# Roadmap for Photonics with 2D Materials

F. Javier García de Abajo,* D. N. Basov,* Frank H. L. Koppens,* Lorenzo Orsini, Matteo Ceccanti, Sebastián Castilla, Lorenzo Cavicchi, Marco Polini,* P. A. D. Gonçalves,* A. T. Costa, N. M. R. Peres, N. Asger Mortensen, Sathwik Bharadwaj, Zubin Jacob,* P. J. Schuck,* A. N. Pasupathy, Milan Delor, M. K. Liu, Aitor Mugarza,* Pablo Merino, Marc G. Cuxart, Emigdio Chávez-Angel, Martin Svec, Luiz H. G. Tizei,* Florian Dirnberger, Hui Deng, Christian Schneider, Vinod Menon,* Thorsten Deilmann,* Alexey Chernikov, Kristian S. Thygesen, Yohannes Abate,* Mauricio Terrones, Vinod K. Sangwan, Mark C. Hersam, Leo Yu, Xueqi Chen, Tony F. Heinz, Puneet Murthy, Martin Kroner, Tomasz Smolenski, Deepankur Thureja, Thibault Chervy,* Armando Genco, Chiara Trovatello, Giulio Cerullo, Stefano Dal Conte,* Daniel Timmer, Antonietta De Sio, Christoph Lienau,* Nianze Shang, Hao Hong, Kaihui Liu, Zhipei Sun,* Lee A. Rozema,* Philip Walther, Andrea Alù,* Michele Cotrufo, Raquel Queiroz, X.-Y. Zhu, Joel D. Cox,* Eduardo J. C. Dias, Álvaro Rodríguez Echarri, Fadil Iyikanat, Andrea Marini, Paul Herrmann, Nele Tornow, Sebastian Klimmer, Jan Wilhelm, Giancarlo Soavi,* Zeyuan Sun, Shiwei Wu,* Ying Xiong, Oles Matsyshyn, Roshan Krishna Kumar, Justin C. W. Song,* Tomer Bucher, Alexey Gorlach, Shai Tsesses, Ido Kaminer,* Julian Schwab, Florian Mangold, Harald Giessen,* M. Sánchez Sánchez, D. K. Efetov, T. Low, G. Gómez-Santos, T. Stauber,* Gonzalo Álvarez-Pérez, Jiahua Duan, Luis Martín-Moreno, Alexander Paarmann, Joshua D. Caldwell,* Alexey Y. Nikitin, Pablo Alonso-González,* Niclas S. Mueller, Valentyn Volkov, Deep Jariwala, Timur Shegai, Jorik van de Groep,* Alexandra Boltasseva,* Igor V. Bondarev, Vladimir M. Shalaev,* Jeffrey Simon, Colton Fruhling, Guangzhen Shen, Dino Novko, Shijing Tan, Bing Wang, Hrvoje Petek,* Vahagn Mkhitaryan,* Renwen Yu, Alejandro Manjavacas, J. Enrique Ortega, Xu Cheng, Ruijuan Tian, Dong Mao, Dries Van Thourhout, Xuetao Gan, Qing Dai,* Aaron Sternbach,* You Zhou, Mohammad Hafezi, Dmitrii Litvinov, Magdalena Grzeszczyk, Kostya S. Novoselov, Maciej Koperski,* Sotirios Papadopoulos, Lukas Novotny,* Leonardo Viti, Miriam Serena Vitiello,* Nathan D. Cottam, Benjamin T. Dewes, Oleg Makarovsky, Amalia Patanè,* Yihao Song, Mingyang Cai, Jiazhen Chen, Doron Naveh, Houk Jang, Suji Park, Fengnian Xia,* Philipp K. Jenke, Josip Bajo, Benjamin Braun, Kenneth S. Burch,* Liuyan Zhao, and Xiaodong Xu

**ABSTRACT:** Triggered by advances in atomic-layer exfoliation and growth techniques, along with the identification of a wide range of extraordinary physical properties in self-standing films consisting of one or a few atomic layers, two-dimensional (2D) materials such as graphene, transition metal dichalcogenides (TMDs), and other van der Waals (vdW) crystals now constitute a broad research field expanding in multiple directions through the combination of layer stacking and twisting, nanofabrication, surface-science methods, and integration into nanostructured environments. Photonics encompasses a multidisciplinary subset of those directions, where 2D materials contribute remarkable nonlinearities, long-lived and ultraconfined polaritons, strong excitons, topological and chiral effects, susceptibility to external stimuli, accessibility, robustness, and a completely new range of photonic materials based on layer stacking, gating, and the formation of moiré patterns. These properties are being leveraged to develop applications in electro-optical modulation, light emission and detection, imaging and metasurfaces, integrated optics, sensing, and quantum physics across a broad spectral range extending from the far infrared to the ultraviolet, as well as enabling hybridization with spin and momentum textures of electronic band structures and magnetic degrees of freedom. The rapid expansion of photonics with 2D materials as a dynamic research arena is yielding breakthroughs, which this Roadmap summarizes while identifying challenges and opportunities for future goals and how to meet them through a wide collection of topical sections prepared by leading practitioners.

## 1. INTRODUCTION

**F. Javier García de Abajo,[1,2,*] D. N. Basov,[3] and Frank H. L. Koppens[1,2]**
[1]ICFO-Institut de Ciencies Fotoniques, The Barcelona Institute of Science and Technology, 08860 Castelldefels, Barcelona, Spain
[2]ICREA-Institució Catalana de Recerca i Estudis Avançats, Passeig Lluís Companys 23, 08010 Barcelona, Spain
[3]Department of Physics, Columbia University, 1150 Amsterdam Avenue, New York, New York 10027, United States
***Corresponding author.** Email: javier.garciadeabajo@nanophotonics.es

Intense research on two-dimensional (2D) materials was sparked in 2004 by the isolation of single-layer graphene and the discovery of its unique photonic properties.[1] This breakthrough introduced the technique of exfoliation, which was quickly applied to other materials composed of atomic layers held together by relatively weak van der Waals (vdW) forces. By stacking these layers together, heterostructures were created,[2] and the introduction of a relative twist angle between the layers produced moiré patterns, further expanding the range of properties of 2D materials.[3] These techniques, along with the broad variety of exfoliable materials, form the backbone of the rapidly growing field of 2D materials. Because of the exceptional optical properties of these materials, photonics occupies a prominent position within this field, where they have generated high expectations for the discovery of new phenomena and the development of groundbreaking applications.

As illustrated in Figure 1, 2D-material heterostructures are becoming increasingly relevant in different areas of photonics. This is due in part to their ability to host polaritons with a range of appealing properties:

- *Strong Optical Confinement*. The in-plane wavelength of polaritonic modes is much smaller than that of light,[4,5] enabling small, laterally structured 2D materials to be optically resonant.[6]
- *Broad spectral range*. 2D polaritons, and material excitations in general, span a wide range of frequencies, extending from the terahertz (THz) domain[7] to the visible[8] and ultraviolet (UV), materialized in the form of plasmons (e.g., in graphene[9] and thin metals[10]), phonons (e.g., in ionic crystals[11,12] such as hexagonal boron nitride (hBN) and $MoO_3$), and excitons (e.g., in transition metal dichalcogenides[13] (TMDs)).

- *Long Lifetime*. Graphene plasmons and phonon polaritons in the mid-infrared (mid-IR) exhibit quality factors (lifetime divided by optical period) of several hundred,[5] while TMD excitons in the visible range can have linewidths of microelectronvolts.[13]
- *Strong Field Enhancement*. When polaritons are excited by external illumination, their associated electric field can be enhanced by several orders of magnitude relative to the incident light.[14]
- *Strong Optoelectronic Response*. Materials like graphene[15,16] and TMDs[17] show a dramatic change in the optical response under electrical doping (e.g., *via* gating), which can switch polaritons on and off or shift them beyond their linewidth.[9,17]
- *High Susceptibility to External Perturbations*. Thermal heating,[17,18] mechanical stretching,[19] magnetic fields,[20] and interactions with additional neighboring materials[21] can also induce substantial, unity-order changes in the linear and nonlinear optical responses of 2D materials.
- *Large Nonlinear Response*. 2D materials exhibit remarkable intrinsic nonlinear responses, with graphene[22] and TMDs[23] featuring nonlinear susceptibilities that rival those of the best available nonlinear materials.

We adopt a broad interpretation of the term *polaritons*, referring to excitations in the material that involve induced charges interacting with light. Depending on the nature of the charges, we consider various types of excitations, such as plasmons, phonons, or excitons, that hybridize with light to produce plasmon, phonon, or exciton polaritons. Polaritons permeate the optical properties of 2D materials, with key features outlined above, which are being leveraged for applications in electro-optical modulation, light emission and detection, imaging and metasurfaces, integrated optics, sensing, and quantum physics across a broad spectral range extending from the far IR to the UV. Additionally, they enable hybridization with spin and momentum textures of electronic band structures.

Noteworthy properties of 2D materials include easy spatial access and precise placement. Moreover, their strong in-plane atomic binding makes them both robust and well-suited for nanofabrication.[2] The atomic structures of 2D materials are well defined, as are the structures of certain atomic defects, which can be leveraged to create quantum optical emitters.[24] Moreover, the electronic band structures of these materials can be both passively[25] and actively[26] modified, providing a powerful interplay between photonics, electronics, and magnetism. The ultrafast dynamics that arises from these interactions configures a rich and fertile research playground.

It should be noted that 2D materials possess quantum properties that are absent in their bulk counterparts. Many of them are not merely thinned-down versions of common three-dimensional (3D) compounds but are electronically richer and different. In this regard, photonics is only beginning to harness the full potential of the rich electronic physics inherent in 2D materials.

These topics are further explored in different sections of this Roadmap, organized into eight areas. However, several sections overlap significantly with more than one of these categories:

- *Fundamentals of 2D Polaritons*. We present an overview of 2D polaritons (Section 2), their generation, detection, and strong coupling in hyperbolic materials (Section 3), plasmonics in twisted 2D materials (Section 4), nonlocal effects in these excitations (Section 5), and the understanding of the optical responses of 2D materials in terms of a Maxwell Hamiltonian (Section 6).
- *Characterization Techniques*. Nanoscopy through near-field optical probes (Section 7) is a driving force in this field, while tip-enhanced nanoscopies (Section 8) and free-electron spectromicroscopy (Section 9) are also powerful sources of unique insights.
- *Excitons and 2D Semiconductors*. 2D exciton polaritons are discussed in Section 10, tunable excitons and trions are analyzed in Section 11, their optical emission properties are covered in Section 12, and the engineered confinement of these excitations is presented in Section 13.
- *Nonlinear and Ultrafast Optical Phenomena*. Ultrafast dynamics in 2D materials present unique characteristics that lead to extraordinary nonlinear optical properties. These areas are discussed with emphasis on ultrafast dynamics in 2D semiconductors (Section 14), quantum-coherent coupling (Section 15), nonlinear photonics (Section 16), nonlinear generation of entangled light (Section 17), nonlinear polaritonics (Section 18), nonlinear valleytronics (Section 19), and nonlinear magneto-optics (Section 20).

- *Chirality, Singularities, Geometric Phases, and Moiré Systems.* This category includes discussions on quantum geometric photonics (Section 21), optical wave singularities (Section 22), topological polaritonics (Section 23), chirality in moiré systems (Section 24), and light control in twisted (Section 25) and anisotropic (Section 26) 2D systems.
- *Metasurfaces and Emerging Materials.* Metasurfaces[27] are a natural application of 2D materials, here discussed in connection with their coupling to photonic structures (Section 27) and a new class of active structures (Section 28). We also present analyses of several classes of emerging materials: the so-called transdimensional materials (Section 29), 2D inorganic compounds known as MXenes (Section 30), microscopic insights into photoelectronic responses exemplified by black phosphorous (Section 31), and plasmons in few-atomic-layer metal films for optical-field confinement and modulation (Section 32).
- *Applications: Integrated Photonics.* This emerging technology can benefit from the participation of 2D materials, as discussed in Section 33. In particular, 2D materials can serve as interconnects (Section 34) and exhibit unique electro-optical capabilities (Section 35).
- *Applications: Light Emission and Detection.* Light emission from 2D materials is discussed for single-photon generation (Section 36), tunneling devices (Section 37), and far-infrared sources (Section 38), whereas light detection is analyzed in the UV (Section 39) and the infrared (IR) (Section 40) spectral ranges.
- *Applications: Disruptive Directions.* Potential applications in quantum information processing are discussed in Section 41 and 2D magnetic photonics is the subject of Section 42.

A concluding Section 43 further discusses future trends in the field. This wide selection of topics aims to provide an overview of the current state of the art and emerging opportunities to guide research in nanophotonics with 2D materials. However, the list is not exhaustive, as this is a rapidly evolving field in which new directions and developments are constantly emerging.

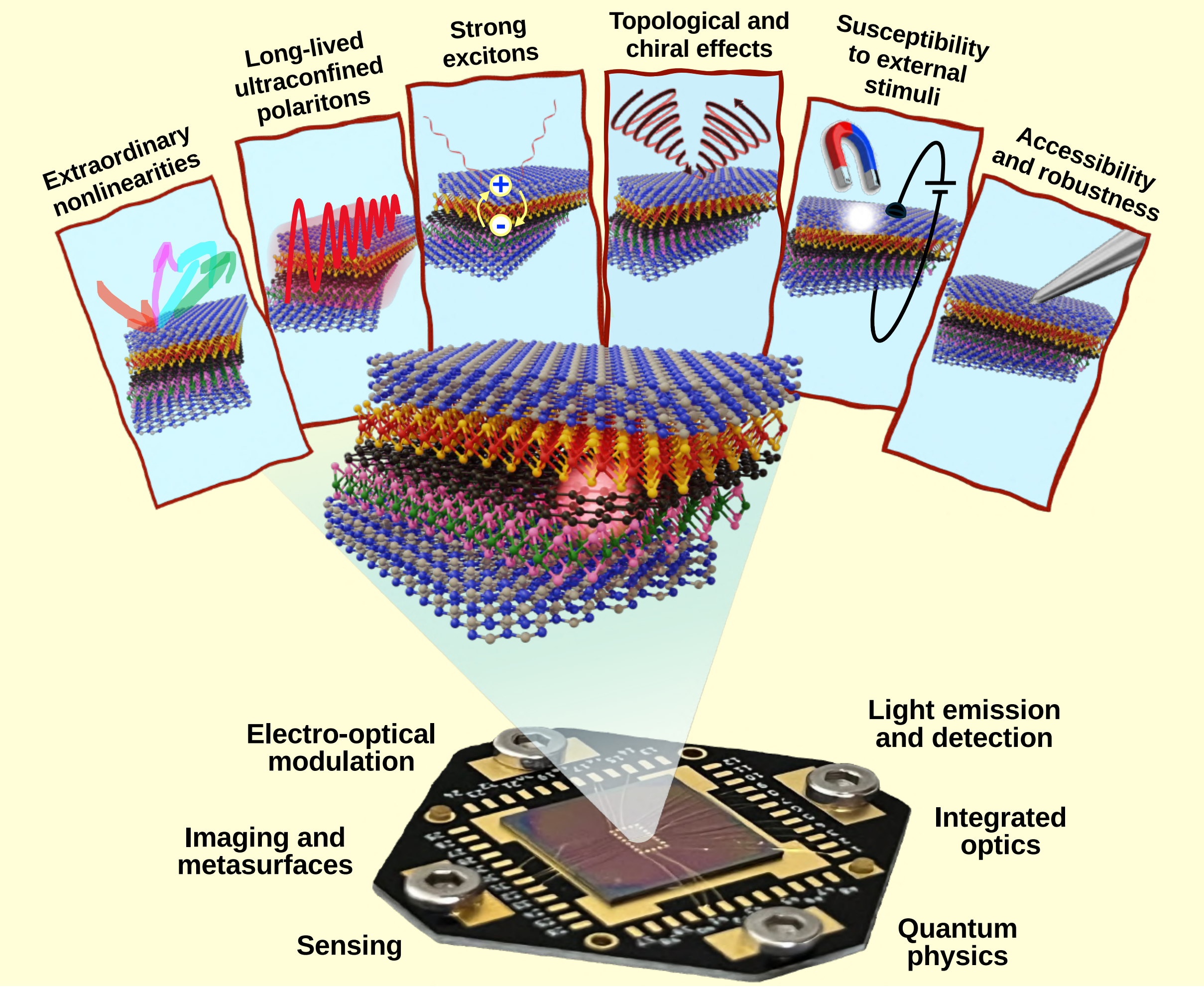

**Figure 1.** Chameleonic properties and uses of 2D materials. Driven by their exceptional properties and unique physical behavior (top), 2D materials are unlocking disruptive applications in photonics (bottom), as discussed in this Roadmap.

**Fundamentals of 2D Polaritonics**

This block describes the fundamentals of polaritons in 2D materials, which are understood here in a broad sense of material excitation modes that hybridize with light. This definition is in the spirit of Hopfield's pioneering work,[28] although the photonic component of 2D polaritons decreases with increasing in-plane wave vector, moving away from the light cone. In this sense, 2D polaritons permeate many other aspects of nanophotonics with 2D materials.

## 2. OVERVIEW OF 2D POLARITONICS

**F. Javier García de Abajo**[1,2,*]
[1]ICFO-Institut de Ciencies Fotoniques, The Barcelona Institute of Science and Technology, 08860 Castelldefels, Barcelona, Spain
[2]ICREA-Institució Catalana de Recerca i Estudis Avançats, Passeig Lluís Companys 23, 08010 Barcelona, Spain
***Corresponding author.** Email: javier.garciadeabajo@nanophotonics.es

### 2.1 Local Description of 2D Polaritons

Polaritons are self-sustained charge oscillations mediated by electromagnetic interactions in different materials.[28] While nonlocality plays a substantial role in 2D plasmons when their wavelength is comparable to the Fermi wavelength (e.g., in extrinsically decorated[29] or laterally patterned[30] graphene, down to the molecular level[31]), the optical response of atomically thin 2D materials can be generally described through frequency-dependent surface conductivities $\sigma(\omega)$, ignoring nonlocal and out-of-plane polarization effects.[32] In these materials, polaritons emerge under the condition $\mathrm{Im}\{\sigma(\omega)\} > 0$ as optical surface modes of p (TM) polarization, revealed through the poles of the corresponding Fresnel reflection coefficient outside the light cone[32] and characterized by the dispersion relation $k_\parallel = i\omega\bar{\epsilon}/4\pi\sigma(\omega)$, where $k_\parallel$ is the in-plane wave vector and $\bar{\epsilon}$ the average environment permittivity (Figure 2a). This description remains valid for material films with small thickness $d$ compared to the in-plane polariton wavelength $\lambda_{\rm polariton} = 2\pi/\mathrm{Re}\{k_\parallel\}$. Because the light wave vector vanishes in the electrostatic limit, the out-of-plane wave vector $k_\perp$ must satisfy $k_\perp^2 + k_\parallel^2 = 0$ (i.e., $k_\perp = \pm i k_\parallel$), therefore confining the mode to a region of vertical extension $(1/2\pi)\lambda_{\rm polariton}$ (Figure 2a) around the material (for $1/e$ decay of the field intensity). This result is universal and applies regardless of material composition and environmental permittivities.

A Lorentzian fit to the surface conductivity (Figure 2b) provides a generic way of describing 2D polaritons and, when plugged into the dispersion relation above, leads to the conclusion that the polariton wavelength is much smaller than the photon wavelength at the same frequency (Figure 2b). The conductivity incorporates a Drude weight $\omega_D$, an intrinsic resonance $\omega_g$ for phonons and excitons, and a phenomenological lifetime $\tau$, whose values depend on the type of polariton and material, as indicated in Figure 2c. From a photonics viewpoint, the nature of the induced charge associated with polaritons (delocalized electrons in plasmons, atomic vibrations in phonons, bound electron–hole pairs in excitons) is fully encapsulated in these parameters. This approach can be equally applied to in-plane anisotropic materials,[33] where $\sigma(\omega)$ becomes a $2 \times 2$ tensor.

The polariton frequencies in laterally patterned structures satisfy the expression[32] $\omega = i\sigma(\omega)D/\eta\bar{\epsilon}$, where $D$ is a characteristic distance of the structure (e.g., a ribbon width or a disk diameter), whereas $\eta$ is an electrostatic eigenvalue that depends on 2D morphology but not on size and composition (e.g., $\eta \approx -0.073$ for the fundamental dipole of a circular disk or $\eta \approx -0.069$ for the transverse dipole of a ribbon, see ref 34).

When stacking 2D layers forming a heterostructure of small total thickness compared with the polariton wavelength, the mode dispersion is also given by the expressions above using a single conductivity that captures all layers. In particular, heterostructures made of vdW materials undergo a minor overlap of their electronic wave functions, and consequently, the conductivities of the individual layers $\sigma_j(\omega)$ remain nearly unperturbed from their pristine values, so we can write the conductivity of the heterostructure simply as a sum over layers $\sigma(\omega) = \sum_j \sigma_j(\omega)$. This zero-thickness approximation (ZTA) works extremely well to explain the dispersion of hybrid modes involving different types of polaritons (e.g., hBN phonons and graphene plasmons[35]). However, it should be noted that the ZTA fails to describe moiré structures with substantial electronic hybridization (see Sections 4 and 24) or when the film thickness is no longer small compared with the polariton wavelength (e.g., in hyperbolic materials, Sections 3, 25, and 26).

## 2.2 Polaritons by Design

During the last decades of the XX century, the development of ultrahigh vacuum (UHV) technology enabled the synthesis of atomically thin layers and vertical heterostructures[10,36,37] through epitaxial growth. The so-generated crystal-quality layers are generally fragile and require UHV to maintain structural stability, thus limiting their applicability to photonics. Spatial placement and lateral patterning (e.g., *via* nanolithography) pose additional challenges that demand material-specific solutions, as recently exemplified through the demonstration of laterally confined plasmons in ultrathin crystalline silver films.[10,37]

This panorama experienced a radical change two decades ago with the isolation of single-layer graphene *via* mechanical exfoliation.[1] Currently, exfoliation and stacking[2] are widely adopted as powerful nanofabrication techniques on par with lithography, generating an impressive number of unprecedented 2D material structures based on vdW single- or few-atomic layers. This approach does not require UHV and benefits from vdW attraction between layers that push polluting molecules to the edges of the exfoliated islands, thus automatically cleaning the interstitial region between the layers.[2] Combined with self-assembly and nanopatterning, these methods configure a comprehensive suite of tools to fabricate previously inaccessible 2D materials of varied geometry and composition (Figure 2d).

The small thickness of 2D materials makes them extremely sensitive to external actions, such as chemical and electrical doping. In particular, long-lived graphene plasmons can be switched on/off and strongly modulated by gating the material,[9] decorating it with donor molecules,[38] or exposing it to magnetic fields.[20] Likewise, electrical gating or heating can modulate the exciton lineshapes of TMDs beyond their pristine linewidths.[17] Mechanical stretching[19] and heating[17] (e.g., *via* optical pumping, see Section 17) also affect the polariton characteristics dramatically. These properties of 2D materials provide great flexibility for engineering structures with designated polaritonic properties (i.e., on-demand parameters $\omega_D$, $\omega_g$, and $\tau$ for $\sigma(\omega)$, see Figure 2b).

### 2.3 Challenges and Opportunities

Figure 2e summarizes several appealing opportunities associated with 2D materials. Besides the structural robustness inherited from the use of vdW layers and the flexibility in polaritonic design, they cover spectral ranges extending from the mid-IR (e.g., hBN phonons and graphene plasmons) to the visible (e.g., TMD excitons and noble metal plasmons), they feature long lifetimes (e.g., vdW material phonons, graphene plasmons, and TMD excitons), they display deep-subwavelength confinement (Figure 2b), and they are susceptible to external stimuli (Figure 2d).

These extraordinary attributes of 2D materials have triggered the imagination of researchers over the last two decades, opening several exciting possibilities that are still facing severe challenges. Here is a short selection of them (Figure 2f):

- *Photon-Free Integrated Polaritonics (PFIP)*. Electrical generation and detection of polaritons are on the agenda to produce optical sensing and signal processing devices with small footprints and no external light sources or detectors needed. Given the fact that 2D polaritons are strongly confined and well described in the electrostatic limit, photons are only involved in a perturbative manner in this approach. The generation and detection of polaritons with spectral sensitivity define an appealing challenge in PFIP.

- *Polaritonic Sensing*. As a byproduct of PFIP, polaritonic sensing could rely on the electrical emission and detection of polaritons to reveal the presence of analytes dispersed on a 2D material region on which they propagate. The dream of all-electron polaritonic sensing is still on the agenda of nanophotonics with 2D materials.

- *Polaritonic Laser*. Long-lived polaritons and the strong Purcell factors observed near polariton-supporting 2D materials constitute a good basis for materializing a nanoscale polaritonic laser analogous to the spaser.[39] A bottleneck in this research frontier is the design of quantum emitters that can undergo population inversion in the mid-IR spectral range where phonon polaritons and graphene plasmons emerge.

- *Nonlinear Nanophotonics*. The relatively large nonlinear response of 2D materials and the strong confinement of 2D polaritons provide an excellent platform for realizing all-optical devices, as discussed in Sections 16, 17, and 18.

- *Single-Photon Emitters*. Quantum light generation from localized sources is emerging as a solid research direction (see Section 36). The production of identical photons and polaritons remains a challenge that would enable the realization of quantum interference within integrated nanoscale environments (i.e., a nanoscale version of plasmonic Hong–Ou–Mandel interference[40]).

- *Integrated Quantum Optics/Sensing*. The realization of quantum-optics devices in an integrated solid-state platform also remains a challenge[9] that demands the development and controlled placement of quantum emitters enhanced by the extraordinarily large Purcell factors associated with ultraconfined 2D polaritons.

- *Small-Footprint Photodetectors*. Several designs have been proposed for IR photodetection using 2D materials and featuring relatively small footprints,[41] particularly in the mid-IR (see Sections 39 and 40). Further developments in this area could explore atomic-scale structures in combination with optical funnels.

These items define an exciting research agenda with high potential for technological impact, stimulating the exploration of intriguing material properties and the discovery of novel phenomena associated with the two-dimensionality of these systems (see Sections 5 and 7).

(a)

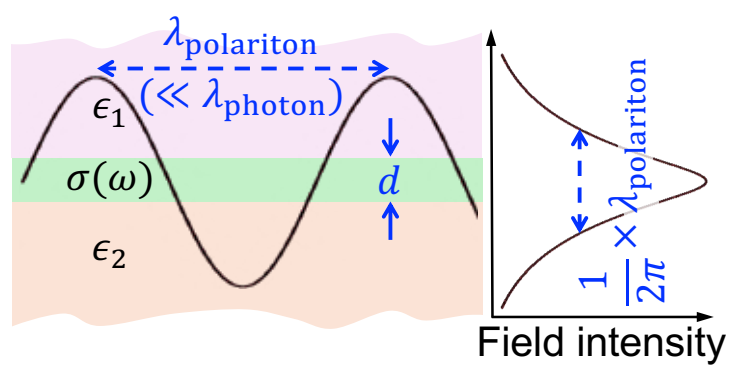

(b)

$$\sigma(\omega) = \frac{e^2}{\hbar} \frac{i\omega_D}{\omega + i\tau^{-1} - \omega_g}$$

$$\frac{\lambda_{\text{polariton}}}{\lambda_{\text{photon}}} = \frac{2\pi}{\bar{\epsilon}c} \text{Im}\{\sigma(\omega)\} \approx \frac{2\pi\alpha}{\bar{\epsilon}} \text{Re}\left\{\frac{\omega_D}{\omega + i\tau^{-1} - \omega_g}\right\} \ll 1$$

(c)

| Type of polariton | Plasmon 2DEG | Plasmon Dirac | Phonon | Exciton |
|---|---|---|---|---|
| Material | thin metals | graphene, TIs | hBN, $MoO_3$ | TMDs |
| $\omega_D$ | $\sim v_F k_F$ | $\sim v_F k_F$ | $f_{ph}\omega_g$ | $f_{ex}\omega_g$ |
| $\omega_g$ | 0 | 0 | 10–100s meV | 1–2 eV |
| $\tau$ | ~ 10s fs | ~ ps | ~ ps | ~ ns |

(d) **Design of 2D polaritonic materials**

**By geometry/composition**
- growth (e.g., epitaxial)
- stacking + twisting
- self-assembly
- nanopatterning

**By external actions**
- chemical doping
- electrical gating
- magnetic/electric DC fields
- strain/mechanical deformation
- optical pumping / thermal heating

(e) **Opportunities**
- a vast suite of **robust** materials
- **on-demand** polaritonic properties
- **broad spectral** ranges
- **low** optical **losses**
- **strong** spatial field **confinement**
- **easy access** and spatial manipulation
- **high sensitivity** to DC EM fields

(f) **Challenges**
- **photon-free** integrated polaritonics
- polaritonic **sensing**
- polaritonic **lasing**
- **nonlinear** nanophotonics
- **single-photon** emitters
- integrated **quantum optics/sensing**
- small-footprint **photodetectors**

**Figure 2**. Overview of 2D polaritons. (a) Universal scaling of the out-of-plane confinement length $(1/2\pi)\lambda_{\text{polariton}}$ in the local electrostatic limit for polaritons in planar films of small thickness $d$ compared with the their wavelength $\lambda_{\text{polariton}}$, sandwiched by media of arbitrary permittivities $\epsilon_1$ and $\epsilon_2$. (b) General parametrization of the frequency-dependent surface conductivity $\sigma(\omega)$ in terms of a Drude weight $\omega_D$, an intrinsic resonance frequency $\omega_g$, and a phenomenological lifetime $\tau$. The polariton-to-photon wavelength ratio, expressed in terms of the fine-structure constant $\alpha \approx 1/137$ and the average environmental permittivity $\bar{\epsilon} = (\epsilon_1 + \epsilon_2)/2$, is much smaller than 1 for actual material parameters. (c) Range of optical parameters in common types of 2D polaritonic materials. We have $\omega_g$=0 for plasmons, while $\omega_D$ scales like the product of the Fermi velocity $v_F$ and the Fermi wave vector $k_F$ in conducting materials. For phonons/excitons, $\omega_D$ is expressed as a fraction $f_{ph}/f_{ex}$ of $\omega_g$. (d) Experimental techniques available to synthesize 2D polaritonic materials based on geometry/composition and the application of external actions such as doping, electromagnetic fields, or heating. (e,f) Selection of challenges and opportunities associated with 2D polaritonics, as discussed in more detail in the main text.

## 3. GENERATION, DETECTION, AND STRONG COUPLING OF DEEP SUBWAVELENGTH HYPERBOLIC POLARITONS IN VAN DER WAALS MATERIALS

**Lorenzo Orsini,[1] Matteo Ceccanti,[1] Sebastián Castilla,[1] and Frank L. H. Koppens[1,2,*]**
[1]ICFO-Institut de Ciencies Fotoniques, The Barcelona Institute of Science and Technology, 08860 Castelldefels, Barcelona, Spain
[2]ICREA-Institució Catalana de Recerca i Estudis Avançats, Passeig Lluís Companys 23, 08010 Barcelona, Spain
***Corresponding author.** Email: frank.koppens@icfo.eu

### 3.1 Introduction

Hyperbolic materials are a special class of anisotropic media characterized by a dielectric permittivity tensor that has both positive and negative principal components within the same frequency range. Their discovery and exploration have been a significant breakthrough in nanophotonics for their ability to host hyperbolic polaritons, deep subwavelength electromagnetic waves with asymptotically large wave vectors, and extreme directional dependence. These features of hyperbolic polaritons can be readily visualized by comparing the isofrequency surfaces of an isotropic dielectric with those calculated for a hyperbolic material. Figure 3a shows the spherical isofrequency surface of an isotropic material, indicating that light can propagate with a finite wavelength in all directions. In contrast, as shown in Figure 3b, the isofrequency surface of a hyperbolic material takes on a more complex, open shape, allowing light to propagate with asymptotically small wavelengths in ways that would not be possible in conventional materials.

Pivotal to this field are 2D polar dielectric materials, which are ideal for studying hyperbolic phenomena due to their crystalline structure. The layered crystalline structure, characterized by strong in-plane covalent bonds and weaker interlayer vdW forces, causes the material's optical phonons to have different energies in different directions, resulting in an anisotropy of the permittivity tensor. When a polar hyperbolic material is thinned into slabs, the propagation of light is guided, and the propagating modes are referred to as hyperbolic phonon polaritons (HPhPs). This behavior has catalyzed research into vdW materials, particularly thin flakes of hBN[11] and α-phase molybdenum trioxide ($\alpha$-$MoO_3$),[12] which stand out as the leading choice for advancing hyperbolic Nanophotonics.

Earlier studies have demonstrated the potential of controlling the hyperbolic phonon polariton wavelength with atomic layer precision, as well as the possibility of designing a large variety of heterostructures where variations in thickness, composition, stacking order, and twist angle can drastically alter the properties of the polaritons. For instance, control of the hyperbolic phonon polariton's group velocity and wavelength can be obtained when the hyperbolic material is placed in contact with any kind of substrate.[42] This proximity effect becomes even more pronounced in heterostructures composed of various hyperbolic materials, leading to remarkable optical phenomena such as polaritonic canalization,[43] isofrequency surface topological transitions,[44] and negative reflection.[45] Furthermore, active control of HPhPs has been demonstrated by coupling hyperbolic media with active materials such as graphene[46], enabling electric tunability of the hyperbolic plasmon phonon polariton wavelength.

### 3.2 Hyperbolic Nanophotonics

Recent research on photonic crystals,[6,47] cavities,[48,49,50] and other optical components[51] created through the direct patterning of hyperbolic materials has expanded beyond translationally invariant systems. Structuring hyperbolic media into customized photonic devices provides a direct method for obtaining specific optical properties. For instance, engineering photonic bandgaps and resonances allows for precise spectral filtering and facilitates strong coupling in light–matter interactions.

The drawback of this approach is the extreme sensitivity of hyperbolic phonon polaritons to imperfections, which, due to the limitations of current fabrication technologies, constrains the quality of nanophotonic devices. For this reason, further advancements have been made using an alternative and more effective approach to spatially modulating HPhPs properties, which involves patterning the substrate beneath or above the hyperbolic material rather than the material itself. In recent years, a substantial number of studies have explored indirectly patterned devices to investigate the reflection and refraction properties of both active[52] and passive[53,54,55,56] nanoscaled lenses and interfaces. Tunable plasmonic lattices have been developed[57] that can be seamlessly coupled to hBN HPhPs, effectively creating tunable hyperbolic photonic crystals. Significant progress has also been made with gold-based indirectly patterned lattices[58] and nanoresonators,[59] which, owing to the use of focused-ion-beam lithography, record-breaking quality factors have been achieved while maintaining tight volume confinement. Moreover, on the same platform, topologically nontrivial one-dimensional (1D) HPhPs lattices have been demonstrated,[60] marking the first realization of a topological nanophotonic system in hyperbolic media. The latter finding suggests a potential extension of these concepts to 2D systems, enabling the exploration of photonic Chern insulators and photonic valley Hall systems. This would allow guided modes to propagate in topologically protected edge states, thereby extending the robustness against disorder. These advances could enable the development of miniaturized photonic isolators, diodes, and logic circuits, and may lead to entirely new concepts for communication systems, optical transistors, and optical information processing.

### 3.3 Future goals

#### 3.3.1 *Active Hyperbolic Devices for Detection and Emission*

By leveraging the unique capabilities of hyperbolic vdW heterostructures, we can now envision an active platform where hyperbolic phonon polaritons are electronically generated, actively or passively manipulated, and detected. As illustrated in Figure 3, this concept integrates the three key elements of generation, manipulation, and detection.

Inspired by the work of Guo *et al.*,[61] which demonstrated that HPhPs can be electrically generated by driving a high-bias current through a simple hBN-encapsulated graphene channel (Figure 3c), this platform could move beyond outcoupling polaritons to the far field. Instead, the generated HPhPs can be guided from the emission site to a manipulation area using directly patterned hBN waveguides, photonic crystal waveguides, or even topologically protected channels. In the manipulation stage, HPhPs could interact with other polaritons *via* nanoscale interferometers, while being focused and redirected using nanophotonic lenses and interfaces or filtered through photonic crystals. These elements, whether passive or active, could also enable applications such as molecular sensing and strong light–matter interactions exploiting the strong electromagnetic field of polaritons (see Figure 3d). Finally, at the detection stage, as shown in Figure 3e, HPhPs can be converted into electronic signals through photo-thermoelectric effect,[62] enabling the retrieval of the processed information. This seamless integration of generation, manipulation, and detection could offer a pathway to functional nanophotonic circuits for a variety of advanced optical and sensing applications, eliminating the need for external sources or signal detection units and enabling full on-chip integration.

### 3.3.2 *Strong Coupling with Hyperbolic Polaritons in Nanophotonics*

Another promising feature of ultrastrong optical field compression is the strong enhancement of light–matter interactions, leading to strong-to-ultrastrong coupling and extreme Purcell factors.[63] Moreover, it has been shown that strongly confined polaritons carry extreme field gradients with wavelengths that reach spatial variations in the electronic wave functions. This leads to novel phenomena that break the limits of the dipole approximation[64,65] or even fundamental topological invariances.[66]

These enhanced coupled systems exhibit single molecule detection,[67,68] improved photovoltaic response,[69] and extreme phenomena in the strong[70] and ultrastrong[71] coupling regimes. Examples include the modification of the quantum Hall in split ring resonators coupled to a 2D electron gas[72] and the macroscopic modification of ferromagnetism in YBCO nanoparticles coupled to a plasmonic substrate.[73] The strong-coupling regime has also been realized by using hyperbolic polaritons coupled to vibrational modes of molecular layers,[74] demonstrating that even the bare presence of the unpatterned hyperbolic material could achieve the few molecules strong coupling.[75] Hyperbolic phonon polaritons have also demonstrated strong coupling with magnetic excitation in graphene,[76] where magneto-excitons were tuned to be resonant with the polaritons, causing the opening of a gap in the dispersion. This experiment revealed contributions to the dispersion arising from Landau transitions that are forbidden by dipole selection rules, showing that the high momentum carried by hyperbolic polaritons lead to the breakdown of the dipole approximation.

Theoretical studies of such systems[77] have demonstrated that subwavelength polaritons require a multimodal nonlocal treatment of the Hamiltonian to explain the behavior of the coupled system, revealing unprecedented regimes of super-strong coupling, where each magneto-exciton is strongly coupled to many hyperbolic modes. These studies pave the way for describing many-body systems coupled to confined hyperbolic polaritons, where strongly correlated electron–photon physics emerges due to intense and rapidly varying electromagnetic fields interacting with wave functions that exhibit comparable spatial gradients. This coupling can lead to macroscopic changes in the electron liquid, such as Fermi velocity renormalization,[78] or induce exotic states like cavity-mediated fractional magnetic phases[79] and cavity-induced superconductivity through modified electron–electron interactions[80].

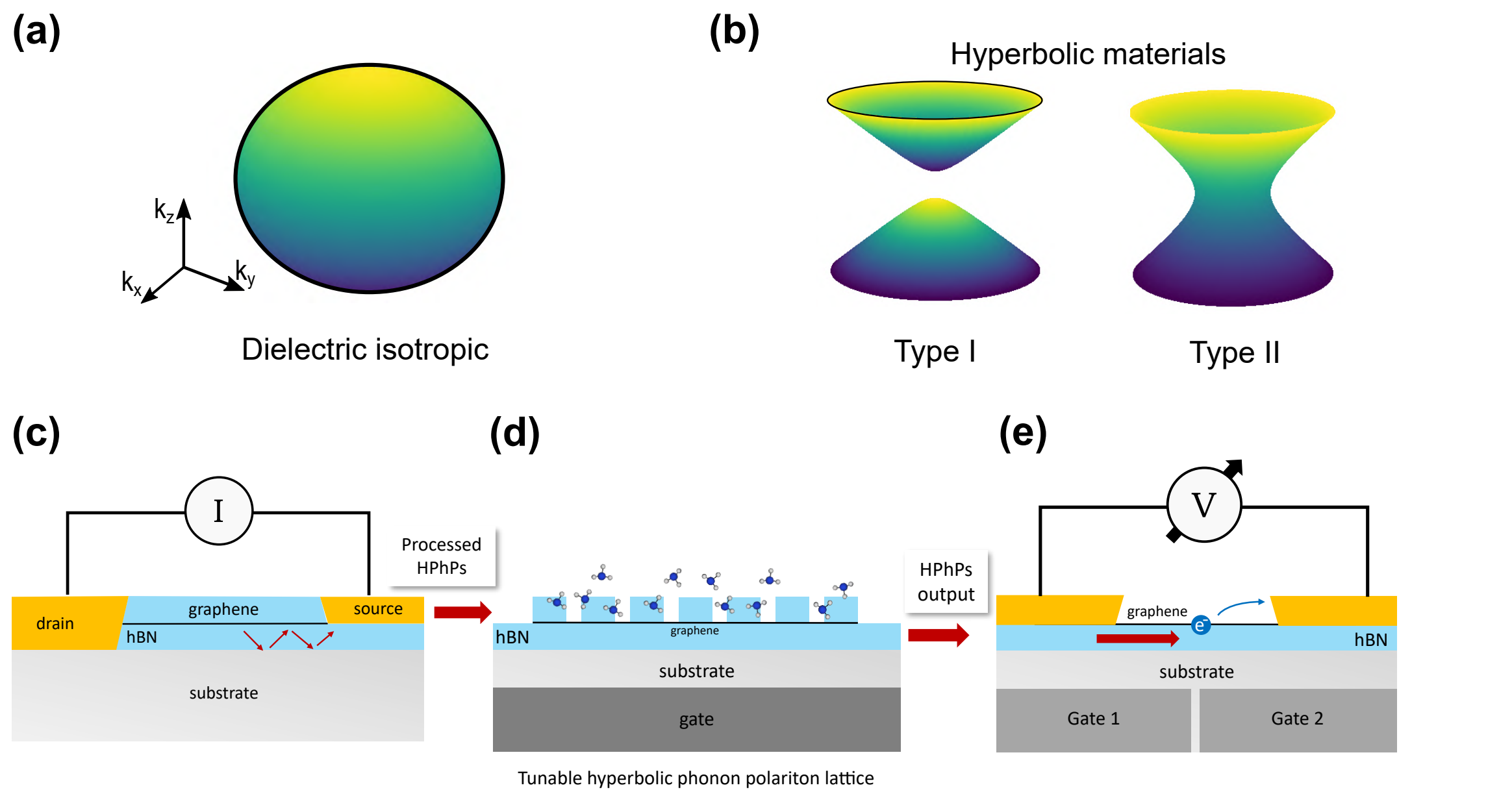

**Figure 3.** Illustration of isofrequency contours and the hyperbolic on-chip nanophotonic platform. (a) Isofrequency surface of a dielectric isotropic medium. (b) Isofrequency surface of a Type I hyperbolic material that is characterized by $\epsilon_{xx} = \epsilon_{yy} > 0$ and $\epsilon_z < 0$. Notably, for any pair of coordinates $(k_x, k_y)$, it is always possible to find a point on the normal surface. (c) On-chip nanophotonic-platform generation: A high-bias current is driven through a graphene channel encapsulated with hBN, producing hyperbolic phonon polaritons (HPhPs) through electroluminescence. The schematic represents the cross section of a configuration in which HPhPs bypass the source contact and propagate for further processing. (d) On-chip nanophotonic-platform manipulation: A tunable polaritonic or topological crystal for molecule detection. The schematic illustrates the cross section of an hBN/graphene/hBN polaritonic lattice (hyperlattice) in which the polaritonic density of states can be electrically tuned by adjusting the gate voltage. This concept could enable molecular selectivity based on their interaction with the crystal lattice. (e) On-chip nanophotonic-platform detection: A graphene-based photo-thermoelectric detector. The schematic illustrates the processed HPhPs incoming from one side of the device being absorbed at the graphene PN junction defined by a split-gate structure. This absorption generates a photo-thermoelectric response, converting the polaritonic signal into an electrical output.

## 4. FUNDAMENTALS OF PLASMONS IN TWISTED 2D MATERIALS

**Lorenzo Cavicchi[1,2] and Marco Polini[2,3,*]**
[1]Scuola Normale Superiore, Piazza dei Cavalieri 7, I-56126 Pisa, Italy
[2]Dipartimento di Fisica dell'Università di Pisa, Largo Bruno Pontecorvo 3, I-56127 Pisa, Italy
[3]ICFO-Institut de Ciencies Fotoniques, The Barcelona Institute of Science and Technology, 08860 Castelldefels, Barcelona, Spain
***Corresponding author.** Email: marco.polini@icloud.com

Plasmonics is a branch of optoelectronics that revolves around the excitation and manipulation of collective modes, known as *plasmons*, which arise from the interaction between electromagnetic fields and free electrons, typically at metallic or conductive interfaces. At the beginning of the 21st century, plasmons were seriously considered as potential candidates for revolutionizing technology, promising a new generation of superfast computer chips and ultrasensitive molecular detectors.[81]

But what is a plasmon? At a fundamental level, a plasmon is a pole of the density–density response function[82] $\chi_{\mathbf{G},\mathbf{G}'}(\mathbf{q},\omega)$, which is a causal (i.e., retarded) linear response function. Such a singularity, physically denoting a self-sustained oscillation of the electron density, emerges for certain values of the momentum $\hbar\mathbf{q}$ and energy $\hbar\omega$ transfer. The position of the pole in the $\mathbf{q}$ and $\omega$ plane, $\hbar\omega_p(\mathbf{q})$, is called the plasmon dispersion. In a crystal, the density–density response function also depends on two reciprocal lattice vectors, $\mathbf{G}$ and $\mathbf{G}'$.[82] This matrix form in reciprocal lattice space is crucial to capture crystalline local field effects (LFEs).

In general, the causal density–density response function can be written in terms of a more fundamental building block, which is dubbed *proper density–density response function*. The name stems from the fact this daunting object is the sum of all the proper Feynman diagrams,[82] from zero order up to *arbitrary* order in the strength of the electron–electron (*e–e*) interaction. Finding a reasonably good approximation that describes the poles of $\chi_{\mathbf{G},\mathbf{G}'}(\mathbf{q},\omega)$ is therefore extremely challenging. Almost 99% of papers in the literature focuses on a specific limit, the so-called *local limit*, in which one sets $\mathbf{G}=\mathbf{G}'=\mathbf{0}$ (thereby ignoring LFEs) and takes the long-wavelength $\mathbf{q}\to 0$ limit. In this limit, the plasmon dispersion can be found analytically within the Random Phase Approximation (RPA),[82] which turns out to be totally equivalent to a standard approach followed by the community of plasmonic scientists. In this plasmonic approach, one combines Maxwell equations with a constitutive equation that links the current to the electric field through the aid of the local (frequency-dependent) conductivity $\sigma(\omega)$. In crystals, one needs to consider the classical intraband Drude contribution to $\sigma(\omega)$ as well as the quantum-mechanical interband contribution. Plasmons obtained in this local limit are therefore completely described by $\sigma(\omega)$ (see Section 2). One finds that, in the long-wavelength limit, $\omega_p(\mathbf{q})$ tends to the well-known plasma frequency in 3D electron systems while it behaves as $q=\sqrt{|\mathbf{q}|}$ in 2D electron systems.[82] While these results remain foundational to plasmonics, modern experimental techniques such as scattering-type scanning near-field optical microscopy (s-SNOM)[83] have made it possible to explore regimes where $q/k_{\mathrm{F}}$ is on the order of unity, $k_{\mathrm{F}}$ being the Fermi wavenumber of the electron system. The reach of this deeply nonlocal regime calls for theoretical approaches that greatly transcend the RPA/local approximation, which is valid only for $q/k_{\mathrm{F}}\ll 1$. s-SNOM, indeed, enables the coupling of far-field light with short-wavelength collective modes through a nanometric metallized tip. The fast development of this great experimental investigation tools was facilitated by the advent of atomically thin 2D materials.

In recent years, in fact, 2D materials, particularly graphene,[84] have emerged as promising platforms for plasmonic applications due to their unique electronic and optical properties. Graphene, a single layer of Carbon atoms arranged in a honeycomb lattice, exhibits exceptional electrical conductivity, tunable carrier density, and strong light–matter interactions. Unlike traditional plasmonic materials such as metals, graphene supports long-lived and highly confined plasmons that can be tuned *via* electrostatic gating or optical pumping.[85] Beyond graphene, other 2D materials, such as TMDs, have also been extensively studied. Their unique semiconducting and excitonic properties make them key materials for the development of innovative electronic and optoelectronic devices.

Research on graphene and related 2D crystals has been, and continues to be, a prominent area of study in condensed matter physics and materials science. Shortly after the isolation of 2D materials in monolayer form, it was realized that their 2D nature enables them to be reassembled into custom heterostructures, known as vdW heterostructures, with precisely tailored layer sequences.[84] This new fabrication technique has unveiled virtually limitless possibilities for engineering novel layered materials by stacking 2D materials such as graphene, TMDs, and many others.

A particularly exciting development is the advent of *twisted* 2D materials, where two or more layers are stacked with a slight rotational misalignment. These structures exhibit highly tunable electronic and optical properties due to the moiré superlattice stemming from the relative twist between monolayers.[86] Remarkably, twisted bilayers have demonstrated to be excellent platforms to explore a plethora of exotic phenomena. In twisted bilayer graphene (TBG) superconductivity.[87] correlated insulating states,[88] strange metal behavior,[89] and topological phases[90] have been observed at a specific twist angle referred to as *magic angle*.[25] In this regime, electronic bands near the charge neutrality point become extremely flat enhancing electronic correlations. Similar studies in twisted bilayer semiconductors, like TMDs, have been equally successful.[91] These systems host moiré excitonic bands, unconventional superconductivity (e.g., in twisted bilayer $WSe_2$[92]) and fractional Chern insulating states (e.g., in twisted bilayer $MoTe_2$[93]).

These breakthrough discoveries in the context of electronic properties are expected to have a counterpart in the realm of plasmonics. Plasmons, being highly sensitive to the band structure of their host 2D materials, are powerful tools for investigating electronic properties, even within the local approximation. Beyond the local approximation and at low densities, however, plasmons are extremely sensitive to profound many-body effects. A simple yet illustrative example is given by systems with Rashba spin–orbit coupling (SOC) where Galilean invariance is broken. In this case, plasmons calculated beyond the RPA are a sensitive tool for measuring effects of *e–e* interactions on the carriers' effective mass and SOC strength.[94]

Looking ahead, the exciting landscape of exotic phases in twisted 2D materials highlights the need for new experimental techniques and more reliable, deeply nonlocal theories of collective excitations.

## 4.1 Current State of the Art

The plasmonic properties of 2D electron systems have been studied since the 1980s. These are known to exhibit a center-of-mass (COM) plasmon, which, in the local limit, disperses as $q$. In twisted 2D materials, the moiré superlattice significantly alters the plasmonic spectrum with respect to that of a simple, single-band 2D electron system, endowing plasmons with exotic properties such low losses, chirality, and topology.

The moiré superlattice reshapes the plasmonic bands even at the single-particle level.[95] This transformation introduces a series of higher-energy, quasi-flat *interband* plasmons accompanying the standard COM plasmon. These modes were first theoretically predicted in TBG[96] and subsequently detected experimentally.[97] In twisted 2D materials tuned to the magic angle, also the highly dispersive COM plasmon morphs into a dispersionless, undamped mode. This phenomenon, originally predicted in TBG,[98] has also been discovered in other twisted systems such as twisted $MoS_2$ and twisted double bilayer graphene.[99]

An intriguing feature of twisted bilayers is the fact that they also harbor *acoustic* plasmons, absent in their monolayer counterparts.[100] Sharp collective modes with an acoustic dispersion arise in multilayer structures, their dispersion providing insights on the interlayer coupling strength and posing severe bounds on the underlying many-body theory. The electromagnetic field associated with acoustic plasmons is highly confined within the layered structure, offering potential applications in strong and ultrastrong light–matter interactions.[101]

Additionally, the structural chirality of twisted 2D materials further enriches their plasmonic behavior. The chiral nature of the underlying atomically thin crystals[102] imparts a longitudinal magnetic moment to the COM plasmon, endowing it with a unique chiral character.[103] These chiral plasmon modes not only propagate with inherent chiral content but also enhance the chirality of the near field, as demonstrated in TBG. Also, the chiral nature of twisted 2D materials offers exciting opportunities for developing novel photonic devices, as these materials could pave the way for the fabrication of chiral sensors.[104]

Lastly, it is worth recalling that in the realm of plasmonics also the concepts of Berry phase and Berry curvature have played an important role. These intrinsic features underlie a variety of phenomena, including electric polarization, anomalous Hall effects, and quantum Hall states. For plasmons, a new class of modes, driven by the interplay between Berry curvature and *e–e* interactions, has been predicted in 2D metallic systems. These chiral plasmonic modes exhibit nonreciprocal energy dispersions at zero magnetic field and offer advanced tools for diagnosing topological band structures.[105]

## 4.2 Challenges and Future Goals

The field of twisted 2D materials has grown rapidly in recent years, presenting new challenges to the scientific community, and leaving many open questions.

Although not directly related to photonics, one of the most compelling challenges is the microscopic explanation of unconventional superconductivity in twisted 2D materials. Despite numerous theoretical proposals, the nature of the superconducting phase in these systems remains elusive. A related but distinct challenge involves understanding the ground state of broken-symmetry phases in twisted 2D materials, which emerge from strong *e–e* interactions in the flat-band regime.[106] Addressing these issues will require the development of novel probing techniques and theoretical approaches.

Furthermore, the nontrivial topology in momentum and real space in twisted 2D materials, particularly quantum geometry, represents a rapidly expanding area of research. While the study of how quantum geometric properties influence macroscopic observables and quantum phases is still in its early stages, advancing our understanding of topology and its effects on collective phenomena remains a major goal for the future.

Looking ahead, a thorough investigation of collective modes in correlated phases of matter—such as unconventional superconductors, correlated insulators, strange metal phases, and topological phases—is expected to shed light on these fascinating phenomena. Twisted 2D materials have proven to be an ideal platform for such exploration, offering a versatile nanoscale laboratory where these phases can coexist. Figure 4 summarizes the list of future challenges alongside their corresponding most promising twisted 2D materials, where we anticipate the most favorable outcomes in plasmonics.

### 4.3 Suggested Directions to Meet Goals

The ability to engineer plasmonic resonances in twisted systems opens new opportunities for manipulating light–matter interactions and exploring exotic quantum phenomena. Theoretical and experimental studies on the plasmonic properties and, more broadly, the collective modes of twisted 2D materials and their correlated phases are highly relevant for both fundamental and applied research, advancing our understanding of exotic quantum phases in these materials.

Experimental efforts are needed to investigate highly nonlocal excitations. To achieve this, the implementation of new exotic photonic couplers will be essential. For instance, advancements in tip engineering for s-SNOM devices are expected to play a crucial role. Innovative tip designs with specific symmetries could be vital for detecting chiral plasmonic modes and launching multipolar excitations. The latter are particularly important for detecting exotic magnetoroton excitations (akin to gravitons), which lack an electrical dipole moment.[107]

From a theoretical perspective, studying the effects of nonlocality on the optical properties of these systems could provide valuable insights into many-body interactions[29] and topological phenomena.[108] For example, analyzing the fully nonlocal density–density response function of a crystal, including crystalline LFEs, has proven useful in identifying correlated phases in twisted TMDs through their plasmonic spectra.[109] Furthermore, plasmons and other collective modes, such as magnetorotons, can deepen our understanding of exotic topological phases like fractional Chern insulators, which have been recently observed in twisted $MoTe_2$.[93] While the exact ground state of these phases remains uncertain, their stabilization appears to be linked to real-space orbital Skyrme textures arising from the moiré superlattice in twisted TMDs.[110] Notably, plasmons in twisted $MoTe_2$ have been proposed as proxies for the real-space topology induced by these Skyrme textures.[111]

Although these findings are promising, the presence of flat bands and strong *e*–*e* interactions characteristic of these systems necessitates nonperturbative approaches for studying their collective modes. A comprehensive description of these modes including nonlocal effects, will likely require a combination of advanced analytical many-body techniques and numerical methods (such as exact diagonalization and quantum Monte Carlo).

The study of plasmons in twisted 2D materials bridges fundamental physics and applied sciences, offering insights into quantum phenomena and opportunities for technological innovation. By addressing current challenges and exploring these suggested directions, we may be able to unlock the full potential of these materials, paving the way for breakthroughs in quantum materials science.

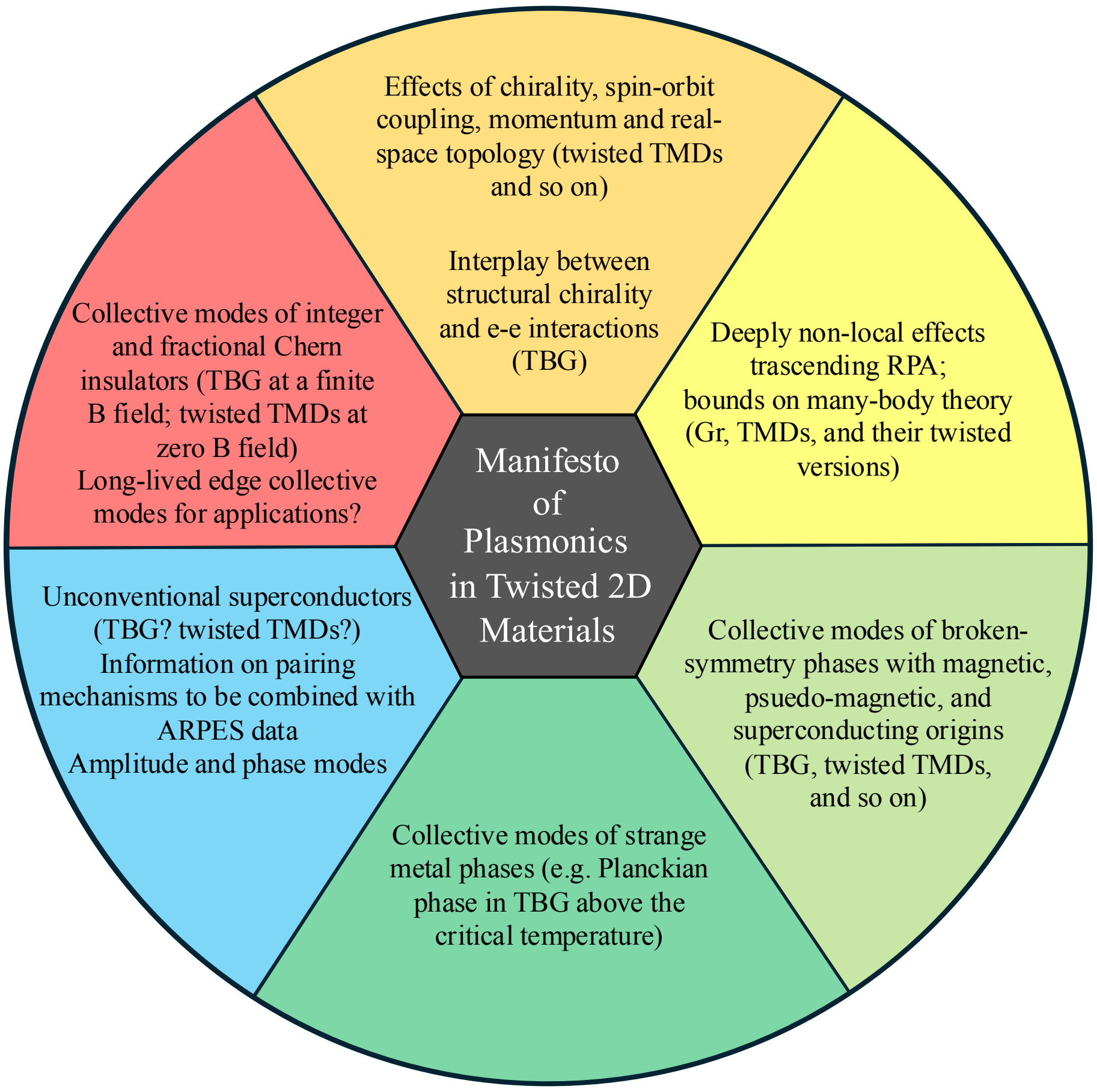

**Figure 4.** Manifesto of plasmonics in twisted 2D materials. We list future challenges and potential outcomes of plasmonics applied to twisted 2D materials.

## 5. 2D POLARITONS FOR PROBING NONLOCAL EFFECTS AND COLLECTIVE MODES IN CONDENSED MATTER

**P. A. D. Gonçalves,[1,*] A. T. Costa,[2] N. M. R. Peres,[3,4] and N. Asger Mortensen[4,5]**

[1]ICFO-Institut de Ciencies Fotoniques, The Barcelona Institute of Science and Technology, Castelldefels (Barcelona) 08860, Spain

[2]International Iberian Nanotechnology Laboratory (INL), Av Mestre José Veiga, 4715-330 Braga, Portugal

[3]Centro de Física (CF-UM-UP) and Departamento de Física, Universidade do Minho, P-4710-057 Braga, Portugal

[4]POLIMA—Center for Polariton-driven Light–Matter Interactions, University of Southern Denmark, Campusvej 55, DK-5230 Odense M, Denmark

[5]Danish Institute for Advanced Study, University of Southern Denmark, Campusvej 55, DK-5230 Odense M, Denmark

***Corresponding author.** Email: padgo@mci.sdu.dk

### 5.1 Current State of the Art

The interaction between electromagnetic radiation and charged particles is a central and recurring problem in condensed-matter physics and nanophotonics. Notably, this interaction can be significantly enhanced by exploiting polaritons, formed through the hybridization of photons with collective oscillations of polarization charges in materials[112] (e.g., plasmons, optical phonons, excitons). Polaritons enable the confinement of electromagnetic fields below the diffraction limit, leading to enhanced light–matter interactions.[113] Over the past few decades, this property has been exploited for various purposes, including the realization of subwavelength nanophotonic circuitry and the control of emitter dynamics. More recently, the advent of 2D materials—such as graphene, hBN, and TMDs—and related vdW heterostructures[63,114] (vdWH) has significantly expanded the range of possibilities for polariton-empowered phenomena. This expansion has been primarily driven by the strong light–matter interactions enabled by polaritons in atomically thin materials,[115] which span the electromagnetic spectrum from the THz to the UV range.[116] While polaritons in 2D materials have been studied in settings similar to those previously addressed with conventional 3D materials, often with superior performance, the strong light–matter interactions promoted by 2D materials offer new opportunities to probe the multifaceted and complex nanoscale response of material nanostructures in unprecedented ways (Figure 5a). This enticing potential is the focus of this section.

#### 5.1.1 *Quantum Nonlocal Effects in Graphene*

Over the last decade, plasmons in graphene-based platforms have attracted significant interest due to the remarkable properties of graphene plasmons (GPs), including strong field confinement, relatively low losses (especially when encapsulated in hBN[4,5] and electrical tunability[32,114]) Among the various configurations where GPs have been exploited, the realization of plasmons in graphene–dielectric–metal (GDM) heterostructures (Figure 5b), where screening from the nearby metal reshapes the GPs dispersion, endowing these plasmons with extremely large wave vectors at small graphene–metal separations,[29,117,118,119,120] has revealed a new paradigm in graphene plasmonics. Such plasmons are dubbed acoustic graphene plasmons (AGPs) due to their nearly linear dispersion relation, and are capable of achieving in-plane wave vectors $q$ approaching the Fermi wave vector $k_{\mathrm{F}}$, while the plasmon group velocity is slowed down toward the Fermi velocity. This feature has dramatic consequences, enabling AGPs to effectively probe both the frequency and momentum dependence (i.e., nonlocality) of graphene's conductivity. Using a scattering-type scanning near-field optical microscope (s-SNOM), Lundeberg *et al.*[29] have experimentally mapped the nonlocal optical conductivity of graphene through measurements of the AGPs' dispersion in GDM structures, revealing nonlocal and many-body effects beyond the random-phase approximation, including Fermi-velocity renormalization and compressibility corrections due to electron–electron interactions.

#### 5.1.2 *Probing Collective Modes in Superconductors*

Beyond plasmon polaritons, the interaction between electromagnetic radiation and (quasi)particles in strongly correlated matter gives rise to a wide range of collective phenomena,[121] including collective excitations in superconductors[119,122,123,124] and in magnetic systems.[125,126,127,128] Such excitations often correspond to otherwise *dark* modes, meaning they do not typically couple to far-field electromagnetic radiation. In this context too, the highly confined graphene plasmons excited in a nearby graphene sheet play a crucial role in probing these collective modes. Additionally, the optical response of a superconductor is inherently nonlocal[129] due to the finite size of the Cooper pairs. Superconductors support a plethora of collective modes,[130] ranging from collective excitations of the amplitude (phase) of the order parameter—i.e., the Higgs (Nambu–Goldstone) modes—to the Bardasis–Schrieffer mode, arising from fluctuations of the subdominant order parameter in an s-wave superconductor. Taking the Higgs mode as concrete example, within linear-response theory, its coupling to electromagnetic radiation vanishes in the local limit ($q \rightarrow 0$), as it is proportional to the magnitude of the in-plane wave vector. For this reason, the Higgs amplitude mode has primarily been studied using intense THz fields in the nonlinear regime.[131,132] However, it has been theoretically proposed that the large wave vectors associated with screened plasmons in a graphene–dielectric–superconductor configuration enable their coupling to the Higgs mode,[122] and that such an interaction can be detected using standard near-field optical microscopy techniques.

The coupling is enhanced for 2D superconductors (2DSCs), in the sense that it leads to larger Rabi-like mode splittings. Furthermore, combining graphene with 2DSCs allows for a variety of configurations, involving one or more graphene and 2DSC layers separated by thin insulating materials. These more complex structures can enhance the coupling between graphene plasmons and collective modes in the 2DSC. Figure 5c shows the imaginary part of the p-polarized reflection coefficient for a graphene sheet sandwiched between two 2DSCs (FeSe, with a superconducting gap $2\Delta = 26$ meV), calculated at $\hbar\omega \simeq 18$ meV (approximately the energy of the Bardasis–Schrieffer mode). The figure shows three peaks: the first two correspond to the hybridization of the graphene plasmon with the 2DSCs' Bardasis–Schriffer mode, whose splitting decreases with the increasing Fermi energy of graphene (because the GP's dispersion is steeper and intercepts the Bardasis–Schrieffer mode at smaller wave vectors ($q \propto \omega^2/E_{\mathrm{F}}$)), whereas the remaining peak at larger wave vectors is associated with the plasmon–Cooper-pair mode. The large wave vector ($q \approx 100\ \mu\mathrm{m}^{-1}$) is due to the hybridization of the two plasmon–Cooper-pair modes (one from each 2DSC) leading to an acoustic-like mode. This implies larger wave vectors in comparison with the individual, bare mode of an individual 2DSC layer.

5.1.3 *Interrogating the Quantum Surface Response of Metals*

Nonlocal effects in conventional plasmonics are inherently difficult to probe because the Fermi wave vector of 3D metals is typically much larger than the one of surface plasmons. With this backdrop, ultraconfined graphene plasmons, and AGPs in particular, constitute a viable route to circumvent this obstacle, due to the large wave vectors attained by AGPs in GDM stacks.[118,119,133,134] Figure 5d illustrates this point, where the quantum surface response of a nearby metal—encoded by the Feibelman $d_\perp$-parameter,[120,135] characterizing the centroid of the induced density—can be inferred from the its influence on the dispersion relation of AGPs.[134] Strikingly, the characteristic electronic length scale, on the order of Å, can be clearly discerned through substantial spectral shifts.[134] For a given wave vector, Re $d_\perp > 0$ (Re $d_\perp < 0$) leads to a redshift (blueshift) of the AGP frequency, while the Im $d_\perp$ contributes to increased losses originating from surface-assisted excitation of electron–hole pairs in the metal. Incidentally, while the Feibelman $d_\perp$-parameter at optical and near-IR wavelengths can be probed using metal-based plasmonics[136] (i.e., $\omega \lesssim \omega_\mathrm{p}$), AGPs enable the characterization of the low-frequency, static ($\omega \to 0$) metallic quantum response, $d_\perp(\omega \simeq 0)$. Such knowledge holds significance for incorporating quantum corrections to the classical electrostatic image theory of particle–surface interaction[135] as well as to the vdW interaction[137] affecting atoms or molecules near metal surfaces.

5.1.4 *Probing Magnetism in Condensed-Matter Systems*

An intriguing approach for exciting magnons in magnetic systems involves their coupling with plasmons (Figure 5e). A magnon represents the quantum of collective spin excitation in magnets, which, in classical terms, can be visualized as spins precessing around their equilibrium direction at a characteristic frequency, $\omega_\mathrm{magnon}$. This coupling mechanism offers a powerful platform for investigating and manipulating light–matter interactions at the nanoscale, opening up exciting opportunities for fundamental research.[138] While much of the progress in this area has been achieved within the framework of linear optics, recent advances in nonlinear optical phenomena present a promising new direction for exploration. These developments have catalyzed the emergence of novel fields such as nonlinear magnonics, which hold significant potential for transformative applications.[139]

5.1.5 *Other Possibilities*

The possibilities outlined in Sections 5.1.1–5.1.4 represent just the tip of the iceberg, with numerous additional possibilities to probe collective excitations in matter using different types of 2D polaritons, supported by the rapidly expanding library of atomically thin materials. Furthermore, the recent discovery of novel physical phenomena in twisted graphene bilayers and twisted metasurfaces has fuelled new prospects for light–matter interactions. These include, for instance, the realization of topological states, photonic flat bands, and transitions between localized and delocalized states of light.[140]

### 5.2 Challenges, Future Goals, and Suggested Directions to Meet These Goals

While some of the opportunities and theoretical proposals discussed above require advancements on the experimental front, others are well within reach using current state-of-the-art techniques. For instance, guided by the theoretical framework introduced in ref 134, the metallic quantum surface response can be characterized using the same s-SNOM approach employed in previous nanoimaging studies of GDM structures.[117,29] Similarly, collective modes in superconductors could be detected by measuring their hybridization with acoustic-like graphene plasmons, for example, through cryogenic SNOM[5,141,142] (especially when incorporating high-$T_\mathrm{c}$ superconductors, where the higher critical temperature reduces cooling requirements). On the theoretical front, advancements are needed to establish a general and practical framework that accounts for both nonlocal and nonlinear effects in the electrodynamics of superconductors.

Another exciting prospect involves exploiting polaritons in vdW heterostructures that combine graphene with superconducting magic-angle twisted bilayer graphene—or other moiré materials—to map the system's optoelectronic response. Such studies could provide valuable insights into the physics underlying the exotic phenomena observed in moiré materials.[143]

Recent advancements in 2D magnetic materials, characterized by strong magnetic anisotropy energy (the presence of an easy axis) and significant spin–orbit coupling, have opened new avenues for studying magnon–plasmon interactions. These materials exhibit gapped magnon spectra, which align well with the energy and momentum scales of graphene plasmons, making them ideal for probing such interactions. Additionally, many 2D magnetic materials possess hexagonal real-space lattices with two atoms per unit cell, a symmetry that gives rise to two magnonic bands, including one at higher frequencies that features a distinctive gap at the Dirac point. The magnonic gap in these systems has the potential to host chiral topological magnetic modes,[144] which are particularly intriguing due to their ability to support robust, dissipation-resistant edge states. Since these modes emerge at the edge of the Brillouin zone, they are not susceptible to probing by graphene plasmons. However, we can consider a scenario where two twisted 2D magnetic layers are present. In this case, the twist-induced superlattice structure leads to a significantly reduced Brillouin zone compared to that of a single layer.[145] As a result, the chiral modes could be probed using the large wave vectors associated with graphene plasmon modes, opening up new avenues for studying topological excitations in 2D magnetic materials. Manipulating the twist-angle is also a viable route to altermagnetism,[146] a new form of collinear magnetic order where the atomic spins adopt a Néel arrangement (like in an antiferromagnet), thus having zero total magnetization, but whose electronic bands are spin polarized (like in a ferromagnet) over sizeable portions of the Brillouin zone. Using the standard Holstein–Primakoff transformation, the mathematical description of twisted magnetic bilayers follows a similar approach to that of their fermionic counterparts, such as graphene layers. Therefore, it is plausible to describe twisted magnetic bilayers as tight-binding magnons, along with the corresponding spin order in each layer.

Graphene plasmons have a magnetic field that lies parallel to the plane of the material. As a result, this field is perpendicular to the magnetic moments of materials with perpendicular magnetization, such as $CrI_3$, $Fe_3GeTe_2$, and other vdW magnets, enabling the excitation of spin waves. In this context, polaritons form through Zeeman coupling of the magnetic moment to the magnetic field of the graphene plasmons. Another mechanism for plasmon–magnon-polariton formation involves the coupling of the electric field of 2D plasmons to the lattice degrees of freedom of the magnetic material. In this scenario, there is an indirect coupling where the magnon hopping is modulated by the electric field, since the former depends on the relative distance between atoms. This unexplored possibility could pave the way for the study of electromagnons in hybrid graphene–2D-magnetic systems.

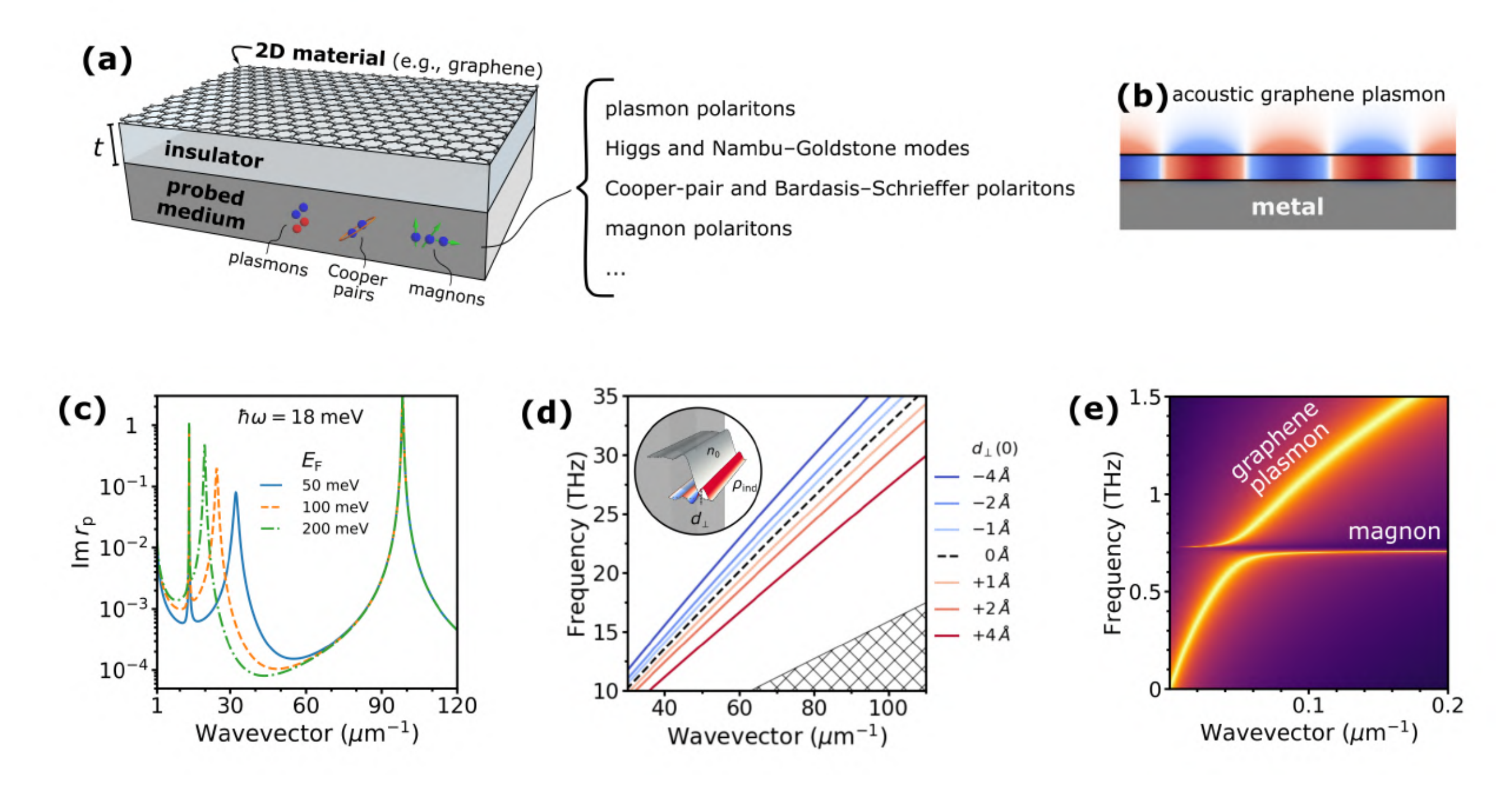

**Figure 5.** Harnessing ultraconfined graphene plasmons to probe collective excitations and strongly correlated phenomena in condensed-matter systems. (a) Schematic of a 2D material (here represented by graphene) in the vicinity of a probing material, separated by a thin insulating slab. The field provided by the 2D polaritons can couple to quasiparticles in the nearby medium, thereby probing the response and collective excitations of the material. (b) Illustration of the out-of-plane component of the electric field associated with acoustic graphene plasmons (AGPs) in a graphene–dielectric–metal (GDM) configuration. (c) Collective modes, emerging as features in Im $r_p$, of a 2DSC–G–2DSC heterostructure. The 2D superconductors (2DSC) is FeSe, with a superconducting gap of 23 meV at zero temperature and supporting a Bardasis–Schrieffer mode at 18.2 meV. See ref 124 for details. (d) Dispersion relation of AGPs in a GDM configuration. The metallic quantum surface response (encoded through the Feibelman $d_\perp$-parameter; see inset) leads to sizable spectral shifts, underscoring the potential of AGPs to infer quantum nonlocal effects in nanoplasmonics. Adapted with permission from ref 134 (Copyright 2021 Springer Nature). (e) Dispersion relation exhibiting the hybridization between graphene plasmons and the uniform ($q = 0$) magnon mode of monolayer $CrI_3$ in a heterostructure composed of three single layers: graphene–hBN–$CrI_3$. Adapted with permission from ref 126 (Copyright 2023 American Chemical Society).

## 6. MAXWELL HAMILTONIAN IN 2D MATERIALS

**Sathwik Bharadwaj[1] and Zubin Jacob[2,*]**
[1]Elmore Family School of Electrical and Computer Engineering, Birck Nanotechnology Center and Purdue Quantum Science and Engineering Institute, Purdue University, West Lafayette, Indiana 47907, United States
[2]Elmore Family School of Electrical and Computer Engineering, Birck Nanotechnology Center and Purdue Quantum Science and Engineering Institute, Purdue University, West Lafayette, Indiana 47907, United States
***Corresponding author.** Email: zjacob@purdue.edu

### 6.1 Introduction

Nanophotonics—the study of light–matter interaction at length scales smaller than the wavelength of radiation—has widespread applications ranging from plasmonic waveguiding[147] and topological photonic crystals[148] to super-lensing.[149] Nanophotonic structures are artificial materials made up of two or more dielectric constituents. The wave dynamics governing these structures can be effectively described by the classical Maxwell equations in continuous media. In contrast, it has been a long-standing problem for more than 40 years to understand the dynamics of optical waves within a crystalline material at the lattice level.[150] For the past century, manifestations of the Schrödinger, Dirac, and Frohlich Hamiltonians have pushed the frontier in the discovery of numerous electronic and phononic phases of 2D materials. It is pertinent to ask what role crystal symmetries play in light–matter interactions and whether a description shift emerges at the sub-nanometer regime in crystalline materials. To address this, we have introduced a Maxwell-Hamiltonian theory of matter combined with a quantum theory of deep microscopic optical polarization.

The concept of Maxwell Hamiltonian was first introduced by Michael Berry to explain the origin of geometric phases in helical optical fibers.[151] We note that the Maxwell-Hamiltonian formulation in free space has also been considered in the past to comprehend the connection between photons and massless fermions in the Dirac equation. However, only recently the Maxwell Hamiltonian has attracted renewed attention in condensed matter to predict deep microscopic electrodynamic phases of matter.[152,153] The Maxwell-Hamiltonian framework satisfies $\mathrm{U}(1)$ gauge symmetry and $\mathrm{SO}(3)$ spin symmetry, hence presenting a unique approach for treating the interaction of materials and electromagnetic fields.

Two-dimensional materials have traditionally served as the primary platform to investigate new topological phases of matter. Even at its inception, the topological phase was first proposed by Haldane[154] on a graphene monolayer with time-reversal breaking *via* next-nearest neighbor hopping. This is an example of a Chern phase $C \in \mathbb{Z}$, and a nontrivial Thouless–Kohmoto–Nightingale–den Nijs (TKNN) invariant. The Kane–Mele model,[155] which is based on spin–orbit coupling in graphene, demonstrates a quantum spin Hall phase $\nu \in \mathbb{Z}_2$, without breaking time-reversal symmetry. Both the quantum hall conductivity $\sigma_{xy} = Ce^2/h$, and spin hall conductivity $\sigma_{xy} = \nu e/2\pi$, are manifested from the 2D Dirac Hamiltonian, and are electrostatic observables at zero frequency $\omega = 0$ and zero momentum $\mathbf{q} = 0$. Since electromagnetic fluctuations in matter span over all frequencies $\omega \neq 0$ and momenta $\mathbf{q} \neq 0$, one must look beyond these paradigms to characterize the topological photonic phases of 2D materials. In this regard, the Maxwell-Hamiltonian framework provides a foundation for the discovery of the topological electromagnetic phases in 2D materials, laying the groundwork for a novel optical classification of condensed matter (Figure 6).

### 6.2 Current State of the Art

In 2D materials, we can strictly focus on transverse-magnetic (TM) polaritonic waves. In this case, the magnetic field $H_z$ is perpendicular to the plane of propagation $\mathbf{q} = q_x\hat{\mathbf{x}} + q_y\hat{\mathbf{y}}$. The Maxwell-Hamiltonian equation in 2D materials can be written as

$$\boldsymbol{H}_M(\mathbf{q}) \cdot \mathbf{f_q}(\mathbf{r}, \omega) = \frac{\omega}{c}\, \mathbf{g_q}(\mathbf{r}, \omega), \tag{6.1}$$

where

$$\mathbf{f_q} = \begin{bmatrix} E_x \\ E_y \\ H_z \end{bmatrix}$$

and

$$\mathbf{g_q} = \begin{bmatrix} D_x \\ D_y \\ B_z \end{bmatrix}$$

define the photon wave function, and the Maxwell Hamiltonian is given by

$$\mathbf{H}_M(\mathbf{q}) = q_x S_x + q_y S_y. \tag{6.2}$$

Here, $q_i = -i\,\partial_\mathrm{i}$ is the photon momentum operator along the direction $i = x, y$. The Maxwell Hamiltonian describes the projection of momentum into the photon spin-1 operators $S_i$. The spin-1 components are given by

$$S_x = \begin{bmatrix} 0 & 0 & 0 \\ 0 & 0 & 1 \\ 0 & 1 & 0 \end{bmatrix}, \qquad S_y = \begin{bmatrix} 0 & 0 & -1 \\ 0 & 0 & 0 \\ -1 & 0 & 0 \end{bmatrix}, \qquad S_z = \begin{bmatrix} 0 & -i & 0 \\ i & 0 & 0 \\ 0 & 0 & 0 \end{bmatrix}. \tag{6.3}$$

The photon spin-1 operators satisfy the angular momentum algebra $[S_i, S_j] = i\epsilon_{ijk}S_k$. We observe that, analogously to the Dirac Hamiltonian for spin-1/2 quasiparticles, the Maxwell Hamiltonian depends on the spin-1 behavior of photons. However, unlike the Dirac Hamiltonian, the time reversal operator $\mathcal{T}$ in the Maxwell-Hamiltonian formalism satisfies the relation $\mathcal{T}^2 = +\mathbb{I}_3$, reflecting the bosonic nature of photons.

6.2.1 *Effective Photon Mass in 2D Gyrotropic Materials*

In 2D gyrotropic materials, $\mathcal{T}$-symmetry is broken,[156] and one can show that the effective photon mass $m_{\text{eff}}^{\text{ph}} \neq 0$ is finite. In this sense, the gyrotropy can also be thought of as a high-frequency equivalence of the DC Hall coefficient. By solving the continuum Maxwell Hamiltonian for a 2D interface between positive and negative gyrotropic layers, it has shown that the system can host unique unidirectional Maxwellian spin waves. These waves are photonic analogs of Jackiw–Rebbi waves[157] (electron waves at the interface between a positive and negative mass), as described by Dirac-Hamiltonian solutions. These Maxwellian spin waves are fundamentally different from surface plasmon polaritons or magnetic edge plasmons, which are present at the interface of a medium with opposite signs for the dielectric constant.

6.2.2 *Maxwell Hamiltonian in 2D Crystalline Materials*

In a material lattice, the Maxwell Hamiltonian is modulated by a periodic optical response, hence the eigen-fields take a vectorial Bloch form, given by $\mathbf{f_q}(\mathbf{r},\omega) = e^{i\mathbf{q}\cdot\mathbf{r}} \sum_{\mathbf{G}} \boldsymbol{\mathcal{U}}_{\mathbf{G}}(\mathbf{q},\omega)e^{i\mathbf{G}\cdot\mathbf{r}}$, where $\mathbf{G}$ runs over reciprocal lattice vectors. Within a linear response framework, the deep microscopic displacement field $\mathbf{g_q}(\mathbf{r},\omega)$ can be expressed as

$$\mathbf{g_q}(\mathbf{r},\omega) = e^{i\mathbf{q}\cdot\mathbf{r}} \sum_{\mathbf{G},\mathbf{G}'} \boldsymbol{\chi}_{\mathbf{GG}'}(\mathbf{q},\omega) \cdot \boldsymbol{\mathcal{U}}_{\mathbf{G}'}\, e^{i\mathbf{G}\cdot\mathbf{r}}, \tag{6.4}$$

where $\boldsymbol{\chi}_{GG'}(\mathbf{q},\omega)$ is the deep microscopic optical response tensor, which includes permittivity, permeability, and magneto-electric coupling. The tensor $\boldsymbol{\chi}_{\mathbf{GG}'}(\mathbf{q},\omega)$ is invariant under spatial and temporal crystal symmetry operations. Combining eqs 6.1 and 6.4, we observe that the Maxwell-Hamiltonian equation of matter belongs to the class of generalized nonlinear eigenproblems. Since $\boldsymbol{\chi}_{\mathbf{GG}'}(\mathbf{q},\omega)$ plays the role of the Green function of the optical polarization density, it naturally possesses a topological invariant that is conserved under continuous deformations, which we define as the optical $N$-invariant.[158]

6.2.3 *Optical N-invariant of 2D Viscous Hall Fluid*

The optical $N$-invariant is obtained by computing the winding number of the microscopic optical response tensor and is distinct from topological invariants known from electronic band theory. We have recently shown that this topological quantum number is captured by the spatiotemporal dispersion of optical constants, which has been defined as the deep microscopic optical band structure of a material.[159,160] We emphasize that the optical $N$-invariant introduces a completely new classification of 2D materials based on the topological electromagnetic phase of matter.[161]

The first candidates for this new class of topological phase ($N \neq 0$) have been identified as hydrodynamic electron fluids in strongly correlated 2D materials such as graphene.[158] The fundamental physical mechanism responsible for this topological electromagnetic classification is the Hall viscosity $\eta_H$, which adds a nonlocal component to the Hall conductivity.[162] In the bulk magnetoplasmons of graphene, the optical $N$-invariant encodes the vorticity of spin-1 Néel-type skyrmions,[163] which are different from skyrmions in plasmonic lattices or photonic crystals. This enables the measurement of the optical $N$-invariant experimentally using magnetic field repulsion resembling the Meissner effect. Further, the optical $N$-phase exhibits gapless unidirectional edge plasmons that are immune to backscattering, which can be used to develop an ultra-subwavelength broadband topological hydrodynamic circulator. A topological circulator based on optical $N$-plasmons can function as an essential component for information routing and interfacing quantum-classical computing systems.[164]

6.2.4 *Deep Microscopic Electrodynamic Dispersion of Materials*

Within the framework of the Hamiltonian band theory of solids, extensive research has been undertaken on obtaining the dispersion for various electronic, phononic, and magnonic excitations in 2D and bulk materials. However, only artificial materials such as photonic crystals and metamaterials have been used to analyze the photonic dispersion and the corresponding field confinement. The Maxwell-Hamiltonian formalism combined with the quantum theory of optical polarization provides a pathway to obtain the deep microscopic electrodynamic dispersion of natural materials interacting with the photon field. It has recently been shown that this formalism predicts the existence of anomalous Maxwellian waves confined with sub-nanometer effective wavelengths even in well-studied materials like bulk silicon.[152] These waves occur in the spectral region where propagating waves are conventionally forbidden in the macroscopic electromagnetic framework. These findings demonstrate that natural media can support a variety of yet-to-be-discovered electromagnetic phases beyond the nanoscale limit.

### 6.3 Challenges, Future Goals, and Suggested Directions to Meet These Goals

The findings presented here highlight the significance of photonic phases of matter as well as the need for the development of first-principles-based computational techniques to reveal novel effects associated with the Maxwell Hamiltonian.

One of the primary challenges in the experimental discovery of these phenomena is the development of appropriate probes capable of detecting the Maxwellian waves hidden within a crystal lattice. Traditional pump-probe spectroscopy, DC transport measurements, and ellipsometry measurements yield the optical conductivity at the far-field limit ($q \to 0$), which significantly limits the spatial resolution required to separate the optical response at the lattice level. The constant development of new near-field tools for light–matter interaction establishes an exciting frontier for investigating the optical response in the extreme momentum regime. We anticipate that THz scanning near-field optical microscopy,[165] microwave impedance microscopy,[166] and momentum-resolved electron energy-loss spectroscopy[167] are best suited to probe these phenomena. These techniques allow one to overcome the Abbe limit and achieve subwavelength real-space resolution.

Further, the newly discovered class of emergent 2D moiré materials[86] are particularly well-suited to host deep microscopic Maxwellian waves. The large effective lattice constant ($a_{\text{moiré}} \sim 10$ nm) of moiré materials is a distinguishing feature that sets them apart from conventional 2D materials ($a < 1$ nm). Hence, one can anticipate significant spatial variation of microscopic fields even within a unit cell of moiré materials. When compared to typical bulk metallic structures, moiré materials can offer stronger photon confinements due to their strong electron correlations and lower dimensions. The discovery of Maxwellian phases in 2D moiré materials further ensures the pathway for designing high-performance devices with spatial dimensions three orders of magnitude smaller than the current limits.

In conclusion, the Maxwell-Hamiltonian framework integrates the fields of 2D materials and photonics to span a novel area of research in materials science.

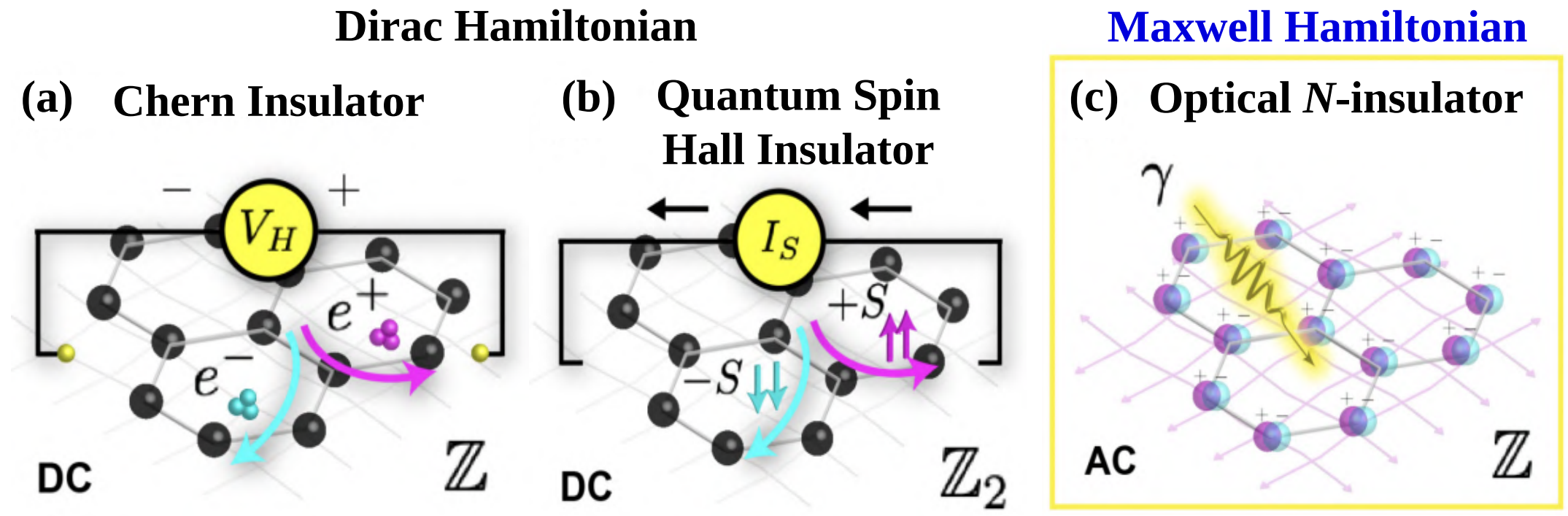

**Figure 6**. Topological phases in 2D materials. (a) A Chern insulator is related to charge transport and generates a voltage $V_H$ across the sample. (b) A quantum spin Hall insulator is connected to spin transport, which results in a current $I_s$ across the sample. Both Chern and quantum spin Hall insulators are manifestations of the 2D Dirac Hamiltonian. (c) In contrast, in certain 2D materials, an effective photon mass term can enter the Maxwell Hamiltonian arising from viscous hydrodynamic light–matter interactions. This leads to an optical topological number: $N$-invariant which captures the winding number of the microscopic optical response tensor. Adapted with permission from ref 161 (CC BY 4.0. International license).

**Characterization Techniques**

## 7. 2D MATERIALS SEEN THROUGH THE LENS OF MULTI-MESSENGER NANOIMAGING

**D. N. Basov,[1,*] P. J. Schuck,[1] A. N. Pasupathy,[1] Milan Delor,[1] and M. K. Liu[2,3]**
[1]Department of Physics, Columbia University, 1150 Amsterdam Avenue, New York, New York 10027, United States
[2]Department of Physics and Astronomy, Stony Brook University, Stony Brook, NY 11794
[3]National Synchrotron Light Source II, Brookhaven National Laboratory; Upton, New York 11973, USA.
***Corresponding author.** Email: db3056@columbia.edu

"Seeing is knowing": this maxim coined by Plato in the *Theaetetus* does apply to the outsized impact that imaging methods are playing in contemporary science and engineering. There is plenty to see in 2D materials, especially if one applies modern scanning probe methods capable of resolving in real space the structure and attributes over a broad range of length scales spanning from atomic arrangements/reconstructions to mesoscale properties of integrated devices. Since the advent of atomic force and scanning tunneling microscopes (AFM and STM), dozens of imaging modalities have been implemented based on these two foundational platforms (Figure 7). A major advance attained in the last two decades is in the implementation of deeply sub-diffraction nano-optical imaging empowered by the coupling of lasers to the AFM (or STM) scanning platforms. Scanning probe images provide powerful visualization of structural, compositional, electronic, optical and magnetic properties information in a form of 2D maps produced by a specific contrast mechanism. Spatial patterns uncovered in images often decode the underlying physics and functions (*vide infra*). Because each scanning probe modality carries its own *message* about the complex materials under investigation, it is instructive to acquire collocated maps with multiple imaging modalities applied to the same spatial region. This task of multi-messenger nanoimaging[168] guided the development of instrumentation and analysis tools carried out in research groups of the co-authors of this perspective and many other groups around the globe.

### 7.1 Current State of the Art

Figure 7 displays a diorama of AFM-based scanning probe methods, including a variety of nano-optical tools. A notable achievement of nano-IR imaging is the visualization of propagating plasmon polaritons, initially on the surface of the prototypical 2D material graphene, accomplished using scattering-type scanning near field optical microscopes (s-SNOM) operating with IR lasers.[15,16] Polaritons are quantum mechanical superpositions of photon states with elementary excitations in solids. 2D materials host many forms of polaritons produced by hybridizations of electronic, lattice, spin, and excitonic processes with photons of a matching frequency.[112] Nano-IR imaging experiments with traveling plasmon polaritons[15,16] paved the way for the systematic exploration of numerous other types of polaritons hosted by 2D materials, including phonon polaritons[169] and exciton polaritons.[8,170] Polaritonic imaging enables an entirely novel approach toward optical studies of the physical phenomena in materials. Traditional far-field optical methods are focused on *colors* and derive insights from the analysis of spectra of emitted, absorbed, reflected, and transmitted radiation. Polaritonic imaging infers physics insights from the examination of shapes and line forms produced by these light–matter standing waves and their rich interference patterns.

#### 7.1.1 *On Spatial Resolution*

The spatial resolution of nanoimaging methods is commonly governed by the radius of the apex of the scanning probe tips reaching single-digit nanometers in many AFM applications. Nano-optical experiments typically require metal coatings on the tips to enhance near-field interaction, and these duller tips yield a spatial resolution of a few tens of nm. Likewise, magnetic force microscopy (MFM) requires ferromagnetic metal coatings that also limit the spatial resolution down to ~10–50 nm; the spatial resolution of MFM is also reduced by the intricacies of the contrast mechanism governed by long-range magnetic forces. Nano-optical and MFM signals can be acquired with the same tip, enabling several concomitant imaging modalities: topography, multi-color optical scattering amplitude, and magnetic susceptibility.[168] In this fashion, multi-messenger nanoimaging allows one to uncover correlations between structural, electronic, and magnetic properties. Nano-terahertz (nano-THz)[165] or nano-photoluminescence experiments also require specialized tips. Counterintuitively, nano-optical experiments can visualize features and textures that are much smaller than the tip diameter. For example, polaritonic standing waves display contrast due to reflections and interference prompted by atomically narrow domain walls in bilayer and multilayer twisted 2D materials, even in maps recorded with metalized tips of ~10 nm diameter.[171] Advances in hardware and signal processing now allow one to visualize a few nm features in twisted 2D bilayers hosting reconstructed moiré patterns using very long-wavelength microwave range imaging.[172]

7.1.2 *On Moiré Reconstructions*

Moiré superlattices in twisted 2D materials can serve as a unique control knob of this class of materials, enabling the control of electronic, excitonic, magnetic, and superconducting properties. These moiré heterostructures have emerged as capable quantum simulators of condensed matter physics.[143] Direct visualization of physical processes in moiré superlattices is imperative to understand and fully realize the opportunities offered by the twist angle control. Specifically, STM data uncovered distinct electronic properties associated with multiple possible atomic arrangements of the two layers in moiré superlattices.[173] Furthermore, information of energy landscapes in bi-layer and multilayer vdW materials can be extracted directly from the analysis of the shapes of the reconstructed pattern *via* the tool of *moiré metrology*.[174] Nano-IR imaging was utilized to explore the electrodynamics of flat bands in twisted graphene near the so-called magic angle, a regime known for hosting superconductivity.[97] Apart from superconductivity, ferroelectricity is yet another notable emergent state in moiré superlattices of 2D materials. A combination of piezo-force, Kelvin force and nano-optical imaging offered a detailed characterization of ferroelectricity emerging at the interface of twisted 2D materials.[175,176,177] Finally, moiré superlattices in the magnetic 2D material $CrI_3$ exhibit intricate textures comprised of ferromagnetic and antiferromagnetic domains, which can be visualized by single-spin quantum magnetometry also known as NV-center imaging.[178]

7.1.2 *Nano-Optical Imaging of Semiconducting 2D Materials*

The exploration of semiconducting 2D materials has benefited from the advances of the nanoimaging versions of light emission experiments including nano-photoluminescence (nano-PL),[179] nano-electro-luminescence (nano-EL),[180] and nano-second-harmonic generation (nano-SHG).[181,182] These methods have already attained sensitivity to emission characteristics down to single monolayer and even single defect[180] levels. The STM platform is likewise well suited to collect both PL and EL spectra and images.

7.1.3 *In Operando Nanoimaging and Nano-Spectroscopy*

One appeal of multi-messenger imaging based on AFM/STM platforms is that these methods empower *in operando* experiments, particularly in the regime where the properties of 2D materials are controlled by external stimuli such as electric fields[15,16] magnetic field[183] chemical dopants[38] photoexcitation[18,184] strain or high-density electrical currents.[185,186] Furthermore, nano-photocurrent imaging[187] stands out for its ability to directly visualize current pathways, which can offer insights into the physics of flat bands and topology in 2D materials structures.

7.1.4 *On Spatiotemporal Imaging*

Nano-optical imaging experiments carried out with femtosecond lasers provide insight into the spatiotemporal evolution of the properties of 2D materials. Notably, these experiments can be carried out under both equilibrium and nonequilibrium conditions. The latter are performed in a pump-probe mode. Here, a pump laser photoexcites the studied 2D material over a large area of a diffraction-limited spot. The probe signal is collected locally from the area underneath the tip of a nano-probe apparatus. By adjusting the time delay between the pump and probe pulses one gains access to the evolution of the system response pixel by pixel.[188] One novelty of nano-pump-probe experiments is that it is feasible to visualize propagating plasmonic or polaritonic waves in nonequilibrium regime when the electronic system of a studied material is agitated by the pump pulse.[18,184,189] This latter capability enabled imaging of the plasmonic response in regimes of high carrier densities of hot electrons created by photoexcitation.[18,184] The equilibrium version of spatiotemporal experiments stems from the ability to track polaritonic motion not only in space but also in time.[75,190] Polaritonic worldlines that are directly observable in space–time metrology experiments facilitated the exploration of electronic interactions in graphene over a broad range of gate voltages, from nearly charge neutral state to highly conducting regimes.[190] The current record in spatiotemporal imaging, picometer length scales and femtosecond timescales, has been achieved at THz frequencies by capitalizing on atomic nonlinearities of mesoscopic near fields.[191]

7.1.5 *Electronic and Photonic Phenomena at Interfaces of 2D Materials*

Multi-messenger nanoimaging is a unique resource for visualizations of the emerging properties at interfaces. For example, 2D layers with distinct work functions reveal prominent charge transfer leading to the formation of conducting interfaces. The effect is purely non-volatile: no applied voltage is required to promote highly conducting metallic interfaces. The emerging metallicity of charge transfer interfaces was quantified through a combination of nano-IR and STM imaging.[192] The charge transfer forms atomically abrupt boundaries that are required, for example, for the implementation of nano-photonic versions of whispering gallery resonators.[193] Additionally, interfaces of hyperbolic 2D materials display nonintuitive optical phenomena including negative reflection[54] and negative refraction,[45,194] which have been visualized by tracking polaritonic motion across interfaces.

7.1.6 *Probing Topology in Electronic and Photonic Phenomena*

2D materials display a rich variety of effects rooted in topology. These effects can be classified into two broad categories: electronic/magnetic and photonic. Both categories of the topological effects can be readily visualized by multi-messenger methods. A hallmark of topological phenomena is the emergence of edge states at the boundary between regions with distinct topological invariants. A recent tour-de-force experiment carried out with microwave nanoimaging has visualized topological edge states in a fractional Chern insulator twisted $MoTe_2$.[195] Earlier nano-IR experiments with polaritonic nano-structures have identified a protocol for the extraction of topological charges from nanoimaging data and confirmed fractional topological charges for phonon polariton vortices in hexagonal boron nitride initiated with circularly polarized light.[196]

### 7.2 Challenges and Future Goals

Over the last two decades, versions of multi-messenger nanoimaging combining nano-optical modalities with other scanning probe tools advanced from sparse demonstrations to an indispensable research tool in many laboratories around the globe. Notably, nano-optical instruments now equip many IR/optical beamlines at synchrotron facilities, offering broad access to users requiring multi-messenger nanoimaging data for their research. The library of nanoimaging modalities continues to expand. We anticipate that expansion of nano-optics in spatially resolved version of 2D spectroscopy,[197] platforms combining advanced multidimensional far-field and near-field imaging, and experiments operating with entangled photons are on the horizon. At the moment, not all available modalities have been implemented under the conditions required for the exploration of condensed matter physics: cryogenic operation and high magnetic fields. Nano-emission experiments remain particularly sparse despite a broad scope of opportunities offered by emission experiments across the electromagnetic spectrum, especially in far-IR and THz ranges.[198] STM electroluminescence experiments offer good opportunities in this regard since they naturally confine the excitation to the atomic scale.[199] Electron tunneling also offers the possibility to directly measure electronic structure changes induced by strong light–matter interactions, and the possibility of observing the electronic component of polaritonic waves in the future.

Data science methods are poised to provide monumental advantages for the analysis of multi-messenger imaging data. The discussion in the previous paragraph shows that multi-messenger datasets are fundamentally multidimensional in nature. Manual processing can only uncover some of the most generic trends. So far, the use of data science in multi-messenger imaging remained rather limited despite remarkable successes of artificial intelligence (AI) approaches in other areas of research. Apart from the analysis, data science methods offer advantages for accelerated measurements[200,201] and data acquisition that are yet to be fully exploited in multi-messenger imaging.

"Not seeing is not knowing" posited Plato. Multi-messenger imaging will continue to deliver new knowledge by allowing practitioners in this bourgeoning field to see the previously unseen and uncover the previously unknown.

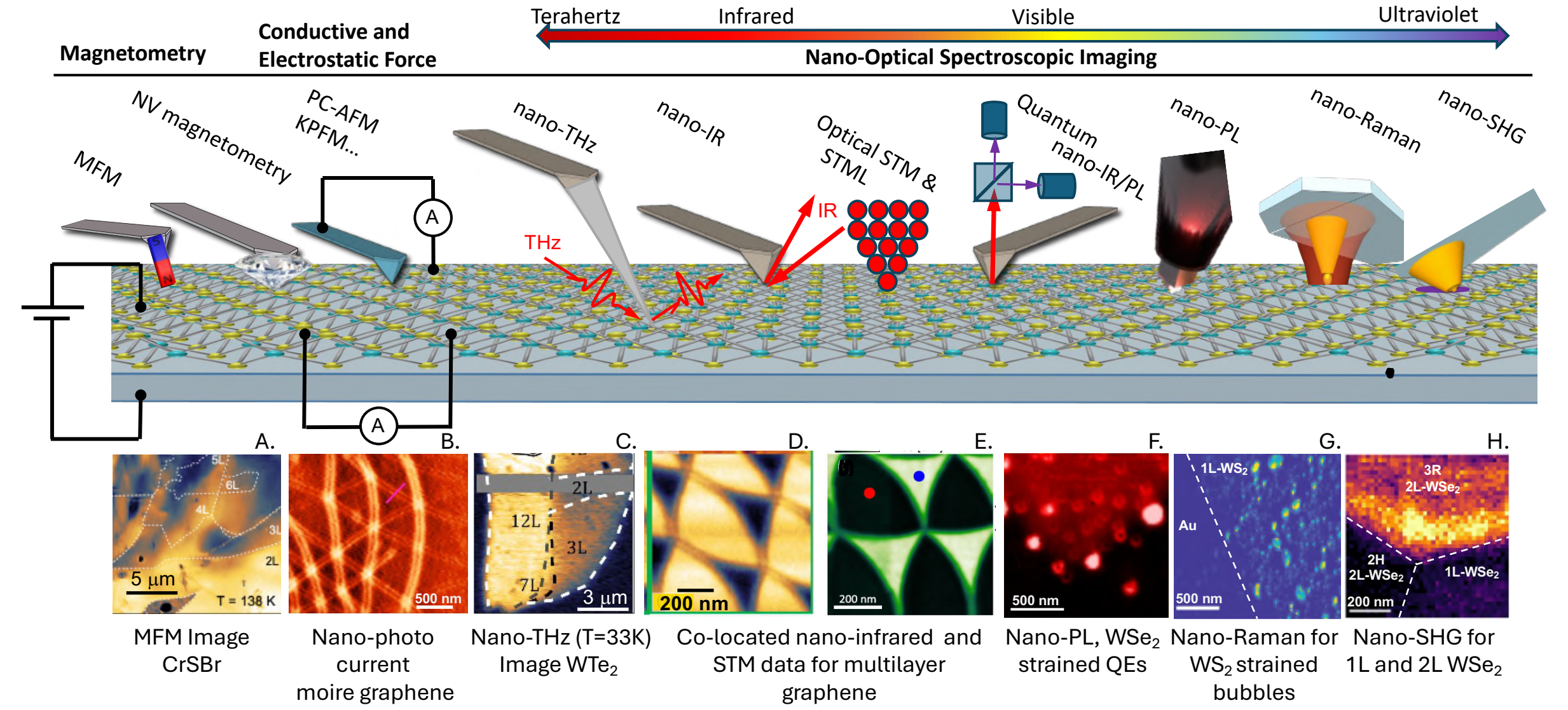

**Figure 7.** Multi-messenger scanning probe nanoimaging is based on AFM and STM platforms. Top panel: Schematic representation of scanning probe imaging modalities commonly applied to the exploration of 2D materials and their heterostructures. Bottom panel: Examples of images of different classes of 2D materials reported by co-authors. From left to right: A) MFM imaging of atomically layered magnetism in semiconducting CrSBr;[202] B) Nano-photocurrent in minimally twisted graphene;[203] C) Nano-THz response of a few layer $WTe_2$ microcrystals;[204] D and E) Co-located nano-IR and STM studies of nonlocal relaxation dynamics in twisted trilayer graphene moiré superlattices;[205] F) Nano-PL image of a strained nanobubbles in a $WSe_2$ monolayer on Au substrate;[206] G) Nano-Raman image of $WS_2$ monolayer bubbles on a Au substrate;[207] H) Nano-SHG characterization of layer stacking in $WSe_2$.[182]

## 8. TIP-ENHANCED NANOSCOPIES FOR ATOMIC-SCALE PHOTONICS IN 2D MATERIALS

**Aitor Mugarza,[1,2,*] Pablo Merino,[3] Marc G. Cuxart,[1] Emigdio Chávez-Angel,[1] and Martin Svec[4,5]**
[1]Catalan Institute of Nanoscience and Nanotechnology (ICN2), CSIC and BIST, Bellaterra, 08193 Barcelona, Spain
[2]ICREA—Institució Catalana de Recerca i Estudis Avançats, 08010 Barcelona, Spain
[3]Material Science Institute of Madrid (ICMM-CSIC), 28049, Madrid, Spain
[4]Institute of Physics, Czech Academy of Sciences, Praha 6 CZ16200, Czech Republic
[5]Institute of Organic Chemistry and Biochemistry, Czech Academy of Sciences, Praha 6 CZ16000, Czech Republic
***Corresponding author.** Email: aitor.mugarza@icn2.cat

### 8.1 State of the Art

Recent advancements in the field of tip-enhanced nanoscopies have opened new avenues for studying light–matter interactions and manipulating quasiparticles down to the atomic scale. These techniques typically involve positioning a sharp metallic tip at nanometer-scale distances from the surface of the material under investigation, while illuminating it with highly focused light. This configuration creates a nanocavity that can significantly enhance and localize electromagnetic fields to volumes well below the diffraction limit. It also facilitates the exchange of photons between the near and far-field, allowing for the coupling of external light sources to the nanocavity or guiding light generated beneath the tip to optical detectors.

Depending on the specific systems and processes being examined, a variety of excitation light sources can be used, spanning from THz to UV wavelengths (1 meV-10 eV). Many of these techniques employ conventional methodologies for photonic absorption, emission, and scattering, with some of the most common being scattering-type scanning near-field optical microscopy (s-SNOM), tip-enhanced Raman spectroscopy (TERS), and tip-enhanced photoluminescence (TEPL). The incorporation of a conducting tip further expands the methodological toolkit, enabling techniques that rely on electronic excitation or detection. For instance, tunneling electrons induced by static or pulsed bias voltages can generate localized electroluminescence, as seen in scanning tunneling microscope-induced luminescence (STML) and incident photons can generate photocurrent that is collected by the tip, enabling near-field photocurrent nanoscopy. Additionally, forces induced by photoexcitation can be measured in IR atomic force microscopy (IR-AFM). The family of tip-enhanced nanoscopies, and the main physical magnitudes and, through them, quasiparticles that can be probed, are schematically summarized in Figure 8.

In a broad, nonsystematic categorization, these techniques can be grouped by the energy range of the excitations involved. This leads to two main families: one suited for studying excitations from a few meV to approximately 100 meV, while the other is more appropriate for examining excitations in the range of eV.

Low-energy excitations include quasiparticles such as phonons, polarons, phonon polaritons, or plasmon polaritons in some 2D materials such as graphene. They can be probed by techniques that use radiation in the same energy scale (THz to IR), as in s-SNOM, its nano-Fourier transform IR spectroscopy (nano-FTIR) variant or IR-AFM/STM. s-SNOM has been extensively applied in the study of 2D materials. Gate-tunable surface-plasmon polaritons (SPPs) have been mapped with resolutions of a few nanometers on graphene devices[15] and hyperbolic SPPs on engineered hBN metasurfaces.[208] Spectral information can be acquired using the nano-FTIR variant for hyperspectral imaging, as shown for phonon polaritons in photonic hBN crystals.[47] Since the first real space visualization of the moiré superlattice acting as a photonic crystal for graphene SPPs in 2018,[171] this technique and its variants have been pivotal in the field of twist-optics. Remarkable demonstrations are the detection of local enhancement of the photo-thermoelectric effect at stacking domain walls,[187] or the 1D channeling of highly localized photons assisted by their strong coupling to phonons in $MoO_3$ bilayers.[44]

Inelastic scattering of high-energy visible radiation can also provide insights into low-energy modes, as exemplified by Raman spectroscopy. While this technique is predominantly used to study phonons, it can also address other low-energy excitations, such as optical polarons and magnons. In its tip-enhanced mode, TERS has been employed to spatially resolve phenomena like the localization of strain solitons at stacking domain boundaries in graphene bilayers.[209] However, most TERS studies involving 2D materials have been conducted in conjunction with TEPL, obtaining correlated information on vibrational and optoelectronic properties. This combined approach allows us to gain insights into phenomena such as doping and strain[210] or interlayer interactions[211] from the optical response of TMD.

Higher-energy excitations, such as excitons, can be effectively probed using TEPL and STML, utilizing visible excitation and detection. TEPL represents the tip-enhanced variant of conventional photoluminescence, while STML employs highly localized tunneling currents to induce electroluminescence. The spatial resolution of these techniques is essential for investigating light emission from heterostructure interfaces, superlattices, or single emitters of nanoscopic dimensions. An illustrative example is the TEPL study of exciton diffusion across $MoSe_2/WSe_2$ lateral junctions, where the observed unidirectional flow mimics thermal Kapitza resistance.[212] Another important application is the ability to discriminate between intrinsic exciton emission of pristine materials and that generated by defects or impurities, which requires spatial resolution well below the diffraction limit. Measurements conducted under ambient conditions have achieved resolutions of 10 nanometers, sufficient to resolve defects such as bubbles and wrinkles.[213] However, detecting emission from single atomic vacancies and impurities requires advancing optical nanoscopies to ultrahigh vacuum (UHV) and cryogenic temperatures. These controlled conditions, along with the development of picocavities that terminate in single-atom protrusions, have enabled the characterization of photonic phenomena at the atomic scale using STML, TEPL, and TERS.[214] While many of these advanced approaches have been applied to single molecules, only a few exemplary experiments investigating 2D materials with atomic resolution exist. These include atomic-scale variations in luminescence from a $MoS_2$/few-layer graphene heterostructure,[215] emission from single vacancies in $WS_2$,[180] and valley excitons in $WSe_2$.[216]

A significant advantage of optical excitation over electronic excitation is the ability to access ultrafast phenomena. Therefore, the use of radio frequency electric pulses limits time resolution to the nanoseconds in time-resolved (TR)-STML.[217] In contrast, the bottleneck for TR-TEPL is on the detector side. The integration of fast single-photon avalanche diodes (SPADs) in the experiments, combined with deconvolution methods, has achieved a resolution of 5 ps.[218] TR-STML and TR-TEPL also differ substantially in the primary type of driving mechanism, and consequently in the subsequent cascade of events and temporal scales that could be explored in a system. While TR-STML is based on charge carrier injection and inherently involves the intervals between the individual electron transfers, TR-TEPL can excite states without electron exchange and probe the exciton lifetimes directly. In the field of TERS, although time-resolved measurements have not been demonstrated yet, static TERS data have already been obtained with 500 fs pulsed lasers.[219] In parallel, THz-STM has evolved, bringing picometer-femtosecond spatio-temporal resolution into the research areas of ultrafast photoexcitation[220] and radiative decay phenomena.[221] This method relies on inducing transient electron tunnelling using sub-cycle THz pulses. It can also be expanded to mid-IR wavelengths that provide higher time resolution, as evidenced by measurements of sub-ps dynamics in $MoTe_2$ related to photocarrier-induced band renormalization.[222] Finally, a variant of this sub-cycle technique that uses two-pulse mixing has achieved sub-fs resolution and has enabled the imaging of quantum electronic coherence in molecular orbitals.[223]

In addition to studying transients, special types of time correlated single photon counting can be conducted to analyze the statistics of a photon source, as the Hanbury Brown and Twiss interferometry. By measuring and binning the time delays between many two-photon emission events, the second-order correlation function can be evaluated to determine the type of emitter—whether it is a single-, two-photon or a stochastic source depending on the observation of photon bunching or antibunching.[224] The technique can be further extended to study temporal order of different photon emissions in a quantum emitter with various charge and excitation states and to extract the transition rates involved in its many-body dynamics.[225] The state-of-the-art time resolution in these experiments is constrained by the jitter time of the SPAD detectors, typically achieving resolutions of only a few tens of picoseconds.

Despite proof-of-principle demonstrations of sub-picosecond electron dynamics in materials like $MoTe_2$[222] and the excitonic insulator candidate $Ta_2NiSe_5$,[226] time-resolved nanoscopies have yet to be systematically applied to study dynamics in 2D materials.

### 8.2 Challenges and Future Goals

One of the primary ongoing efforts is focused on studying light–matter interactions in atomically engineered materials at the ultimate spatiotemporal limits of picometers and attoseconds. Concurrently experiments are advancing in laboratories worldwide to characterize emission statistics, directionality, coherence, chirality, and the integration of these characteristics into spintronic and electronic devices. However, numerous challenges accompany these ambitious goals. In the following, we will address both the overarching challenges of tip-enhanced nanoscopies and the specific technical and fundamental issues related to characterizing photonic phenomena in 2D materials.

A primary objective is to expand the pool of quasiparticles that can be probed within the near-field of a nanocavity. Beyond plasmons, excitons, polarons, and phonons, other quasiparticles—such as magnons, Cooper pairs, spinons, skyrmions, and solitons—may also be manipulated using optomechanical or optoelectronic methods involving sharp probes and electromagnetic fields. Controlling the transduction between light and other state variables at the nanometer scale could open new avenues for innovative multivariable quantum technologies. Developing local nanocavity techniques for addressing new quasiparticles optically would enable studies in their genuine nanoscale environments and at their characteristic scales.

Quasiparticles such as the excitons, which are routinely studied by a range of tip-enhanced techniques, can reveal new phenomena when harnessed in engineered 2D superstructures. Imprinting metagate electrostatic potentials to manipulate the energy, lifetime, and localization of excitons can lead to excitonic superlattices that exhibit exotic many-body phenomena. Metagates can be achieved, for instance, by twisting layers, directly nanostructuring the 2D layer, or by interfacing with nanostructures. A recent study utilized local photocurrent detected by STM to map in-plane charge transfer excitons arranged by the moiré potential landscape in twisted $WS_2$.[227] Similar investigations could be performed at 1 nm-scale superlattices by using bottom-up nanostructures as atomically precise metagates. At this scale comparable to the characteristic size of excitons in 2D semiconductors, strong inter-exciton interactions are anticipated. Additionally, the rather unexplored charge transfer excitons in lateral

heterostructures—the 1D analogues of interlayer excitons—could be studied by taking advantage of the lateral resolution capabilities of the tip-enhanced nanoscopies.

Chirality effects can be investigated using circularly polarized light, as recently shown in the probing of valley excitons in a decoupled $WSe_2$ monolayer by STML.[216] Similar studies could be extended to TEPL and TERS to shed light on the origins of gyrotropic electronic order[228] or on chiral-selective magnon-phonon coupling.[229] These approaches would allow for the correlation of polarized emission with simultaneous maps of electronic and spin density of states, achievable through STM.

Furthermore, the numerous degrees of freedom offered by scanning probe methodologies will enable the control of exotic optoelectronic phenomena through local and external stimuli. Recent advancements have demonstrated the gating of emission from single quantum emitters *via* current injection;[230] similar atomic-scale switching can be achieved using local tip-induced electric or magnetic fields. The generation of single photons on demand and the emission of heralded entangled photons of various wavelengths is now at hand, along with superradiant sources that exceed the quantum efficiencies of their individual constituents.[231] These can be engineered by using the atomically precise, deterministic capability of STM to manipulate matter.

The broad range of the electromagnetic spectrum offers a wealth of physical processes that can be characterized and potentially controlled. In the coming years, we envision extending TERS and TEPL to the NIR and UV regimes, a yet unexplored territory in most tip-enhanced setups. This extension will facilitate the study of 2D materials with narrow bandgaps, such as black phosphorus, $PdSe_2$, and $PtSe_2$, to explore their applications in telecommunications and theragnostics, as well as enabling investigations of UV photochemistry at the nanoscale. Additionally, it will allow for resonant TERS and TEPL studies across a broad range of wavelengths.

Accessing the near-field induced fast excitonic decays, electronic transients, or photoexcitation dynamics requires achieving time resolution in the sub-picosecond regime. Although significant advancements have been made in time-resolved nanoscopic methodologies, their application to the study of ultrafast dynamics in 2D materials is still in its infancy. The same holds true for time correlation studies of quantum emitters, such as point defects in 2D materials. In this case, expanding the studies to include multiphoton events and their correlations will provide valuable insights into the nature of the emitters, energy transfer between color centers, and cascade emission phenomena.

### 8.3 Suggested Directions to Meet These Goals

In this section, we will outline our perspective on potential directions to address some of the key challenges discussed above. These suggestions will encompass both materials and device fabrication, as well as anticipated advancements in the techniques themselves.

Fundamental studies of the coupling between photons and other quasiparticles will greatly benefit from correlative measurements. Although scanning probe microscopies can provide simultaneous electron, force, and optical spectroscopy, complementary information is often necessary. Correlative studies using techniques such as angle-resolved photoemission spectroscopy (ARPES), X-ray absorption spectroscopy (XAS), and its magnetic circular dichroism (XMCD) variant can yield a comprehensive, atomistic picture of the system under investigation. Achieving this requires the development of strategies for the colocalization of micrometer scale regions and the transfer of samples between different instruments under UHV conditions. In this regard, initiatives aimed at creating multi-technique platforms for correlative characterization, such as the InCAEM facility at ALBA, are beginning to emerge.[232]

The application of sub-nanometer optical nanoscopies in 2D materials is often limited by sample quality. Atomic scale investigations of surfaces always require stringent conditions in terms of sample preparation and measuring conditions, typically involving UHV and cryogenic temperatures, and in most cases the use of STM as local probe. The need of conductive samples for this technique clashes with the need of decoupled systems to prevent the quenching of radiative processes and preserve their intrinsic properties, thus adding further challenges to sample preparation. While vdW epitaxy on inert, low-conductivity substrates such as graphene on SiC allow for *in situ* growth and characterization of monolayers in UHV for some 2D materials,[180] more versatile approaches for twisted layers and heterostructures, as well as the implementation of gate and source/drain electrodes, primarily rely on top-down approaches transferring flakes onto insulating layers. In thick insulating buffers, as the ones needed for gating, metallic electrodes need to be contacted to the 2D material to facilitate electronic transport. The main challenges for conducting atomic scale nanoscopy in these systems include the stringent cleanliness requirements that far exceed those for macroscopic measurements, the precise positioning of the tip on the flake with micrometer accuracy, and the fabrication of efficient tunneling electron paths. Cleanliness can be enhanced by developing glovebox and UHV sample transfer protocols. For accurate and reproducible navigation, high magnification capabilities provided by the embedded lens systems of optical nanoscopes can be utilized, along with strategic sample marking. Appropriate metallic contacts can be achieved using patterned multilayer graphene. Several research groups are beginning to overcome these barriers,[215,216,227] and we anticipate that their pioneering results will inspire the community.

Engineering nano- and picocavities is a challenging endeavor that has proven essential for advancing our fundamental understanding of quantum states confined in nanocavities. Plasmonic nano- and pico-cavities formed at the STM junction offer a controlled and versatile platform for this purpose, thanks to their intrinsic atomic-scale precision in tip positioning and tip apex geometry. Pioneering studies have achieved several significant milestones, including the realization of directional emission through atomic-scale cavity engineering,[233] the control of plasmonic modes by shaping Fabry–Pérot resonators at the tip,[234] and the quantification of coupling strength and Purcell factors derived from localized Fano resonances.[235] Additionally, the contributions of near-field effects[218] and tunneling current[236] on exciton lifetimes can also be simultaneously assessed. Tip engineering can also be used as a strategy to eliminate far-field contributions, for instance by employing nanowire waveguides attached to the tip that allows for the separation of excitation and detection positions.[237] The near-field enhancement can alternatively be tailored, by modulating the amplitude and phase of incoming photons.[238] Progress in the design and fabrication of tailored SPM-based plasmonic cavities will open avenues for manipulating light at subwavelength scales. Tip reproducibility being a challenge even for commercial tips, developing protocol standards[239] will be crucial for advancing in this direction.

The extension of TERS and TEPL to the NIR and UV regimes also represents a challenge from the tip engineering side. The necessity for plasmonic tips that operate in the UV and deep-UV ranges requires the exploration of materials beyond the commonly used Au and Ag. Mg, Al, Ga, or Rh offer large field-enhancement factors and localized surface plasmon resonances in the UV regime. Notably, Rh stands out for its combination of a high field-enhancement factor and a low oxidation rate, and tips fabricated from this material have already been successfully developed.[240]

In terms of time resolution, reaching picosecond resolution is already at hand through several methods. While these techniques can select specific electronic excitations and investigate their relaxation dynamics, they provide limited spectroscopic information. Incorporating picosecond-gated spectroscopy would enhance ultrafast techniques by adding the ability to reveal sequenced, correlated phenomena, such as energy transfer in coupled chromophores and color centers, or cascade emission at atomic resolution. This could be realized by implementing streak camera detectors. Furthermore, advancing beyond second-order correlation in photon statistics would yield valuable insights into multiphoton quantum states characterized by superposition, entanglement, or nonclassical correlations. This could be accomplished by measuring the photon-number distribution using a true photon-number-resolving detector, such as a streak camera, or through manifold coincidence counting, in the spirit of the Hanbury Brown and Twiss experiment.

In conclusion, the integration of tip-enhanced nanoscopies into the field of photonics for 2D materials has already proven crucial for addressing significant challenges, yet it remains in its infancy, with highly promising prospects ahead. Key milestones, such as the exploration of new photon-quasiparticle interactions, operando studies of vdW optoelectronic devices, or the implementation of picometer-attosecond resolution techniques, are anticipated in the coming years. The application of these techniques—particularly in moiré superlattices, single-atom emitters, lateral heterostructures, and other vdW engineered materials—will provide the atomistic insights necessary for an in-depth understanding of the interactions of light and quasiparticles in these complex energy landscapes.

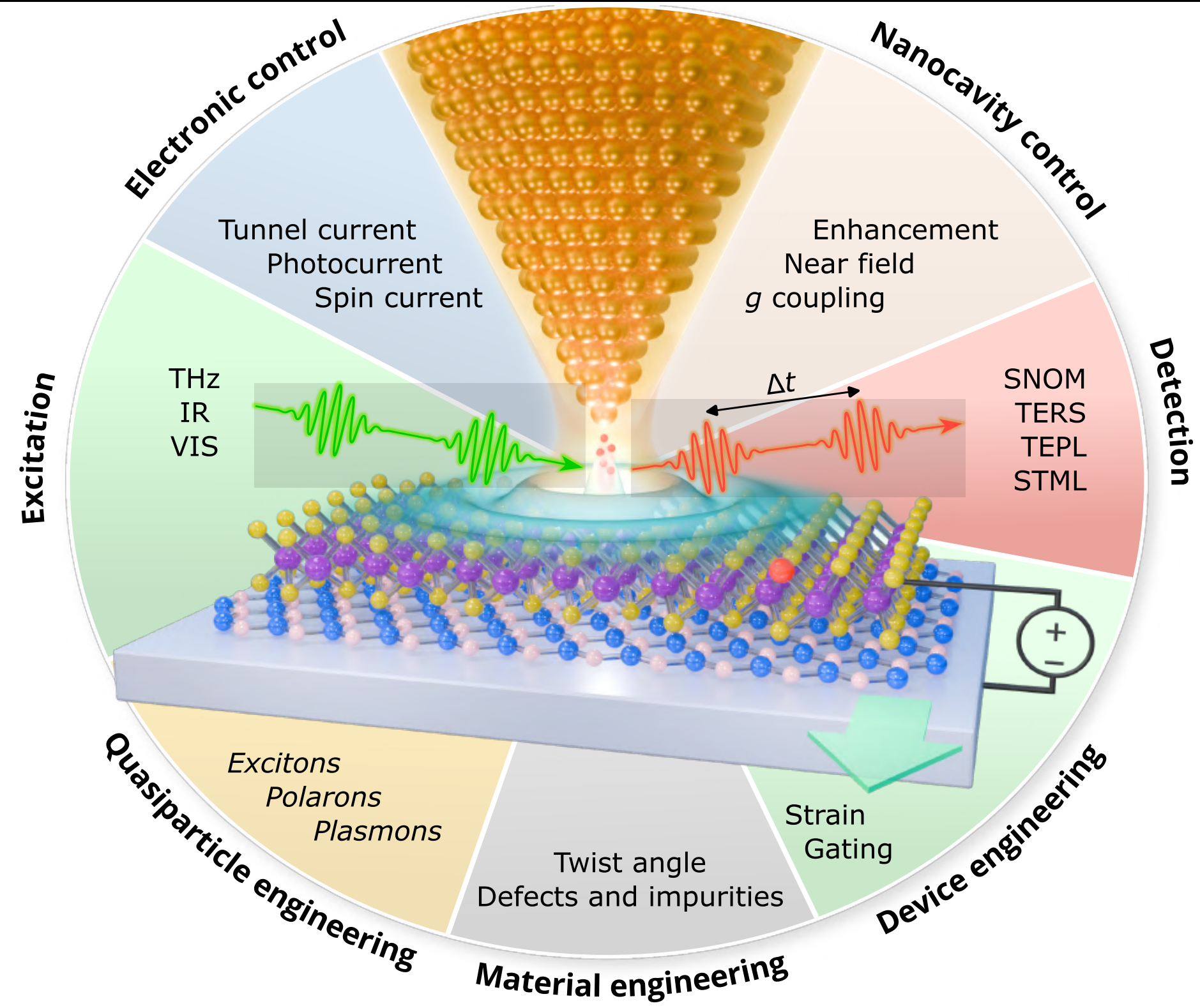

**Figure 8**. Schematic illustration of tip-enhanced nanoscopies applied to the study of photonics in 2D materials. The tip–sample nanocavity can confine electromagnetic fields and couple near- to far-field photons, and vice versa, enabling local probing of optical and electronic excitations with atomic-scale spatial resolution. This can be done by using photons or tunneling electrons for excitation and detection. The intrinsic surface sensitivity and sub-nanometer spatial resolution of tip-enhanced techniques make them especially suited for studying and controlling light–matter interaction phenomena in 2D systems. The spatial resolution is particularly useful for the study of systems exhibiting nanometer-scale lateral modulations or inhomogeneities such as moiré superlattices, nanostructures, lateral heterostructures, or defects. The application of external stimuli such as strain, electrical gating, and magnetic fields can be incorporated for the manipulation of the local properties and correlating it with mesoscopic properties *via in operando* studies. Incorporating fast detection or pump-probe schemes adds time resolution to study quasiparticle dynamics with the same spatial resolution. All these elements converge in a toolbox of techniques that enable deterministic characterization and manipulation of photonic phenomena at the ultimate spatiotemporal limits of picometers and attoseconds.

## 9. ELECTRON MICROSCOPY AND SPECTROSCOPY FOR 2D MATERIALS

**Luiz H. G. Tizei**[1,*]
[1]Université Paris-Saclay, CNRS, Laboratoire de Physique des Solides, 91405 Orsay, France
***Corresponding author.** Email: luiz.galvao-tizei@universite-paris-saclay.fr

### 9.1 Introduction

Electron microscopy has become an indispensable tool for investigating the properties of 2D materials at the atomic scale. The structure, morphology, chemistry, vibrational response, and optics of these materials have been extensively studied down to the atomic level using electron microscopy (EM) and spectroscopy. Since the initial production of 2D materials, EM contributed to different aspects of their understanding. Imaging and diffraction can map deformation in 2D monolayers;[241] electron energy-loss spectroscopy[242] (EELS) (Figure 9a) and energy-dispersive X-ray (EDS) spectroscopy[243] allow for the chemical identification of individual atoms in 2D layers. Vibrational EELS can map the vibrational response of a single Si atom on a graphene monolayer.[244] EELS and cathodoluminescence (CL) have been shown to measure spectra similar to those of optical absorption and photoluminescence in hBN-encapsulated $WS_2$[245,246] (Figure 9b). The objective of this section is not to be an extensive review of electron microscopy and spectroscopy for 2D materials, but to give a perspective of which technical developments could help solve standing questions for these materials.

### 9.2 Structure, Chemistry, and Electronic Structure

EM has been crucial in understanding 2D materials and their vdW heterostructures. For example, diffraction and imaging allowed the measurement of the roughness of 2D layers.[247,248] These techniques have also been used to study the influence of vdW stacking ambient conditions on interlayer spacing.[249] Furthermore, they have enabled the mapping of strain distribution with sub-nanometer precision for large fields of view.[241] The chemistry and electronic structure of defects in graphene and hBN have been probed by spectroscopy in EM.[250,251]

Advanced imaging techniques such as ptycography, coupled with structural and imaging simulation, might be a way to measure the interlayer spacing in 2D material heterobilayers and more complex vdW heterostructures, including when moiré patterns are formed. This approach could help determine how interlayer spacing affects the electronic band structure. In this respect, ptychography has already been used to measure the interlayer distance in bi- and tri-layer graphene.[252]

A standing problem in 2D materials is the identification of atomic structures that can operate as single-photon emitters (SPEs). Specifically, in TMDs, SPEs are linked to the presence of defects and strain, which funnel excitons toward the defects, leading to localized photon emission.[253]

Coupling EM with other methods, such as photoluminescence, can be quite fruitful. Furthermore, the improvement of detectors and microscopes to optimize 4D-STEM and ptychography has significantly increased spatial resolution and data throughput, opening new avenues for correlative microscopy.

### 9.3 Nano-Optics

CL and EELS have been extensively used to study the optical properties of materials at the nanoscale, including 2D materials.[254,245,246] These techniques probe excitonic transitions in TMD monolayers. Probing these transitions with EELS at nanometer spatial resolution only became possible with the development of modern electron monochromators, which allow for spectral resolution in the 5-30 meV range.[255] For CL, the spatial resolution is limited by the diffusion and drift of excitations, while for EELS, it is limited by the delocalization of inelastic scattering. Both types of spectroscopies have been used to study excitonic transitions in 2D materials, revealing nanoscale phenomena. For example, the modification of the charged-to-neutral exciton emission intensity down to tens of nanometers has been reported in hBN-encapsulated $WS_2$. A SPE in hBN at 4.1 eV was detected using a Hanbury Brown and Twiss interferometer coupled to a CL system.[256] The spatial modulation of $MoS_2$ neutral excitons due to the moiré structure formed in a $MoS_2/WSe_2$ bilayer was probed by EELS.[257]

Finally, these spectroscopies can be implemented with temporal resolution, especially CL (time-resolved CL, TR-CL). This allows for the detection of decay traces from the emitter in 2D materials with sub-nanosecond temporal resolution.[258]

The spectroscopic signatures in EELS and CL, such as changes in plasmon energy or phonon population, can be used to measure the temperature of materials at high spatial resolution.[259] This capability has also been applied to 2D materials.[260] Recently, a synchronized laser injection-EELS experiment with an electron event-based detector has demonstrated the possibility of measuring the spectroscopic signature of thin materials, including 2D layers, as a function of temperature with nanosecond and nanometer resolutions.[261]

These experiments are mostly well-established (except for the time-resolved ones) in the EM community and currently present in many laboratories and universities around the world. A promising next step is to perform these experiments under different stimuli: electrical biasing, controlled magnetic field, variable temperature, and laser irradiation, among others. Specifically, for 2D materials, *in situ* electron spectroscopies might be extremely valuable, as they would allow for experimental approaches similar to those used in more standard optical ones. For example, doping and internal electric field control using electrostatic potentials in a variable temperature setup is a way to identify neutral and charged excitonic states.

### 9.4 New Electron Spectroscopies

New electron microscopy and spectroscopy techniques are in constant development, aiming to expand the range of measurable physical quantities and enable new types of experiments that reveal deeper insights into materials. Here, we shall describe three recent ones that might have a strong impact in the study of 2D materials. First, electron–photon (EELS-CL) coincidence experiments have recently been revisited, based on new event-based electron detectors (Timepix3).[262,263] The resulting technique is a nanometric analog of photoluminescence excitation (PLE) spectroscopy. It allows one to measure the relative quantum efficiency of light-emitting materials in a large spectral range (the same available for EELS). This technique, tentatively called CL excitation spectroscopy (CLE), can identify the excitation pathways that lead to light emission following an electron inelastic scattering event. It can also measure temporal decay profiles of the emitter similarly to TR-CL. For 2D materials, it might be the best way to understand the detailed mechanisms of how CL is generated, specifically in vdW heterostructures with TMDs. Specifically, it might be a way to pinpoint why energy shifts have been reported between PL and CL of the same devices.[254]

Another electron spectroscopy method that has been developed over the last 20 years is photon-induced near-field electron microscopy (PINEM). This technique involves coupling a laser field to a fast electron mediated by the near field of a nanostructure.[264] With PINEM, the structural dynamics of thin materials can be followed down to the picosecond temporal scale.[265] Also, electron-photon coupling with chirped optical pulses mediated by a thin $MoS_2$ film has been reported.[266]

Related to PINEM is another spectroscopic method called electron energy-gain spectroscopy[267] (EEGS). EEGS can achieve spectral resolutions much better than the energy linewidth of the electron beam, limited only by the laser linewidth used. For 2D materials, an evident application of EEGS is the measurement of absorption linewidths in materials such as TMD monolayers. These materials have intrinsic linewidths much smaller than those currently achievable in EELS measurements (above 2.5 meV in the best cases, and typically 5 meV in state-of-the-art microscopes).

Finally, a recently theoretically proposed improvement of EELS will allow the measurement of signatures with degrees of freedom analogous to light polarization: phase-shaped EELS (PEELS)[268] In this method, an electron beam with a specific phase distribution is scattered by a material. A measurement of the scattered beam under certain conditions allows for the identification and distinction of modes that would usually be missed in standard EELS.[268] Modes beyond the dipolar approximation, inaccessible to optical experiments, can also be detected.[269] Applications for TMD materials are evident, where polarization degrees of freedom play a key role in spin-valley states. PEELS could therefore provide unique insights into the nanoscale optical and electronic properties of these materials.

## 9.5 Conclusions

The perspectives for electron spectroscopies to study 2D materials are very promising, especially given the improved spectral resolution approaching those achievable in optical experiments. Moreover, some key recent advances might be very beneficial for experiments with 2D materials. These include improved vacuum quality in the sample chamber, reaching a truly clean $10^{-10}$ torr range, and hBN encapsulation. New scanning methods and detection approaches to minimize the required dose will help reduce electron-beam-related sample damage.[270,271,272,273] Finally, greater automation and standardization of microscopes and related equipment (sample holders, apertures, *in situ* capabilities) are needed. This is crucial for reproducibility, increased throughput of data acquisition, and easier design and development of new experiments. Ultimately, such advancements could democratize these powerful experimental methods, making them accessible to the entire scientific community, beyond a small group of specialists.

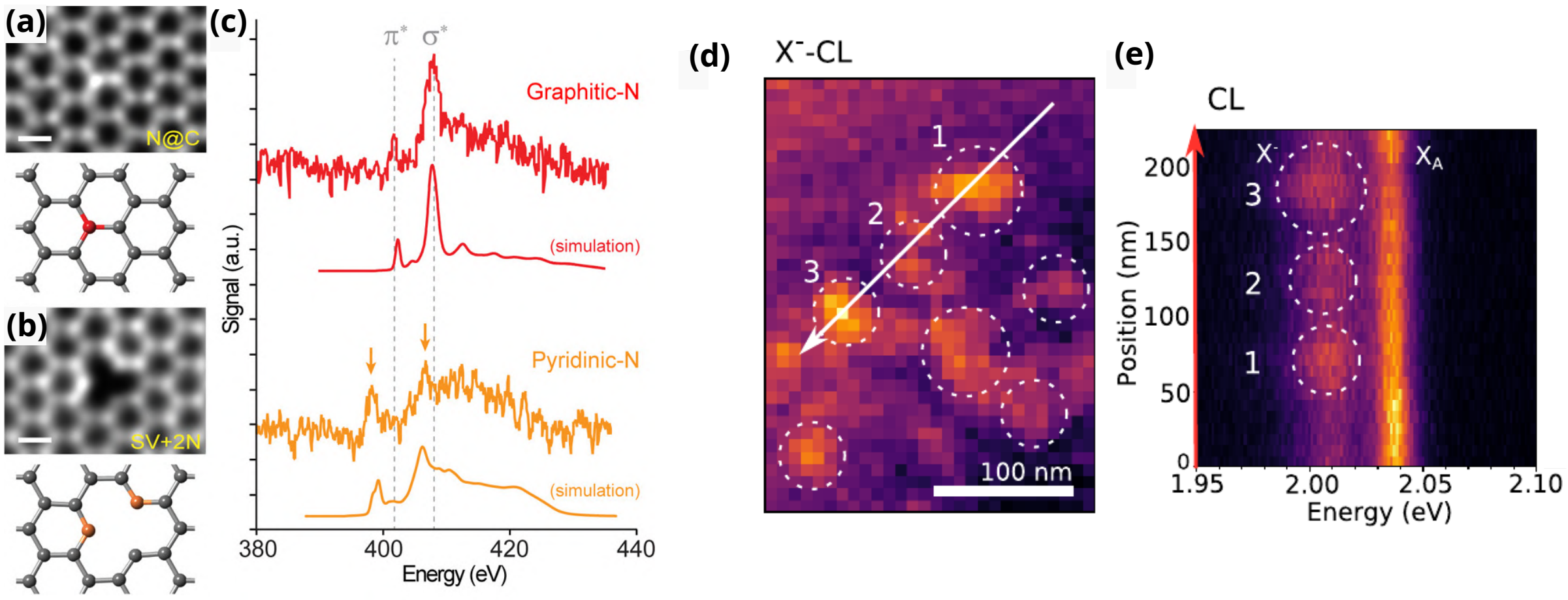

**Figure 9.** Chemistry, electronic structure, and optics of 2D materials with fast electron spectroscopies. (a-b) Electron-microscopy images and model for nitrogen atoms in graphene in different atomic configurations. (c) Electron energy-loss spectroscopy (EELS) core-loss spectra and simulation for the nitrogen K edge of atoms in the graphitic and pyridinic configurations. (d) $WS_2$ trion ($X^-$) cathodoluminescence (CL) intensity map. High-intensity spots are numbered and marked by dashed circles to be compared with (e). (e) Spectral profile along the arrow in (d), showing how the trion-to-neutral exciton CL emission evolves on the local maxima. Panels (a-c) are adapted with permission from ref 241 (Copyright 2018 American Association for the Advancement of Science); Panels (d-e) are adapted with permission from ref 244 (Copyright 2020 American Association for the Advancement of Science).

**Excitons and 2D Semiconductors**

# 10. EXCITON POLARITONS IN 2D MATERIALS

**Florian Dirnberger,[1] Hui Deng,[2] Christian Schneider,[3] and Vinod Menon[4,5,*]**
[1]TUM School of Natural Sciences, Physics Department, Technical University of Munich, 85748 Garching, Germany
[2]Physics Department, University of Michigan, 450 Church Street, Ann Arbor, Michigan 48109, USA
[3]Institut für Physik, Fakultät V, Carl von Ossietzky Universität Oldenburg, 26129 Oldenburg, Germany
[4]Department of Physics, City College of New York, 160 Convent Ave., New York, NY 100031, USA
[5]Department of Physics, The Graduate Center, City University of New York, New York, NY 10016 USA
***Corresponding author.** Email: vmenon@ccny.cuny.edu

## 10.1 Current State of the Art

### 10.1.1 *Introduction*

Nearly seventy years after the hybrid light–matter quasiparticles called exciton polaritons (EPs) were postulated, and decades into their research in 3D semiconductors and quantum-well cavities, strong exciton–photon coupling was demonstrated in a monolayer 2D semiconductor.[274] Since then, research on strong light–matter coupling in 2D materials has been thriving.[275] The large exciton oscillator strength in these materials eases formerly more stringent requirements for high cavity quality factors while binding energies reaching hundreds of millielectronvolts ensure that excitons remain stable excitations at elevated temperatures. With 2D materials, strong light–matter coupling has become a staple of solid-state optics now routinely achieved at room temperature. Initial experiments with planar microcavity structures were soon followed by demonstrations of strong coupling using external mirror cavities[276] and cavities made of plasmonic or dielectric lattice structures.[277,278,279,280] Optical excitations occasionally referred to as self-hybridized exciton polaritons,[281] which are formed by excitons strongly coupling to the photons confined inside bare bulk crystals, were also observed. Figure 10 shows a schematic illustration of the different material types, resonator structures, and intriguing phenomena that have been observed in the context of EPs in 2D materials.

### 10.1.2 *Materials*

A common experimental approach to obtain strong light–matter interaction is to incorporate a monolayer semiconductor from the family of TMDs into an optical cavity. Such monolayers are typically encapsulated between hBN to avoid detrimental effects produced by the substrate. Besides the advantage of realizing Rabi splitting energies of several tens of millielectronvolts, these materials host exciton complexes such as trions, with reasonable binding energy and oscillator strength, which can also be coupled to cavity photons.[282] In the limit of high doping, the trion state transitions to the Fermi polaron picture[283] where novel effects have been predicted.[284]

More recently, EP studies employ twisted hetero- or homo-bilayers of TMD layers to modulate the potential landscape into a superlattice that can induce collective excitonic behavior. When coupled to a cavity, the resulting moiré exciton polaritons combine the large oscillator strength of the direct excitons in TMDs with a strong nonlinearity resulting from the quantum-dot nature of each moiré cell.[285] Such a system circumvents the tradeoff between exciton binding energy and exciton nonlinearity, thereby providing a pathway to strongly nonlinear and robust polaritons. Additionally, pristine bilayer TMDs are of high interest for polaritonics. $MoTe_2$ bilayers feature a direct bandgap with high-oscillator-strength excitons in the near IR,[286] and interlayer excitons with a permanent dipole moment emerge in bilayer $MoS_2$[287,288] facilitating dipolar interactions, which can enhance the polariton nonlinearity.[289]

A relatively new direction for EP research is currently emerging from the discovery of excitons in vdW materials with highly correlated degrees of freedom. An archetypal platform for polariton physics from this class of materials is the antiferromagnetic correlated insulator *$NiPS_3$*. Its excitons[290] enabled the first experimental demonstration of strong exciton–photon coupling in a magnetic vdW material.[291] The study found that many properties of EPs are intimately linked to magnetic order. Prominent magneto-optic effects also characterize the layered magnetic semiconductor CrSBr,[292] a material recently attracting considerable attention. Its excitons couple directly to magnetic order,[293] while supporting self-hybridized polaritons[294] and pronounced interactions with magnons.[295] Besides magnetic materials, ferroelectric vdW crystals, such as Sn*,* further enrich the palette of exciton polaritons in ferroic materials,[296] with few-layer TMD crystals of rhombohedral stacking remaining a promising candidate to functionalize EPs with ferroelectric properties.[297] Lastly, the recent emergence of large-bandgap monolayer topological insulators with strongly pronounced exciton resonances, such as Bismuthene, could compose an interesting platform to explore the physics of quantum spin Hall polaritons.[298]

### 10.1.3 *Phenomena*

As light–matter hybrids, EPs obtain properties from both excitons and photons. This is why, from a photonics perspective, they often exhibit exceptionally large nonlinearities which are primarily inherited from the excitons' exchange coupling and a nonlinear saturation of the oscillator strength. Such properties are highly sought after in the quest for reaching the important quantum blockade regime.[299] Polaritons offer pronounced nonlinearity from the excitonic–, and long coherence from the photonic component. Efforts to enhance nonlinearity for reaching the blockade regime in the weak pump limit have included the use of Rydberg states of excitons, moiré EPs, trion polaritons and dipolar EPs.[285,289,300,301] Nonlinearities also play a key role in the context of polariton Bose–Einstein condensation, an important platform for reservoir computing and potentially low-threshold lasing.[302] Polariton condensation and the emergence of coherence was recently reported in TMD based EPs.[303,304,305] From their excitonic component, EPs also adopt symmetries of the underlying crystal lattice. Their emission in TMD monolayers, for instance, shows intriguing polarization properties that reflect the locking of exciton spin and valley degrees of freedom.[306] As a direct consequence, the emergent valley-momentum coupling in optical cavities has the effect that photons with different helicity propagate in opposite directions.[307] In addition, it was recently demonstrated that the transport of EPs in TMDs can be further manipulated *via* the topological order of the optical cavity structure.[308] Owing to the large binding energy of excitons in TMDs, they also host excitonic complexes, including trions and Fermi polarons in the limit of small and large doping, respectively. Polariton states realized using these charged excitonic systems are especially interesting from the standpoint of nonlinear interaction owing to Coulomb interaction.[283]

Interesting EP properties also result from the magneto-optic effects of 2D magnetic materials. By inducing linear or circular dichroism, magnetic order can impart optical polarization selection rules onto EPs. Magnetic order can also be a powerful control knob to alter the electronic band structure. In CrSBr, the changes in the electronic structure that occur when the magnetic layer configuration is forced from an antiparallel into a parallel configuration can shift excitonic resonances by tens of millielectronvolts[293], inducing considerable magnetochromism. When CrSBr excitons are strongly coupled with cavity modes exhibiting a different polarization dichroism, the resulting polariton modes have polarization properties that are distinct from either that of the exciton or the photon but are controllable *via* both and in turn by a magnetic field.[309]

## 10.2 Challenges and Future Goals

### 10.2.1 *Crystal Size*

Exciton polaritons in 2D materials could bring many benefits to optics applications. Technologically, however, these benefits are currently still outweighed by the limited size and homogeneity of exfoliated 2D crystals, which typically do not exceed a few tens of micrometers. For most photonic devices, this size needs to reach millimeter and centimeter scales. Continuing efforts directed at achieving wafer-scale growth and passivation of TMD monolayers, as well as metal tape-assisted exfoliation and passivation techniques[310] already report significant progress. Thus, while the primary challenge toward technological applications with EPs in 2D materials remains to increase lateral crystal sizes, recent advances are promising.

### 10.2.2 *Nonlinearity*

A major goal for strong light–matter coupling with 2D materials is realizing polariton blockade, potentially enabling single photon nonlinearity and switching. Although large nonlinear responses have been observed in TMD based EPs, the nonlinear interaction strengths measured are well within the classical limit. Currently, approaches to enhance this further *via* cavity engineering, reducing excitonic linewidths or field effects are being pursued.

### 10.2.3 *Coupled Degrees of Freedom*

The primary challenge with EPs in correlated materials is to obtain a better understanding of the microscopic interactions involved in the coupling of excitons, photons, phonons, and magnons. Of particular interest are polariton Bose–Einstein condensates in antiferromagnetic crystals, since magnons in these materials often exhibit very high frequencies. These condensates could be interesting for (quantum) transduction. Interactions between magnons with ~THz frequency and exciton-polariton condensates might be utilized to modulate the condensate's coherent emission on ultrafast timescales. Despite much smaller magnon frequencies, the antiferromagnetic semiconductor CrSBr is a highly promising candidate to demonstrate this effect. In addition, ferroelectric materials could provide new degrees of electrical control over polariton laser emission.

### 10.2.4 *Coupling Electron & Photon Transport*

Transports of photons and charges are usually two independent processes studied in different types of experiments. EPs induced by trion/Fermi polaron resonances may however provide means of efficiently connecting them. Similar phenomena are expected for EPs and polaron polaritons,[284] in either a thermal or condensed state, where coupling with photons generally helps overcome local disorders for both ballistic and coherent transport, with the perspective to reach a polariton-mediated regime of superconductivity.[311]

## 10.3 Suggested Directions to Meet These Goals

Utilizing strong light–matter interactions in 2D materials offers compelling advantages over conventional optics in many regards. Realizing their potential is directly linked to achieving a consistent fabrication of high-quality 2D materials with technologically meaningful lateral extensions. A promising path toward this goal may lie with recent efforts in printing 2D materials.[312]

A direction for fundamental research is to continue exploring new materials with excitonic resonances suitable for achieving strong light–matter coupling. Especially tapping into the vast pot of materials with strongly correlated degrees of freedom should provide insight into interactions between different quasiparticle excitations and new functionalities for optics. Of interest in this regard are optically anisotropic materials, and multiferroics promising a large degree of control over optical properties by external voltages and magnetic fields. Ongoing efforts in increasing nonlinearities could also benefit substantially from exploring new materials and quasiparticle interactions. From the cavity engineering point of view, a promising direction currently pursued is using chiral cavities for breaking symmetries and allowing new functionalities of optical devices.

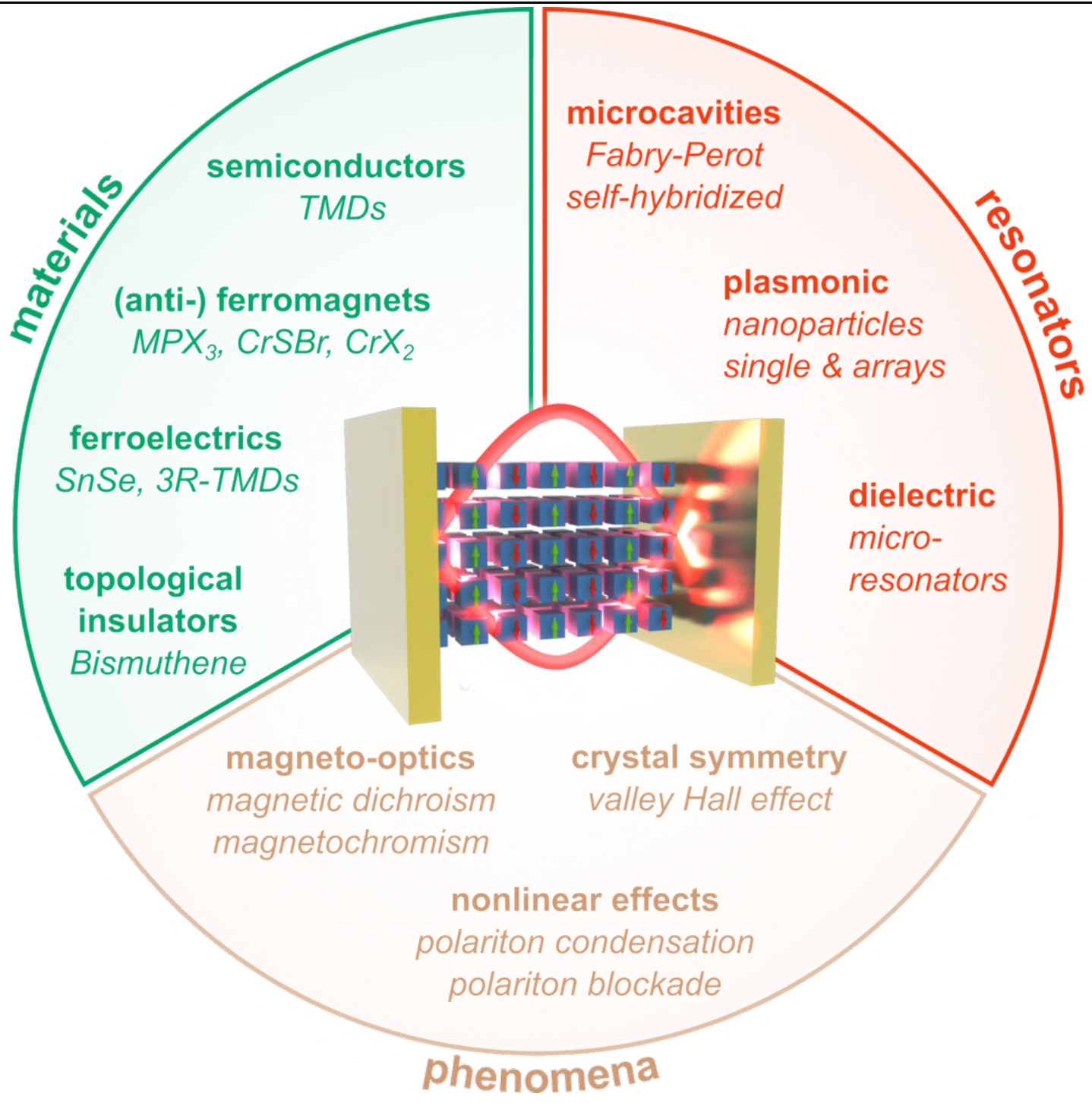

**Figure 10.** Study of strong exciton–photon coupling in 2D vdW crystals combines different classes of materials with optical resonators to investigate intriguing phenomena related to optics and photonics.

## 11. TUNABLE EXCITONS AND TRIONS IN TMD STRUCTURES

**Thorsten Deilmann,[1,*] Alexey Chernikov,[2] and Kristian S. Thygesen[3]**
[1]Institut für Festkörpertheorie, Universität Münster, 48149 Münster, Germany
[2]Institute of Applied Physics and Würzburg-Dresden Cluster of Excellence ct.qmat, TUD Dresden University of Technology, 01062 Dresden, Germany
[3]Computational Atomic-Scale Materials Design (CAMD), Department of Physics, Technical University of Denmark, 2800 Lyngby, Denmark
***Corresponding author.** Email: thorsten.deilmann@uni-muenster.de

### 11.1. Current State of the Art

Mono- and few-layer structures of semiconducting TMDs host strongly bound excitons with binding energies of several hundred meV.[313] These binding energies are much larger than those of their bulk counterparts due to quantum confinement and reduced dielectric screening leading to strong electron–hole Coulomb attraction. In the presence of free charge carriers the neutral excitons may capture a free electron or hole and form charged excitons called trions (see Figure 11a).[314,315] An alternative description increasingly used at higher doping densities is that of the Fermi polaron corresponding to the exciton being dressed by the excitations of the Fermi sea.[283,316] Typical binding energies of the extra charge carrier to the exciton in monolayer TMDs are a few tens meV making them observable even at room temperature for sufficiently high doping. Experimentally, excitons, trions, and Fermi polarons can be probed by optical absorption and emission-type measurements as well as a variety of nonlinear and time-resolved techniques.[313] Theoretical calculations of excitons can be based on *ab initio* many-body methods like GW and Bethe-Salpeter equation (BSE),[317] analytical effective Hamiltonian approaches,[315] or employing the Bloch equations methods.[318] Altogether, the stability of the excitonic complexes combined with design flexibility of TMD structures and their tunability renders them excellent platforms for exploration and manipulation of fundamental exciton physics as well as for optoelectronic and excitonic devices operating up to room temperature.

Historically, the interest in excitons in TMD materials was motivated by the demonstration of bright, luminescent excitons in monolayer $MoS_2$ in 2010.[319,320] Theoretical and experimental studies of the exciton properties quickly followed (see Wang *et al.*[313] for an overview) demonstrating both large binding energies and deviations from the hydrogen-like excited states series in contrast to more traditional quantum well systems. The non-Hydrogenic exciton series originates from the nonlocal nature of the dielectric function of a 2D material.[321,322] In most experimental setups the TMD structures are either placed on insulating substrates or encapsulated in other 2D materials (typically hBN). The chemically inert nature of the TMDs limits the formation of chemical bonds with the surrounding materials. However, the long range Coulomb interaction inside the TMD can be significantly affected by the dielectric environment, which in turn can renormalize the electronic bandgap and exciton binding energies.[317] In fact, by varying the dielectric constant of the substrate, it is possible to tune not only the size of the exciton binding energy,[323] but also the qualitative form of the exciton series.[324]

The lowest bright excitons in TMDs consist of transitions close to the degenerate K and K' points of the hexagonal Brillouin zone. Due to the spin–orbit coupling, degenerate excitons at K and K' have different total angular momenta. As a result, the inequivalent K and K' valleys can be selectively excited using circularly polarized light—an effect that motivated propositions for valleytronic applications using valley index as a carrier of quantum information.[325] The bright excitons have vanishing angular momentum ($l$=0), finite g-factors and diamagnetic shifts increasing with the principal quantum number of excited states. This leads to a fan-like splitting in a magnetic field, see Figure 11e.[326] Of particular importance, especially for W-based TMD monolayers, are spin- and momentum-dark states involving K/K' transitions or those to the $\Gamma$ or $Q(\Lambda)$ points that can be lower in energy compared to bright ones (see Figure 11d). The ordering and precise splitting of the dark and bright excitons is determined by a competition of exchange interaction and spin–orbit coupling.[327,328] These excitons can be probed by light *via* phonon-assisted processes and play an important role for exciton dynamics (see also Figure 11c). In close analogy to bright excitons the dark states can also bind additional charges and form trion states. Moreover, both the energies and the nature of the excitonic wave functions can be strongly affected by externally tunable strain potentials opening up interesting opportunities for basic science and applications, see Figure 11d.[19]

If different layers are stacked, the excitons are no longer confined to the same layer and interlayer excitons may form. A general TMD heterostructure thus typically hosts both intra- and interlayer states. The latter are challenging to probe in reflection or transmission because of the small spatial overlap of the electron and hole, albeit possible in selected cases.[329,330] When intralayer and interlayer excitons are close in energy they may even hybridize to form a mixed interlayer exciton with both intra- and interlayer character (Figure 11b and f). Such states combine a significant oscillator strength with an out-of-plane dipole moment allowing them to be tuned by a vertical electric field.[287,331] In analogy to the intralayer excitons, negative or positive interlayer trions may form depending on the doping level.[332,333] Moreover, in structures containing more than two layers, dipolar interlayer excitons may couple and form quadrupolar excitons in symmetric trilayers.[334] In general, in-plane lattice mismatch and finite rotation angles will lead to moiré or atomically reconstructed structures upon stacking of TMDs depending on the particular combinations[335] including trion complexes.[336] In this case, the effects discussed above remain largely valid, yet, crucially, are also subject to the spatial modulation with an substantial impact of their band structure including predictions of topological edge states.[337] Particularly interesting are the practical consequence of the external screening mentioned above. This allows the use of the excitons as dielectric sensor for a rich variety of *electronic* many-particle correlations including Wigner crystals,[338] Mott insulating,[339] or even superconducting states[340] among many others.

### 11.2. Challenges, Future Goals, and Suggested Directions

While the physics of excitons and trions in TMDs are generally well explored, intriguing avenues of research remain for future studies. Below we briefly mention a selection of outstanding challenges and opportunities for both theory and experiment. From a theoretical perspective we have identified three main challenges in the description and understanding of excitonic properties of 2D semiconductors:

### 11.2.1 *Complex Structures*

These include vertical stacks with two or more lattice mismatched or rotationally misaligned layers such as moiré or reconstructed superlattices. A material-realistic representation of such structures often requires simulation cells with several thousand atoms, which is challenging even with standard density functional theory (DFT). Moreover, a quantitatively accurate description of band structures and excitons requires beyond-DFT methods that are currently only feasible for systems with less than 100 atoms. Similar challenges face the description of crystal defects (which can couple to and/or trap excitons) and local strain fields (e.g., due to interactions with substrates). One solution to these challenges may be constituted by emerging machine learning interatomic potentials. However, the potentials are currently limited to the prediction of total energies and structures. The prediction of electronic band energies using machine learning seems significantly more difficult and currently does not represent a viable solution.

### 11.2.2 *High-Density Regime*

Most theory and calculations of excitons in TMDs assume a comparatively low density of excitons or trions. However, many phenomena of interest, such as exciton condensation, superfluidity, and Mott transition occur in the high-density regime where exciton-exciton interactions become of crucial importance or the excitons even dissociate into dense correlated carrier gases/liquids. A similarly interesting and unexplored area is that of high doping where the high density of free carriers may lead to the formation of more complex charged quasiparticles beyond the three-particle trions.

### 11.2.3 *Exciton Dynamics*

The important role of dynamics is so far mainly investigated based on parameter-based models. Further progress of *ab initio* approaches in this direction would be highly desirable. This would require the inclusion of electron-phonon coupling alongside electron–hole interactions in combination with real time propagation either within the framework of time-dependent DFT or many-body Green-function theory (the Kadanoff–Baym equations). Especially in combination with the previous discussion of defects, a simplified model may remain required to achieve a better understanding and comparison with experiments, for example, on transport-related quantities.

From the experimental perspective the usual challenges to obtain large area homogeneous or, at least, *structurally well-defined samples* remain. This is particularly important for more complex heterostructures considering instances of stable moiré superlattices but also microscopic and mesoscopic atomic reconstruction. Aside from that, the following challenges and directions can be of interest to resolve and to follow:

### 11.2.4 *Time-Resolved Imaging on Nanometer Scale*

There is already a number of state-of-the art approaches featuring atomic-scale spatial resolution and femtosecond dynamics successfully applied also to TMDs. It would be highly desirable, however, to combine it with time-resolved imaging capabilities to directly monitor the evolution of the excitonic wave functions in real space under a variety of external conditions. The capability to optically track the excitonic response on extremely short time-scales could be then exploited to understand structurally complex scenarios including edge states at in-plane interfaces, interplay of localization and delocalization in superlattice potentials, and, possibly, excitonic quantum transport phenomena. Experimentally this would require independent, simultaneous tracking of the optical response from different sample positions with high spatial and temporal resolution. Recent demonstrations of excitons in photoemission could motivate using focused electron beams both as a pump and probe, potentially combining it with tip-enhanced microscopy and similar methods. While technologically demanding, this could open up exciting opportunities to combine high level many-body theory with closely matching experimental observables.

### 11.2.5 *Dynamics of Electronic Correlations from Excitonic Sensors*

The direct consequence of using excitons as optically accessible detectors for electronic correlations would be the extension of this concept to the time domain. A wide range of interesting scenarios would then become possible including formation, relaxation, and decay of stable and metastable correlated states using pump-probe-based and similar techniques. The excitons themselves could be briefly promoted to states more or less sensitive to specific correlations, including dressing them with free charge carriers or promoting them to highly excited states. This could potentially enable time-dependent sensitivity also to the out-of-plane direction due to the transiently modified long-range interaction from different excited states. By independently controlling the exciton and free carrier densities in the exciton sensing layers would further allow for precise tuning to experimentally desired conditions.

### 11.2.6 *Control of Neutral and Charge Excitons with Low-Energy Photons*

This concept goes back to studies of excitonic quasiparticles using low-frequency microwave[341] and THz[342] radiation. In TMDs this process can be very efficient and fast[343] and should thus allow for transient modification of the exciton composition, their charged state, or possible even spin-, valley- or intra/inter-layer configurations. This could be also exploited for controlling a variety of excitonic complexes or even higher-particle states including electronic correlations probed by the excitons. There are clear experimental challenges considering sample size and geometry as well as the properties of the THz or microwave radiation that can be more difficult to adjust for the atomically thin samples in contrast to the visible spectral range. In addition, the by now routinely used methods of tip-enhanced THz and NIR microscopy would be ideal for implementing similar schemes of transient excitonic control.

Altogether, considering an ever-increasing material base of low-dimensional platforms and their combinations in hybrid heterostructures, these are only a few among many exciting directions for research. Advances are to be expected both in theory and experiment with the common goal of better understanding the properties of Coulomb-bound states in nanostructured matter. In that respect, TMD-based structures are likely to remain on par with other types of optical materials such as conventional semiconductor quantum wells, nanoplatelets, and nanotubes, or even outperform these types as a platform for advancing our understanding of excitonic quasiparticles as a platform for exploring fundamental physics of excitonic quasiparticles.

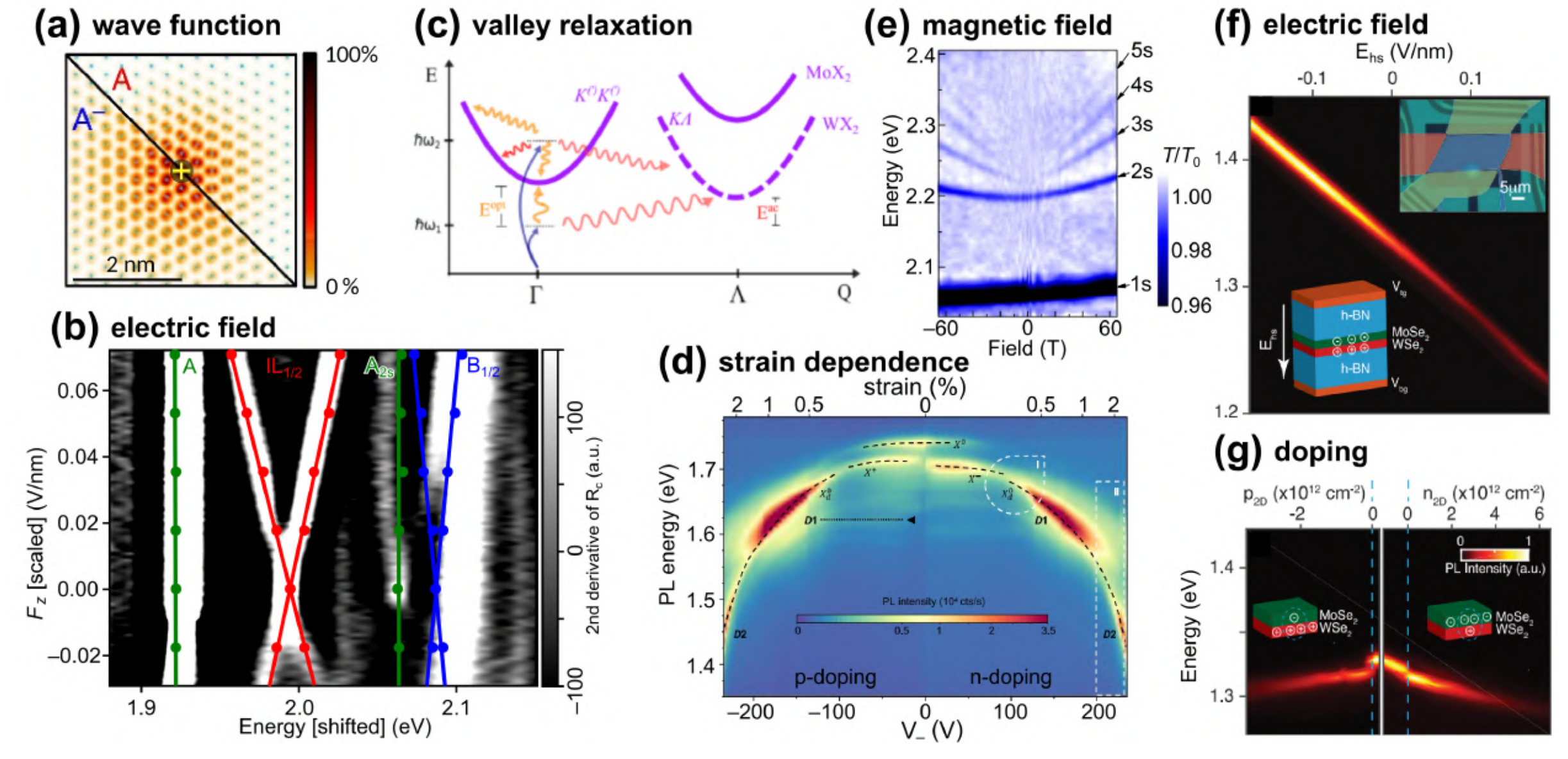

**Figure 11.** Tunable properties of excitons and trions. (a) Exciton wave function of excitons (right top) and trions (left bottom) in monolayer $MoS_2$. The hole is fixed in the center and the (averaged) electron distribution is shown in a top view. Adapted with permission from ref 323 (Copyright 2017 Springer Nature). (b) Exciton energies with respect to an applied vertical electric field in a bilayer $MoS_2$. While intralayer states are almost constant in energy, states with strong interlayer contributions show an X-like shape. The colored data are *ab initio* results overlaying experimental reflectance measurements. Adapted with permission from ref 331 (Copyright 2021 Springer Nature). (c) Representative relaxation channels and momentum dependent excitations in TMDs. While transitions between valence and conduction bands at K have a momentum close to Γ, momentum-indirect transitions involving phonons (arrows) are also possible with a pronounced temperature dependence. Adapted with permission from ref 318 (Copyright 2017 American Physical Society). (d) Strain dependent photoluminescence of $WSe_2$ monolayer featuring bright ($X^0$, X) and dark (D1, D2) excitons. Adapted with permission from ref 19 (Copyright 2022 Springer Nature). (e) Magnetic tuning of the ground and excited states in monolayer $WS_2$.[326] (f) Tuning of interlayer excitons in a $MoSe_2/WSe_2$ heterobilayer with electric field.[332] (g) Corresponding injection of free charges demonstrating *p*- and *n*-type interlayer trions. Panels (f,g) are adapted with permission from ref 332 (Copyright 2019 American Association for the Advancement of Science).

## 12. TUNING OPTICAL EMISSION IN 2D SEMICONDUCTORS VIA DOPING, INTERCALATION, AND ALLOYING

**Yohannes Abate,[1,*] Mauricio Terrones,[2] Vinod K. Sangwan,[3] and Mark C. Hersam[3,4,5]**
[1]Department of Physics and Astronomy, University of Georgia, Athens, Georgia 30602, United States
[2]Department of Physics, The Pennsylvania State University, University Park, PA 16802, USA
[3]Department of Materials Science and Engineering, Northwestern University, Evanston, Illinois 60208, United States
[4]Department of Chemistry, Northwestern University, Evanston, Illinois 60208, United States
[5]Department of Electrical and Computer Engineering, Northwestern University, Evanston, Illinois 60208, United States
***Corresponding author:** yohannes.abate@uga.edu

### 12.1. Current State of the Art

#### 12.1.1. *Introduction*

Controlling material properties at the atomic scale is a long-sought goal in materials science and holds profound implications for science and technology. In a visionary 1959 talk, Richard Feynman famously asked, "What would the properties of materials be if we could really arrange the atoms the way we want them?" With the rise of atomically thin 2D materials hosting quantum phenomena, we are closer than ever to realizing this dream. For example, bound electron–hole pairs (i.e., excitons), which dominate the optical properties of these materials, can be manipulated by many-body effects arising from the dynamic interplay of charge, spin, and moiré superlattice structures. Despite recent advances, achieving complete atomic-level control remains a significant challenge. In this section, we review recent advances that enable the tuning of excitonic quasiparticles at the atomic scale in 2D semiconducting chalcogenides through doping, alloying, and intercalation of vdW crystals (Figure 12a). We correlate these fundamental approaches with advances in characterization methods and resulting applications (Figure 12). Finally, we summarize the remaining challenges for this field and propose a forward-looking roadmap aimed at realizing the full potential of 2D semiconducting chalcogenides for photonic applications.

#### 12.1.2. *Synthesis*

The modification of 2D materials by doping, alloying, and intercalation can be achieved either during or after the growth through bottom-up or top-down approaches. Bottom-up approaches such as chemical vapor deposition (CVD) or metal-organic chemical vapor deposition (MOCVD) build up the materials from atomic or molecular gaseous precursors. In this manner, these techniques are naturally suited to growing doped and alloyed wafer-scale 2D materials by adjusting the gas composition, temperature, substrate, and pressure (Figure 12b.1).[344,345] Top-down approaches involve exfoliating bulk layered materials (Figure 12b.2) that could be doped or alloyed into 2D monolayers or ultrathin nanosheets *via* adhesive tape exfoliation, liquid-phase exfoliation (LPE), or electrochemical intercalation followed by LPE, where intercalated ions expand interlayer spacings to facilitate LPE.[346,347,348,349] Intercalants can also control doping, providing the possibility of exfoliation and doping in one step.[350,351,352] Intercalation is controlled by the magnitude of inter-layer vdW interactions, the electrical conductivity of the bulk crystals, and the size and structure of the intercalants.[349,351,352] Self-intercalating molecules have also been shown to form superlattices within host vdW materials resulting in molecular-level control of doping and emergent properties such as magnetic order.[353] Finally, doping can be realized by electrical gating to study and tailor excitonic properties.[348,349]

#### 12.1.3. *Characterization*

Exciton characterization in alloyed and doped 2D materials typically integrates structural microscopy with diffraction-limited optical microscopy and spectroscopy.[354] Information on dopant and alloy distribution is obtained using various atomic-resolution structural imaging techniques.[355] For example, scanning transmission electron microscopy (STEM) reveals atomically resolved variations in dopant and defect concentrations (Figure 12c.1).[344] Far-field optical methods can analyze the chemical composition and excitonic properties, showing how defects, doping, and alloying impact optical behavior and electronic band structure. Raman imaging characterizes vibrational properties and reveals concentration differences, while photoluminescence (PL) spectroscopy probes excitonic properties and the impact of defects, doping, and alloying on electronic transitions (Figure 12c.1).[356] Conventional far-field optical techniques are, however, diffraction-limited and thus struggle to resolve features smaller than half the illumination wavelength. Various tip-based nanoscopy techniques have been developed to overcome the diffraction limit, enabling optical imaging at the nanoscale. For example, tip-enhanced photoluminescence (TEPL), tip-enhanced Raman spectroscopy (TERS), and scattering-type near-field optical microscopy (s-SNOM) achieve sub-diffraction resolution through enhancement of electric fields at the tip apex, enabling exciton studies at ~10 nm spatial resolution.[357,358] Additionally, theoretical investigations into doping mechanisms, energetics, and electronic structures using first-principles calculations play a significant role in guiding experiments and interpreting experimental results.[359,360]

12.1.4. *Applications*

The potential applications enabled by intercalation, doping, and alloying fall into two broad categories: (i) High-performance photonic technologies such as photodetectors; (ii) Emergence of new phenomena including bandgap engineering,[348] monolayer behavior in bulk,[352] superconductivity,[361] quantum phase transitions,[362] chemochromism in intercalated 2D compounds,[363] and novel spin structures in vdW magnets.[353] In the case of alloying, the bandgap can be tuned to vary the spectral responses of light-emitting diodes, photodetectors, and single photon sources.[364] Substitutional doping facilitates the realization of lateral and vertical p–n heterojunctions that show gate tunability of photonic responses.[365] Through electric-field doping, sub-bandgap trap states in 2D TMDs can be filled, which has resulted in near unity gain in photoluminescence albeit at low fluence.[366] Overall, 2D photodetectors have shown high internal photoconversion gain ($>10^5$) in the visible range, but they are more competitive in the mid-IR where options for bulk semiconductors are limited.[367]

## 12.2. Challenges and Opportunities

12.2.1. *Synthesis*

Challenges in achieving controlled modification of 2D materials by doping, alloying, and intercalation share characteristics with the general challenges for bottom-up and top-down approaches. For example, desired properties from doping compete with unintentional effects from contamination and disorder in 2D materials, so doping-induced tuning of photonic devices assumes the availability of relatively defect-free intrinsic 2D materials.[368] In this regard, five main challenges can be delineated: (i) Production of wafer-scale intrinsic 2D materials with minimum defects; (ii) Controlled dopant density and alloy stoichiometry with minimum strain; (iii) Post-growth patterning of doped and alloyed regions for on-demand production of heterojunctions with ohmic electrical contacts; (iv) Reversible doping strategies such as healing of dopant-induced defect states; (v) Low-temperature growth and transfer strategies for back-end-of-line (BEOL) processing.[369,370] Spatial control of dopant distributions during CVD growth presents additional challenges and opportunities. For example, dopants such as Cr and Fe can be confined to specific regions of hexagonal 2D sheets given the high affinity of chalcogen zig-zag edges to create metal vacancies that can be filled with dopants.[371] For reversible doping, approaches are needed to eliminate synthetic defects after growth.[372] The two main challenges for intercalation are: (1) Avoiding spurious electrochemical reactions with the host crystal; (2) Controlling the density and spatial distribution of the intercalant. For completeness, it should be noted that chemical functionalization of the basal plane can also tune electronic properties for photonic applications as has been described in detail elsewhere.[373]

Future opportunities include: (i) Templated growth to achieve highly crystalline, rotationally commensurate materials using epitaxy or remote epitaxy approaches; (ii) Growth of Janus crystals for tunable photonic properties; (3) Direct growth of vdW and moiré heterostructures from different 2D materials.[374] The opportunities in top-down approaches include: (i) Morphotaxial conversion synthesis that can produce 2D counterparts of materials that do not exist as vdW crystals;[375] (ii) Tunable magnetic properties in doped 2D materials such as vanadium-doped $WSe_2$.[376] Post-growth processing with proton and ion irradiation in controlled plasma environments can further tailor the spatial distribution of doped regions, preferably in a manner that minimizes perturbation of the underlying crystalline lattice.

12.2.2. *Characterization*

Although tip-based nanoscopy methods (such as those mentioned above) are increasingly used to characterize excitons and related quasiparticles, it has been challenging to achieve the spatial resolution needed to probe exciton emission at the length scale of the Bohr radius (1–2 nm). Moreover, most TEPL studies are confined to ambient conditions, whereas a low-temperature setup could reveal intrinsic excitonic properties with enhanced PL intensity without homogeneous broadening, thereby quantifying the effect of dopants on excitonic structures and recombination processes. By combining cryogenic s-SNOM with TEPL, for example, demodulation can be achieved for the tip-scattered PL signal at high harmonics. Recent efforts have also correlated electronic structure from low-temperature atomically resolved scanning tunneling microscopy (STM) with tip-induced excitons and their luminescence spectra, mapping structure-property relationships at nanometer-scale spatial resolution (<5 nm).[215] Coupling ultrafast laser excitation within nanoscopy methods will further quantify competing recombination pathways involving excitons, dopants, defects, and vdW heterojunctions (Figure 12c.2). In particular, recent advances hold promises for atomic-scale near-field spectroscopy and imaging such as optical tunneling emission microscopy[191] and X-ray-excited resonance tunneling that measures the local density of states.[377]

*In situ* high-resolution optical and structural diagnostic techniques are needed for real-time monitoring of crystallinity, defect states, and electronic structures during growth in order to provide insights into defect formation and photochemical pathways that can be tailed for controlled doping and alloying (Figure 12e). For example, *in situ* characterization of TMDs can capture critical kinetic information often missed in post-growth interrogation, offering insights into the intermediate complex formation, grain boundary stitching, defect dynamics, phase transitions, and chemical reactions. These insights will help address key challenges in understanding how growth conditions affect material structure, enabling precise control over properties. Correlated measurements using high-resolution spatial, spectral, and temporally resolved spectroscopy and microscopy in a multimodal setup under ultrahigh vacuum and cryogenic temperatures will facilitate a comprehensive analysis of optical and optoelectronic properties, which will facilitate optimization for photonic applications. Finally, integrating advanced machine learning methodologies with large-scale datasets can enable the extraction of nuanced insights into excitonic dynamics, acceleration of data analysis, and a deeper understanding of the physical mechanisms governing complex systems. Including device performance metrics in machine learning analyses would further refine our understanding of multifaceted excitonic phenomena that play performance-limiting roles in photonic devices (Figure 12e).

12.2.3. *Applications*

Current challenges in realizing viable 2D photonic technologies include: (i) Minimizing disorder in wafer-scale, monolayer films; (ii) Controlled doping and alloying; (iii) Effective passivation and contact engineering. For example, the electrochemical intercalation of individual micron-sized black phosphorus flakes has shown monolayer behavior, but this scheme needs to be generalized to other wafer-scale crystals such as TMDs.[352] As an example of molecular-level control, left-handed or right-handed intercalants (such as R-α-methylbenzylamine) have shown strong chirality-induced spin selection effects in 2D crystals such as $TaS_2$ and $TiS_2$ (Figure 12d.2).[378] Further progress in this direction that achieves long-range lattice order in interacted species would be of particular interest for tailoring the optospintronic properties of magnetic vdW semiconductors.[379] In additive manufacturing approaches, printed 2D films require a high fraction of monolayers in inks and efficient charge transport in percolating films.[380] Solution-based densification of monolayers into bulk crystals has been demonstrated, but these strategies are not likely to preserve monolayer behavior without spacer layers provided by covalently tethered molecular adlayers.[381] While electrostatic doping in all-2D stacks using micron-sized graphene, hexagonal boron nitride, and TMC flakes have been shown to minimize disorder for high-quality 2D photonic devices (Figure 12b.1), the challenge is to generalize these approaches to wafer-level photonic applications.[378,379] Finally, post-growth substitutional doping, contact engineering, encapsulation, and BEOL integration with readout integrated circuits are of high interest for photonic applications, but these methodologies need to be achieved without introducing additional disorder.

Doping and alloying impact several photonic applications including single-photon sources, IR photodetectors, electro-optical switches relying on strong nonlinear optical processes, and light-emitting devices (Figure 12e).[364,382] All of these technologies will benefit from resolving the materials science challenges mentioned above. For example, a recently demonstrated megasonication approach enhanced the monolayer fraction in solution-processed $MoS_2$ inks in a manner that preserves monolayer character in restacked $MoS_2$ flakes, ultimately leading to electroluminescence from bulk films (Figure 12d.2).[383] Monolayer behavior in intercalated crystals is also promising for quantum emitters where localized strain can funnel excitons to increase quantum yield. Effective passivation strategies are also essential for emerging 2D magnetic semiconductors that couple excitonic quasiparticles with spin states.[379] Combining two different doping strategies could yield further control over photonic properties such as intercalating alloyed TMDs to achieve a greater bandgap tunability. Likewise, combining electric field effects in ambipolar TMDs and heterojunctions between doped monolayers (e.g., Nb-doped p-type $MoS_2$ and Re-doped n-type $MoS_2$) can independently induce electrostatic control and electrical emission of quasiparticles such as trions, biexcitons, and dark indirect excitons.

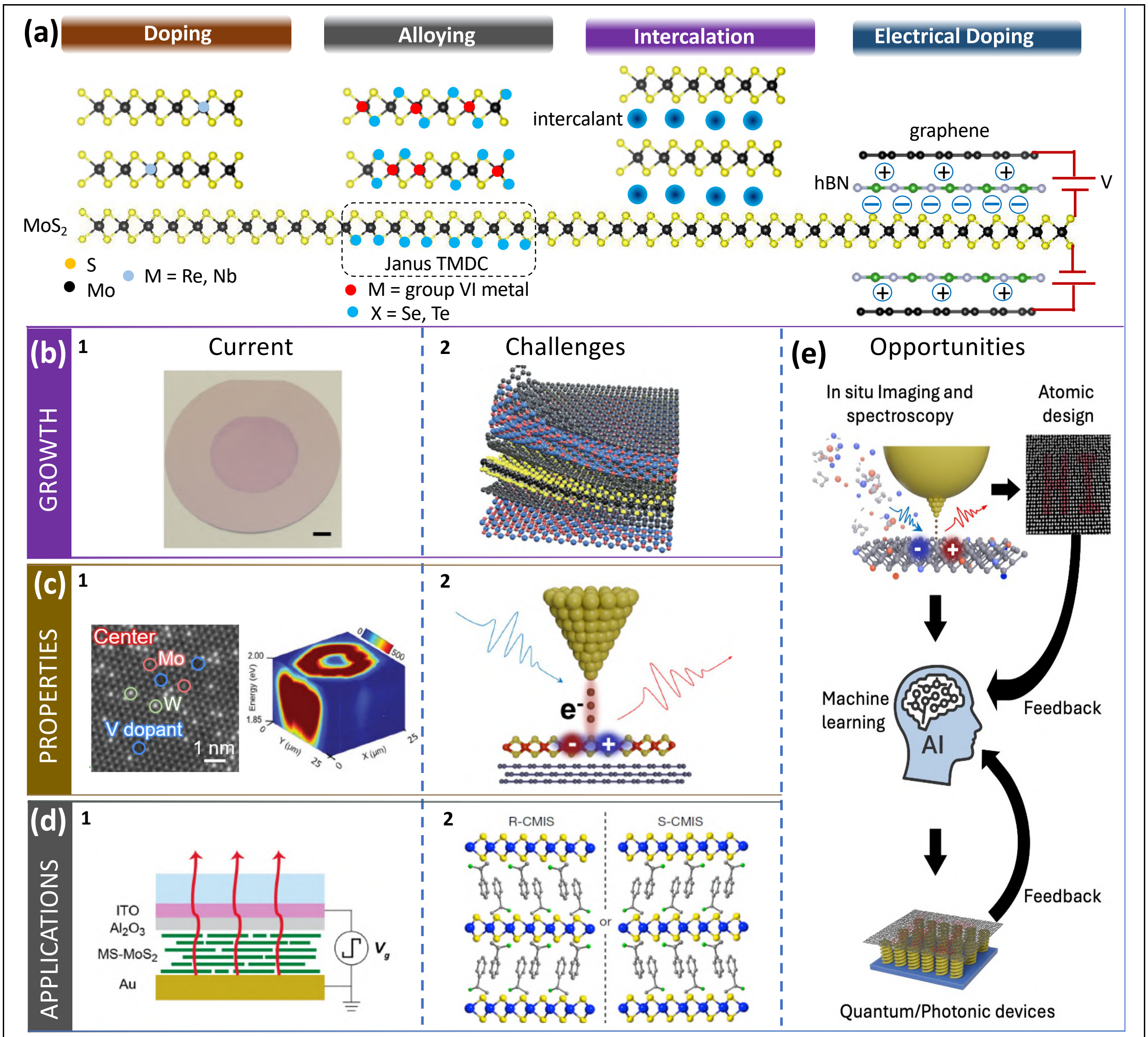

**Figure 12.** Overview of growth, characterization, and applications in doping, alloying, and intercalation of 2D vdW materials. (a) Schematic of structural changes in vdW materials and heterostructures for doping, alloying, and intercalation. (b.1) The current status of growth efforts includes doped and alloyed high-quality wafer-scale 2D materials by bottom-up approaches such as chemical vapor deposition. However, producing wafer-scale 2D materials with minimal defects and achieving controlled dopant density and alloy stoichiometry with minimal

strain is still a challenge. Adapted from ref 384 (Copyright 2021 Springer Nature). (b.2) The remaining challenges include low-temperature growth, reversible doping strategies, and precise post-growth patterning that are critical for applications relying on heterojunctions and BEOL processing. Adapted from ref 385 (Copyright 2016 Springer Nature). (c.1) The current status of characterization methods for quasiparticles is exemplified by STEM (left) and far-field optical spectroscopy techniques (right) that reveal dopant and defect distributions, chemical composition, and excitonic properties in 2D materials. Left and right schematics are adapted from ref 344 (Copyright 2020 American Chemical Society) and ref 356 (Copyright 2024 Wiley), respectively. (c.2) A key challenge in characterizing intercalated, doped, and alloyed samples is achieving the spatial resolution necessary to probe both the atomically resolved crystal structure and the dynamics of excitons at the Bohr radius length scale (1–2 nm) at low temperatures. (d.1) An example of a current application enabled by solution-processed 2D materials is electroluminescence from megasonicated $MoS_2$ inks that retain monolayer direct bandgap character in percolating films. Adapted from ref 383 (Copyright 2023 American Chemical Society). (d.2) Challenges for photonic applications include the ordering of dopants, substitutional species, and intercalated molecules. The specific example depicted here is ordered intercalated chiral molecules in $MoS_2$ that have the potential to realize macroscopic chiral optical response. Adapted from ref 378 (Copyright 2022 Springer Nature). (e) Schematic illustrating opportunities for integrating *in situ* high-resolution imaging and spectroscopy techniques for real-time monitoring of crystallinity, defect states, and electronic structures during growth, which can accelerate the realization of patterned doped regions without disturbing the underlying material lattice. The incorporation of advanced artificial intelligence and machine learning approaches with large-scale datasets will accelerate data analysis, while an integrated feedback mechanism will further refine the understanding of the physical mechanisms influencing device performance.

## 13. ENGINEERED CONFINEMENT OF EXCITONS IN 2D SEMICONDUCTORS

**Leo Yu,[1,2] Xueqi Chen,[2,3] Tony F. Heinz,[1,2] Puneet Murthy,[4] Martin Kroner,[4] Tomasz Smolenski,[5] Deepankur Thureja,[6] and Thibault Chervy[7,*]**
[1]Department of Applied Physics, Stanford University, Stanford, CA 94305, USA
[2]SLAC National Accelerator Laboratory, Menlo Park, CA 94025, USA
[3]Department of Physics, Stanford University, Stanford, CA 94305, USA
[4]Institute for Quantum Electronics, ETH Zürich, Auguste-Piccard-Hof 1, 8093 Zürich, Switzerland
[5]Department of Physics, University of Basel, Klingelbergstrasse 82, 4056 Basel, Switzerland
[6]Department of Physics, Harvard University, Cambridge, MA, USA
[7]NTT Research, Inc. Physics & Informatics Laboratories, 940 Stewart Dr, Sunnyvale, CA 94085, USA
***Corresponding author.** Email: Thibault.chervy@ntt-research.com

### 13.1 Introduction

Excitons—bound electron–hole pairs—dominate the optical response of 2D TMD semiconductors. The reduced Coulomb screening in these materials, together with the high quality of state-of-the-art encapsulated monolayer samples result in robust excitonic states with binding energies on the order of $E_B \sim 0.5$ eV. The states exhibit large oscillator strength and nearly radiatively limited linewidth at low temperatures.[386,387] These features have established 2D TMDs as a promising light–matter interface for photonics applications.[388] However, weak interactions between tightly bound 2D excitons have so far limited their relevance for quantum nonlinear devices. Largely motivated by the possibility of realizing exciton-based quantum photonic devices, several recent works have focused on confining the center-of-mass motion of these bosonic quasiparticles as a pathway to increase their nonlinear response. In this context, the 2D materials toolbox offers a wide variety of possible approaches, including patterning of the semiconductor or surrounding dielectric,[389,390] formation of moiré superlattice potentials,[391] strain-engineering,[392,393] and electrostatic engineering,[394,395,396] among other schemes.

In context of precisely engineering the confinement of exciton, a key recent development is the possibility of tuning the confinement length by lithographic patterning and the use of applied strain and electrostatic fields. These degrees of freedom allow one to reach an optimal confinement scale that exhibits large optical nonlinearity, while simultaneously maintaining sizeable oscillator strength.

In this section, we will focus on two approaches: strain engineering and electrostatic engineering, which constitute two scalable platforms for the control of multiple nonlinear excitonic sites, a key ingredient of future quantum photonics applications.

## 13.2 Strain Confinement of Excitons

### 13.2.1 *State of the Art*

Strong confinement of excitons associated with localized tensile strain was initially shown to produce quantum emitters (QEs) in tungsten-based TMD monolayers.[392,397] This was achieved by deposition of the monolayer on a substrate that had been patterned with pillar arrays (Figure 13a,b). The pillars induce a strain field in the monolayer and localize the QEs with a spatial precision of ~ 100 nm. The deposition was often accompanied by the formation of wrinkles in the monolayer, which may result in variation in the strain profile. This issue has been addressed by decoupling the straining process from the initial monolayer deposition in two subsequent studies.[398,399] In both works, localized strain was created by the monolayer conforming to a depression, rather than a protrusion in the substrate. In one approach, nanoindentation on polymeric substrates was explored[398] and in thermal molding[399] was used in the other. Complementing strain confinement, additional control over the defect density by electron-beam irradiation was utilized in another recent investigation,[13] in which encapsulation of the monolayer in thin dielectric layers improved the spectral quality of the QEs. Further integration also enabled electrically driven QEs.[400]

While $WSe_2$ and $WS_2$ have been the materials of choice in most of the studies of strain-induced QEs, the origin of their (quantum) emission has been a source of discussion. Naïvely, the emission could come from bright excitons that are tightly localized in the localized strain field. Subsequent theoretical analysis,[253] however, attributed the origin to dark excitons that are brightened due to strain-induced mixing with in-gap defect states. Two experimental signatures were considered in the analysis: the spectral gap (> 40 meV) between the QEs and 2D bright excitons, and the (dis-)agreement of the exciton $g$-factor between the QEs and the 2D (bright) dark excitons. With this understanding, in a subsequent work the strain and the defect density were controlled independently in the fabrication process,[13] and it was found that neither factor alone was able to produce QEs. More recently, in one investigation where $WSe_2$ monolayers are incorporated into nanomechanical devices,[19] *in situ* strain tuning allowed the hybridization between dark and localized excitons to be observed directly. Also, correlated photoluminescence imaging and atomic-force microscopy have found smooth strain variation around the QEs, which is considered insufficient to confine bright excitons sufficiently tightly to enable quantum emission.[401] The picture of strain-induced mixing between dark and localized excitons is thus supported by growing experimental evidence.

### 13.2.2 *Challenges, Future Goals, and Suggested Directions to Meet Goals*

One of the major challenges limiting the $WSe_2$ QEs is their spectral quality, especially in terms of the variation of their emission energies. This site-to-site inhomogeneity renders current $WSe_2$ QEs unsuitable for quantum optics applications where photon indistinguishability is essential. While progress has been made to reduce the energy spread,[398,13] it remains much greater than the linewidth of the QEs around 100 μeV.[13] It should be noted that the best $WSe_2$ QE linewidth is also far from the lifetime limit (< 1 μeV). This problem can in principle be solved by increasing the photon decay rate of the QEs *via* Purcell enhancement in photonic structures.[402,403]

This challenge calls for new directions in the research of 2D confined excitons and QEs. One such direction is to start with monolayers with bright (spin-allowed) exciton ground states, thereby eliminating the defect band as a requirement to brighten the dark excitons of $WSe_2$. Progress toward this direction has been made using monolayer $MoSe_2$,[399] where the spin-allowed nature of the QEs was supported by the measured exciton $g$-factor and the lack of a discernible zero-field splitting. The authors also showed an improved method for developing localized emitters that appeared to reduce the QE spectral inhomogeneity down to a few meV from a few tens of meV in other works. Another material that also features bright excitons is $MoTe_2$.[24] The $MoTe_2$ QEs exhibited characteristics of spin-allowed transitions like those of the $MoSe_2$ QEs, but with emission in the telecom wavelength range that is desirable for long-distance quantum communication. Additionally, multilayer GaSe can be tuned toward a direct gap with applied strain.[402] The strain-induced GaSe QEs show distinct emission lines without the lower-energy defect-related bands that accompany $WSe_2$ QEs, but further spectroscopic investigations are needed to determine unambiguously the nature of the relevant optical transition of the QEs.

Another promising direction is to combine strain with layer stacking to induce QEs in heterobilayers (HBLs) that host interlayer excitons. Interlayer excitons possess static dipole moments and energy shifts under an electric field could be used to compensate for spectral variation up to tens of meV. This tunability would, however, typically come at the price of reduced radiative rate of interlayer excitons. This approach does not presume the elimination of defects, as the HBLs reported to show localized interlayer excitons also include the $WSe_2$ layers.[404] We also note that the cause of the localization in this report is not attributed to moiré confinement. Epitaxial HBLs from nearly lattice-matched materials can thus be used to avoid the complexity of moiré structures. A work in this direction has already shown strain-induced interlayer QEs in $MoS_2$/$WSe_2$ HBLs,[405] although Stark was not explicitly demonstrated because of the lack of electrical contacts.

With the engineerability of monolayers, progress has been made in achieving strain confinement, vdW heterostructures, and coupling to photonic structures. We expect further improvement and integration of these capabilities to enable a scalable source of indistinguishable single photons for quantum photonics applications.

## 13.3 Electrical Confinement of Excitons

### 13.3.1 *State of the Art*

Electrostatic confinement of excitons is a promising direction for scalable nonlinearities because of the possibility afforded for *in situ* energy tuning and lithographic patterning. Electrostatic confinement may be achieved in several ways, depending on the nature of the excitonic state involved, and the electric-field distributions considered. Early experiments on III–V materials already reported the electrostatic control of out-of-plane dipolar excitons by applying spatially varying vertical electric fields. In this configuration, the position-dependent Stark shift creates a potential energy landscape for out-of-plane dipolar excitons that can be used to control their center-of-mass motion. Similar results have been obtained in 2D materials, in particular using stacks with type II band alignment where the hole sits in one layer and the electron sits in the other layer.[406] Such hetero-structures generally also exhibit moiré superlattice potentials, which can further confine excitons or interfere with their electrostatic control. Another consideration related to confinement strategies based on interlayer excitons is their aforementioned low radiative rate.

Exciton complexes such as trions with net charge can also be controlled electrostatically through the engineering of spatially varying electric potential. Examples of such approaches include local electrostatic doping using carbon nanotubes,[407] biased AFM tips,[408] and nanostructured gate electrodes.[395,396]

Recently, electrostatic quantum confinement of neutral intra-layer excitons in monolayer TMDs has been achieved using the second-order Stark effect in gate-defined lateral p–i–n junctions.[394] In this technique, the steeply varying in-plane electric field present in the depletion region polarizes and traps excitons, resulting in a 1D confinement potential that follows the natural edges of the gate. Repulsive charge–exciton interactions in the doped regions (repulsive polaron formation) cam further confine the neutral excitons to the region of charge neutrality. Building on this approach, researchers have demonstrated quantum confinement of excitons in a variety of geometries, including quantum dots,[395,396] quantum rings, and arrays[395] by lithographic structuring of gate electrodes (Figure 13c). A salient feature of this approach is its scalability, as the quantum dot positions are lithographically defined, and the energy of individual trapping sites of a lattice can simultaneously be tuned by independent gate electrodes. Simultaneous resonant tuning of three independent quantum dots in the same device has been achieved,[395] demonstrating a route toward large arrays of degenerate solid-state quantum devices (Figure 13d). The energy tuning range of the trapped excitons, limited by their field ionization, reaches 10 meV in state-of-the-art devices at cryogenic temperatures. The 0D nature of such emitters has been demonstrated through the on-off blinking behavior of emission, typical of 0D quantum emitters.[396] Finally, electrical control of valley hybridization of the localized exciton states has been reported.[409] For applications, however, the samples must be optimized to suppress blinking and spectral jumps of the 0D exciton resonance, and photon correlation measurements are needed to confirm their QE nature.

### 13.3.2 *Challenges, Future Goals, and Suggested Directions to Meet Goals*

These advances in electrostatic trapping of excitons pave the way for several new research directions and applications. We identify three main themes (Figure 13e) that will benefit from the techniques introduced above for electrostatic, as well as for strain confinement.

(i) Integration in photonic structures: A central challenge in the field of 2D materials photonics is the realization of highly nonlinear optical media, in which photons coherently propagate while undergoing strong exciton-mediated nonlinearities. A key advantage of 2D materials for realizing this polaritonic platform is their suitability for integration with photonic elements, such as optical cavities, photonic crystals, and waveguides. In particular, interfacing confined excitons with a specifically engineered photonic mode would have implications for the long quest for scalable single photon sources.

(ii) Large-scale control and quantum optical metasurfaces: Another important goal is the upscaling of existing platforms to realize very large arrays of controllable quantum confined excitons. Electrically trapped intra-layer excitons in TMDs hold promise thanks to their controllability with CMOS-compatible voltages and their sizeable optical cross section. With advances in 2D materials growth, monolayers with larger area and high quality are expected to become available. This will open the possibility of integrating these heterostructures in wafer-scale CMOS architectures, enabling a massive scale-up of the demonstrated devices. The large-scale control over quantum confined exciton states offers a new avenue to the design of metasurfaces operating in the quantum nonlinear regime. The ability to control electronic fluctuations in the local environment of trapped excitons, the demonstration of QE statistics, and the management of crosstalk in large-scale devices constitute near-term challenges for this approach.

(iii) Interfacing with the 2D materials ecosystem: A unique feature of 2D materials is that they cover the whole range of electronic, optical, and magnetic properties. This brings up exciting opportunities for electrical confinement of excitons in other vdW systems, such as layer-hybridized excitons in bilayer semiconductors, moiré superlattices, electronic quantum-dots in bilayer graphene, ferromagnetic semiconductors (e.g., CrSBr), and many other possibilities. For instance, gate-control of the charge density in CrSBr[410] could be used to realize electrically trapped excitons with energies sensitively depending on the inter-layer spin alignment. The confined state energy could thus be controlled with small external magnetic fields, providing a new tuning knob for quantum photonics applications. Another exciting direction is the interaction of gate-defined excitonic QDs with their electronic counterparts in bilayer graphene. This may lead to the creation of a spin-photon interface hosted entirely in vdW materials.

The design and control of large-scale nonlinear photonic devices operating in the quantum regime is a tantalizing frontier of 2D semiconductor optics. While new approaches hold promises for the scalable assembly of quantum nonlinear building blocks, material constraints have so far limited the progress toward that goal. Achieving high-quality, large-area encapsulated devices with homogeneous excitonic response over millimeter scales is key for getting such devices out of the research labs. Once such materials become available, interfacing with CMOS technologies, as well as with existing integrated photonics platforms, should enable significant breakthroughs.

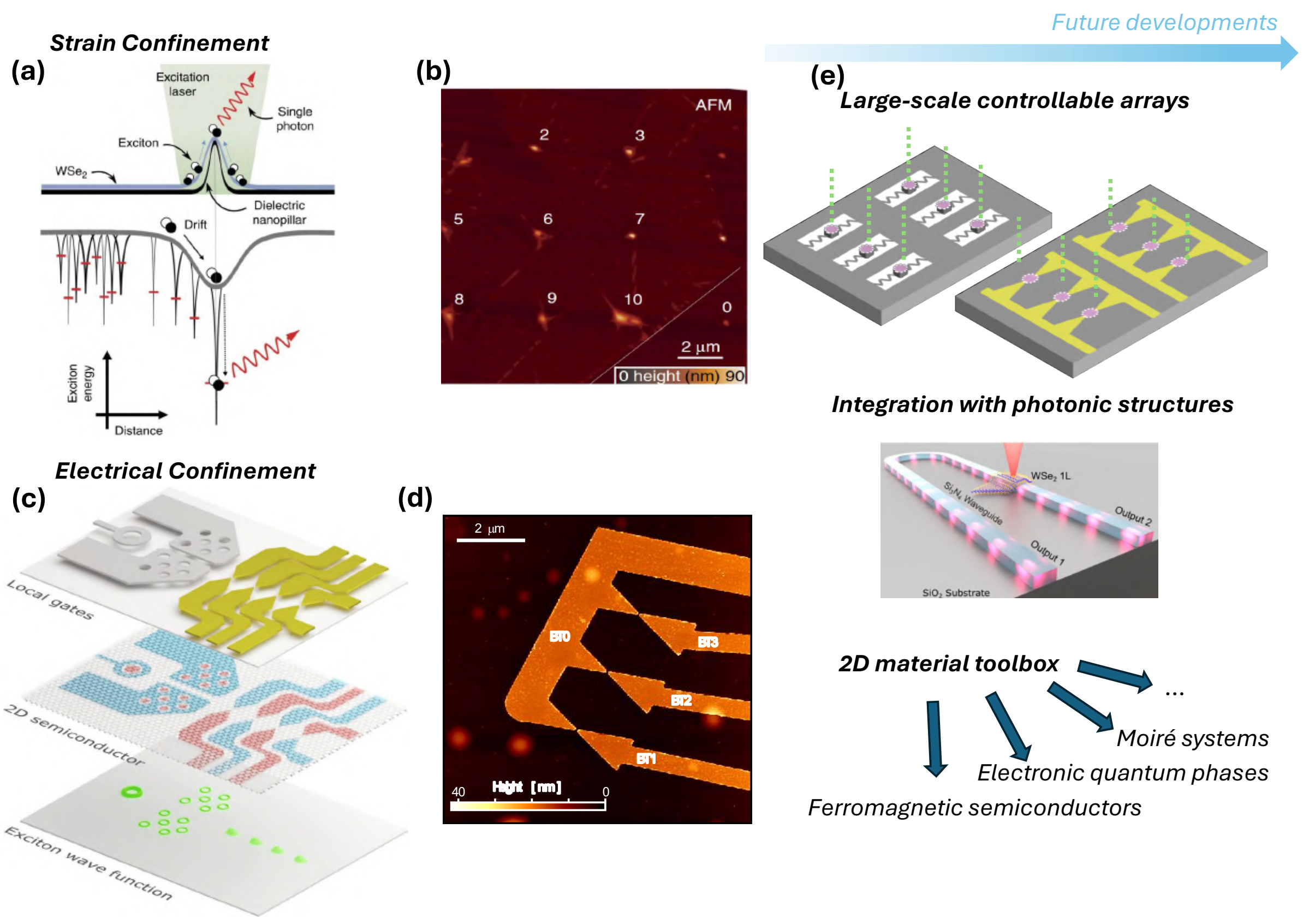

**Figure 13.** Strain and electrostatic engineering for monolayer exciton confinement. (a,b) The principle of strain-induced quantum confinement (a) and representative site-controlled quantum emitters (b). Reproduced with permission from ref 397 (Copyright 2017 Springer Nature). (c,d) Structures for 0D electrostatic quantum confinement of intra-layer excitons with position and energy tunability. Reproduced with permission from ref 395 (Copyright 2024 The American Association for the Advancement of Science). (e) Some future possibilities for scaled-up arrays of controlled QEs, interfacing with photonic integrated circuits, and new interfaces with 2D material platforms. Adapted with permission from ref 411 (Copyright 2021 American Chemical Society).

**Nonlinear and Ultrafast Optical Phenomena**

## 14. ULTRAFAST EXCITON DYNAMICS IN VAN DER WAALS HETEROSTRUCTURES

**Armando Genco,[1] Chiara Trovatello,[1,2] Giulio Cerullo,[1] and Stefano Dal Conte[1,*]**
[1]Politecnico di Milano, Dipartimento di Fisica, Piazza Leonardo da Vinci 32, Milano (MI), Italy
[2]Department of Mechanical Engineering, Columbia University, New York, NY 10027, USA
***Corresponding author.** Email: stefano.dalconte@polimi.it

Two-dimensional TMDs have been at the forefront of research on condensed matter physics due to their unique optoelectronic properties enabled by strong light–matter interaction. Quantum confinement and reduced Coulomb screening make TMDs an ideal playground to study excitons and their many-body complexes (i.e., trions, biexcitons, and other multiparticle states) and to investigate their tunability with external parameters like electric/magnetic fields and the lattice strain. Moreover, the possibility to optically address and manipulate the spin/valley degree of freedom further increased interest in these materials.[412]

Since the advent of 2D materials, out-of-equilibrium spectroscopies have been extensively used to study the mechanisms of formation, relaxation, and recombination of excitons occurring on different timescales, ranging from tens of femtoseconds to nanoseconds. Exciton line-shape analysis performed at different delay times has shown that the transient optical response as measured by pump-probe optical spectroscopy is the result of the interplay between dynamical screening of Coulomb interaction (i.e., renormalization of the quasiparticle bandgap and the binding energy of the exciton), broadening of the exciton peak due to excitation-induced dephasing and reduction of its absorption due to phase-space filling effect.[413] In the time domain, the transient optical signal decays over multiple timescales as a consequence of radiative and nonradiative relaxation processes, scattering with defects, and energy equilibration to the lattice.[414]

A fundamental step forward in the research on 2D materials was the realization of vdW heterostructures made by vertically stacking two or more 2D monolayers without any constraints due to the lattice mismatch.[84] Here, the twist angle acts as an external parameter to tune the optical properties and the exciton landscape. Early optical pump-probe experiments performed on heterostructures with type II band alignment focused on the study of ultrafast electron or hole transfer dynamics occurring between the K/K' valleys, which are weakly hybridized due to their in-plane orbital character.[415] These interlayer scattering processes are extremely fast (i.e., tens to hundreds of femtoseconds)[416] and show a weak dependence on the interlayer twist angle because they are mediated by intermediate scattering through layer-hybridized valleys. Charge separation suppresses the intervalley electron–hole exchange interaction, which is the main cause of fast valley depolarization in TMDs, leading to the formation of spin-polarized holes characterized by valley polarization close to unity and spin/valley lifetime of the order of microseconds.[417] The strong Coulomb interaction binds electrons and holes located in opposite layers of the heterostructure (see Figure 14a), giving rise to interlayer excitons (ILXs) characterized by lower emission energies than their intralayer counterparts and reduced oscillator strength as a result of the small overlap between the electron and hole wave functions. Optical pump-probe and time-resolved photoluminescence spectroscopy measurements have shown that ILXs have an extremely long lifetime (exceeding by several orders of magnitude the lifetime of intralayer excitons) and enhanced valley polarization retention. The ILX formation process has been also investigated by different nonequilibrium spectroscopy techniques.[418] Broadband transient absorption spectroscopy measurements performed on $MoSe_2$/$WSe_2$ heterostructures have reported picosecond formation dynamics of the optically bright ILXs at the K valleys (i.e., a timescale much slower than the build-up time of intralayer excitons). Theoretical calculations have shown that (i) the delayed formation is the result of a phonon-assisted interlayer exciton cascade process of hot ILX populations down to the ground state (see Figure 14b) and (ii) the dynamics of optically bright excitons are strongly affected by optically dark momentum-indirect excitons.[419]

Time- and angle-resolved photoemission spectroscopy (tr-ARPES) techniques have noticeably expanded the research on excitons in TMDs because of their unique ability to combine energy and momentum resolution, also in the time domain, enabling direct tracking of the band structure evolution following photoexcitation. A clear photoemission signature of photoexcited momentum-direct excitons (at the K valleys) has been predicted by theory and reported experimentally. The excitonic signature, located at a lower energy with respect to the single particle band and shifted by an amount equal to the exciton binding energy, has a peculiar energy–momentum dispersion, which is inherited from the valence band. The capability of accessing the exciton dispersion has enabled the retrieval of the excitonic wave function in real space, allowing to estimate its spatial extent.[420] Another advantage of tr-ARPES is the possibility to directly address the formation dynamics of momentum-forbidden dark exciton states (i.e., excitons characterized by finite crystal momentum with an electron and a hole residing on different valleys in the conduction and valence bands) which cannot be detected with all-optical spectroscopy techniques.[421]

Time-resolved momentum microscopy, a variant of tr-ARPES, has also been applied to track intra- and interlayer exciton dynamics in TMD heterostructures.[422] In particular, in the $MoS_2$/$WSe_2$ heterostructure, interlayer coupling results in the hybridization of the Σ valley conduction bands of both $WSe_2$ and $MoS_2$ layers. A direct consequence of the layer hybridization is the formation of the so-called momentum-indirect hybrid excitons (i.e., Coulomb bound electron–hole pairs where the hole resides at the K valley of the $WSe_2$ layer and the electron at the Σ valley is shared between the two layers). Upon photoexcitation of intralayer bright exciton of $MoS_2$, hybrid excitons form almost instantaneously by phonon scattering, and mediate the formation of momentum-direct ILXs at the K point on a slower timescale[423] (see Figure 14c).

The ability to control the orientation between the layers of the heterostructure with increasing precision has led to the systematic study of exciton emission and absorption as a function of the twist angle. The periodic alignment of the atomic registry in the twisted layers of the heterostructure results in an overarching periodic pattern (i.e., the moiré pattern) which modulates the electronic band structure of the heterostructure. The spatial periodicity depends on the relative orientation and the lattice mismatch between the layers. The periodic potential induced by the moiré pattern spatially confines ILXs and gives rise to a series of sharp excitonic resonances (called moiré excitons), spaced in energy and characterized by alternating spin/valley selection rules.[424]

During the last years several experimental studies have tried to capture the effect of the twist angle and the moiré potential on the recombination dynamics of the excitons. Unlike the charge transfer and exciton formation process, the radiative recombination time of ILXs can drastically change in response to small variations of the twist angle.[425] Two mechanisms are responsible for this effect: (i) the growing momentum mismatch between the conduction/valence band states at K which are involved in the formation of the ILXs, and (ii) the localization effect of excitons in the moiré potential. An experimental signature of moiré potential affecting the spatiotemporal diffusion of ILX is reported in refs 426 and 427 (see Figure 14d).

### 14.1 Challenges, Future Goals, and Suggested Directions

Despite the efforts made in recent years to understand, characterize, and control the dynamics of excitons in TMD heterostructures, much still needs to be done to clarify certain aspects and answer new questions. Here we present possible future directions in this research field, and novel experimental techniques that can help in this quest.

#### 14.1.1 *Effect of Moiré Potential on Exciton Dynamics*

The measurements of exciton dynamics in TMD heterostructures as a function of the twist angle, mentioned earlier, have been performed when little was known about the nature of moiré excitons. Only recently a number of studies have shed more light on the nature of these excitons, clarifying how the transition from localized to nonlocalized states can drastically alter their absorption and emission spectra. This was made possible by successfully imaging the moiré periodic domains *via* electron microscope techniques and by correlating them to the optical response of the excitons.[428] Recent studies have also shown that rigid lattice moiré pattern models cannot be applied for small twist angles, and atomic reconstruction must be considered.[429] Calculations have shown that the rearrangement of the atoms results in the appearance of multiple flat bands which affect the excitonic spectrum.[429]

In light of these new results, novel time-resolved optical spectroscopy experiments will be extremely useful to understand how the dynamics of exciton formation and thermalization change when they are spatially confined within the moiré potential or when they are in a delocalized regime. Since spatial localization of excitons is supposed to enhance their mutual interaction, optical pump-probe experiments as a function of the incident fluence will enable to quantify the role of the Auger-type exciton-exciton annihilation process and understand how this process depends on the confinement. Another dynamical aspect of moiré excitons that deserves to be considered is the origin of their spatiotemporal diffusion, which might not be fully compatible with a simple model of exciton hopping between moiré sites. Systematic studies of the role of phonon-mediated, exciton-exciton scattering processes in the diffusion dynamics are needed. The depth of moiré potential and the degree of exciton localization within the potential is another parameter that might affect the diffusion dynamics. Shallow-trapped excitons are expected to diffuse more freely on the crystal than deep-trapped excitons. Experimental verifications of this prediction are still missing.

#### 14.1.2 *Ultrafast Dynamics of Correlated Electronic States in Moiré Superlattices*

One of the big revolutions brought by the advent of moiré materials is the appearance of strongly correlated states of electrons in the periodic potential created by the moiré superlattice. When the Coulomb repulsion energy $U$ of charge carriers is much larger than the kinetic energy term $W$ related to the tunneling between the potential pockets, the system can be described by the Fermi-Hubbard model. This model predicts Mott insulating phases and high-temperature superconductivity.[86] Strongly correlated insulating states have been observed in different TMD moiré systems when $U>W$, such as graphene-gated heterobilayers of $WSe_2/WS_2$, $MoSe_2/WS_2$ and $MoTe_2/WSe_2$, as well as twisted homobilayers of $WSe_2$ and $MoSe_2$.[91] In such systems, the signatures of the electronic Mott insulating phase can be optically probed by monitoring the exciton photoluminescence or reflectance spectrum changing the doping level around integer filling factors of the moiré lattice sites.

Although this new research field is sparking tremendous interest in the scientific community, to date only a handful of studies have aimed at understanding and optically manipulating the ultrafast dynamics of strongly correlated states in moiré materials, with several fundamental questions remaining still open. Very recently, temperature-dependent pump-probe optical experiments on a gated $WSe_2/WS_2$ heterostructure revealed the prominent role of electron-phonon coupling in the collapse of the Mott insulating phase, occurring within few picoseconds, previously mainly attributed to electron–electron interactions. These findings also suggest the presence of a polaron lattice in the TMD moiré system.[430] However, the processes behind the recovery of the Mott insulating state after photoexcitation remain still unclear. Because of the possibility to tune the intercell tunnelling term by the twist angle and the average filling factor of holes and electrons in the moiré cell by the electrostatic gating, moiré TMD heterostructures are an ideal and a novel framework for studying the out-of-equilibrium physics of different correlated states by transient optical spectroscopy techniques. We expect several studies in the future on this topic.

### 14.1.3 *Advances in Ultrafast Spectroscopy Techniques*

Optical pump-probe spectroscopy is a mature technique that has been extensively used to unveil the exciton dynamics in TMD heterostructures. An extension of this technique which employs two phase-locked femtosecond excitation pulses is 2D electronic spectroscopy (2DES). Although less frequently used, this technique has the advantage to disentangle homogeneous and inhomogeneous excitonic linewidth and to combine high temporal and spectral resolution, allowing to retrieve important dynamical properties like exciton and spin-valley coherence time, exciton many-body effects and coherent coupling between excitons.[431] The possibility to extend 2DES improving the spatial resolution to the diffraction limit (in a collinear geometry) will provide the opportunity to study interlayer coherent carrier and energy transfer and the coherent coupling involving intralayer, interlayer and hybridized excitons in aligned and twisted TMD heterostructures.

In close connection with 2DES experiments, multiple pump-probe optical techniques based on the use of delay-controlled phase-locked few-cycle excitation pulses can allow the manipulation of the valley polarization on a timescale shorter than the exciton dephasing time, opening the way to valleytronics in the petahertz regime, as recently proposed in ref 432. In this context, sub-cycle switching of the valley polarization has been recently achieved by photoexcitation with strong THz pulses, paving the way to light-wave valleytronics in TMDs.[433]

As mentioned before, tr-ARPES has been used to further extend the out-of-equilibrium excitonic landscape to finite-momentum (i.e., optically dark) excitons. We envisage that further developments of this technique will add novel insights and bring a deeper understanding of the exciton physics in TMD heterostructures. By improving the spatial resolution (i.e., $< 1$ μm) and the energy resolution (i.e., $< 20$ meV), energy and momentum space studies of exciton dynamics in higher quality TMD heterostructures with moiré superlattice effects will become feasible. High energy resolution combined with 100 fs temporal resolution will also allow one to detect the dynamics of many-particles excitonic states with momentum resolution and to resolve the spin–orbit energy splitting of the conduction band states at the K valleys: this will enable to unambiguously disentangling spin-flip and spin preserving intervalley scattering processes.

Finally, it is worth noting that all the pump-probe spectroscopy measurements discussed so far are performed in the far-field regime, where the limited spatial resolution prevents singling out the response of an individual moiré cell. In this context, the field of ultrafast spectroscopy has recently witnessed disruptive developments entering the near-field regime and combining atomic spatial resolution and sub-picosecond temporal resolution. Such recently developed near-field techniques, including ultrafast scanning tunneling spectroscopy, are extremely promising and could open novel pathways to explore the moiré physics in TMD heterostructures and to image the dynamics of localized moiré excitons in real space.[191]

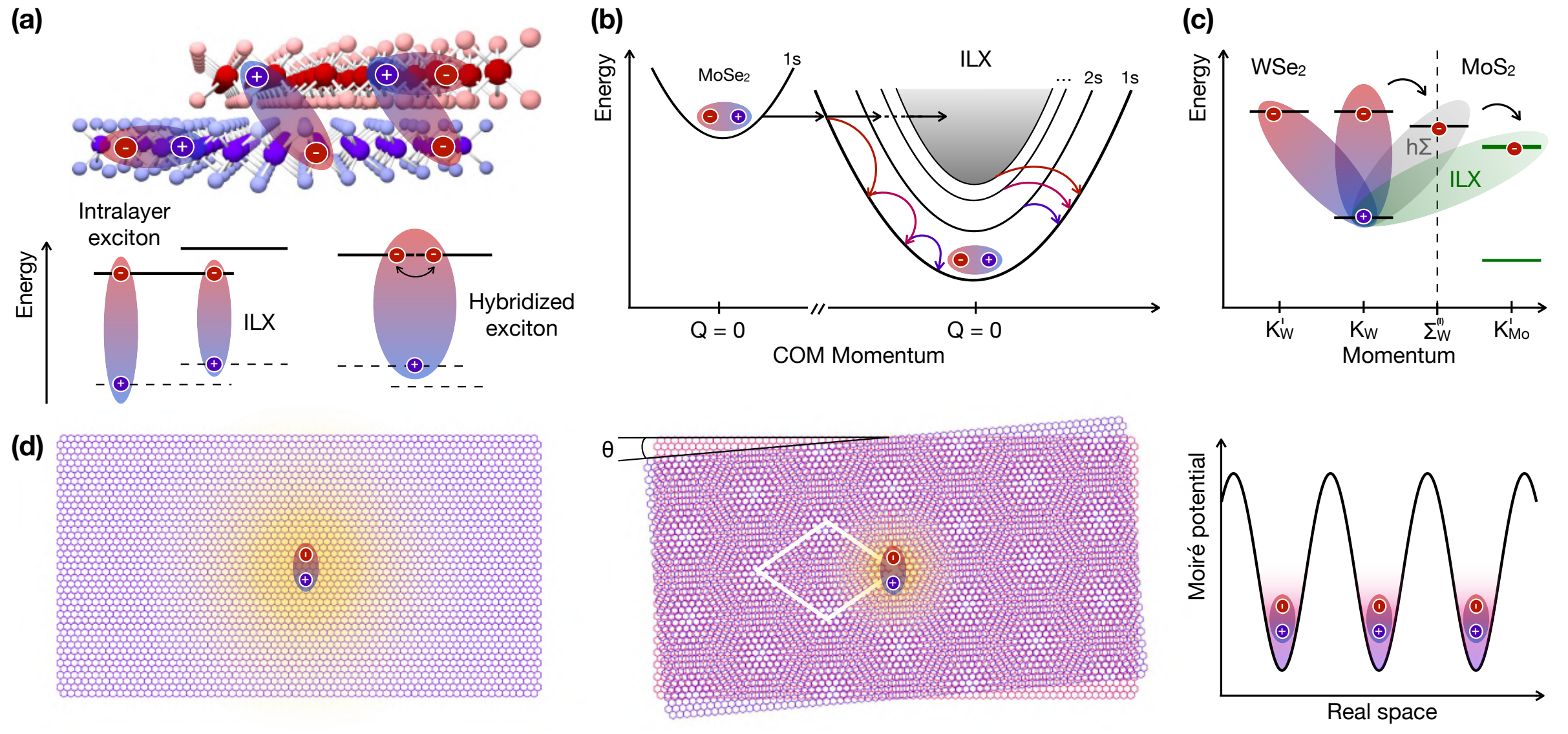

**Figure 14.** Ultrafast optical processes in vdW heterostructures. (a) Sketch of a TMD heterostructure illustrating intralayer, interlayer, and hybrid excitons, along with the simplified representation of the heterostructure band structure with the different band alignments hosting different excitonic species. Hybrid exciton results from resonant interlayer electron tunneling. Adapted with permission from ref 434 (Copyright 2021, American Chemical Society). (b) Graphical representation of the interlayer-exciton (ILX) formation process in a $MoSe_2/WSe_2$ heterostructure upon resonant photoexcitation of the bright exciton in $MoSe_2$. Hole transfer results in the formation of a hot ILX population with finite momentum Q. Hot ILXs quickly relax to the optically bright ILX ground state by exchanging energy and momentum with phonons. Adapted from ref 419 (Creative Commons CC BY). (c) Sketch of the single-particle valleys of $MoS_2/WSe_2$ heterostructures at some high-symmetry points (i.e., K and Σ). After resonant photoexcitation of the intralayer exciton in $MoS_2$, an ILX at the K valleys forms *via* intermediate phonon-mediated scattering with hybrid exciton (hΣ). Adapted with permission from ref 423 (Copyright 2023 IOP Publishing). (d) Exciton diffusion in TMD heterostructures with and without moiré superlattice (left and center panels). θ is the twist angle between the two monolayers. The moiré potential (sketched in the right panel) tends to localize ILXs and to hinder their diffusion.

## 15. QUANTUM-COHERENT COUPLING IN TWO-DIMENSIONAL MATERIALS PROBED BY ULTRAFAST MULTIDIMENSIONAL SPECTROSCOPIES

**Daniel Timmer,[1] Antonietta De Sio,[1,2] and Christoph Lienau[1,2,*]**
[1]Institut für Physik, Carl von Ossietzky Universität, 26129 Oldenburg, Germany
[2]Center for Nanoscale Dynamics (CeNaD), Carl von Ossietzky Universität, 26129 Oldenburg, Germany
***Corresponding author.** Email: christoph.lienau@uni-oldenburg.de

### 15.1. Current State of the Art

Transition metal dichalcogenide monolayers present a fascinating nanoscale laboratory for probing coherent couplings and many-body interactions (MBI). Already their linear optical properties are governed by strong Coulomb correlations, forming excitons with binding energies of up to several hundred meV, stable even at room temperature.[313,435] Their unusual band structure and broken inversion symmetry couples spin and valley degrees of freedom. This makes the optical absorption spectra of different valleys sensitive to the circular polarization of the incident light, opening up the field of valleytronics. Weak dielectric screening enhances the formation of charged quasiparticles with properties that can be controlled by chemical doping and/or electrostatic gating. This weak screening also makes excitons in TMD monolayers particularly sensitive to strain-induced variations in the potential landscape, resulting in exciton localization and the creation of single photon sources[436] with direct applications in quantum communication, metrology, and computing.

Coherent couplings and MBI govern not only the linear but also the nonlinear optical properties of TMDs. Their optical nonlinearities largely arise from MBIs such as excitation-induced dephasing (EID) and resonance energy shifts (EIS).[437] For sufficiently low temperatures, biexcitons and charged exciton complexes appear as new resonances in the nonlinear spectra. Importantly, coherent couplings among quasiparticles can manifest as distinct dynamic oscillations in time-resolved nonlinear experiments. These rich inherent physical phenomena are largely affected by the stacking of multiple TMD layers and by introducing external degrees of freedom. Of particular relevance are couplings to plasmonic nano- or dielectric microcavities, resulting, for example, in polariton lasing or polariton condensation at room temperature. Moreover, TMD bilayers have recently emerged as a tunable moiré system for studying and designing correlated electron physics. In such bilayers, the moiré potential leads to the formation of triangular electron or hole spin lattices exhibiting correlated magnetic phases that can be controlled *via* an external magnetic field.

An important experimental tool to unravel coherent couplings and MBI in TMD-based monolayers and heterostructures is ultrafast multidimensional coherent spectroscopy, specifically 2D electronic spectroscopy (2DES).[437,438] In its simplest implementation, 2DES is a powerful extension of pump-probe spectroscopy that replaces the impulsive excitation of the system under investigation by an excitation with a phase-locked pair of two ultrashort pulses. As in multidimensional nuclear magnetic resonance spectroscopy, 2DES results in energy-energy maps (Figure 15a) that correlate the excitation and detection as a function of the waiting time $T$ between the second pump and the time-delayed probe pulse. To generate these maps, differential absorption or reflection spectra are measured as a function of the time delay $\tau$, the coherence time, between the two pump pulses. Fourier transform of this set of spectra along $\tau$ provides the excitation axis of the 2DES map. Here, two distinct excitation scenarios should be distinguished. If the first pump and the probe pulse interact nonlinearly with the same resonance in the system, this gives rise to a diagonal peak in the map. If, in contrast, first pump and probe interact with different resonances in a nonlinear system, this creates a cross peak provided that the two resonances are interacting. A variation of the waiting time provides access to the quantum dynamics. Again, two different scenarios appear. If both pump pulses interact with the same resonance, the waiting time evolution of the diagonal and cross peaks probes incoherent population relaxation and energy transfer dynamics. If, however, the two pulses interact with different resonances, this gives rise to a coherent quantum beat that oscillates at their difference frequency during the dephasing time of the induced coherence. The induced coherent oscillations of cross and diagonal peaks are distinct signatures of coherent couplings in the investigated quantum system.[439]

The first 2DES of TMD monolayers has been reported by Moody *et al.*[440] in 2015. The work makes use of the ability of 2DES to separate homogeneous and inhomogeneous broadening, revealing narrow homogeneous exciton linewidths in $WSe_2$. This linewidth is shown to be highly sensitive to EID induced by exciton-exciton interactions. This central role of MBI for the optical nonlinearities of TMDs has been further confirmed by semiconductor Bloch equation simulations[441] and recent pump-probe experiments.[413] Zero-quantum 2DES with circularly polarized pulses has directly measured valley coherence of excitons in $WSe_2$, persisting for ~100 fs.[442] Coherent couplings between excitons and trions in monolayer $MoSe_2$ have directly been identified by resolving coherent oscillations of 2DES cross peaks for waiting times of up to 150 fs.[443] The intrinsic trion coherence time and trion valley coherence have also been explored in $MoSe_2$.

Broadband 2DES with pulses covering simultaneously A and B exciton resonances, combined with valley-selective excitation, have demonstrated strong intravalley exchange interaction and coherent mixing between A and B excitons as a consequence of strong spin–orbit couplings in $MoS_2$, confirmed by first-principles calculations of the Bethe-Salpeter equation.[444] A similar experimental approach using helicity-resolved excitation has revealed cross peak formation due to a Coulomb-induced, Dexter-like, intervalley coupling.[445] Very recently, such cross peaks have also been resolved in 2DES studies of $WS_2$, where A and B excitons are detuned by almost 400 meV.[446] Here, coherent intervalley couplings give rise to pronounced 11.5 fs oscillations in pump-probe, persisting for 100 fs at room temperature. Semiconductor Bloch equation simulations confirm these direct signatures of coherent intervalley couplings.

First 2DES studies of TMD bilayers have resolved ultrafast sub-100-fs electron and hole transfer processes in large area $MoS_2/WS_2$ heterostructures.[416] Pioneering 2DES imaging experiments with sub-µm resolution on $MoS_2/WSe_2$ heterostructures have shown that – despite noticeable local strain – interlayer couplings and exciton lifetimes are mostly robust across the sample.[447]

Helicity-resolved double quantum 2DES has allowed for directly observing biexcitons in monolayer $WS_2$ by spectrally separating them from the correlated unbound two-exciton state. [448] A biexciton binding energy of 26 meV has been deduced. In electrostatically doped $MoS_2$ monolayers, both attractive and repulsive Fermi polarons, excitons that are coherently dressed by the surrounding Fermi sea of unbound electrons and/or holes, and their cross peaks have been observed.[449] The dependence of the polaron energy, oscillator strength and linewidth on the Fermi energy of the surrounding electrons has been compared to polaron theory.

Two-dimensional fluorescence-detected micro-spectroscopy has allowed to investigate exciton–phonon coupling in single-layer $MoSe_2$. From a coherent oscillation of the 2DES amplitude with a period of ~2 ps a Huang-Rhys factor of ~1, larger than in most other inorganic semiconductor nanostructures, has been deduced.[450] In a first 2DES study of a microcavity with an embedded $WSe_2$ vdW heterostructure at room temperature, multiple polariton branches have been observed.[451] These arise from exciton-photon-phonon hybridization in the cavity, consistent with the results of a vibronic polariton Hamiltonian model.

### 15.2 Challenges and Future Goals

This short summary of the state-of-the-art highlights the enormous potential of 2DES for uncovering coherent couplings and MBI in TMDs. Such experiments provide a unique tool for probing the time evolution of the density matrix in complex quantum materials, even in the presence of fast decoherence processes in the 100-fs range. This requires broadband and highly sensitive 2DES setups, providing sufficient time resolution. Currently, suitable 2DES systems exist in a handful of laboratories worldwide and, so far, have mainly been applied to studies of TMD monolayers. Extensions of such experiments to (twisted) bilayers, gated and doped samples, and polaritonic systems are just emerging and form important future goals. In particular, 2DES studies of strong coupling phenomena and coherent spin dynamics in magnetic heterostructures hold promise for significant new insight.

Ideally, such experiments should probe quantum dynamics across all relevant regions of the TMD band structure. This presents a substantial experimental challenge, as it requires not only valley-selective (Figure 15b) but also fully momentum-resolved studies of charge carrier dynamics with a time resolution much below the decoherence time. Since the optical and electronic properties of existing TMD samples are substantially affected by local strain variations and defects, it will be crucial to add high spatial resolution to ultrafast 2DES. This will be instrumental for isolating the dynamics of localized quantum emitters and therefore provide valuable insight into the electronic and optical properties of single and entangled photon sources based on TMDs.

The emergence of moiré physics and twistronics (Figure 15c) highlights the need for optical studies with a resolution beyond the superlattice period. With these systems emerging as novel simulators of Hubbard physics, coherent studies of the spin dynamics may offer qualitatively new insight into the underlying coherent couplings. High spatial resolution experiments of quantum dynamics of polaritonic systems are another key future goal for exploring and exploiting the strong coupling between TMDs and external resonator modes (Figure 15d). Such studies may shed light on nonequilibrium dynamics relevant for polariton condensation and lasing.

Plasmonic nanoresonators hold great promise due to the possibility of creating spatially highly localized electromagnetic fields with high local field enhancement. Their strong coupling to TMDs will therefore create spatially strongly inhomogeneous polariton systems enabling real-space quantum transport of optical excitations which is still largely unexplored so far. Importantly, in addition to exploiting vacuum field fluctuations of such resonators to create new quasiparticles, also strong external optical fields, spanning from the THz to the visible and UV range, offer new possibilities for transiently altering the electronic structure of TMD materials. This could facilitate inducing and controlling new types of light-driven quantum dynamics.

### 15.3 Suggested Directions to Meet These Goals

To reach the goals outlined above, it will be crucial to further advance 2DES beyond the state-of-the-art. Further improvements in spectral bandwidth are important to cover a broader range of elementary excitations. The recent discovery of UV interlayer excitons, for example, calls for 2DES experiments covering the full visible range[452] and extending into the UV range. Current 2DES experiments have resolved coherent coupling with energies of up to ~ 400 meV corresponding to oscillation periods in the 10-fs range, which represents the upper limit of what can be reached in current 2DES setups. Improved time resolution will be essential to achieve broader spectral coverage. Combinations of supercontinuum white light generation with light field synthesis[453] and/or pulse shaping have the potential for designing pulses with durations of <3 fs. Yet such pulses have not been applied in 2DES experiments so far.

Equally important are efforts to enhance sensitivity and reduce the acquisition time of 2DES experiments. Here, the implementation of recent developments in ultrafast dual- and multi-comb spectroscopy for ultrafast 2DES seems promising, as they enable rapid optical delay scans without moveable parts and noise suppression.[454,455,456]

Not only the recording but also the interpretation of experimental 2DES data needs further development. This includes, for example, the use of rapidly developing artificial intelligence methods for spectral analysis as well as of automated schemes to compare experiments and simulations based on model Hamiltonians. We emphasize that high-quality 2DES measurements provide an important benchmark for theoretical descriptions of nonequilibrium quantum dynamics in TMDs, using, for example, semiconductor Bloch equation calculations. Improving the merger between theory and experiment will therefore be crucial for future advancements. This is especially true for the development of *ab initio* descriptions which still are in a very early stage.[457]

So far, all the discussed experiments have used all-optical 2DES detection schemes. Time- and angle-resolved photoelectron spectroscopy has provided crucial insight into momentum-resolved incoherent exciton dynamics[458] and interlayer exciton formation.[422] Integrating such momentum-resolved photoelectron detection schemes with ultrafast 2DES may give intriguing new insight into the coherent quantum dynamics in TMDs and their heterostructures. Equally promising are combinations of 2DES with high-resolution microscopy techniques. This includes both further improvements of far-field 2DES microscopy[447] and in particular also the development of entirely new detection schemes providing nanometer-scale spatial resolution and few femtosecond temporal resolution. A promising step in this direction is the coherent 2D optical nanoscopy technique reported in ref 459. More generally, combining 2DES with action-based detection schemes, such as, for example, photocurrent or photoelectron emission microscopy, or with the all-optical readout of the optical field emitted from the junction of a scanning probe tip, may bring 2DES microscopy of TMD materials to the nanometer scale. These advancements will enable deeper and more detailed exploration of spatio-temporal light-driven quantum transport in 2D materials and heterostructures.

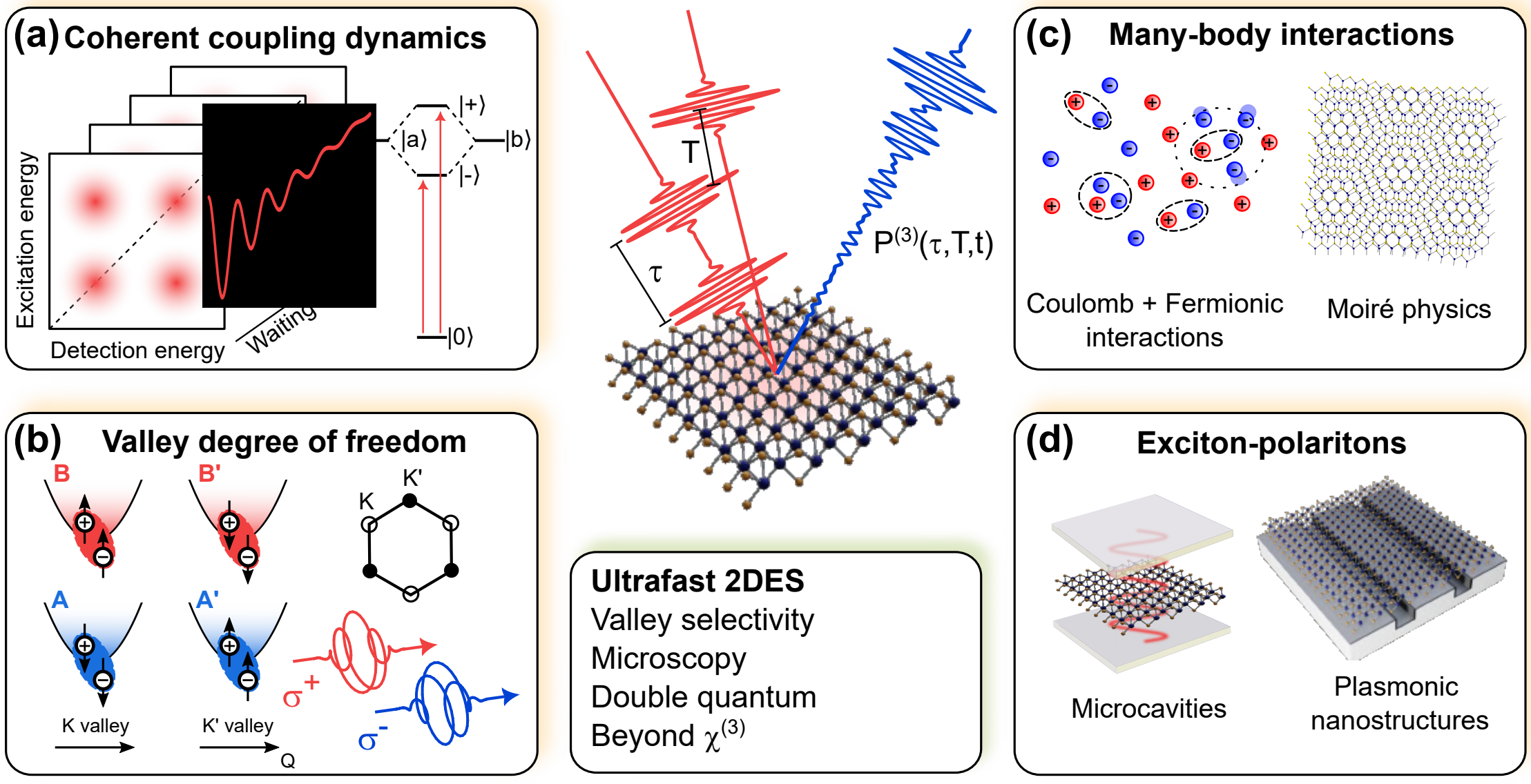

**Figure 15.** Ultrafast 2D electronic spectroscopy of 2D materials. The sample is resonantly excited with a phase-locked pulse pair with delay $\tau$. A probe pulse at waiting time $T$ induces a nonlinear polarization in the material.

(a) Coherent couplings induce temporally oscillating diagonal and cross peaks in the 2DES map. (b) Valley-selective 2DES of A and B excitons using circularly polarized light pulses. (c) Intrinsic many-body interactions result in a rich variety of quasiparticles and dominate their optical properties. Such interactions can be tuned by moiré potentials. (d) New quantum states are created by hybridization with fields in external microcavities and plasmonic nanoresonators.

## 16. 2D MATERIALS-BASED NONLINEAR PHOTONICS

**Nianze Shang,[1] Hao Hong,[2] F. Javier García de Abajo,[3,4] Kaihui Liu,[2] and Zhipei Sun[1,*]**
[1]QTF Centre of Excellence, Department of Electronics and Nanoengineering, Aalto University, Espoo 02150, Finland
[2]State Key Laboratory for Mesoscopic Physics, Frontiers Science Center for Nano-optoelectronics, School of Physics, Peking University, Beijing, 100871, China
[3]ICFO-Institut de Ciencies Fotoniques, The Barcelona Institute of Science and Technology, 08860 Castelldefels, Barcelona, Spain
[4]ICREA-Institució Catalana de Recerca i Estudis Avançats, Passeig Lluís Companys 23, 08010 Barcelona, Spain
***Corresponding author.** Email: zhipei.sun@aalto.fi

### 16.1 Introduction

Two-dimensional materials with third-order and higher-order optical nonlinearities, complementing the capabilities of second-order nonlinear effects, represent a promising frontier in optical science. Despite their atomic-scale thickness, these materials—such as graphene, TMDs, black phosphorus (BP), and their heterostructures[460,461,462]—exhibit remarkable nonlinear optical susceptibilities. Fully harnessing these advanced nonlinearities holds the potential to drive revolutionary advances in next-generation photonic devices, enabling applications across light emission, all-optical modulation, frequency conversion, sensing, computation, and quantum technologies.

### 16.2 Current State of the Art

Unlike second-harmonic generation, which typically requires non-centrosymmetric structures, third- and high-odd-order nonlinear susceptibilities (i.e., $\chi^{(3)}$, $\chi^{(5)}$, …, $\chi^{(2n+1)}$) of materials do not necessitate symmetry breaking, making them broadly applicable for 2D material-based photonic applications. For example, the real part of $\chi^{(3)}$ supports several coherent optical nonlinear frequency generation processes such as third-harmonic generation (THG) and four-wave mixing (FWM). In contrast, the imaginary part of $\chi^{(3)}$ is associated with nonlinear absorption processes arising from light–matter interactions, such as two-photon absorption and saturable absorption. Indeed, recent advancements in various high-order nonlinear optics processes with 2D materials have highlighted their unique properties, including significant high-order susceptibilities, ultrafast response time, and large tunability.

#### 16.2.1 *Third-Harmonic Generation*

THG, primarily governed by the real part of $\chi^{(3)}$, is among the most extensively studied third-order nonlinear processes and has proven to be a versatile, effective, and non-destructive method for characterizing 2D materials. It enables direct measurement of the nonlinear susceptibilities of these materials, providing an approach to discover and evaluate novel materials with high nonlinear efficiency. Moreover, the sensitivity of the $\chi^{(3)}$ tensor to both internal factors (e.g., defects, doping, band structures, excited states) and external conditions (such as substrates, strain, and electric or magnetic fields) enhances the ability of THG to investigate the fine structures and phase transitions of 2D materials,[463] and opens new avenues for efficient frequency generation.[464]

#### 16.2.2 *Four-Wave Mixing*

FWM is a process of third-order nonlinear generation in which photons of two or three different energies are combined to produce one or two new photons. This process has diverse applications in areas such as wavelength conversion, THz generation, optical signal amplification, and imaging. Research on FWM in 2D materials began with graphene[465] and gained momentum as these materials were integrated with various photonic structures, including microring resonators,[466] optical fibers,[467] waveguides,[468] and others. These integrations have unlocked significant potential in 2D materials for supercontinuum generation and frequency comb applications.[469]

16.2.3 *Two-Photon Absorption (TPA)*

Governed by the imaginary part of $\chi^{(3)}$, TPA in 2D materials has emerged as a pivotal phenomenon in the study of their nonlinear response. Taking monolayer $MoS_2$ as an example, TPA is observed to be up to $10^3$ times stronger than in bulk materials, which is attributed to the strong nonlinearity under reduced dimensionality.[470] Just like $MoS_2$ and other TMDs, the strong optical responses and tunable bandgaps make TPA in 2D materials tailorable for applications in imaging systems, ultrafast spectroscopy,[471] and photonic detection applications.[472]

16.2.4 *Saturable Absorption (SA)*

SA in 2D materials is typically explained by Pauli blocking. This phenomenon has been observed across a variety of 2D materials, from graphene to various TMDs, BP, and their heterostructures. Consequently, an increasing number of 2D material-based saturable absorbers have been developed for mode-locked or Q-switched lasers covering the visible to mid-IR ranges,[473] as well as in all-optical modulators.[474]

16.2.5 *High-Harmonic Generation (HHG)*

Higher-order nonlinear processes, such as HHG, have been less explored due to limitations in pump energy and wavelength range for excitation pulses. However, HHG plays a crucial role in generating deep UV wavelengths and attosecond pulses. Graphene has demonstrated remarkable HHG efficiency, attributed to its ultrafast carried relaxation time, which aligns with the THz wave periods used as excitation pulses. HHG effects up to the sixteenth order have also been observed in monolayer TMDs and various members of the 2D materials family.[475,476,477] The strong many-body interactions and unique electronic properties of 2D materials position them as promising candidates for HHG applications, prompting ongoing research into their potential for higher-order nonlinearities.[478]

## 16.3 Challenges and Future Research Directions

16.3.1 *Theoretical Challenges*

One of the primary challenges in nonlinear optics with 2D materials is the limited understanding of their nonlinear optical properties. Accurately modeling nonlinear interactions in 2D materials is complex, and hence, high-order nonlinearity theories become increasingly complex. In addition, traditional nonlinear optical theories often rely on bulk material assumptions that do not readily apply to atomically thin layers. In 2D materials, quantum-confinement, surface, edge, and doping effects become significantly more pronounced, altering their optical behaviors in a non-negligible fashion. Additionally, the presence of many-body interactions and strong excitonic effects introduces further complexities that require advanced computational approaches beyond conventional mean-field theories. These theoretical limitations hinder the predictive modeling of nonlinear responses, complicating efforts to optimize 2D materials and their versatile, artificially stacked heterostructures for targeted applications. Overcoming these obstacles will require the development of novel theoretical frameworks and computational methods that accurately account for the unique quantum and multi-body interactions in 2D systems, ultimately enabling the design of high-performance, application-specific nonlinear optical devices.

16.3.2 *Designing 2D Materials with Higher Nonlinearity*

Various computational methods, such as density functional theory and the Bethe-Salpeter equation, have been extensively employed to guide the prediction of band structures and nonlinear optical coefficients in artificially designed 2D materials, incorporating many-body effects. The progress of machine learning and artificial intelligence (AI) technology makes it even more productive to identify or predict potential candidates of nonlinear 2D materials with the large databases developed so far. Further, advanced experimental approaches, such as atomic engineering to create Janus structures, ion intercalation, and heterostructure design, offer promising routes to develop hybrid materials with high nonlinear efficiency that do not exist in nature. With ongoing experimental measurements, we can anticipate that more 2D materials will be discovered for their high nonlinear optical properties, which will then be integrated with a theoretical framework to deepen our understanding of these materials. As illustrated in Figure 16, the short-term goal is to uncover previously unknown trends that can help predict the nonlinear properties of 2D materials, facilitating the discovery of high-performance materials. Ultimately, we aim to artificially create new nanomaterials for nonlinear devices that demonstrate superior performance.

### 16.3.3 *Light–Matter Interaction Enhancement*

A practical challenge in nonlinear 2D photonics arises from the atomic thickness of 2D materials, which as extensively addressed in previous literature, limits their associated light–matter interaction. With interaction lengths typically around or even less than several nanometers, the resulting nonlinear signal output is constrained despite the high nonlinear susceptibility per unit volume that 2D materials possess. Additionally, the nature of nonlinear processes leads to weaker intensity for higher-order nonlinear signals, compounded by the overall low output. This makes it difficult to implement 2D materials in real photonic circuits that require multi-device communication or cascade structures. One of the key objectives in decades of research on nonlinear 2D photonics is to enhance the nonlinear optical conversion efficiency of 2D materials (e.g., stacking engineering, $\chi^{(n)}$ engineering, field enhancement, Figure 16).

One strategy to enhance light–matter interactions leverages the sensitivity of 2D materials to lattice and electronic structures, enabling the engineering of their nonlinear properties through mechanical, electrical, or optical methods. For example, strain engineering, carrier injection,[479] heterostructure construction—such as 2D-2D, 2D-1D, or 2D-0D moiré superlattices[3,480] with interlayer twisting angles[481]—, and optical techniques like laser writing. It can be foreseen that novel strategies in material, chemical, mechanical, thermal, or other levels will be developed in the future. Another effective approach to increasing light–matter interactions in 2D material photonic devices is through field-intensity enhancement techniques like plasmonic resonances and resonant-microcavity embedding.[482] Integration with new emerging photonic structures—such as photonic crystals, metasurfaces, and bound states in the continuum (BIC)—will also contribute to this objective and create new paradigms for photonic applications.[483] Finally, increasing the light–matter interaction length is the most well-studied and practical approach. Two directions can be followed: (i) In the $x - y$ in-plane, integrating 2D materials with optical fibers, waveguides, photonic crystals, and microring resonators can be highly beneficial.[484,485] Mass-production methods, such as chemical vapor deposition (CVD), can further enhance the scale of these hybrid 2D material photonic structures, leading to improved nonlinear conversion efficiency. (ii) In the $z$ out-of-plane direction, using artificial stacking methods or optimizing stacking phases during growth, it is possible to create thicker 2D material crystals with a thickness of hundreds of nanometers or micrometers, high nonlinear efficiency, and novel phase matching conditions.[486] This trend toward thicker 2D materials is expected to play a significant role in developing efficient photonic devices.

### 16.3.4 *Integration Challenges*

Integration of 2D materials with various photonic devices offers a dual benefit: this integration not only enhances light–matter interactions, as previously discussed, but also maximizes the practical utility of 2D materials in advanced photonic applications. The mature approach of transferring mechanically exfoliated 2D material flakes to photonic structures such as optical waveguides, micro-fibers, and ring-resonators yields successful enhancement of nonlinear performance of 2D materials but is still limited to micrometer-scaled devices. Using CVD-grown 2D materials instead of exfoliation can lead to a significant increase in the device scale accompanied by stronger nonlinear output. Pioneering demonstrations have been made in graphene or TMD-grown photonic crystal fibers, but the precise control of the 2D materials quality, such as uniformity, orientation, and defect density, remains a challenge for future research (e.g., hybrid integration in Figure 16). Designing 2D material-compatible novel photonic structures, like BICs, will also benefit 2D photonics integration. Nevertheless, the integration of 2D materials into photonic circuits can enable novel devices for signal amplification and lasing, contributing to the miniaturization and enhancement of existing technologies. The tunable nature of 2D materials also allows for adaptive optical components, which can dynamically adjust their properties in response to external stimuli, paving the way for smart photonic systems. We believe that improvements in fabrication and integration techniques of 2D materials will continuously broaden their application in photonics and optoelectronics, paving the way for innovations in computing, signal processing, and quantum information science.

16.3.5 *Application Perspectives*

Due to their exceptional high-order nonlinearity, tunable electronic structures, and high damage thresholds, 2D materials are transforming certain applications in photonic devices. For example, 2D materials with outstanding saturable absorption like graphene, TMDs, and BP can serve as mode-lockers in ultrafast lasers. Looking forward, 2D materials and their engineered heterostructures open exciting possibilities for developing compact, tunable, and highly efficient ultrafast laser systems across a broad spectral range, making them invaluable for emerging technologies in quantum communications, light detection and ranging (LiDAR), and ultrafast spectroscopy.

Up until now, attosecond pulses have primarily been generated through HHG processes. Recently, the detection of HHG in atomically thin 2D materials highlights the potential to explore strong-field and attosecond phenomena in lower-dimensional materials. Moreover, recent findings indicate that strong-field-driven phenomena in nanoscale structures often have sub-cycle durations, offering insights into collective electron dynamics and enabling the generation of optical-field-driven currents on an attosecond timescale. Currents induced in graphene by few-cycle optical pulses can be precisely controlled on the attosecond scale,[487] suggesting that field-driven processes, including HHG in graphene and other 2D materials, hold promise for a variety of novel attosecond-science applications, such as sub-cycle photoemission, attosecond metrology, and attosecond control of electronic processes.

Efficient HHG and ultrafast dynamics of 2D materials also make them an ideal platform for new emerging petahertz electronics because of the efficient responses to petahertz-frequency electric fields (1 petahertz = $10^{15}$ Hz). This ability makes such materials particularly promising for faster electronics beyond the current THz range. Materials like graphene, with its zero-bandgap and linear dispersion, are particularly attractive because they can exhibit broadband, high-speed responses across the visible and near-IR spectral ranges, enhancing their utility in petahertz electronics.[488]

Besides the applications mentioned above, we anticipate that high-order nonlinearities of 2D materials will also play significant roles in the field of biosensing, all-optical computing, white light emission, quantum photonics, and numerous interdisciplinary fields, pushing the frontier of 2D photonics forward and beyond.

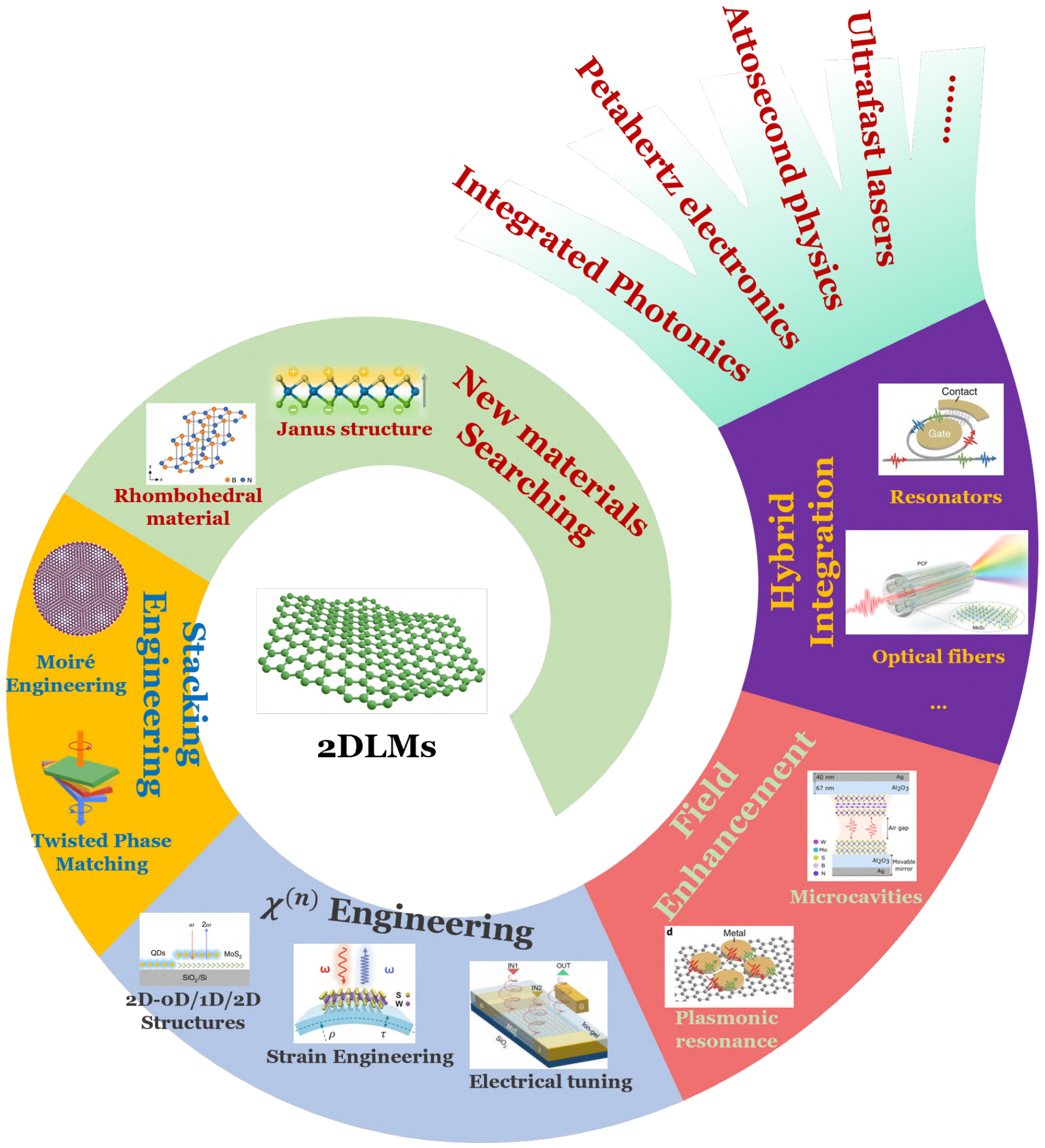

**Figure 16.** The representative path toward next-generation photonics based on high-order nonlinearities in 2D layered materials (2DLMs). Stacking engineering, $\chi^{(n)}$ engineering, field enhancement, and hybrid integration methods can be used for the overall enhancement of the nonlinear conversion efficiency of 2D materials. With new emergent 2D materials exhibiting better nonlinear performance, disruptive advances and revolutionary findings can be foreseen in integrated photonics, attosecond and strong-field physics, petahertz electronics, and ultrafast lasers.

## 17. NONLINEAR OPTICAL GENERATION OF ENTANGLED QUANTUM LIGHT IN 2D MATERIALS

**P. James Schuck,[1,*] Chiara Trovatello,[1,2] Giulio Cerullo,[2] Lee A. Rozema,[3] Philip Walther,[4] Andrea Alù,[5] Andrea Marini,[6] Michele Cotrufo,[7] Milan Delor,[8] Raquel Queiroz,[9] D. N. Basov,[9] X.-Y. Zhu[8]**

[1]Department of Mechanical Engineering, Columbia University, New York, New York 10027, United States
[2]Dipartimento di Fisica, Politecnico di Milano, Milano 20133, Italy
[3]University of Vienna, Faculty of Physics, Vienna Center for Quantum Science and Technology (VCQ) & Research Network Quantum Aspects of Space Time (TURIS), 1090 Vienna, Austria
[4]University of Vienna, Faculty of Physics, Vienna Center for Quantum Science and Technology (VCQ) & Research Network Quantum Aspects of Space Time (TURIS), 1090 Vienna, Austria; Christian Doppler Laboratory for Photonic Quantum Computer, Faculty of Physics, University of Vienna, 1090 Vienna, Austria
[5]Photonics Initiative, Advanced Science Research Center, City University of New York, New York, New York10031, United States; Physics Program, Graduate Center of the City University of New York, New York, New York 10016, United States
[6]Dipartimento di Scienze Fisiche e Chimiche, Università degli Studi dell'Aquila, L'Aquila 67100, Italy
[7]The Institute of Optics, University of Rochester, Rochester, New York 14627, United States
[8]Department of Chemistry, Columbia University, New York, New York 10027, United States
[9]Department of Physics, Columbia University, 1150 Amsterdam Avenue, New York, New York 10027, United States
***Corresponding author.** Email: p.j.schuck@columbia.edu

### 17.1 Current State of the Art

Recent work has revealed the exceptional potential of vdW crystals and device architectures for achieving simple, scalable, and tunable entangled photon-pair production[489,490,491,492]—a direction in the field that is both attractive and challenging.[493] Efforts to date have focused on the generation of photon pairs and squeezed states *via* spontaneous parametric down-conversion (SPDC), a nonlinear process that depends on the second-order nonlinear optical susceptibility $\chi^{(2)}$ of the material. $\chi^{(2)}$ can be intrinsically large in vdW systems, and beyond broken inversion symmetry, it is increasingly linked to a nontrivial electronic structure characterized by a large quantum metric.[494,495] Driven by these sizeable nonlinearities, with $\chi^{(2)}$ reaching values ~10-1,000-fold larger than those in crystals utilized for current state-of-the-art devices, the field of nonlinear optics with layered materials has experienced rapid growth. Innovative vdW-based nanostructures can serve as key components in the miniaturization paradigm of quantum light sources, promising previously unattainable functionalities while offering direct, on-chip integration for applications in quantum information, computing, cryptography, spectroscopy, and sensing, all exploiting the use of entangled photons.

Notably, nonlinear optical processes in atomically thin vdW materials can be achieved over ultrabroad operational bandwidths, with phase-matching-free $\chi^{(2)}$ processes such as second-harmonic generation (SHG), optical parametric oscillation (OPO), and amplification (OPA) being demonstrated at the ultimate monolayer (1L) limit.[496,497] Within the context of quantum- and nano-photonic applications, monolayers are appealing thanks to their straight-forward on-chip integrability, easy realization of multifunctionality from vdW interfaces, direct bandgaps allowing large transmittance at IR communication wavelengths, and much-reduced screening of the Coulomb potential. The latter is responsible for the large oscillator strength from tightly bound excitons, strong many-body interactions and quantum phase transitions, and the ability to modulate many of their properties even at ultrafast timescales.[313]

While featuring a large $\chi^{(2)}$ (a thickness-independent material property), monolayers typically suffer from limited nonlinear frequency-conversion and entangled-light-generation efficiencies owing to the surface-like nonlinear interaction. Intense research has been dedicated to enhancing these efficiencies by integrating vdW crystals into photonic structures such as waveguides, resonators, microcavities, fibers, and metasurfaces,[483,498,499,500] as well as by exploiting layered crystal symmetries like the 3R (rhombohedral) crystal polytype that remain non-centrosymmetric even in the bulk.[490,491] For the latter, vdW crystals that are tens to hundreds of nm thick combine substantial optical path lengths with large $\chi^{(2)}$ to provide the high efficiencies required by real applications.

Within the vdW materials family, ferroelectric layered compounds such as niobium oxide dihalides $NbOX_2$ (X = Cl, Br, and I) have emerged as appealing nonlinear crystals thanks to their anisotropy and large $\chi^{(2)}$ (≈ 200 pm/V).[501] $NbOI_2$ and $NbOCl_2$ can undergo a reversible ferroelectric-to-paraelectric phase transition and a ferroelectric-to-antiferroelectric phase transition, respectively, potentially rendering external electric fields an effective tuning knob for nonlinear optical device applications[502]. These crystals have recently enabled the generation of entangled photon pairs at λ = 800 nm from sub-μm-thick flakes[503] *via* SPDC. Evidence of pair generation was provided by measurements of the second-order correlation function $g^{(2)}(\tau)$ and the pump power-dependent coincidence rate. The coincidence rate shows a quadratic dependence on crystal thickness, demonstrating that phase matching constraints do not play any role in the thickness range explored by these researchers, lying within the material coherence length. Very recently, SPDC[490] has been observed in thin 3R-stacked TMD flakes, with researchers demonstrating broadband generation of maximally polarization-entangled Bell states with fidelity as high as 96% at telecom wavelengths.[492]

For all these known materials, further improvements in efficiency necessitate the use of even thicker crystals, where phase matching requirements come into play. In nonlinear optics, the key phase matching parameter is the coherence length, $l_c$, defined as the path length over which the fundamental and converted frequency waves become completely out-of-phase due to material dispersion. In 3R-$MoS_2$, $l_c$ has now been experimentally measured[491] to be ≈ 500 nm at communication-band-relevant wavelengths (for light normally incident to the layered plane). Knowledge of $l_c$ has now enabled the demonstration of so-called periodically poled (PP) TMD stacks, where 3R-TMD slabs of thickness $l_c$ are vertically stacked, alternating the orientation between each constituent slab by 180 degrees (Figure 17a), achieving the quasi-phase-matching condition. Photon-pair generation has now been observed in PP 3R-$MoS_2$ crystals, demonstrating an increase in the SPDC brightness by ~20× compared to a single crystal slab with thickness equal to $l_c$.[490]

The successful demonstration of high-efficiency SPDC over microscopic thicknesses in PP 3R-TMDs is now motivating further research on quasi-phase-matched vdW structures and devices for entangled photon generation, as well as other novel strategies that may enable tunable, dynamic nonlinear quantum light generation with higher efficiencies and smaller footprints.

### 17.2 Challenges, Future Goals, and Suggested Directions to Meet These Goals

Layered 2D compounds and their heterostructures promise to heavily impact the integration of nonlinear optical functionality with nanoscale photonic devices. A notable advantage of vdW materials is their potential ease of integration, as their direct transfer and stacking onto structured substrates are compatible with existing photonic chip fabrication processes.[485] Still, an open challenge remains the development of strategies for optimizing the coupling between vdW structures and universal photonics platforms.[483,498] To date, researchers have successfully integrated 2D materials with photonic crystals, optical fibers, and optical waveguides, increasing the effective interaction length through cavity mode coupling and evanescent field coupling, respectively.[504,505,506,507,508,509,510] However, more efficient light coupling with the nonlinear material will require sophisticated designs, due in part to the large refractive index mismatch that exists between promising nonlinear vdW materials and silicon oxide/nitride optical elements on chip. The large refractive indexes[511] are beneficial in terms of enhancing light confinement within the material, but also demand precision optical elements for optimizing in- and out-coupling.[512]

A promising approach for both optimizing conversion efficiency and streamlining device design is fabricating photonics elements directly from vdW crystals themselves.[490,491,513,514] Efficient frequency doubling within microscopic 3R-TMD waveguides (Figure 17b) has now been demonstrated,[491,512] and thus a near-future goal is the realization of ultracompact SPDC sources utilizing optimized vdW waveguide geometries. In such structures, a key challenge is attaining phase matching, for example, by birefringence or by designing the waveguide geometric dispersion. To this end, recent work has shown that birefringent phase matching can be achieved for SH generation in 3R-$MoS_2$ waveguides, combining large $\chi^{(2)}$ and optical interaction lengths to reach up to few-percent conversion efficiencies. This strategy should be generalizable to SPDC and other vdW crystals since it exploits the natural anisotropy between their in- and out-of-plane optical susceptibilities. Quantifying the out-of-plane dielectric functions and $\chi^{(2)}$ components of promising 2D materials is therefore critical. Measuring these fundamental material tensors will require advanced characterization methods such as high-resolution ellipsometry and scanning near-field microscopies capable of probing the nanostructured photonic elements.

Since the largest $\chi^{(2)}$ values are often reached under resonant conditions (i.e., when the photon energies of the fundamental and/or nonlinear light match the excitonic resonances in TMDs), understanding the trade-off between field enhancements versus absorption losses in these structures is also important. Similar trade-offs have been explored in nonlinear metasurfaces featuring other polaritonic phenomena,[515] and optimal strategies may be devised for the target nonlinear operation.[516] Key questions include: can full frequency conversion, resulting in depletion of the fundamental wavelength, be achieved before losses dominate? Do polariton states play a role? Self-hybridized polariton states are known to exist in vdW crystal slabs,[517,518] and the larger optical density of states afforded by these and other polariton modes could further boost SPDC entangled photon generation rates. Characterizing these modes will be a crucial future direction, enabled by ongoing developments in high-resolution measurement techniques for real time and space visualization of frequency conversion and propagation.[198,519,520]

The use of vdW metasurfaces, illustrated in Figure 17c, and similar resonating structures promises to bypass phase-matching constraints altogether by further reducing the layer thickness required for achieving efficient entangled-photon generation. Metasurfaces are an excellent platform for boosting the efficiencies of processes such as SHG and SPDC within thin (subwavelength) films[515] while simultaneously controlling the polarization state of the interacting fields, as they can be engineered to support optical modes that simultaneously enhance fields at both fundamental ($\omega_0$) and final frequencies, (e.g., $\omega_0/2$ for SPDC and $2\omega_0$ for SHG). A common approach to implement this paradigm consists of transferring atomically thin layers (down to 1L) onto existing metasurfaces comprised of other materials. However, as noted above, it may be even more desirable to fabricate MSs and resonators directly from vdW crystals. The newly available non-centrosymmetric vdW crystals are appealing for these purposes, thanks to their nonzero bulk nonlinear responses. Moreover, 3R-$MoS_2$ and other TMDs are excellent materials for realizing metasurfaces due to their high refractive indexes[511] in the visible and communication wavelength regimes (~≥4) that reduce design constraints. To date, enhanced nonlinear signals have been experimentally demonstrated in $WS_2$ metasurfaces [521] and $MoS_2$[514] nanodisks. More generally, the trade-off between mode quality factor and operational bandwidth is an important, application-dependent consideration. Current efforts are investigating strategies for tuning these parameters by tailoring geometrical asymmetries that induce modes with controllable quality factors such as quasi-bound states in the continuum.

Looking further ahead, more complex photonic structures can be envisioned. Such future nanostructured devices, for example, multilayered media, nanocavities, nonlinear antenna arrays, and metasurfaces, will be able to further manipulate the phase matching requirements that fundamentally affect SPDC. In this context, the mixing of forward and backward photons in such innovative nanostructures is expected to yield qualitatively new quantum phenomena. The adoption of metasurfaces for orbital angular momentum (OAM) manipulation[522] promises SPDC of spatially structured photons, yielding prospects for a new generation of quantum computation schemes based on qbits (quantum states in a d-dimensional Hilbert space). Combining SPDC with nanosystems and metasurfaces encompassing spin-to-OAM conversion will enable the generation of hyperentangled photon pairs involving richer quantum correlations. Engineered nonlocalities in metasurfaces hold the promise of realizing local structuring of the nonlinear states without sacrificing strong spatial correlations stemming from the underlying lattice resonances,[523] leading to more sophisticated control over the nonlinearly generated waves.[524] Hyperentanglement (i.e., entanglement in multiple degrees of freedom in a quantum system) is promising for a range of potentially groundbreaking applications in quantum information, enhancing the capabilities of quantum computers and quantum networks.[525]

Importantly, 2D materials also offer inherent programmability and tunability, which is expected to yield unprecedented control and manipulation of quantum radiation states in future architectures[493]. Material properties can be tuned, for example, by applying stimuli such as light, mechanical strain, electric and magnetic fields, or local changes to the dielectric environment.[526] The polarizations of the fully entangled states generated in 3R TMDs by SPDC are completely tunable, with constant efficiency, by simply changing the polarization of the incident fundamental beam.[492] In addition, the interlayer twist angle degree of freedom offers a compelling control knob to tune and optimize crystal symmetries, optical nonlinearities and phase matching in vdW homo- and heterostructures.[527,493] By precisely controlling twist and assembly of vdW interfaces, new microscopic 3D nonlinear crystals with unique nonlinear susceptibilities and polarization selectivity will be realized.[493]

Finally, advances on the materials, engineering, and fabrication fronts are needed to accelerate the realization of robust, efficient, compact sources of quantum-entangled photons using 2D materials. Large- or wafer-scale growth of 3R TMD crystals will enable broad, commercial technology development, as will the demonstration of facile, scalable encapsulation methods (adapted, for example, from the solar cell industry) for less-stable vdW crystals. Improvements in robotic and parallel stacking capabilities will greatly enhance sample throughput and push the utility of novel heterostructures beyond the research sector. Similarly, improvements in less-intrusive fabrication methods such as laser patterning[512] will reduce fabrication steps (and costs) while also reducing the need for photoresist or other polymers whose presence contaminates devices and precludes reliable production of complex device architectures. Ultimately, these and other advances will not only improve SPDC-based sources but may lead to the development and widespread adoption of additional, complementary routes for generating complex entangled quantum light states in layered materials, including biexciton cascade and the robust manifestation of superradiant or superfluorescent states (e.g., cooperative emission from a collection of moiré-trapped excitons).[528,529]

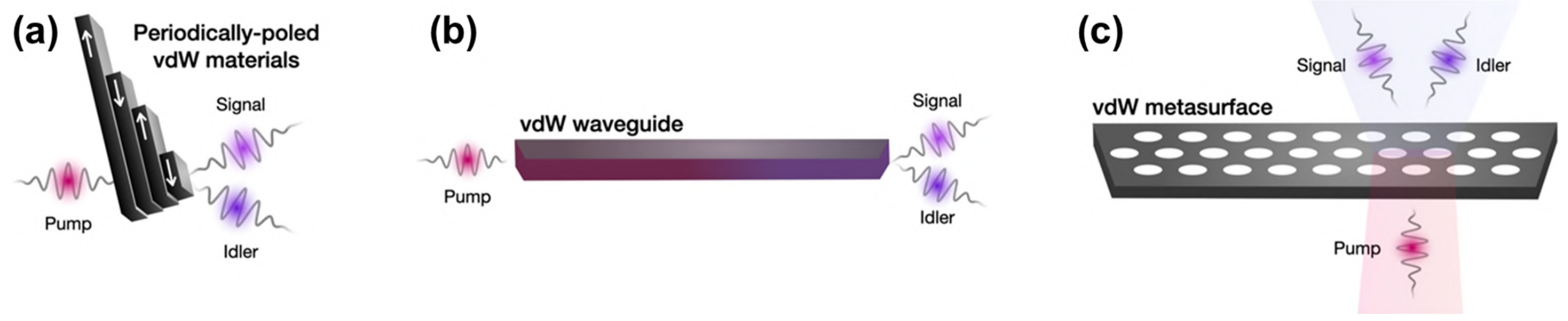

**Figure 17.** Nanoscale platforms for efficient SPDC generation. (a) Periodically poled vdW materials. (b) vdW waveguide. (c) vdW metasurface.

## 18. NONLINEAR POLARITONICS IN 2D MATERIALS

**Joel D. Cox,[1,2,*] Eduardo J. C. Dias,[1] Álvaro Rodríguez Echarri,[3] Fadil Iyikanat,[4] Andrea Marini,[5] and F. Javier García de Abajo[4,6]**
[1]POLIMA—Center for Polariton-driven Light–Matter Interactions, University of Southern Denmark, Campusvej 55, DK-5230 Odense M, Denmark
[2]Danish Institute for Advanced Study, University of Southern Denmark, Campusvej 55, DK-5230 Odense M, Denmark
[3]Max-Born-Institut, 12489 Berlin, Germany
[4]ICFO-Institut de Ciencies Fotoniques, The Barcelona Institute of Science and Technology, 08860 Castelldefels, Barcelona, Spain
[5]Dipartimento di Scienze Fisiche e Chimiche, Università degli Studi dell'Aquila, L'Aquila 67100, Italy
[6]ICREA-Institució Catalana de Recerca i Estudis Avançats, Passeig Lluís Companys 23, 08010 Barcelona, Spain
***Corresponding author.** Email: cox@mci.sdu.dk

### 18.1 Introduction

Nonlinear optics facilitates diverse light-based technological applications while opening new areas of fundamental research (e.g., in attosecond science, high-harmonic-generation (HHG) spectroscopy, spatiotemporal pulse shaping, mode-locked fiber lasers, parametric oscillators, and the generation of nonclassical photon states). In nanophotonics, the enhancement of nonlinear optical phenomena is generally pursued by boosting the local optical field strength,[530] increasing the light–matter interaction time,[531] and engineering the intrinsic nonlinear optical response of materials. The former two strategies can be effectively utilized by polaritons—quasiparticles formed by the hybridization of light with collective dipole-carrying excitations in matter—acting as nanoscale optical resonators that concentrate electromagnetic energy on subwavelength scales.[112] Enhancement of the intrinsic nonlinearity may involve patterning and heterostructuring nanomaterials to modify their optoelectronic properties (e.g., through quantum confinement) or applying external stimuli that actively tune the optical response.

These approaches to expanding nonlinear nanophotonics can be uniquely addressed using 2D materials. In the 2D (atomically thin) limit, polaritons push near-field confinement to extreme levels[63] (see Section 2), enhancing the nonlinear field produced in their host media. In addition, the comparatively small number of charge carriers in 2D materials makes their response highly susceptible to sizeable modulations driven by mild stimuli (e.g., charge-carrier doping, applied static fields, or temperature variations by optical pumping or Joule heating), opening exciting possibilities for active and ultrafast tuning that cannot be accessed in bulk materials. Furthermore, the electronic structures of 2D materials, which govern their polaritonic and nonlinear optical properties, can be more easily customized than their bulk counterparts, for example, by patterning or stacking.

## 18.2 Current State of the Art

### 18.2.1 *Nonlinear Graphene Plasmonics*

The emergence of highly doped graphene as a platform for strongly confined and actively tunable plasmons (e.g., through electrostatic doping), combined with the large intrinsic nonlinear optical response originating from its Dirac electronic dispersion,[465,532] motivates intensive research efforts in nonlinear graphene plasmonics to capitalize on these synergistic effects (Figure 18a). Numerous theoretical studies have predicted relatively strong optical nonlinearities associated with propagating plasmons in extended graphene[533] and localized plasmons in graphene nanostructures.[534] On the experimental side, far-field measurements of graphene nanoribbons have revealed that localized THz plasmons enhance coherent four-wave mixing,[535] while difference-frequency generation has been exploited for all-optical excitation of graphene-plasmon polaritons in extended samples.[536] Aside from the *instantaneous* plasmon-driven nonlinear optical response associated with coherent processes (e.g., harmonic generation and wave mixing), the transient electronic heating in graphene by optical pumping constitutes a strong source of incoherent (*delayed*) nonlinearity that is also boosted by plasmon resonances (Figure 18b).[537] In this regard, the inherently low electronic heat capacity of graphene enables rich and sizeable electron temperature dynamics down to sub-picosecond timescales, enhancing the natural nonlinear optical properties of the carbon monolayer and opening possibilities for the spatiotemporal modulation of its optical response (Figure 18c). Acting as artificial doping in graphene, electron temperature can tune and either enhance or suppress graphene plasmon resonances in an ultrafast manner using intense femtosecond laser pulses.[538] While such transient plasmonic states have been experimentally observed,[18] their potential for nonlinear optics has yet to be fully realized.

### 18.2.2 *Nonlinear Polaritonics*

Beyond graphene, other 2D materials hold significant promise for applications in nonlinear optics. Hexagonal boron nitride naturally supports highly confined phonon polaritons exhibiting long lifetimes and low propagation losses, which make them ideal for enhancing nonlinear interactions at mid-IR frequencies, as explored in first-principles theory[539] and experiments.[540] The polaritonic response in hBN can be tuned across the mid-IR spectrum (e.g., passively by patterning or varying the thickness, and actively through temperature or exposure to DC electric fields), enabling explorations of nonlinear optical effects over a broad spectral range where many other materials exhibit a weak nonlinear optical response.

Transition metal dichalcogenides, such as $MoS_2$, $MoSe_2$, $WS_2$, and $WSe_2$, are semiconductors that, in their monolayer form, display large excitonic binding energies due to reduced dielectric screening in the 2D limit, leading to robust exciton polaritons capable of significantly enhancing nonlinear optical processes. Unlike many materials in which polaritons require low temperatures to be stable, exciton polaritons in TMDs are observable at room temperature,[541] making them practical for real-world nonlinear optical devices. In contrast to graphene or thin metallic films, these materials are non-centrosymmetric in monolayer form and can therefore naturally support strong even-ordered nonlinear processes.[471] Second-order coherent nonlinear optical effects in monolayer TMDs, encompassing second-harmonic generation as well as parametric down-conversion, amplification, and oscillation, present opportunities to generate and manipulate quantum photon states at the nanoscale.[542]

18.2.3 *Nonlinear Nanophotonics in Low-Dimensional Crystals*

Recent explorations of nonlinear plasmonic phenomena in ultrathin crystalline noble metal films show that these materials constitute a paradigmatic example of nonlinear 2D polaritonics. Approaching the 2D limit in ultrathin noble metal films,[543] the plasmonic field enhancement resulting from reduced screening and anharmonic electron motion produced by vertical quantum confinement simultaneously increases as the film thickness decreases (see Figure 18d,e). These complementary behaviors were hypothesized to boost plasmon-driven optical nonlinearity in ultrathin crystalline noble metal films,[544] a prediction that is consistent with recent experimental measurements of second-harmonic generation.[545] Incidentally, the intense ultrafast laser pulses typically used to measure coherent optical nonlinearities can also excite hot electrons that rapidly thermalize to an equilibrium distribution with an elevated electronic temperature, effectively modulating the optical response during a transient period as heat energy is transferred to the lattice. Compared to bulk systems, this photothermal nonlinearity is considerably enhanced in ultrathin metal films with inherently small surface-to-volume ratios and fewer charge carriers per surface area.[538]

## 18.3 Challenges and Future Goals

18.3.1 *Coherent Polariton-Driven Nonlinear Optical Phenomena*

Despite the large body of theoretical research in nonlinear graphene plasmonics, there are relatively fewer experimental demonstrations of proposed phenomena. Measurements of large coherent plasmon-driven nonlinear optical effects (e.g., harmonic generation) could be pursued in doped graphene nanoribbons,[534] offering a simplified geometry that supports electrically tunable and polarization-dependent localized plasmon resonances that can in- and out-couple to free-space illumination. Beyond patterning, leveraging hybrid structures that combine graphene with other photonic materials could enable enhancements in nonlinear responses.

As a complementary direction, many of the concepts developed in nonlinear graphene plasmonics can be extended to polaritons in other emerging 2D materials, which offer unique advantages and appealing properties. For instance, phonons in hBN sustain high-quality resonances that are predicted to drive efficient harmonic generation and optical bistability.[539] Exciton polaritons in TMDs are also long-lived and offer additional valley-dependent degrees of freedom,[546] although they feature low group velocities that render them inadequate to propagate encoded signals. The strong second-order nonlinear response of TMDs is particularly promising for the manipulation of quantum radiation states. Indeed, thanks to the surface-like nonlinear interaction, the phase-matching condition is bypassed, so that photon state generation and manipulation is expected to yield qualitatively new quantum phenomena. The integration of TMDs with conventional plasmonic elements can also boost the overall nonlinear optical response of the hybrid system while enabling subwavelength control over the generated near fields.[547]

Polaritons in 2D materials are also envisioned to overcome the long-standing challenge of triggering nonlinear optical phenomena at the single-photon level. In this context, theory proposals based on the optical Kerr effect have been put forward to achieve unity-order interactions among propagating graphene plasmon polaritons,[548] but unfortunately they rely on extreme nanopatterning and large quality factors that could be difficult to achieve in practical experiments with current fabrication methods.

18.3.2 *Photothermal Nonlinearities and Incoherent Phenomena*

The large photothermal response of graphene presents both challenges and opportunities in the context of nonlinear optics. Intense ultrashort optical pulses are routinely employed in nonlinear optics, which result in sizeable electron heating. This is detrimental to coherent plasmon-driven nonlinear effects, but it plays a central role in driving thermally mediated, incoherent phenomena. Graphene thermoplasmons, which are sustained by temperature-induced carrier dynamics, provide an appealing platform for exploring such effects, enabling optical switching and modulation based on thermal engineering (see Figure 18b).[18]

## 18.4 Suggested Directions for Nonlinear 2D Polaritonics

While plasmon resonances in graphene are typically limited to THz or IR frequencies, they can be boosted to the near-IR or visible domains by extreme doping or nanostructuring. The latter approach holds intriguing possibilities to harness quantum finite size effects in the nonlinear plasmonic response of structures with lateral sizes below the Fermi wavelength ~10 nm.[534] For instance, the large photothermal response of relatively few electrons in small graphene nanoislands is predicted to enable unity-order modulations in plasmon resonances triggered by single-photon absorption (see Figure 18g),[534] while narrow graphene nanoribbons could be optically pumped to support near-IR thermoplasmons that can drive a coherent nonlinear plasmonic response.[549]

In phonon polaritonics, the vibrational origin of the nonlinear optical response allows for larger applied light intensities without boosting optical losses associated with electronic degrees of freedom. Additionally, introducing strong in-plane DC fields can actively modify the nonlinear response, further enhancing the control over the system. Indeed, as shown in Figure 15f, by introducing in-plane DC fields through, for example, lateral gating, significant shifts are predicted to occur in the frequency of phonon polaritons in monolayer hBN, exceeding the spectral widths of the modes.[539]

The design of atomically thin heterostructures comprised of materials with complementary optoelectronic properties can open new paths to enhance nonlinear light–matter interactions at the nanoscale. Polaritons in 2D materials interfacing with noble metal films (behaving as perfect conductors) can hybridize with their mirror images to form so-called image polaritons,[550] reaching extreme light-focusing regimes for nonlinear optics. This concept has been demonstrated for graphene–dielectric–metal heterostructures,[22] and should similarly produce large polariton-driven optical nonlinearity in other 2D systems. In a related approach, the strong coupling of excitons in TMDs with optical cavity modes can be leveraged for nonlinear light–matter interactions such as saturable and multiphoton absorption.[546,551] Turning to the intrinsic nonlinear optical response of 2D materials, moiré electronic-structure engineering offers an appealing route to tailor anharmonic electron motion,[481] with polaritons supported by twisted 2D materials or nearby metal–dielectric interfaces enhancing nonlinear light–matter interactions.

In quantum nonlinear optics, recent theory proposals to generate entangled photon pairs directly in the guided modes of optical waveguides show promise for emerging quantum photonic technologies. These strategies can be mapped directly to waveguided 2D polaritons in nanophotonic architectures (e.g., the optical generation of plasmon-polariton pairs in graphene nanoribbons *via* spontaneous parametric down-conversion[552]). In an alternative scheme, free electrons that impinge normally to a 2D waveguide can excite guided polaritons that are entangled in energy and momentum, and heralded by measuring the energy loss of undeflected electrons.[553]

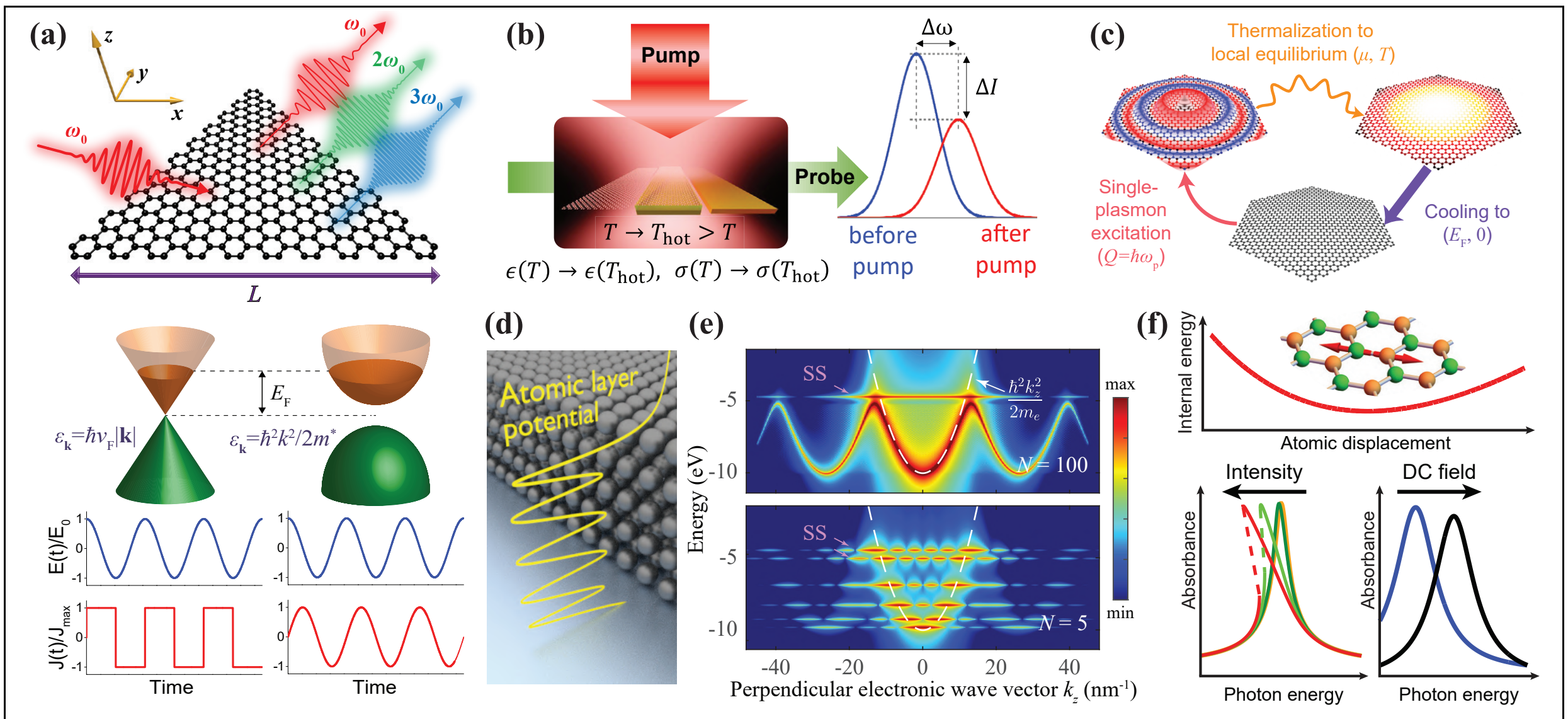

**Figure 18.** Nonlinear polaritonics in 2D materials. Nonlinear polaritonics in 2D materials. (a) Intense high-harmonic generation from graphene nanostructures (upper part) is predicted to arise from the combination of plasmonic field enhancement and anharmonic electron motion associated with the conical band structure of graphene (left-lower part), the latter attribute leading to an intrinsically nonlinear optical response that contrasts the harmonic response exhibited by a parabolic electron dispersion (right-lower part). Adapted with permission

from refs 554 and 555 (Copyright 2014 and 2017 Springer Nature). (b,c) The large transient photothermal response of graphene and ultrathin metals can be harnessed to trigger incoherent nonlinear light–matter interactions that enable the creation of transient optical responses (b) and optical switching at the single- of few-photon level (c). Adapted with permission from refs 538 (Copyright 2020 Springer Nature) and 556 (Copyright 2019 American Chemical Society). (d,e) In crystalline ultrathin metal films (d), anharmonicity can be engineered by vertical quantum confinement to enhance the nonlinear response of plasmon polaritons (e). Adapted with permission from ref 544 (Copyright 2021 de Gruyter). (f) The nonlinear response of phonon polaritons in hBN originates from the anharmonicity of the potential landscape under atomic displacements (upper part), which can produce non-perturbative effects (lower left), and enables electrical modulation of the phonon frequency through wave mixing with a static field in the V/nm range (lower right). Adapted with permission from ref 539 (Copyright 2021 American Chemical Society).

## 19. NONLINEAR VALLEYTRONICS AND ALL-OPTICAL PROBE OF BROKEN TIME-REVERSAL SYMMETRY IN 2D MATERIALS

**Paul Herrmann,[1] Nele Tornow,[1] Sebastian Klimmer,[1,2] Jan Wilhelm,[3,4] and Giancarlo Soavi[1,5,*]**
[1]Institute of Solid State Physics, Friedrich Schiller University Jena, Helmholtzweg 5, 07743 Jena, Germany
[2]ARC Centre of Excellence for Transformative Meta-Optical Systems, Department of Electronic Materials Engineering, Research School of Physics, The Australian National University, Canberra, ACT, 2601, Australia
[3]Regensburg Center for Ultrafast Nanoscopy (RUN), University of Regensburg, 93040 Regensburg, Germany
[4]Institute of Theoretical Physics, University of Regensburg, 93053 Regensburg, Germany
[5]Abbe Center of Photonics, Friedrich Schiller University Jena, Albert-Einstein-Straße 6, 07745 Jena, Germany
***Corresponding author.** Email: giancarlo.soavi@uni-jena.de

### 19.1 State of the Art

Monolayer TMDs are direct-gap semiconductors with energetically degenerate minima (maxima) of the conduction (valence) band at the ±K points of the Brillouin zone (i.e., the ±K valleys). The combination of broken space-inversion symmetry (SIS) $E_{\uparrow}(+K) \neq E_{\uparrow}(-K)$ and preserved time-reversal symmetry (TRS) $E_{\uparrow}(+K) = E_{\downarrow}(-K)$ imposes that electrons residing in opposite valleys have opposite spins, leading to the so-called spin-valley locking.[557] The deterministic control of the valley information naturally leads to the field of valleytronics, which has a dual scientific and technological relevance. From one viewpoint, the valley-contrasting Berry curvature and Berry phase make TMDs the ideal platform to study fundamental processes unexplored in standard bulk semiconductors, such as the valley Hall effect[558] and the valley-exclusive Bloch–Siegert shift.[559] From another viewpoint, valleys in TMDs represent a unique platform for information technology, with the possibility to write and read the spin-valley information with all-optical methods, potentially increasing the operation speed by orders of magnitude compared to standard electronic devices.

All-optical valleytronic operations are ultimately defined by the selection rules of light–valley interactions,[557] which in TMDs lead to the coupling of circularly polarized light of opposite helicity to the opposite valleys. This can be used, for instance, to generate a valley imbalance either by a real excited state population with linear[560] (see Figure 19a) and two-photon absorption,[561] or by a valley selective bandgap modulation with the coherent optical Stark and/or Bloch–Siegert shifts[559] (see Figure 19b). For a real excited state population, the readout of the valley information can be achieved either by helicity-resolved photoluminescence (PL, see Figure 19c), or by optical Kerr rotation measurements (see Figure 19d). Note that optical Kerr rotation is able to probe both an imbalance in real carriers residing in the valleys[562] and a coherent bandgap modulation.[563] Recently, new methods based on nonlinear optics have emerged as an alternative and powerful approach to measure the valley degree of freedom in TMDs.

To date, nonlinear valleytronics has only been realized in two configurations:

(1) The measurement of a rotation angle in the emitted second harmonic (SH) polarization upon creation of a valley imbalance.[564,565,566,567] Here, the valley imbalance can be generated by real excited states carriers,[564] off-resonant coherent bandgap modulation,[565] or external magnetic fields.[566] This method is equivalent to the nonlinear Kerr rotation, which has been used in the past to probe bulk magnetic materials.[567]

(2) The detection of a deviation in the ratio of the SH intensity for circularly *versus* linearly polarized excitation.[568] This approach builds on the following observation: ±K resonant second-harmonic generation (SHG) in TMDs is dictated by the $C_{3h}$ wave vector group of the ±K valleys, rather than the $D_{3h}$ wave vector group of the Γ point. The generation of a valley imbalance can then be traced back to a specific modulation of the different elements of the $C_{3h}$ second-order nonlinear susceptibility tensor.

Another interesting aspect of nonlinear valleytronics is that, in TMD monolayers, a valley imbalance breaks TRS. Thus, nonlinear optics can serve as the ultimate tool to probe changes in symmetries (both SIS and TRS) in a crystal and, consequently, the field of nonlinear valleytronics can be further generalized to the field of all-optical detection of broken TRS. The most significant limitation of SHG as a universal probe of broken TRS is undoubtedly its limited applicability since strong SH within the electric-dipole approximation is only possible in crystals with broken SIS. While electric-quadrupole SH, which is possible in any material, can be employed to probe TRS breaking,[569] this approach is experimentally challenging. The electric-quadrupole SH is orders of magnitude weaker than the electric-dipole SH, it requires an oblique angle of incidence, and a larger amount of polarization scans for multiple angles of incidence.[569]

Finally, the development of these ultrafast methods to probe broken TRS also calls for new and alternative all-optical methods to break TRS. To date, the most common methods have been the valley-Zeeman splitting *via the* application of strong external magnetic fields[566] and the introduction of a carrier imbalance between the ±K valleys by circular/elliptical one-photon absorption.[560] To reach the boundaries in terms of operation speed, all-optical valleytronics can now rely on two additional approaches for ultrafast bandgap engineering: (1) asymmetric bandgap opening by the valley-selective optical Stark effect and Bloch-Siegert shifts[559,565] and (2) application of a trefoil topological field, tailored to the crystal symmetry.[570]

## 19.2 Challenges and Future Goals

Looking ahead, we identify five main challenges for the field of nonlinear valleytronics and all-optical probing of broken TRS, which we further divide into fundamental and technological challenges.

### 19.2.1 *Fundamental Challenges*

*I.* The most fundamental and important challenge for this field is arguably the necessity to find a universal, ultrafast, and noninvasive probe of broken TRS. Here, with *universal* we refer to a method that can be applied to any crystal, regardless of SIS. As discussed, electric-dipole SH applies only to a limited set of materials (those with broken SIS), while electric-quadrupole SH is experimentally challenging. The goal is to find a simple rule within the realm of nonlinear optics that allows to unambiguously tell if a system displays broken TRS, analogous to SHG as the universal probe of broken SIS.

*II.* In addition to experimental endeavours, further development of the underlying theories is required to understand and guide experiments. The field of topology, which is fundamentally linked to symmetry, is the primary focus in this regard. The inseparable interplay of topology and TRS was demonstrated in condensed matter physics, for example, in the light-induced anomalous Hall effect in graphene[571] (see Figure 19f), and has also been extended to other areas such as photonics.[572] For this reason, future methods for the all-optical probe of broken TRS should develop alongside new theories that allow to explain experimental findings on the basis of topological quantities, such as Berry curvature, Berry phase, and the Chern number.[573]

### 19.2.2 *Technological Challenges*

*III.* The first technological challenge for valleytronic devices is miniaturization. While 2D materials already represent the ultimate limit of device thickness, the minimum possible lateral dimensions are yet unknown. We identify two factors that can have a major impact on the lateral dimension: (1) for all-optical operations, the lateral size of a device is limited by the spot-size of the driving/reading light pulse; (2) the valleys as an electronic property of 2D materials are derived from the ansatz of periodic Bloch-states. When shrinking the lateral size of a device, this might turn out to be increasingly inaccurate.

*IV.* The shortest switching time possible to achieve with all-optical valleytronics is ultimately given by the optical cycle. In turn, ultrashort pulses have an ultra-broadband spectral coverage, dictated by the Fourier theorem. This limits the applicability of nonlinear valleytronics, which is in large parts predicated on ±K resonant excitation/detection.[565,568] Therefore, ultrashort pulses might not be able to adequately address the desired states due to their broad energy spectrum.

*V.* Being in an early stage, nonlinear valleytronics requires a precise calibration of the ultrafast methods employed for TRS breaking, namely the amplitude of the optical Stark and Bloch-Siegert shifts, and the resulting asymmetric bandgap opening, the effective bandgap change upon application of a topological field, and the impact of the aforementioned processes on the different elements of the nonlinear susceptibility tensor.

### 19.3 Directions to Meet Goals

Building on the challenges and future goals mentioned above, we envision new methods and theoretical approaches for the future of nonlinear valleytronics.

From our point of view, the most promising candidate as a universal probe of broken TRS is third-harmonic generation (THG), in particular third-harmonic (TH) Kerr rotation (see Figure 19f). THG combines universality (i.e., applicability to any material regardless of SIS) with a simple experimental realization, especially in comparison to the electric-quadrupole SH. To date, TH Kerr rotation has been applied to measure magnetic phases in EuO,[574] and we believe it is now the time to fully unlock the potential of this technique by applying it to layered materials and related heterostructures.

For the challenges associated with the lateral size of devices, we foresee solutions based on local field tip-enhancement (see Figure 19e), similar to what has been done, for example, in Raman or PL,[357] combined with nanofabrication and patterning down to a few nanometers.

Topology has been identified in linear Hall currents *via* the Chern number[575,] and even- and odd-order nonlinear responses *via* the Berry phase and Berry curvature,[573] respectively. However, the current model connecting NLO with topology does not include excitonic effects. Furthermore, the topological nature of NLO remains yet to be explored experimentally We believe that a combined effort is crucial for advancing the field. From our point of view, a unification of the Chern number with the conservation of angular momentum of crystals and light appears promising.

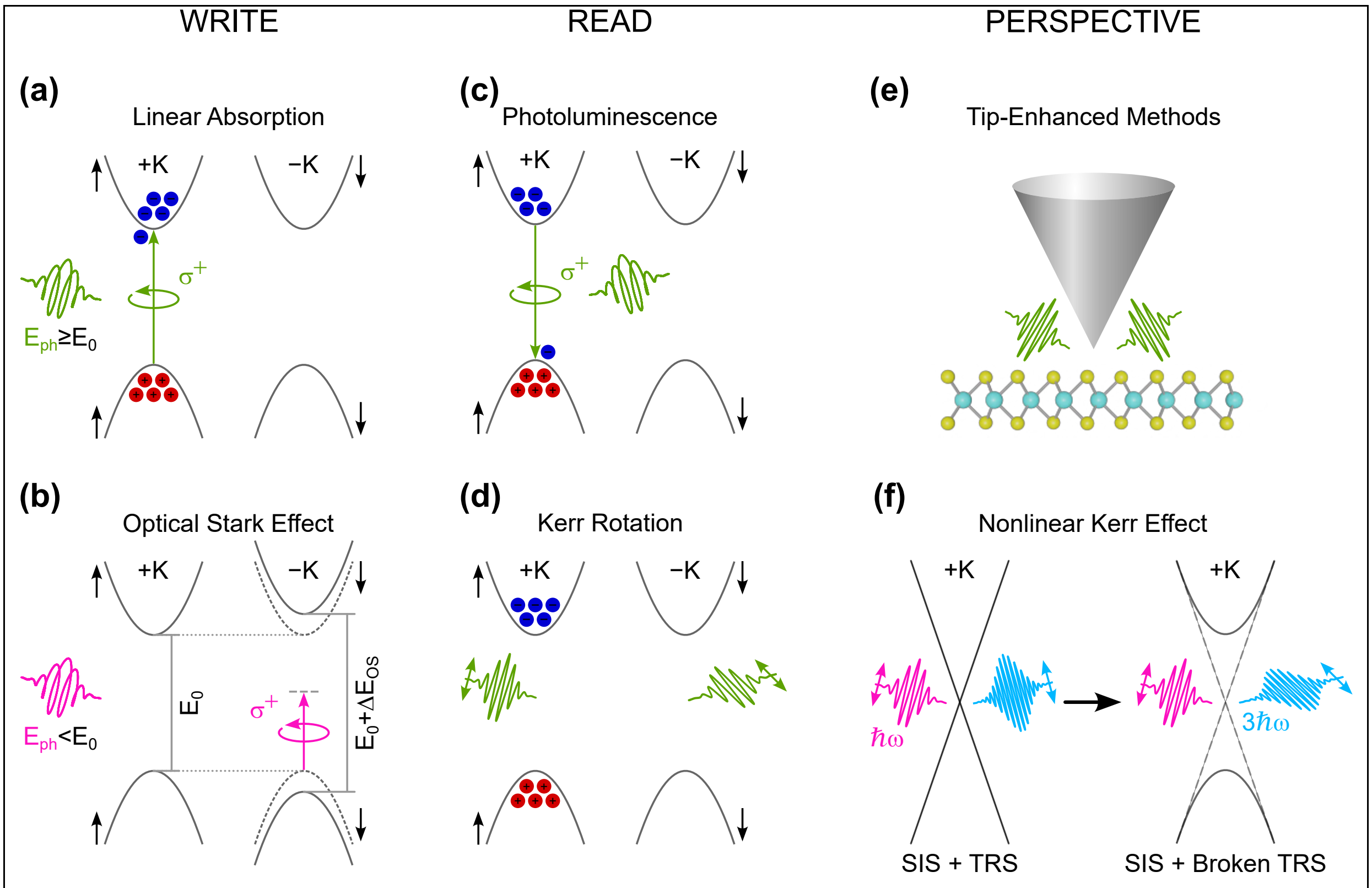

**Figure 19.** Schematic of all-optical valleytronic operations and time-reversal symmetry (TRS) modulation/detection. A valley imbalance can be generated by resonant one-photon absorption of circularly polarized light (a) or the coherent valley-selective optical Stark effect (b). The valley imbalance is typically measured *via* helicity-resolved PL (c) or ultrafast resonant Kerr rotation (both linear and nonlinear) (d). Future valleytronic devices could use local field enhancement *via* metallic tips (e) and potentially probe broken TRS even in centrosymmetric crystals by nonlinear third-harmonic Kerr rotation.

## 20. NONLINEAR MAGNETO-OPTICS IN 2D MAGNETIC MATERIALS

**Zeyuan Sun[1] and Shiwei Wu[1,2,*]**
[1]State Key Laboratory of Surface Physics, Key Laboratory of Micro and Nano Photonic Structures (MOE), and Department of Physics, Fudan University, Shanghai 200433, China
[2]Institute for Nanoelectronic Devices and Quantum Computing, and Zhangjiang Fudan International Innovation Center, Fudan University, Shanghai 200433, China
***Corresponding author.** Email: swwu@fudan.edu.cn

### 20.1 Current State of the Art

The discovery of 2D magnetic materials such as $CrI_3$ and $Cr_2Ge_2Te_6$ has greatly benefited from linear magneto-optical effects including Kerr rotation and reflective magneto-circular dichroism (RMCD).[576,577] These linear magneto-optical effects are sensitive to nonzero total magnetization of the materials and, thus, effective in detecting ferromagnetic states or materials. In contrast, the nonlinear magneto-optical effects such as magnetic second-harmonic generation (SHG) are sensitive to the breaking of spatial inversion symmetry in spin lattices under study.[578,579] This spatial inversion symmetry breaking often occurs in antiferromagnetic materials or at the surface of ferromagnetic materials.

Indeed, the very first observation of magnetic SHG was done on a Fe(110) surface under ultrahigh vacuum conditions by Reif *et al.* in 1991.[580] Consistent with the theoretical prediction by Pan *et al.* in 1989,[581] the intensity of observed SHG changed with the reversal of the magnetization. Later, magnetic SHG was detected in the antiferromagnetic bulk crystal $Cr_2O_3$ by Fiebig *et al.* in 1994.[582] Since then, magnetic SHG has become a powerful tool to study multiferroic orders in functional oxide materials.[578] However, the second-order nonlinear optical susceptibility arising from magnetic orders in these bulk materials is typically weak, in comparison to that contributed by the crystallographic structures with broken spatial inversion symmetry. Surprisingly, it was found that the layered antiferromagnetic states in bilayer $CrI_3$ yield giant magnetic SHG,[583] with nonlinear susceptibility approximately three orders of magnitude larger than that of $Cr_2O_3$, and even ten orders of magnitude larger than that at Fe(110) surface. This discovery was soon followed by the observation of significant magnetic SHG signals in other 2D antiferromagnetic materials, including $MnPS_3$,[584] $MnPSe_3$,[585] and CrSBr.[586]

Similar to conventional SHG from crystallographic structures, magnetic SHG is also subject to the symmetry constraints, and the corresponding response is described by the second-order nonlinear optical susceptibilities $\boldsymbol{\chi}^{(2)}$ (Figure 20). If the long-range magnetic order breaks the spatial inversion symmetry, $\boldsymbol{\chi}^{(2)}$ becomes nonzero under the electric-dipole approximation. Note that this SHG comes from the symmetry breaking of spin-polarized electronic structures, and thus it is electric dipole allowed and differs from the magnetic dipole contribution. The specific form of $\boldsymbol{\chi}^{(2)}$ is determined by the symmetry of magnetic orders and has been systematically tabulated according to the magnetic point groups.[587] Because the magnetic orders can reverse their direction under time-reversal operation, the tensor elements in $\boldsymbol{\chi}^{(2)}$ are often time-noninvariant and referred to as c-type, in contrast to the conventional i-type time-invariant $\boldsymbol{\chi}^{(2)}$ that arises from crystallographic structures.

Taking bilayer $CrI_3$ as an example, the layered antiferromagnetic structures in its ground state break the spatial inversion symmetry, although the crystallographic structure is centrosymmetric.[583] The corresponding $\boldsymbol{\chi}^{(2)}$ is a third-rank tensor as an odd function of its Néel vector $\boldsymbol{L}$, meaning that the sign of tensor elements in $\boldsymbol{\chi}^{(2)}$ changes with the direction of $\boldsymbol{L}$. If the magnetization of bilayer $CrI_3$ is forced to align in the same direction by an external magnetic field, the magnetic symmetry becomes centrosymmetric, and the time-noninvariant c-type $\boldsymbol{\chi}^{(2)}$ vanishes. Similarly, when the sample temperature is increased above the Néel temperature $T_c$, the c-type $\boldsymbol{\chi}^{(2)}$ also vanishes due to the disappearance of long-range magnetic order. These symmetry-governed phenomena were exactly observed in the nonlinear magneto-optical measurements.

Because of the resonant optical excitation and enhanced light–matter interaction in atomically thin 2D materials, the magnetic SHG in bilayer $CrI_3$ turns out to be strong, comparable to those time-invariant i-type SHG observed in non-centrosymmetric crystals such as $MoS_2$[588,589] and GaSe.[590] By conducting the polarization-resolved SHG measurements, the anisotropic patterns dictated by $\boldsymbol{\chi}^{(2)}$ provide an experimental means to determine the magnetic symmetry. The beauty of this SHG technique is that the revealed symmetry would be independent of the excitation wavelength used in the experiments, although the SHG intensity and pattern may change. Since the magnetic structure in bilayer $CrI_3$ is registered with the crystal lattice as in many magnetic materials, the measurement of magnetic symmetry can also reflect the crystallographic symmetry. In bilayer $CrI_3$, the structure and symmetry are highly sensitive to the stacking order at the vdW interface. Magnetic SHG has thus helped to unveil the mysterious stacking structure and elucidate the interfacial magnetic coupling.

As Louis Néel famously remarked, antiferromagnetic materials are extremely interesting. Unfortunately, few experimental techniques are available for probing 2D antiferromagnets. Because of the atomically thin thickness, neutron scattering becomes impractical despite its powerfulness for studying rich magnetic structures in bulk crystals. Meanwhile, linear magneto-optical tools and some emerging novel scanning probes such as SQUID[591] and NV-center magnetometry[592] mostly work for ferromagnetic materials but are ineffective for antiferromagnetic systems with zero total magnetization. In this context, magnetic SHG can nicely fit in the niche, particularly for its noninvasive optical microscopic capability with spatial resolution down to the diffraction limit (roughly 1 micron).

Very recently, we developed a phase-resolved SHG microscopy that is compatible with cryogenic and high vacuum conditions.[593] By applying this technique to layered antiferromagnet CrSBr few layers, we were able to resolve the magnetic polymorphs, which have the same magnetization but different magnetic structures.[594] Because these magnetic polymorphs are counterparts by spatial inversion or time reversal operations, their corresponding $\boldsymbol{\chi}^{(2)}$ from magnetic structures exhibit a sign change that can be measured by phase-sensitive SHG as a π phase shift. Furthermore, an interesting magnetic layer-sharing effect between laterally interconnected bilayer and tetralayer was observed, manifesting the strong lateral exchange interaction within a ferromagnetic intralayer. With the extreme sensitivity of magnetic

SHG to antiferromagnetism, more intriguing magnetic properties and effects are thus being observed in 2D antiferromagnets, which hopefully could lead to some functional spintronic or opto-spintronic devices.

### 20.2 Problems and Challenges

The study of nonlinear optics in solid materials often starts with the known crystal structures and symmetries, as exemplified in textbooks. Based on the symmetry analysis, one can deduce the corresponding nonlinear optical susceptibilities and predict the expected responses and phenomena. However, in many practical scenarios, prior knowledge of structure and symmetry is absent, posing significant challenges for analysis. With the addition of nonlinear magneto-optical effects in magnetic materials, the *reverse* analysis is even more challenging and involves multiple possibilities to include or exclude. A notable example is the case of polar magnets, whose crystallographic structures are non-centrosymmetric.[595,596] Even in the ferromagnetic states, the magnetic symmetry is no longer centrosymmetric, leading to contributions to the SHG from both the crystal and magnetic structures. The coherent superposition of both contributions complicates the experimental observations. Additionally, surface and interface effects are also worth mentioning as they intrinsically break the spatial inversion symmetry and may contribute to the SHG signal.[580] Sometimes, higher-order contributions such as the electric quadrupolar term are also appreciable.[597]

Moreover, 2D materials are more susceptible to external controls such as carrier doping, electric field, magnetic field, tensile strain and hydrostatic pressure. Therefore, these tuning knobs provide unique means to decipher the origin of nonlinear optical response and develop advanced nonlinear photonic devices. For nonlinear magneto-optical effects, the most relevant control is the external magnetic field. Yet under the external magnetic field, the Faraday rotation through optical components such as microscope objective is appreciable, which can reach as high as 6°/T·cm at wavelength of 400 nm for fused silica.[566] Therefore, special care must be taken during nonlinear optical measurements under magnetic field, particularly in the linear polarization basis. Furthermore, current SHG measurements under magnetic field are limited to normal incidence due to the constraints by bore size of superconducting magnet. This restriction means that only 8 out of the 27 elements in the third-rank nonlinear tensor can be effectively obtained. Thus, there is a pressing need to develop advanced nonlinear optical microscopy with *z*-polarization capabilities that can operate at low temperature and magnetic fields.

### 20.1 Future Directions and Goals

Moving forward, nonlinear optical studies of 2D magnetic materials offer substantial research opportunities and potential applications. While current studies mainly focus on the characterization of static magnetic structure, exploring ultrafast spin dynamics through nonlinear optical pump-probe techniques offers a key direction. For example, pump-probe SHG is expected to provide exceptional sensitivity in 2D antiferromagnetism with atomic thickness, making it ideal for investigating phenomena such as magnon dynamics and related quasiparticle interactions.

In addition, the development of nonreciprocal nonlinear magneto-optical devices based on 2D magnets offers exciting possibilities. These devices can be fabricated by stacking different vdW materials through dry transfer methods, enabling the construction of integrated 2D opto-spintronic circuits. The tuning of magnetism will not be limited to magnetic fields but can also be achieved through various means such as charge doping, electric field, optical field, and external strain. The modulation of nonlinear magneto-optical responses lays the foundation for coherent multiphoton applications such as quantum computing, neural networks, and artificial intelligence.

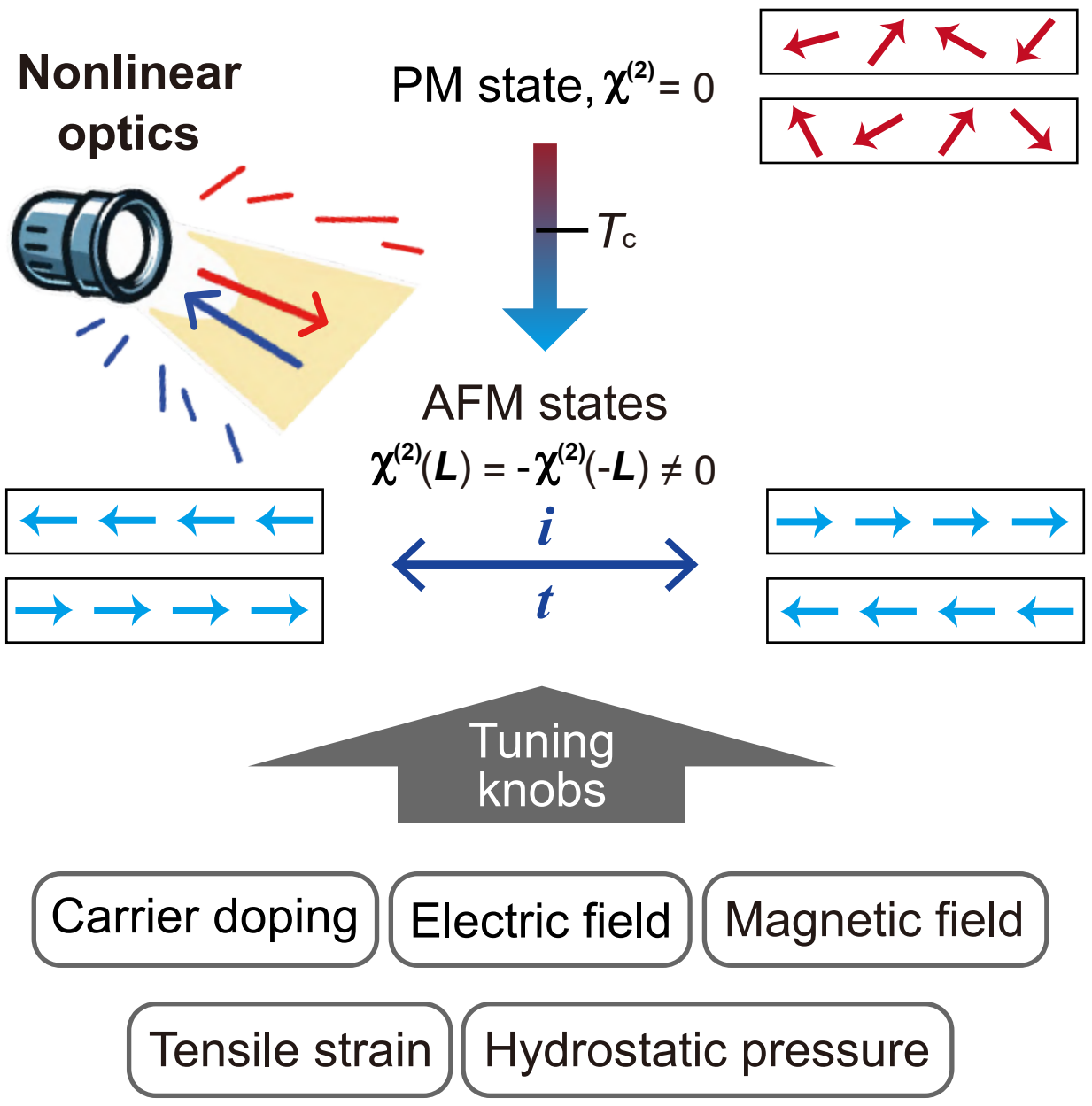

**Figure 20.** Symmetry-sensitive nonlinear optics as a powerful tool for probing 2D magnets under external stimuli. PM and AFM represent paramagnetic and antiferromagnetic, respectively. $T_c$ denotes the Néel temperature of the magnetic phase transition. The symbols ***i*** and ***t*** indicate spatial inversion and time reversal operations, respectively.

**Chirality, Singularities, Geometric Phases, and Moiré Systems**

# 21. QUANTUM GEOMETRIC PHOTONICS IN 2D MATERIALS

**Ying Xiong,[1,2] Oles Matsyshyn,[1] Roshan Krishna Kumar,[3,4] and Justin C. W. Song[1,*]**
[1]Division of Physics and Applied Physics, School of Physical and Mathematical Sciences, Nanyang Technological University, Singapore 637371
[2]Science, Mathematics and Technology (SMT), Singapore University of Technology and Design, 8 Somapah Road, 487372 Singapore
[3]Catalan Institute of Nanoscience and Nanotechnology (ICN2), CSIC and BIST, Campus UAB, Bellaterra 08193, Barcelona, Spain
[4]ICFO - Institut de Ciencies Fotoniques, The Barcelona Institute of Science and Technology, 08860 Castelldefels (Barcelona), Spain.
***Corresponding author.** Email: justinsong@ntu.edu.sg

## 21.1 Introduction

Unlike conventional optoelectronics that are dominated by free-particle-like dynamics and scattering, a new range of optoelectronic responses are emerging that directly rely on the winding of the Bloch wave functions in quantum materials—termed quantum geometry. Such quantum geometric optoelectronic responses have now recently exploded in 2D material. For instance, valley circular dichroism and a light-induced Hall effect can be controlled by the Bloch band Berry curvature;[598] nonreciprocal propagation of collective modes (such as plasmons) can be controlled by the quantum metric dipole.[599]

Recent attention, however, has focused on bulk quantum geometric photocurrents induced even in the absence of a p–n junction. In these, photocurrents flow in a direction determined by a material's intrinsically broken centrosymmetry. Unlike built-in fields that control the conventional photovoltaic effect, or temperature gradients and the Seebeck coefficient for the photothermoelectric effect, bulk photocurrents are now understood to strongly depend on the intrinsic properties of Bloch wave functions[573] (see Figure 21). For instance, when electrons are photoexcited in non-centrosymmetric materials, light-induced coherences during the interband transition from a valence-to-conduction band change a material's electric polarization. Such a continuous rate of change of electric polarization produces a shift photocurrent. Shift photocurrents have been related to a range of quantum geometric properties, including a Pacharatnam–Berry phase[600] and a Hermitian connection,[601] and contribute to a bulk photocurrent response for linearly polarized irradiation in non-magnetic materials. Similarly, under circularly polarized irradiation, bulk photocurrents can be induced that depend on an interband Berry curvature termed injection photocurrents.

## 21.2 State of the Field

Due to the pronounced and tunable quantum geometry in 2D materials, large bulk photovoltaic photocurrents have recently proliferated in a diversity of 2D systems that range from $WSe_2$/black phosphorous heterostructures,[602] monolayer $MoS_2$,[603] and twisted bilayer graphene.[604] These 2D systems have stood apart from those found in traditional bulk photovoltaics due to the immense tunability of 2D materials: interface configuration and dipoles can be directly customized by stacking,[602] quantum geometry and band structure tuned by twist[604] or strain,[603] and Fermi energy can be tuned *in situ* by gate voltage.[604]

### 21.2.1 *Metallic Quantum Geometric Optoelectronics*

Going beyond interband transitions, an emerging range of optoelectronic processes are being explored that are *intraband* in nature. A key example are second-order nonlinearities enabled by the Berry curvature dipole[605] that produce a rectified photocurrent or second-harmonic generation from the dynamics of Bloch electrons on a metallic Fermi surface (see right panels of Figure 21). These have gained recent attention due to the pronounced Berry curvature dipole-induced nonlinear susceptibilities found in 2D materials[606] that are gate tunable as well as their quantum geometric origin. The latter quantum geometric origin is particularly striking since in non-magnetic materials, second-order nonlinearities cannot arise from the simple classical group velocity dynamics of electrons about the Fermi surface even when inversion symmetry is broken; instead, they arise from quantum geometric corrections such as the anomalous velocity.[605] Second-order nonlinearities arising from the classical group velocity of electrons typically only arise when time-reversal symmetry is broken, for example, in magnetic materials. Beyond the berry curvature dipole, a growing zoo of nonlinear intraband processes have emerged that include other intrinsic contributions such as the Berry connection polarizability[607] (see, for example, ref 608 for an experiment in the heterostructure $MnBi_2Te_4$ on black phosphorous), intrinsic Fermi surface contributions,[609] extrinsic scattering such as skew scattering nonlinearities[610,611] (see, for example, ref 612 for an experiment in graphene/hBN), and mixed extrinsic-intrinsic contributions such as anomalous skew scattering.[613]

While most experimental studies in metals have focused on (near DC) low-frequency operation, many of these intraband metallic nonlinearities persist to higher frequencies. Taking a characteristic electron scattering time of about a picosecond in a good metal, these nonlinearities can persist into the THz frequency range providing new tools for *metallic optoelectronics* without p–n junctions. As a result, metallic optoelectronic devices operating in the THz have been recently proposed that include energy harvesters,[614] as well as electro-optic modulators[610,614,615,616] that modify absorption and reflection by an applied in-plane current.

Electro-optic modulation is particularly exciting since it can provide tools for nonreciprocal THz devices. For large enough applied electric fields, these metallic nonlinearities have even been envisioned to produce the preliminary conditions for THz amplification[614,615] without population inversion in a (p–n) junction-less metal. Note that for large electric fields, high-order nonlinear electro-optic effects can ensue; for example, beyond the linear electro-optic modulation (i.e., Pockels effect), higher-order effects such as second-order electro-optic modulation (e.g., nonlinear Kerr effect) can manifest.[616] Systematically taking into account the range of mechanisms responsible for electro-optic modulation (and, more generally, in the use of quantum geometric nonlinearities) in realistic materials remains an ongoing challenge in the field—implementing these promises to reveal new types of electro-optic modulation. Indeed, the metallic Fermi surface can become strongly modified in the presence of periodic driving[617] to produce a rich out-of-equilibrium phenomenology that can depend on the details of relaxation.

Beyond the charge degree of freedom, there has been recent attention focused on using quantum geometry to control other degrees of freedom. In layered materials, quantum geometry can grant optical control over the layer polarization in both chiral[618] and achiral[619] stacks of 2D materials; in a similar fashion, it can enable the manipulation of magnetization and spin.[620]

## 21.3 Challenges and Goals

21.3.1 *Dynamical Control and Light-induced Ferroic Phases*

Light can be used to dynamically control band structures and electronic phases. Indeed, recent experiments in ref 571 have demonstrated light control of a Floquet–Bloch band structure and anomalous Hall effect; and these have only persisted for ultrafast timescales. Pushing these to longer timescales remains an ongoing challenge. One emerging strategy is to exploit the effect of optically induced polarization or magnetization that induce feedback on the underlying band structures. For instance, a light induced layer polarization in rhombohedral trilayer graphene can boost the electric potential drop across the layers transforming screening (the reduction of applied electric fields across the layers) into anti-screening (the amplification of applied electric fields across the layers).[621] Strikingly, this feedback can even sustain dynamical symmetry broken phases such as a light-induced ferroelectric-like state in rhombohedral trilayer graphene[621] or a light-induced ferromagnet in disks of graphene[622]—both are examples of symmetry breaking induced by the quantum geometric dynamics of electrons. Using such optical control to sustain broken symmetry phases as well as a tool for optically controlling the dielectric environment of stacked materials is a new, and as yet experimentally unrealized, vista for future growth.

21.3.2 *Interacting Quantum Geometric Photocurrents*

More generally, the role of electron interactions in modifying fundamental quantum geometric optoelectronic functionality in 2D materials (such as photocurrent) is an ongoing intense area of research. The pronounced quantum geometric photocurrents in twisted moiré materials make it a particularly striking example. For instance, Hartree corrections to the band structure of twisted bilayer graphene can induce appreciable changes to the shift photocurrent;[623] these corrections have been recently used to explain unconventional filling-dependent photovoltage close to charge neutrality in twisted bilayer graphene.[604]

Even more signs that electron interactions may play a dominant and essential role in photocurrent of 2D materials are beginning to show up in experiments in 2D materials. For example, THz photocurrent measurements in magic angle twisted bilayer graphene heterostructures have revealed multiple sign changes in bulk photocurrent as filling is tuned across well-defined fractions ¼, ½, and ¾ of a moiré unit cell.[604] This mirrors the Dirac cascades of magic angle TBG. These measurements simultaneously indicate that photocurrent can be used as a probe of the interacting ground state but also highlight the urgent need for a comprehensive understanding of quantum geometric nonlinearities in interacting systems.

Another sign of the importance of interaction effects is the bulk photocurrent in TMD heterostructures such as $WSe_2$/black phosphorus[602] or strained $MoS_2$ monolayers[603] where photocurrent is induced for frequencies that corresponded to exciton peaks. Importantly, even as excitons are charge neutral, they enable to enhance charge photocurrents[624] and occur below the nominal single-particle bandgap. Surprisingly, photocurrent in unstrained TMD systems were found to be highly suppressed[602,603] even as symmetry allows for them to develop.[603] In contrast, straining the TMDs dramatically activates the photocurrent. Understanding the detailed mechanisms for this suppression in pristine samples remains a challenge in the field.[603]

## 21.4 Future Directions

21.4.1 *Nanophotonic Quantum Geometric Optoelectronics*

Harnessing quantum geometric optoelectronics have largely focused on materials with pronounced quantum geometry whereby symmetry (e.g., centrosymmetry or time-reversal symmetry) is broken intrinsically in the material. These have typically utilized electro-magnetic fields in free space where the interaction of light and matter takes on an electric dipole character. A different strategy is emerging that involves exploiting nanophotonic techniques where the spatial profile of electromagnetic fields either activate or enhance quantum geometric optoelectronics. By accessing the small wavelengths available in a proximal plasmonic material (e.g., graphene), pronounced quantum geometric (polariton-drag) photocurrents can be activated in a target quantum material placed on top of the plasmonic material[625,626] even if the target quantum material possesses high symmetry. For example, the finite wave vector of polaritons can activate circular shift and linear injection type photocurrents in a nonmagnetic and inversion symmetric material.[626] Such photocurrents track higher-order quantum geometry (e.g., shift current dipole[625] or even a momentum-resolved quantum geometry[626]). Indeed, such finite-wave vector quantum geometric optoelectronics are currently being explored even in a single contiguous stack (e.g., twisted spiral multilayer $WS_2$)[627] and represent a promising new direction for enhanced quantum geometric nonlinearities.

21.4.2 *Stack Engineered Optoelectronics*

Another emerging direction is the use of stack engineering to control optoelectronic response. For example, by twisting adjacent layers of graphene together to form a chiral stack,[102] optical circular dichroism was induced; similar dichroism can be activated even at low frequencies arising from the quantum geometric properties (e.g., an in-plane magnetic moment) of a metallic Fermi surface in such a chiral material. Similarly, bulk photocurrents can be sensitive to AB or BA atomic stacking arrangements in gapped bilayer graphene[600] even as the dispersion relations for AB or BA stacked gapped bilayer graphene are the same; instead AB/BA stacking arrangements are imprinted into the shift vector—a property of electronic wave function. By tailoring stacks rationally engineered in the $z$ direction,[627] artificial materials with pronounced quantum geometric optoelectronic responses can be envisioned.

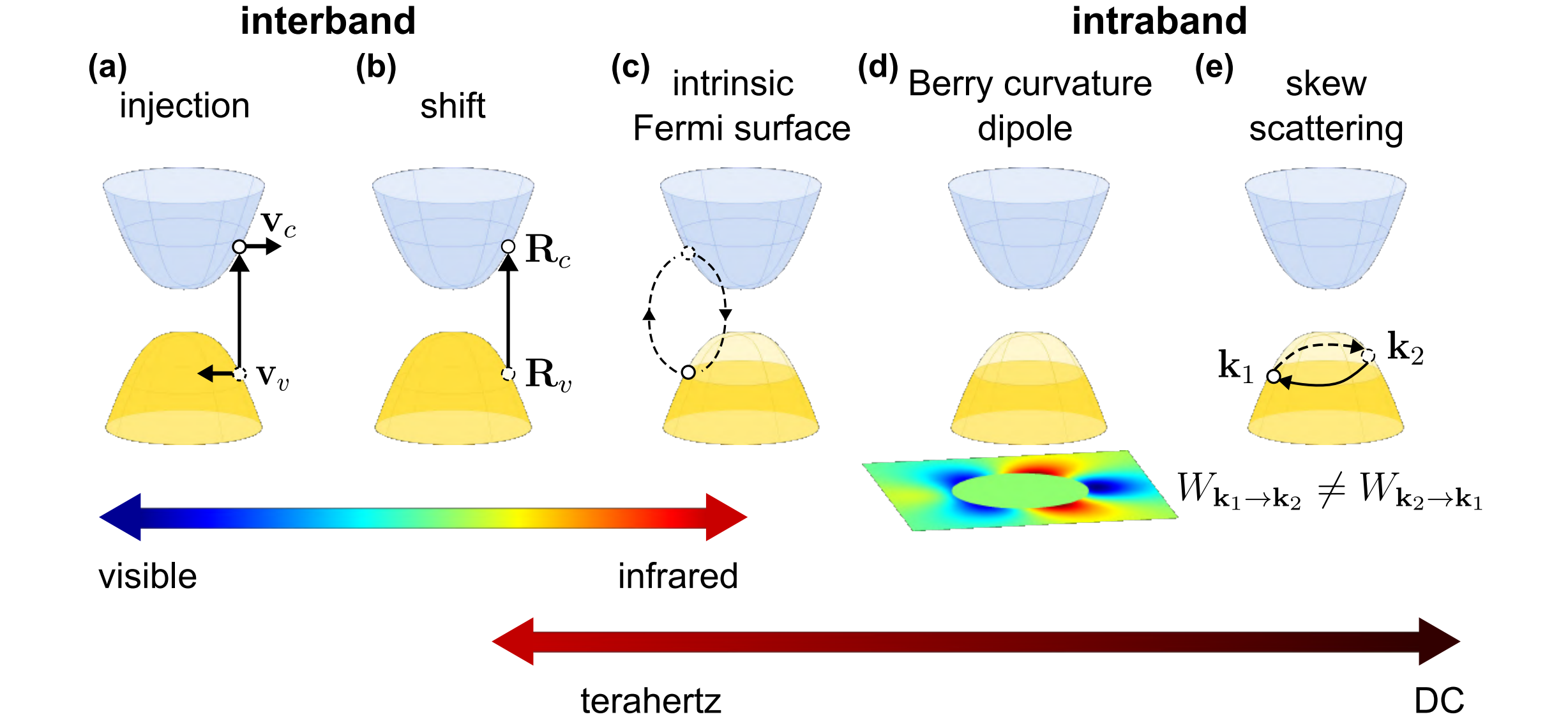

**Figure 21.** Bulk quantum geometric photocurrents. (a,b) Injection (a) and shift photocurrents (b) arise from the change in electron group velocity ($\mathbf{v}_c$ and $\mathbf{v}_v$ in the conduction and valence bands) and real space displacement ($\mathbf{R}_c$ and $\mathbf{R}_v$ in conduction and valence bands) when electrons are photoexcited from valence to conduction bands in non-centrosymmetric materials. (c-e) In the presence of a Fermi surface, a variety of other mechanisms become activated including (c) an intrinsic Fermi surface effect (arising from virtual interband processes), (d) a Berry curvature dipole (inset; asymmetric distribution of Berry curvature across a Fermi surface), and (e) skew scattering contributions (produced when the transition rates $W$ between Bloch states at momenta $\mathbf{k}_{1,2}$ become imbalanced). These mechanisms provide means for achieving a type of optoelectronics based on bulk non-centrosymmetric metals.

## 22. SINGULARITIES OF RANDOM WAVES IN 2D MATERIAL PLATFORMS

**Tomer Bucher,[1] Alexey Gorlach,[2] Shai Tsesses,[3] and Ido Kaminer[4,*]**
[1]Andrea & Erna Viterbi Department of Electrical and Computer Engineering, Technion–Israel Institute of Technology, 3200003 Haifa, Israel
[2]Department of Physics, MIT, Cambridge, MA, US.
***Corresponding author.** Email: ido.kaminer@gmail.com

### 22.1 State of the Field

Wave singularities, such as vortices and field dislocations, refer to points or lines in a wave field where intensity vanishes, and the wave phase is undefined. These singularities have been a central topic in all areas of physics due to their universal properties across different systems. Since the pioneering works by Berry and Nye on singularities and dislocations in electromagnetic waves,[628,629] various types of singularities have been demonstrated using light, such as Möbius rings,[630] knots,[631] skyrmions,[632] and hopfions,[633] with each of these holding a conserved topological charge, a quantized property representing the underlying structure of the singularity (Figure 22a,b).

The topological charge can imbue light with well-defined physical traits, such as angular momentum,[634,635] as first demonstrated in the phase of propagating optical beams,[636] making optical singularities a versatile tool for many applications. Such applications include control over light–matter interactions with either bound[637,638,639] or free[640,641,642] electrons; super-resolution imaging techniques;[643] and the encoding of classical[644] and quantum[645] information.

Singularities in polariton waves, and specifically surface plasmon polaritons, provided exceptional degrees of tunability of the singularities' dynamics and singularity-pairs creation and annihilation[646] events, reaching exotic regimes for light–matter interaction.[647] In such systems, the spin–orbit interaction of light[648] was shown to alter the physical properties of singularities, leading to the creation of unique vector singularities in the electric field[649,650] and to the time-resolved observation of the rotation of light.[651]

Of particular interest are singularities in polariton waves propagating in 2D materials, such as graphene and hBN. Such materials have exceptional tuneability—via biasing[15,16] or the construction of heterostructures[44]—facilitating unique optical properties, like hyperbolic dispersion,[652] manipulation of magnetic fields,[653] and low group velocities with long lifetime.[11] The tunability of these properties in 2D materials provides novel ways of controlling light–matter interactions and altering the dynamics of optical singularities.[654]

### 22.2 Challenges and Goals

One of the universal characteristics of singularities that has proven challenging to study experimentally is their correlations and statistical properties in random waves,[655] which were studied theoretically at length.[656,657] The vortex patterns emerging in ensembles of singularities exhibit spatial correlations (Figure 22c) that share similarities with liquid-like distributions,[657] hinting at an underlying organizing principle in random wave systems. While theoretical frameworks have been developed to describe such singularity correlations,[656] capturing their evolution and interactions remains challenging. Experimental investigations on random waves[657,658] provided insights into spatial correlations of vortices, but no work so far observed their dynamical correlations (Figure 22d).

2D materials offer a robust platform to study such collective dynamics of optical singularities, with unique properties that support advanced light–matter interaction. For example, the high tunability and slow group velocities enable the study of dynamics in ensembles of singularities, and the signatures of creation-annihilation events.[654] Leveraging the high tunability of 2D materials could provide access to new properties of singularity statistics, for example, their dynamics in the presence of nonlinearity,[539] especially by employing advanced fabrication methods to sculpt 2D-polariton cavities.[59] Recent advancements in ultrafast electron microscopy[264] enable the detection of low-intensity fields,[659] which is especially important for resolving the singularities and their dynamics accurately because the field near every singularity quickly drops to zero. Enhancing detection sensitivity is especially important in creation and annihilation events, which are hard to accurately pinpoint, and can allow observing and analysing otherwise inaccessible phenomena in sensitive materials that cannot withstand strong fields.

### 22.3 Future Directions and Conclusions

Looking forward, heterostructure engineering[2] could control material phase transitions, and study how the emergence of topological order may affect the statistics of singularities emerging by their light–matter interaction. Controlling the dispersion of unique structures like molybdenum trioxide[44] may induce topological transitions not only in the wave dispersion but also in the behavior of vector singularities, similar to topological phase transitions.[660] By modulating both in-plane and out-of-plane interactions,[2] these heterostructures allow for tunable electronic band structures achieving phase transitions that will change the statistics and correlations of singularities in a controllable manner. A complementary approach could be altering the coupling of the electric field to the magnetic field, which will incorporate interesting statistics and correlations among singularities.

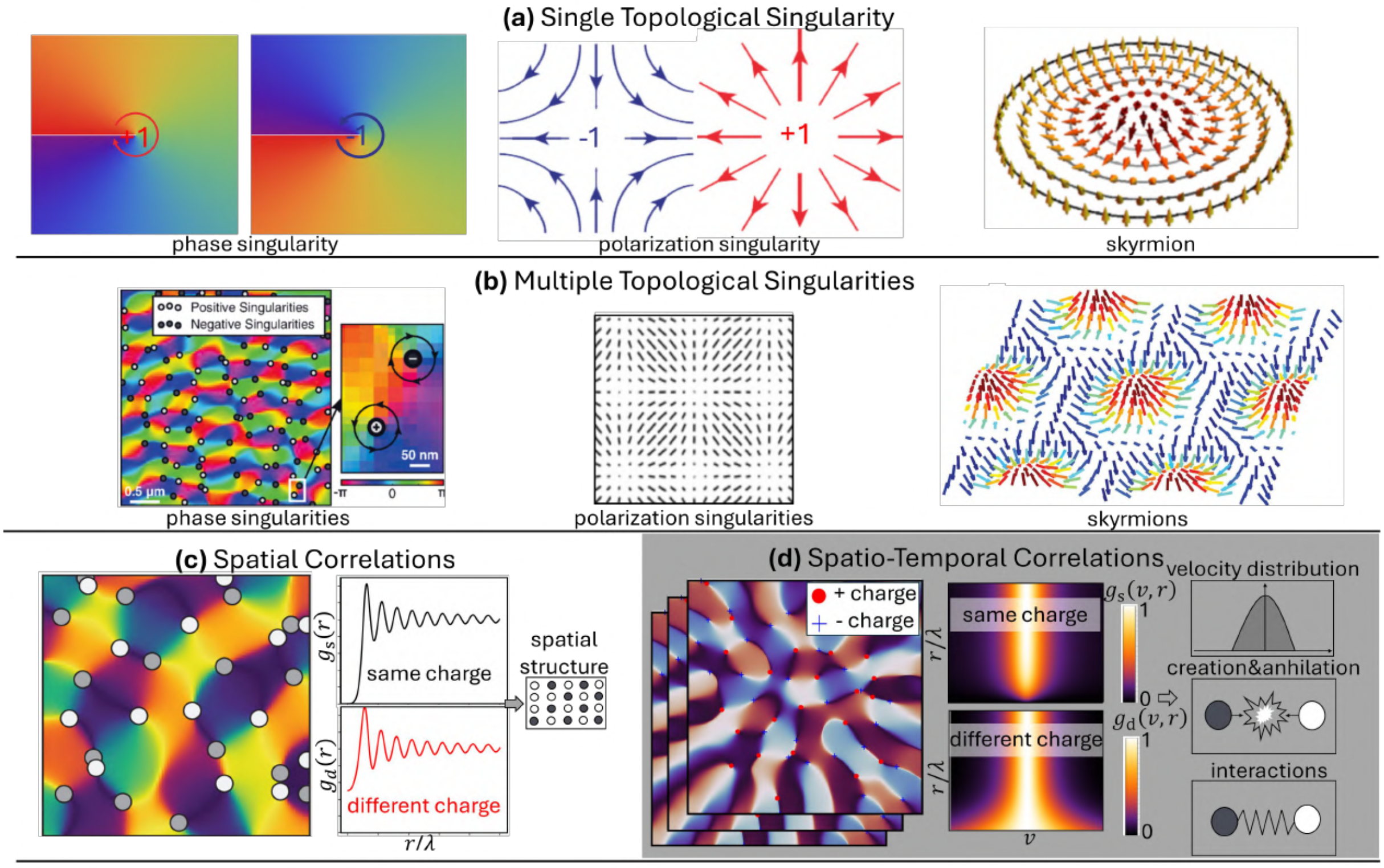

**Figure 22.** Overview of optical singularities and their correlations. (a) Different types of optical singularities and their experimental demonstrations including phase singularities, polarization singularities, and more complex types of singularities such as knots and skyrmions. (b) A vast range of systems show natural occurrences of multiple singularities, which can interact and annihilate in pairs, conserving their combined topological charge. Adapted with permission from refs 649 and 657 (Copyright 2020 American Association for the Advancement of Science and 2016 American Physical Society). (c) Ensembles of phase singularities emerging from the interference of random waves have spatial correlations as theoretically predicted in ref 656 and experimentally demonstrated in ref 657. Adapted with permission from ref 657 (Copyright 2016 American Physical Society). (d) The interference of random wave packets results in spatiotemporal dynamics of phase singularities, characterized by the creation and annihilation of vortex pairs. The spatiotemporal correlations of these phase singularities remain an underexplored area, warranting further investigation.

## 23. 2D TOPOLOGICAL POLARITONICS

**Julian Schwab,[1] Florian Mangold,[1] and Harald Giessen[1,*]**
[1]4th Physics Institute, Research Center SCoPE, and Integrated Quantum Science and Technology Center, University of Stuttgart; 70569 Stuttgart, Germany
***Corresponding author.** Email: giessen@pi4.uni-stuttgart.de

The advent of topological plasmonics has led to the discovery of plasmonic skyrmions[649] and merons.[661] Methods such as scanning near-field optical microscopy (SNOM) and two-photon photoemission electron microscopy (PEEM) gave the necessary spatial resolution in the sub-10 nm range to explore skyrmions and merons. By combining PEEM with ultrafast laser sources with pulse durations of around 16 fs, it was also possible to reveal the ultrafast dynamics of these excitations.[650,662]

Skyrmions and merons are 3D topological defects on a 2D plane, which can exhibit particle-like properties.[663] They can be characterized by certain vector field properties such as the skyrmion number $S$, the polarity $P$, and the vorticity $V$, which are defined as

$$S = \oiint_A s\, dA = \frac{1}{4\pi} \oiint_A \mathbf{e} \cdot \left( \frac{\partial \mathbf{e}}{\partial x} \times \frac{\partial \mathbf{e}}{\partial y} \right) \mathrm{d}x\mathrm{d}y = P \cdot V,$$

$$P = -\frac{1}{2} \int_0^R \frac{\partial e_z}{\partial r} \mathrm{d}r,$$

$$V = \frac{1}{2\pi} \int_0^{2\pi} \frac{\partial e_\phi}{\partial \phi} \mathrm{d}\phi,$$

respectively. The polarity $P$ gives the number of rotations that the out-of-plane component of a vector carries out when going along the radius from the center of the defect to the boundary of the topological structure. The vorticity $V$ gives the number of rotations that the in-plane component of a vector carries out when going azimuthally around the defect of the topological structure. The skyrmion number is given as the product of polarity and vorticity, $S = P \cdot V$. Skyrmions, which exhibit $S = 1$, have been demonstrated in a variety of different experimental realizations, both in free space as well as in plasmonic systems.[664] Just recently, also higher-order integer skyrmion numbers in the form of skyrmion bags have been discovered in plasmonics.[665] Fractional skyrmion numbers lead to so-called merons, and different meronic structures have also been found, in free space as well as in plasmonic systems.[650,661] Not only the electric vector field can be analyzed regarding its structure, but also magnetization,[666] magnetic field components of the electromagnetic field, spin angular momentum $\mathbf{S} = \mathrm{Im}\{\epsilon\, \mathbf{E}^* \times \mathbf{E} + \mu\, \mathbf{H}^* \times \mathbf{H}\}/(2\omega)$, and, recently, also the Poynting vector of the free-space electromagnetic field[667] have exhibited such topological structures.

Du *et al.*[668] and Zheludev *et al.*[669] have pointed out that certain features in these topological vector field structures can be much smaller than the wavelength of the excitations, since – in contrast to variations of the intensity – spatial changes of the direction of the electromagnetic field are not limited by diffraction. In particular, Du *et al.* pointed out that spin texture features can be deeply subwavelength and reported spin reversals on length scales as small as $\lambda/60$ = 10 nm. This could make topological light excitations an attractive candidate for high-resolution imaging and precision metrology.

In plasmonics, the so-called long-range surface-plasmon polaritons (SPPs) are a common subject of investigation. These are evanescent surface waves at the boundaries between a thick metal and a dielectric, for example the gold-air interface of a gold flake. However, when going toward thin gold films, in the range of 20 nm or thinner, it is possible to excite surface plasmon polaritons on both sides of the thin metallic film using light impinging from above.[670,671] Hybridization between those two SPPs takes place, forming an odd and even coupled mode. The odd mode is nonradiative and hence exhibits a long propagation distance, typically for gold in the range of a few hundred μm, hence it is termed long-range SPP, whereas the even mode is radiative and hence has only a limited propagation distance of a few μm. However, the dispersion of this long-range SPP is very close to the light line, resulting in a SPP wavelength that is very close to the excitation light wavelength. Yet the dispersion of the short-range SPP is more plasmon-like, thus giving a slope that deviates strongly from the light line (it is much flatter) and exhibits hence a higher effective refractive index, leading to compression of the SPP wavelength. Values as short as 176 nm have been measured experimentally for incident light at 800 nm,[670] yielding a compression factor of around 4.5. In graphene, the ultimate 2D material, Koppens *et al.* have demonstrated surface plasmon polaritons that exhibited a compression factor of up to 150 ($\lambda_{\mathrm{SPP}}$ = 70 nm for 10.5 μm light wavelength),[4,9,15,16] which has led plasmonic skyrmions with wavelength compression factors of up to 100.[672]

Therefore, it is very exciting to study other polaritonic excitations.[63] Especially, phonon polaritons have been known to exhibit extreme dispersion values that can give effective refractive indices of over 100. Hallmark systems include SiC[673,674,675] or hBN.[169] Here, we would like to expand the concept of topological excitations to phonon polaritons in thin and 2D solid-state materials, with the possibility to utilize their strong dispersion for extreme wavelength compression and hence ultradeep subwavelength structured light features.

Figure 23a depicts the dispersion relation of surface phonon polaritons (SPhPs) in SiC membranes.[676] As the phonon (i.e., the oscillatory vibration of the atoms in the solid lattice) can occur transversely and longitudinally with respect to the phonon propagation vector, the phonon dispersion is split into a transverse and a longitudinal optical branch. In the forbidden zone between the transverse and the longitudinal phonon resonances $\omega_{\mathrm{TO}} = 797\ \mathrm{cm}^{-1}$ and $\omega_{\mathrm{LO}} = 973\ \mathrm{cm}^{-1}$ (Reststrahlenband), the permittivity is negative and leads to an evanescent character of plane waves in the medium. Therefore, localized and propagating SPhP modes can be supported analogous to SPP resonances in metals. As visible in Figure 23a, the phonon polariton dispersion of the odd mode exhibits extreme sublinear behavior for larger wavenumbers and deviates strongly from the light line, allowing for smaller SPhP wavelengths. This effect is even stronger for thinner polaritonic media, ideally for 2D materials.

Figure 23b displays the oscillatory behavior of the different atoms in a polar lattice for a phonon polariton mode, yielding a dipole moment to which the light field can couple. Furthermore, the two possible solutions of Maxwell's equations for SPhP propagation on the upper and lower interface are indicated, yielding an odd and even solution. We would like to focus here onto the odd solution as the even mode is either very close to the light line or has a propagation length of $L/\lambda_{\mathrm{SPhP}}$ below unity.

Recently, Mancini *et al.*[677] have utilized the sublinear SPhP dispersion on a 200 nm SiC membrane to tune the phonon polariton wavelength over an extremely wide range, ranging from 1.5 μm to 10 μm for incident light wavelengths of around 11 μm. Using this extreme sublinear dispersion, they were able to demonstrate multiplication of the orbital angular momentum of phonon polaritons. In fact, they could achieve OAM tuning from $l = 1$ to $l = 7$ when tuning the laser merely from 880 to 924 $\mathrm{cm}^{-1}$ (i.e., a relative tuning of only 5% of the wavelength). They obtained a wavelength compression factor as large as 7.

Here, we suggest to utilize SPhPs to realize strongly compressed topological excitations, such as phonon-polaritonic merons and skyrmions. We theoretically predict that much higher compression factors of the SPhP wavelength should be attainable using thinner SiC membranes as visualized in Figure 23a and suggest using sub-100 nm thick SiC membranes with suitable fabricated boundary structures to launch the SPhPs. In fact, we devise using concentric rings, or hexagonally shaped boundaries and excite them using circularly polarized light for skyrmions to appear in the spin angular momentum. Alternatively, Archimedean spirals with an opening of $\lambda_{\mathrm{SPhP}}$ ($l = 1$), or hexagonally shaped boundaries with an offset shift of $\lambda_{\mathrm{SPhP}}/6$ for each hexagonal border can be used to attain skyrmions in the electric field. Likewise, merons can be obtained analogous to previous approaches. Using transmission SNOM with impinging light of the necessary polarization for a tunable mid-IR laser, the out-of-plane component of the electric field could be detected. Using $\mathrm{div}\,\mathbf{E} = 0$ from Maxwell's equation, the in-plane components can subsequently be derived, yielding the complete phonon-polaritonic vector field. This way, for a 200 nm SiC membrane, phonon polaritonic skyrmions with a SPhP wavelength of 800 nm at an incident wavelength of 10.8 μm (924 $\mathrm{cm}^{-1}$) and a compression factor of 13.5 should be attainable. This compression increases rapidly for thinner membranes of 50 nm (10 nm) to factors of 27 (135). This would yield the possibility of obtaining deep subwavelength spin texture features in phonon polaritonic topological excitations, resulting in sub-100 nm sized features for an incident wavelength of over 10 μm!

Figure 23c illustrates such a phonon polaritonic skyrmion vector field in 3D view and in a side-cut view. Note that the length scale is determined by the SPhP wavelength and could thus be on a deep sub-micrometer scale.

One possibility would be to use hBN, $MoO_3$, or other vdW materials which are in the few nm thickness range and would give extreme confinement way beyond 100, leading to structured light features in the sub-100 nm range with laser wavelengths in the mid-IR spectral region. In h-BN, we would find isotropic 2D dispersion and hence isotropic skyrmions, merons, and such, whereas in $\alpha$-$MoO_3$, one could expect elliptical skyrmions, as the dispersion in the 2D plane is anisotropic. This is very unique and can only be found in 2D materials with strong anisotropy.[678] An exciting perspective is coupling of such phonon polaritonic structured light to matter excitations (OAM light or light with nontrivial topological vector textures to chiral material excitations, for example, in Weyl semimetals).[111]

As phonon polaritons are inherently nonlinear, one could also use this concept to try generation of nonlinear topological excitations, for example, solitons, so that eventually skyrmion–kyrmion interaction becomes possible and could resemble DMI interaction as in magnetic skyrmions.

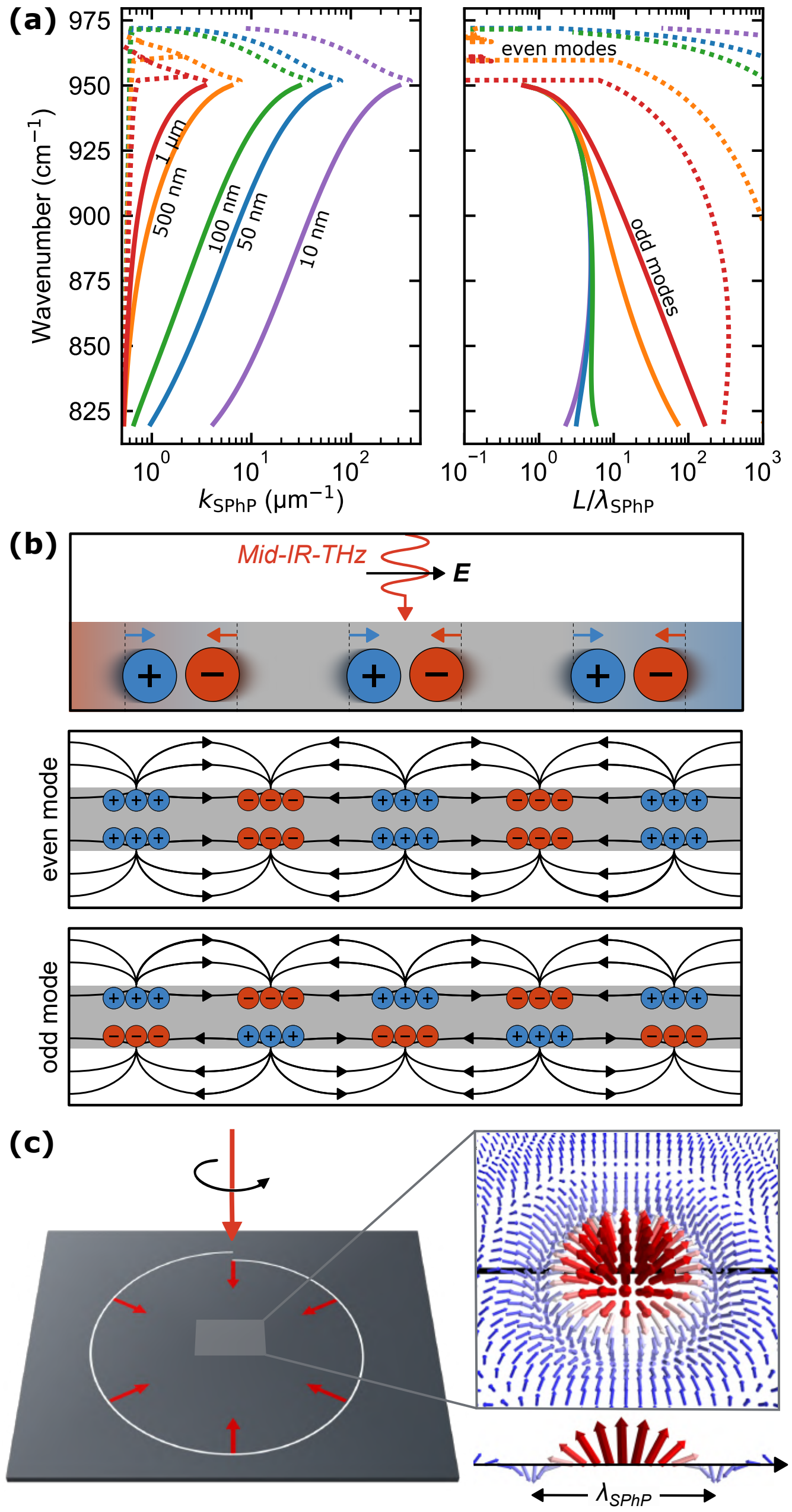

**Figure 23.** Surface-phonon polaritonic (SPhP) skyrmions and other topological features with varying spatial compression. (a) Theoretical SPhP dispersion for different SiC membrane thicknesses from 10 nm to 1 µm. Displayed are the odd and even modes, calculated using the method outlined in ref 677. On the left, the excitation wavenumber is plotted as a function of SPhP wave vector and on the right the excitation wavenumber is given as a function of the normalized propagation length $L/\lambda_{SPhP}$. There are two even modes, one of them very close to the light line and the other one with propagation length $L < \lambda_{SPhP}$. The odd mode features high propagation lengths and strong wavelength compression. Such wavelength compression increases rapidly for thinner membranes. (b) Schematic representation of collective atom displacements in polar dielectrics as the foundations of even (odd) SPhP modes that are illustrated in the center (bottom) image. (c) Exemplary implantation of a skyrmion in the SPhP electric field. SPhPs are launched from a spiral structure with an opening of $\lambda_{SPhP}$, which is illuminated by circularly polarized light in the mid-IR. The SPhPs interfere in the center of the structure, creating a skyrmion vector texture in the electric field, as depicted in the vector plot and the 2D cut on the right. The size of the skyrmion is determined by the SPhP wavelength and can be compressed strongly in comparison to the excitation wavelength by using thin SiC membranes.

## 24. CHIRALITY IN MOIRÉ SYSTEMS

**M. Sánchez Sánchez,[1] D. K. Efetov,[2,3] T. Low,[4] G. Gómez-Santos,[5] and T. Stauber[1,*]**
[1]Instituto de Ciencia de Materiales de Madrid (ICMM-CSIC), E-28049 Madrid, Spain
[2]Fakultät für Physik, Ludwig-MaximiliansUniversität, D-80799 München, Germany
[3]Munich Center for Quantum Science and Technology (MCQST), D-80799 München, Germany
[4]Department of Electrical & Computer Engineering, University of Minnesota, Minneapolis, Minnesota 55455, USA
[5]Departamento de Física de la Materia Condensada, Instituto Nicolás Cabrera and Condensed Matter Physics Center (IFIMAC), Universidad Autónoma de Madrid, E-28049 Madrid, Spain
***Corresponding author.** Email: tobias.stauber@csic.es

### 24.1 Introduction

Properly designed metasurfaces offer control over amplitude, phase, and/or polarization state of light at a certain frequency range. Including more layers, with or without direct coupling between them, can further increase the control and also open up more functionalities. Multilayer metaoptics thus comprises two or more independent metasurfaces that are designed to operate as sequential optical elements.[27]

Usually, these metasurfaces are separated by a few wavelengths of the incoming light and the optical elements are bulky and difficult to couple. Moiré materials may provide a viable alternative as their layers are only a few Ångströms apart and display a variety of correlated phases.[679] This atomistic separation leads to quantum-mechanical coupling between the layers and may give rise to novel chiral properties that cause intriguing optical phenomena such as circular dichroism (CD) (see ref 102). Furthermore, the CD shows a resonance at transitions of states that are maximally hybridized between the two layers, depending on the twist angle (see Figure 24a).

In this section, we summarize the current state of the art regarding chiral moiré systems and possible applications. We also discuss strategies to enhanced the chiral response in moiré multilayers. We close with future goals and experimental directions.

### 24.2 Current State of the Art

Polarization control is a crucial aspect of optoelectronic devices, and twisted 2D material stacks, including twisted bilayer graphene (TBG), offer new avenues for tunable polarization optics. Twisted vdW materials, either using quantum interlayer coupling[103,680,681] or in-plane birefringence,[682] can be used to create polarization transformers, such as wave plates and polarizers. By adjusting the twist angle and doping levels in TBG, the eigenstates of the system's Jones matrix—which describes the polarization transformations of incident light—can be modified. This leads to the development of tunable wave plates that can function as elliptic or linear birefringent elements, useful for controlling the polarization state of light. These properties highlight the potential of TBG-based devices in polarimetry and Stokes polarimetry, where multiple polarization states can be measured to infer unknown polarization characteristics of light. The high degree of tunability in TBG makes it a promising candidate for reconfigurable photonic circuits and polarization-sensitive detectors[683] in advanced optical communication systems.

TBG offers the possibility of rotating the polarization plane without breaking time-reversal symmetry since the two enantiomers can be distinguished by their relative rotation between the two layers—to the left or to the right. It thus displays a circular dichroism without the presence of a magnetic field, which becomes largest for optical frequencies that correspond to transitions close to the van Hove singularities (i.e., where the hybridization between the two layers is largest). However, chiral effects are also expected in the static (intraband) regime, and plasmons, with its confined wavelength, generally allow for substantial enhancement in light–matter interactions.[671]

Plasmonic excitations display chiral effects in TBG as the electric dipole oscillations are linked to magnetic dipole oscillations. This coupling is either parallel (+) or antiparallel (−) as longitudinal current densities in one sheet give rise to transverse current densities in the other sheet, see Figure 24b. Interestingly, the sign (+/−) not only depends on whether the twist angle is clockwise or anti-clockwise, but also whether the twist angle is less or greater than the *magic angle* at which flat bands emerge.[684] Nevertheless, the dispersion of bulk plasmonic modes is only slightly changed from the usual square-root behavior as for chiral effects retardation needs to be included.[685] These effects lead to a spin component of the confined plasmonic near field that also points in the propagation direction (see Figure 24c). We finally note that the energy of plasmonic edge modes in TBG depends on chirality even in the nonretarded regime.[686]

TBG's chiral sensing capabilities provide an attractive platform for biosensing and enantioselective photochemistry (see Figure 24b).[104,687,688,689,690] Especially, the interaction between plasmons and electromagnetic waves in TBG can be harnessed to detect small changes in the optical properties of surrounding materials, making it a powerful tool as a label-free biosensor. The ability to enhance chiral interactions through external tuning mechanisms, such as electrostatic doping or changing the twist angle, makes TBG an ideal platform for testing these ideas.

### 24.3 Challenges

The constitutive equations for an isotropic moiré bilayer such as TBG are given by

$$\begin{pmatrix} \mathbf{p} \\ \mathbf{m} \end{pmatrix} = \begin{pmatrix} \alpha & -G \\ G & \chi \end{pmatrix} \begin{pmatrix} \mathbf{E} \\ \mathbf{B} \end{pmatrix}, \tag{24.1}$$

where the electric and magnetic moments $\mathbf{p}$ and $\mathbf{m}$ are usually directly related to the electric and magnetic fields $\mathbf{E}$ and $\mathbf{B}$ *via* the electric polarizability $\alpha$ and magnetic susceptibility $\chi$. In chiral systems, though, also the electromagnetic response $G$ enters.

The chiral response is linear with the interlayer separation ($G \sim d_\perp$) and, thus, small due to the atomistic spacing ($d_\perp \approx 3.4$ Å). Systems with (substantially) larger interlayer separation can principally be ruled out since the chiral effects require a quantum mechanical coupling between the layers which is exponentially reduced for larger distances. However, we may explore several stacks of the same interlayer separation and look for enhanced chirality in moiré multilayers.

The constitutive relations in eq 24.1 are obtained from linear response theory for the layer-resolved sheet-current density $\mathbf{j}^i$ that is induced by layer-resolved in-plane electric fields $\mathbf{E}^i$:

$$\mathbf{j}^i(\omega) = \sum_{i'=1\ldots N} \boldsymbol{\sigma}^{i,i'}(\omega)\mathbf{E}^{i'}(\omega)\,, \tag{24.2}$$

where $\boldsymbol{\sigma}^{i,i'}$ denote $2 \times 2$-matrices for the layer-resolved conductivity tensors. In eq 24.2, we have generalized the response function to $N$ layers with $i, i' = 1 \ldots N$ and the conductivity tensors are further related to each other through the electronic structure and geometry of the system. Note that, in this framework, chirality is described by the nonlocal response with respect to the layers (i.e., it is encoded in the nondiagonal matrix elements of $\boldsymbol{\sigma}^{i,i'}$ with $i \neq i$). A detailed analysis of this equation should be able to guide experimentalists to exploit chiral responses in moiré multilayers.[691]

Apart from this theoretical challenge, an experimental challenge is to contact the layers separately which seems feasible if the twisted graphene sheets have different dimensions. This would open up the possibility to observe the layer-contrasted Hall effect and a chiral supercurrent in twisted bilayer,[692] and we expect more complex response patterns for three, four, or even more layers especially when normal modes close to resonances are excited.

### 24.4 Future Goals

The circular dichroism in moiré multilayers scales with the number of layers. However, mechanical stacking of these moiré multilayers to a thickness comparable to the optical wavelength is not feasible experimentally. Recently, it was demonstrated that continuously twisted chiral TMD superstructures can be grown experimentally (see Figure 24d).[693,694] This breakthrough suggests a potential pathway for producing moiré multilayers with thickness approaching the light or plasmon wavelengths.

Also, recent interest in emerging optical responses arising from the quantum geometry of Bloch electrons in moiré multilayers warrants further examination. Magnetoelectric responses can further arise from the Berry curvature dipole, the magnetic moment texture of Bloch electrons, or the concomitant presence of these two.[695] Optoelectronic responses that probe the Berry curvatures and their dipoles in moiré graphene bilayers have been studied experimentally.[683,696] Their emerging magnetoelectric responses present a compelling avenue for further experimental investigation.

Finally, chiral molecules are widespread, constituting a significant fraction of pharmaceuticals, flavorants, and biological building blocks. Using light–matter interactions to drive chiral-specific chemistry is appealing for potentially enhanced selectivity and tunability. Chiral graphene platforms can selectively activate bonds in only one enantiomer of a racemic mixture. Achieving enantioselectivity in molecular excitation is challenging because the asymmetry in light absorption between chiral pairs is minimal. However, chiral plasmons in twisted bilayers generate near-field electromagnetic waves with significantly larger asymmetry. These chiral polaritonic platforms would also allow for precise tuning of plasmonic resonances through electrostatic doping and twist-angle control, enabling alignment with the vibrational modes of the target chiral molecules.

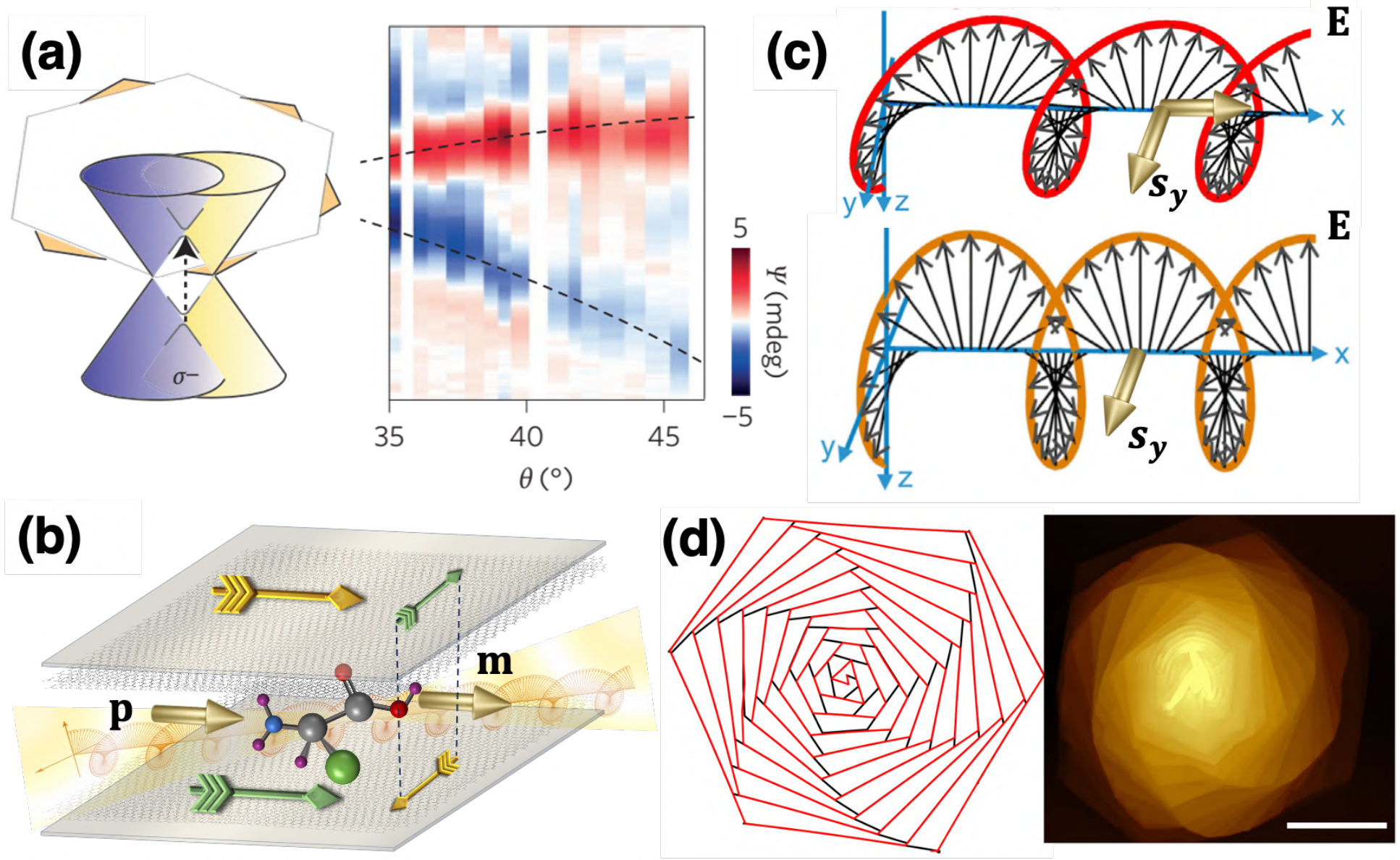

**Figure 24**. Chiral response in moiré multilayers. (a) Ellipticity or circular dichroism spectra due to the resonant interband transition at the M point as function of the twist angle observed in twisted bilayer graphene. Adapted with permission from ref 102 (Copyright 2016 Springer Nature) (b) Schematic chiral cavity consisting of moiré systems with enhanced chiral near field suitable for sensing. Adapted with permission from ref 687 (Copyright 2020 American Chemical Society). The electric moment **p** due to longitudinal current densities is locked to a magnetic moment **m** due to transverse current densities. (c) The electric field associated with plasmon excitations propagating in the $x$ direction together with its helicity/spin. For ordinary plasmons, the spin points in the $y$ direction, whereas for chiral plasmons there is also a component in the $x$ direction. Adapted with permission from ref 685 (Copyright 2020 American Physical Society). (d) Schematic view of supertwisted multilayers and its AFM image of a representative supertwisted $WS_2$ sample with continuously twisting layers (scale bar 4 μm). Adapted with permission from ref 694 (Copyright 2024 Springer Nature).

# 25. TWISTOPTICS: CONTROLLING LIGHT AT THE NANOSCALE WITH TWISTED 2D LAYERS

**Gonzalo Álvarez-Pérez,[1,2,3] Jiahua Duan,[4,5] Andrea Alù,[6] Tony Low,[7,8] Luis Martín-Moreno,[9,10] Alexander Paarmann,[11] Joshua D. Caldwell,[12] Alexey Y. Nikitin,[13,14] and Pablo Alonso-González[1,2,*]**

[1]Department of Physics, University of Oviedo, Oviedo, Spain.
[2]Center of Research on Nanomaterials and Nanotechnology, CINN (CSIC-Universidad de Oviedo), El Entrego, Spain.
[3]Istituto Italiano di Tecnologia, Center for Biomolecular Nanotechnologies, Via Barsanti 14, 73010 Arnesano, Italy
[4]Centre for Interdisciplinary Science of Optical Quantum and NEMS Integration, School of Physics, Beijing Institute of Technology, Beijing, China.
[5]Center for Quantum Physics, Key Laboratory of Advanced Optoelectronic Quantum Architecture and Measurement (MOE), School of Physics, Beijing Institute of Technology, Beijing, China.
[6]Photonics Initiative, Advanced Science Research Center, City University of New York, 85 St. Nicholas Terrace, 10031, New York, USA
[7]Department of Electrical and Computer Engineering, University of Minnesota, Minneapolis, Minnesota 55455, USA
[8]Department of Physics, University of Minnesota, Minneapolis, Minnesota 55455, USA
[9]Instituto de Nanociencia y Materiales de Aragón (INMA), CSIC-Universidad de Zaragoza, Zaragoza 50009, Spain
[10]Departamento de Física de la Materia Condensada, Universidad de Zaragoza, Zaragoza 50009, Spain
[11]Fritz Haber Institute of the Max Planck Society, Berlin, Germany.
[12]Vanderbilt University, Nashville, TN, USA
[13]Donostia International Physics Center (DIPC), Donostia, San Sebastian, Spain.
[14]IKERBASQUE, Basque Foundation for Science, Bilbao, Spain.
***Corresponding author.** Email: pabloalonso@uniovi.es

## 25.1 Introduction and State of the Art

One of the most attractive features of vdW crystals is their straightforward mechanical exfoliation into extremely thin layers, potentially down to atomic-scale thicknesses. Mechanical exfoliation enables the production of nanoscale materials with minimal defects, without the need for conventional nanofabrication techniques like lithography or milling, which always introduce optical losses to some extent. Moreover, vdW layers can be stacked, regardless of their lattice parameters, to form homo- or heterostructures with unique properties, allowing even layers with significantly different lattice constants to be easily overlaid. This feature not only adds versatility and precise control over the resulting structure but also enables stacking at arbitrary twist angles. This unique twisting-and-stacking technique was crucial for the discovery of unconventional superconductivity[87] and correlated insulating states[88] in twisted bilayer graphene, a landmark in condensed matter physics research in the last decade that initiated the field of *twistronics*.

In 2020, an analogous breakthrough in nanophotonics occurred when four independent research groups experimentally demonstrated that the propagation of polaritons could be significantly altered by stacking flakes of in-plane hyperbolic materials[43,44,697,698] at different twist angles. The studies showed that, when two such flakes are stacked with their optical axes not aligned, hyperbolic phonon polaritons (HPhPs) in the top flake may propagate along in-plane directions that are different from those in the bottom flake. Since the electromagnetic field of the modes in both layers partially overlaps, they can hybridize, with the strength of the coupling depending on the relative orientation of the polaritonic propagation directions, which are related to the opening angle of each layer's hyperbolic isofrequency curve (IFC, a cross section of the polariton dispersion at a fixed frequency). This coupling results in hybridized HPhPs with intriguing physical properties: the propagation of PhPs in the bilayer can change dramatically based on the twist angle. Specifically, as the twist angle increases, the IFC of the coupled system undergoes a transition from a hyperbolic shape to an elliptical one. This marks a topological transition from an open to a closed curve, which can happen as a function of both the twist angle and the frequency.[699] At the specific twist angle for which this transition occurs—referred to as *photonic magic angle* in analogy to its electronic counterpart in twistronics—, the IFC is bound to exhibit a flat band, resulting in highly collimated propagation at a single, well-defined propagation direction. Such flat IFC is associated to the *canalization* regime, meaning that the wave can propagate without geometrical decay, that is, the propagation is diffraction-less. While topological transition and canalization of polaritons can be observed in hyperbolic metasurfaces at a specific frequency around the metasurface resonance, the twisted geometry offers the opportunity to tune these effects across a broad range of frequencies, simply by twisting. Such broadband canalization phenomenon was directly observed *via* nano-IR imaging of PhPs in twisted α-$MoO_3$ slabs, initiating the field of *twistoptics* (Figure 25). In contrast to twistronics, the slabs in twistoptics are typically tens to hundreds of nanometers thick.

Nevertheless, twisted bilayers support a single predetermined magic angle at a given frequency, that is, the propagation direction of canalized polaritons cannot be controlled in the same device without changing the twist angle. Another work in 2023 demonstrated that, by adding a third layer, the resulting device can support multiple photonic magic angles at a given frequency, enabling in-plane, all-angle polariton canalization.[700] Moreover, magic angles in twisted α-$MoO_3$ trilayers can be made robust to frequency variations, a phenomenon known as *broadband canalization* of polaritons.

Apart from polariton canalization effects, it is also possible to realize platforms and devices where the polaritonic properties become dependent on the twist angle. For example, placing an α-$MoO_3$ slab on an array of sufficiently separated metallic ribbons allows Fabry–Pérot nanoresonators to be uniquely tuned by simply rotating the host crystal.[701] The principles of twistoptics can also be extended to polaritonic crystals:[702,703,704,705] rotating an α-$MoO_3$ layer on a periodic hole array fabricated in a metallic layer enables the selective excitation of Bloch modes, an effect that can also be achieved by patterning the hole array directly onto the α-$MoO_3$ flakes.

Beyond directing the flow of energy in the near-field, twisted stacks of anisotropic vdW materials are also promising for manipulating optical properties in the far-field. For instance, electrostatically doped twisted stacks can control the light polarization state,[682] since they can function as birefringent waveplates or polarizers with tunable degrees of non-normality. This, in turn, gives access to a plethora of polarization transformers, which are important in applications in information processing, telecommunications, and spectroscopy.

However, twisting 2D materials supporting HPhPs for light control in the near field and the far field is not the only approach to twistoptics. Alternatively, moiré photonics leverages periodic stacking misalignments in twisted vdW layers to create moiré patterns, offering a novel platform for engineering optical responses and exploring salient optical phenomena[706] (see Section 24). For instance, second-harmonic generation can be modulated by a factor of 50 by changing the alignment of two hBN layers.[707] This was shown by *in operando* AFM rotation of the top hBN layer. Moreover, intriguing optical effects have also been observed in moiré patterns of twisted bilayer graphene, including interband plasmons driven by transitions within the moiré superlattice[97] as well as hyperbolic enhancements of photocurrent patterns.[187,203] Furthermore, moiré patterns can also create confined conducting channels that can enable a *quantum* polaritonic crystal for graphene plasmon polaritons,[171] a novel approach to manipulate and route the flow of energy at the nanoscale.

Additionally, due to the quantum coupling between twisted atomic layers, light–matter interactions in these structures are inherently chiral, offering a promising platform for chiral polaritons at the nanoscale. Chiral plasmons have been demonstrated to possess longitudinal spin in addition to the conventional transverse spin of plasmons,[685] opening up avenues for novel spin–orbit light–matter interactions. With the extensive range of atomically thin materials available, twisted vdW heterostructures can also serve as fundamental building blocks for advancing chiral plasmonics research, such as chiral sensing of molecular enantiomers.[687] Since different enantiomers can have drastically different effects (for instance, one enantiomer of a drug may be therapeutic, while the other could be inactive or harmful), their accurate detection is essential for drug development, food safety, and environmental monitoring.

### 25.2 Challenges and Opportunities, and Suggested Directions to Meet These Goals

Many current challenges and opportunities in twistoptics revolve around the development of more mechanisms to actively tune its features through external stimuli (see Section 26). Major advancements could be triggered by the ability to dynamically control canalized polaritons through external stimuli. This is essential not only for implementing twistoptics in nano-optical technologies, but also for exploring fundamental phenomena. A promising approach was theoretically predicted[708] and experimentally demonstrated:[709] an electrical tuning mechanism in twisted bilayer α-$MoO_3$ through hybridization with gated-graphene plasmon polaritons. By adjusting the doping level of the graphene layer, polariton canalization can be tuned on demand, enabling dynamic control over topology, propagation, and confinement. Also, as many of the materials of interest, such as α-$MoO_3$, are semiconductors, carrier injection *via* photoexcitation and gating could also be employed. This is key for the implementation of polariton canalization in optoelectronic devices, allowing for reconfigurable, high-speed, and energy-efficient photonic circuits, sensors, and modulators. Moreover, mechanical tuning constitutes an interesting alternative to electrical control mechanisms: by using an AFM tip to push the vdW layers, the alignment between them—and consequently the optical response of the whole system—could be dynamically modified, as demonstrated in twistronics studies on twisted bilayer graphene.[710] This approach would be particularly interesting when using s-SNOM tips, as they would allow for an *in operando* tunable response. In contrast to twisted bilayer graphene, the large adhesion forces among the α-$MoO_3$ layers, as well as between the α-$MoO_3$ and traditional substrates in nanophotonics, such as $SiO_2$, make it challenging to perform this operation in a reproducible and controllable way. A complementary alternative bridging electrical and mechanical control mechanisms to actively tune the twist angle involves using piezoelectric and microelectromechanical systems (MEMS), which allow precise control over the twist angle and vertical separation between multiple twisted layers, as well as other parameters like lateral translation, stretching, tilting, and shearing.[711] This flexibility opens up a wide range of photonic applications. Phase-change materials also present an encouraging platform for dynamic control in twistoptics, whether through temperature-dependent metal-to-insulator transitions or by creating rewritable nanostructures.

Additionally, several challenges in twistoptics are similar to those encountered in the PhP field, where twistoptics has been most extensively explored. One of the challenges is to extend the polariton frequency ranges. Namely, while twistoptics has already been extended to the THz regime,[712] further extension and tunability could be achieved by intercalating atoms of different materials, thereby passively modifying the underlying collective excitations. Notably, the discovery of hyperbolic plasmon polaritons in the visible spectrum opens the door to advance twistoptics also into the visible range. In addition, leveraging twistoptics in materials with even greater optical anisotropy constitutes a growing interest in nanophotonics. In this regard, materials with a monoclinic crystal structure, like β-$Ga_2O_3$ and $CdWO_4$, which host hyperbolic shear polaritons, could unlock new extraordinary optical features. In fact, recent research has also shown that by placing a thin α-$MoO_3$ layer on a monoclinic β-$Ga_2O_3$ substrate, it is possible to engineer an IFC that is approximately linear along the asymptotes of the hyperbolic dispersion while suppressing half of the IFC, leading to *unidirectional ray propagation.*[713] This effect can also be achieved by combining two α-$MoO_3$ layers with significantly different thicknesses. The potential can be even greater in lower-symmetry triclinic systems, such as $ReS_2$, where real-space imaging of polaritons have remained elusive so far. The search for low-symmetry crystals has become a central pursuit in nano-optics, as demonstrated by the intense study of strongly anisotropic PhPs in vdW slabs, heterostructures, and bulk crystals in recent years (see Section 26).

Like any other 2D-material-related field, twistoptics also faces challenges associated with the fabrication of high-quality 2D layers. Currently, the best quality flakes are produced through mechanical exfoliation, which yields small flakes with limited reproducibility—posing a hurdle for twistoptics to reach industry requirements. Advances in growth processes, such as CVD and other techniques, could help standardize the fabrication of twisted samples. Another challenge lies in the creation of efficient polariton antennas and emitters within the samples, as well as nanoscale detectors and absorbers, crucial for integration into on-chip devices, and which are largely missing as well. While gold antennas and the s-SNOM tip have been used as launchers, both present disadvantages: the former lack broadband capabilities, and the latter cannot be integrated into future devices.

As the number of parameters in twisted stacks increases, designing precise propagation for specific outcomes in nanophotonics is becoming more challenging. Although analytical methods and approximations are available, inverse design is emerging as a powerful alternative to traditional intuition-based approaches. This shift has led to recent efforts in applying machine learning and advanced data processing to nanophotonics. Inverse design has already proven successful in tasks such as beam steering in twisted layers[714] and designing other nanophotonic, plasmonic, and metamaterial structures.

Many of the applications employing polaritons—such as polaritonic chemistry, which involves coupling to molecular vibrations and could lead to advances in sensing and catalysis—stand to gain from the development of twistoptics, for instance by the capability of guiding heat with canalized polaritons and without the need for patterned structures.[715] In fact, the diffractionless nature of canalized polaritons makes them particularly promising for achieving extremely directional thermal emission exceeding classical blackbody radiation limits. This could unlock precise control over heat flow, which would be useful in thermal imaging and targeted cooling applications, such as efficient cooling solutions in nanoscale devices. Additionally, phonon polaritons offer an ideal platform for strong coupling with molecules. Investigating vibrational strong coupling in the canalized regime could lead to advancements in sensing and security. Similarly, because polariton canalization enables the collimation of the many different wave vectors characteristic of hyperbolic materials into a single direction, resulting in a very high density of states along it, applications such as enhancing the coupling between quantum emitters and facilitating long-range dipole-dipole interactions could as well be boosted by the advancement of twistoptics. This has significant implications for fields such as quantum information processing through coupling to quantum emitters, with potential impacts on entanglement and information technologies. Advances in polariton canalization could also trigger the development of super- and hyper-lenses that benefit from the extremely collimated behavior to achieve unprecedented resolution, driving advances in nanoimaging. On a more fundamental level, exploring deeply subwavelength, topologically protected polaritons in twisted crystals could provide new insights, as could the design of extremely small nanocavities based on the noncollinearity of the wave vector and the group velocity for the largest part of the momenta distribution of canalized polaritons.

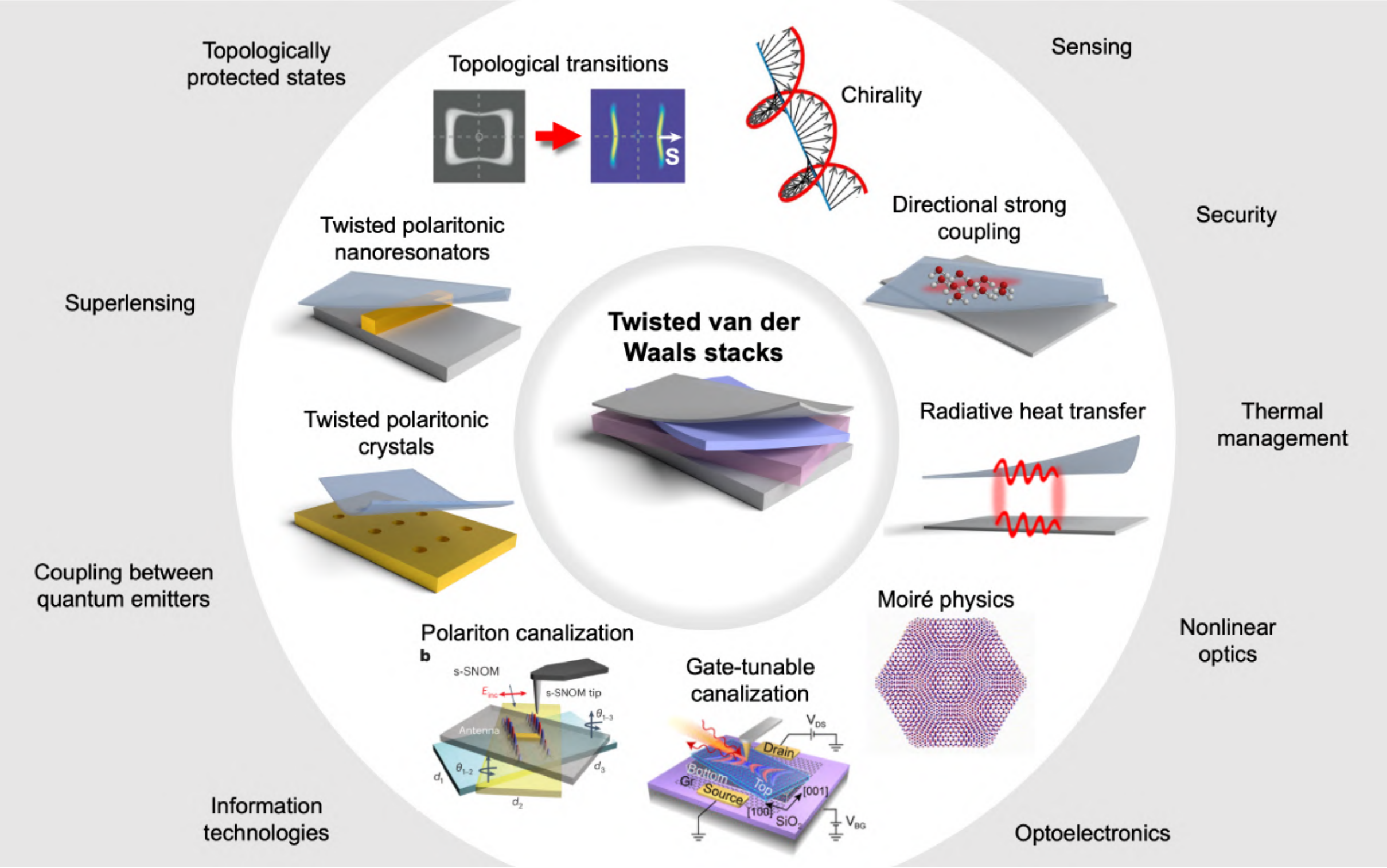

**Figure 25.** Roadmap for twistoptics. Current and potential functionalities in twistoptics (gate-tunable polariton canalization,[709] moiré physics,[706] twisted polaritonic crystals,[702] topological transitions[700]) that could trigger a wide array of applications in nanophotonics (sensing, information technologies, radiative heat transfer, optoelectronics). The "Chirality" sketch is adapted with permission from ref 685 (Copyright 2020 American Physical Society); the "Gate-tunable canalization" panel is adapted with permission from ref 709 (Copyright 2023 American Chemical Society).

## 26. OPTICAL ANISOTROPY OF 2D VAN DER WAALS MATERIALS

**Niclas S. Mueller,[1] Alexander Paarmann,[1] Pablo Alonso-González,[2,3] Alexey Y. Nikitin,[4,5] Andrea Alú,[6,7] Tony Low,[8,9] Valentyn Volkov,[10] Luis Martín-Moreno,[11,12] and Joshua D. Caldwell[13,*]**
[1]Fritz-Haber-Institut der Max-Planck-Gesellschaft, Faradayweg 4-6, 14195 Berlin, Germany
[2]Department of Physics, University of Oviedo, Oviedo 33006, Spain
[3]Center of Research on Nanomaterials and Nanotechnology, CINN (CSIC-Universidad de Oviedo), El Entrego 33940, Spain
[4]Donostia International Physics Center, Manuel de Lardizabal Pasealekua 4, Donostia/San Sebastián 20018, Spain
[5]IKERBASQUE, Basque Foundation for Science, Bilbao 48013, Spain
[6]Photonics Initiative, Advanced Science Research Center, City University of New York, New York, NY, USA
[7]Physics Program, Graduate Center, City University of New York, New York, NY, USA
[8]Department of Electrical and Computer Engineering, University of Minnesota, Minneapolis, Minnesota 55455, USA
[9]School of Physics and Astronomy, University of Minnesota, Minneapolis, Minnesota 55455, USA
[10]Emerging Technologies Research Center, XPANCEO, Internet City, Emmay Tower, Dubai, United Arab Emirates
[11]Instituto de Nanociencia y Materiales de Aragón (INMA), CSIC-Universidad de Zaragoza, Zaragoza 50009, Spain
[12]Departamento de Física de la Materia Condensada, Universidad de Zaragoza, Zaragoza 50009, Spain
[13]Department of Mechanical Engineering, Vanderbilt University, Nashville, TN, 37235 USA
***Corresponding author.** Email: josh.caldwell@vanderbilt.edu

Anisotropy of materials is at the heart of many optical phenomena.[63,671,716,717,718] It leads to birefringence and hyperbolicity, which are key effects that enable light manipulation and nanophotonic applications. Two-dimensional vdW materials intrinsically possess anisotropic crystal structures, owing to their strong in-plane covalent bonds and weak out-of-plane vdW interaction. This gives rise to unique anisotropic optical properties that make them promising building blocks for flat optical components and nanoscale waveguides.[63,671,716,717,718] Many functionalities arise from polaritons, which are formed through the strong coupling between electromagnetic (EM) fields and dipolar excitations such as phonons, plasmons, or excitons.[63,671,716,717,718] The primary motivation for studying polaritons resides in their ability to enhance light–matter interactions at the nanoscale, particularly when these interactions are not only anisotropic, but they are so anisotropic that the real parts of different permittivity tensor components have opposite signs—the hallmark of hyperbolicity.[63,671,716,717,718] This gives rise to hyperbolic polaritons that are associated with open, hyperbolic isofrequency contours in momentum space, enabling large electromagnetic energy confinement within narrow, directional channels with a large photonic density of states. Hyperbolic wave propagation was initially introduced in the context of artificial materials. In many 2D vdW materials, these hyperbolic modes instead arise naturally due to their strong anisotropy, without the need for artificial structuring. This makes them highly attractive for nanophotonic on-chip components, such as flat lenses, resonators, or waveguides. The vast range of vdW materials, of which only a few have been explored so far optically, opens thus a large tool-box to exploit extreme anisotropy for optoelectronic applications.

### 26.1 Current State of the Art

The expansion of research into 2D materials has led to an ever-broadening class of vdW materials with different degrees of structural asymmetries (Figure 26a).[716,717,719]

#### 26.1.1 *Hexagonal*

Graphene, hBN, and TMDs, are hexagonal vdW materials (Figure 26a, first column). This high symmetry leads to a uniaxial optical response that is isotropic in-plane but can be largely anisotropic out-of-plane. One consequence is a giant optical birefringence, which can reach $\Delta n$~3 for TMDs in the visible (VIS), and $\Delta n$~1.6 and ~0.8 for hBN in the IR and THz spectral range, enabling extreme forms of light manipulation.[511,720] The large index contrast is associated with highly directional material resonances driven by phonons, excitons or plasmons. The resonances are often so pronounced that the associated permittivity component becomes negative, giving rise to hyperbolic polaritons.[63,716,717] A prime example are hyperbolic phonon polaritons in hBN at IR wavelengths, which propagate as rays in the material bulk.[11] The ray direction is perpendicular to the asymptotes of the hyperbolic isofrequency contour, which gives access to very large wave vectors associated with highly confined fields and extreme light–matter interactions. This has been exploited for hyperlensing, ultraslow light propagation and extreme compression of the polaritonic wavelength. [11,63,671,716] In general, a finite thickness of vdW materials is required to exploit the out-of-plane uniaxial anisotropy. Therefore, further symmetry reductions are required to achieve in-plane, 2D anisotropic responses.

26.1.2 *Orthorhombic*

Significant advances toward 2D optical anisotropy have been made in the past few years with materials with orthorhombic crystal symmetry, such as α-$MoO_3$, $WTe_2$ and CrSBr (Figure 26a, second column).[716,717,720,721] These materials have a biaxial response leading to in-plane birefringence, linear dichroism and hyperbolicity. Particularly noteworthy has been the discovery and control of in-plane hyperbolic phonon polaritons in α-$MoO_3$ and α-$V_2O_5$.[716,717] Using near-field microscopy, researchers have visualized concave wave-fronts for these polaritons, occurring along specific in-plane directions. These enable flat optical lenses, diffraction-free propagation across interfaces and peculiar effects such as negative reflection and refraction. The isofrequency contours of these modes undergo topological transitions between elliptical (closed) and hyperbolic (open) shapes, enabling exotic phenomena like canalization. Furthermore, hyperbolic plasmon polaritons in $WTe_2$ show intriguing tunability[717]. Here, the combination of intraband (free carrier) and interband (valence-conduction) transitions produce an inductive and capacitive response in orthogonal directions, leading to an in-plane hyperbolic dispersion.[722] Recently, in-plane anisotropic exciton polaritons were discovered in CrSBr in the near-IR range at low temperatures (Section 10).[294] The material is particularly interesting because of its antiferromagnetic response that opens a route to tune polaritons with magnetic fields (Section 42).

26.1.3 *Monoclinic, triclinic*

A further reduction of symmetry to monoclinic and triclinic crystals gives rise to hyperbolic shear polaritons (Figure 26a, third column).[716,717] These exhibit a peculiar in-plane anisotropic propagation, arising from off-diagonal elements in the dielectric permittivity tensor. In addition, the optical axes are no longer aligned with the crystal frame and rotate with excitation frequency changes, offering additional tunability.[716,723] Hyperbolic shear polaritons were first observed in β-$Ga_2O_3$ and $CdWO_4$, which are non-vdW materials, but can be exfoliated as thin films.[716] Recent experiments with plasmon polaritons in monoclinic $MoOCl_2$ and excitons in triclinic $ReSe_2$ give hope that such shear phenomena may be leveraged soon to tailor light propagation in 2D vdW materials.[723,724]

## 26.2 Challenges and Future Directions

The field of optically anisotropic vdW materials and planar hyperbolic polaritons is set for future growth, driven by advances in material discovery and polaritonic engineering. Several challenges and promising directions stand out (Figure 26b,c):

26.2.1 *Materials Search*

Given the vast range of vdW materials that have been exfoliated from their known bulk crystal forms, but have not been characterized optically, we expect a surge of research in polariton phenomena.[716,717,720,725,726] Many of these materials belong to orthorhombic, monoclinic or triclinic crystal classes, which underlines the importance of understanding their optical anisotropy in order to utilize them in photonic applications. It is currently challenging for ellipsometry to extract the full frequency-dependent components of the permittivity tensor, in particular for the out-of-plane response. Thus, further developments for full nano-optical characterization are needed. Furthermore, predictive *ab initio* modelling of the optical properties, including material losses, would benefit materials searches. A material database that reports crystal structures along with the frequency-dependent permittivity tensors would be highly desirable. Furthermore, expanding the range of vdW materials supporting hyperbolic polaritons in the mid-IR molecular fingerprinting region, beyond hBN, would be desirable for sensing applications.

26.2.2 *Characterization*

The prime tool for characterizing polariton propagation and nano-optical phenomena is scattering-type, scanning near-field optical microscopy (s-SNOM), which has been mostly used in conjunction with tunable continuous-wave lasers to access frequency-dependent properties[63,716,717] (Section 7). This works well in the mid- and near-IR spectral ranges, but is technically challenging at VIS frequencies for stability reasons and in the far-IR and THz ranges because of the lack of narrowband lasers. The latter long-wavelength spectral ranges are currently only accessible with advanced light sources, such as free-electron lasers or synchrotron sources. Recent developments in narrowband difference-frequency generation down to $\omega \approx 500\ cm^{-1}$ promises to widen the range of materials accessible with table-top lasers, but there is still a need for narrowband light sources to bridge the THz gap. A notable emerging capability is time-resolved nanoscopy with ultrafast laser systems, opening a route toward spatiotemporal characterization and temporal photonic control.[727,728] Several experiments have achieved all-optical activation of polaritonic states by exploiting transiently excited carriers, which enabled, for example, transient hyperbolicity in $WSe_2$.[728] Another notable recent development is cryo-SNOM, which is crucial for the characterization of functional materials with phase transitions and enables characterization with greatly reduced material losses.[170]

26.2.3 *Fabrication and scalability*

For scalability and device integration it is crucial to fabricate large-area highly crystalline vdW materials of defined thickness. The currently most popular approach of exfoliation is nondeterministic and not scalable, although recent advances with metal-assisted exfoliation offer a route to large-area mono- and few-layers.[729] A promising approach to large areas and scalability is chemical vapor deposition (CVD), but recent experiments showed that maintaining high-quality polariton propagation is challenging.[730] An unexplored opportunity for CVD is to employ isotopically pure precursors, which have greatly improved phonon polariton properties in exfoliated crystals.[11] Another device challenge is to pattern vdW materials with lithography or focused ion beam milling without damaging their crystal structure. Recent approaches of engineering polaritons through a patterned substrate instead of structuring the material itself are promising.[731] Ultimately it would be desirable to grow vdW materials on a nanostructured template directly.

26.2.4 *Tunability*

A major challenge for polaritons at large is that many interesting optical phenomena, such as hyperbolicity, only occur within narrow spectral ranges that are set by material resonances. A wide range of approaches has been explored to achieve tunability.[63,716,717,718] For plasmon polaritons, the plasma frequency is shifted through charge carrier doping and gating; an approach that also carries over to other material excitations when hybridized with plasmons, as demonstrated for hBN/graphene and $MoO_3$/graphene heterostructures.[718] For exciton polaritons a key advantage, but also a challenge, is their dependence on excitation density and temperature, offering tunability of frequency and oscillator strength, and even condensation and lasing when tailoring their dispersion (Section 10). Active tuning of phonon polaritons is much more difficult, as their spectral range is set by the vibrational frequencies of the lattice. Isotopic substitution, intercalation and heterostructures of different materials and isotopes have been demonstrated as concepts to design and shift phonon resonances. A particularly promising approach to actively induce and tune anisotropy and hyperbolicity is strain, which is so far largely unexplored experimentally.[732] Other emergent fields are twist optics and polaritonic metasurfaces, which are discussed in dedicated Sections 25 and 27.

26.2.5 *Flat and on-chip photonic devices*

Optical anisotropy and hyperbolicity in vdW materials provide unique tools to realize flat optical devices and on-chip photonic components.[716,717] The giant birefringence and linear dichroism of biaxial vdW materials are ideally suited to realize ultra-flat polarization optics and light manipulators.[511,721] These flat optical components can be stacked to combine different functionalities in compact and miniaturized devices. In addition, hyperbolic polaritons pave the way for on-chip light processing through their unique propagation and confinement. Several key concepts and components have been realized recently, including flat lenses, bend-free refraction, and directional propagation, that occur on deep subwavelength scales.[716] Transformation optics and inverse design are powerful approaches to expand the search for new designs. The strong and tunable confinement of polaritons also offers a route to frequency multiplexing and conversion in nanoscale waveguides, bridging functionalities of different spectral ranges. Importantly, the implementation of on-chip devices relies on interfaces between electronic and optical components (Section 34).

26.2.6 *Sensing*

Sensing applications can benefit from the extreme field confinement of hyperbolic polaritons, particularly in the mid- and far-IR molecular fingerprinting regions.[717] This enhances sensitivity to environmental changes, making hyperbolic media ideal for chemical detection and biosensing. The enhancement of molecular fingerprint spectra mainly occurs through the hybridization and strong coupling of IR vibrations with phonon polaritons.[717,718] Recent experiments, demonstrating deeply subwavelength confinement down to $10^{-7}/\lambda_0^3$ mode volumes, give hope that the sensitivity can be pushed to the single-molecule level.[59]

26.2.7 *Thermal applications*

Another promising and emerging field is to exploit material anisotropy and polaritons in the IR for thermal emission and transport.[717] Coherent and directional thermal light sources have been realized based on hyperbolic metasurfaces and phonon polaritons in structured 3D polar crystals.[733,734] It will be interesting to realize similar concepts through the natural optical anisotropy of vdW crystals.[735] Hyperbolic phonon polaritons are expected to enable super-Planckian thermal emission, surpassing the black-body limit, which could lead to innovations in thermophotovoltaics, radiative cooling, and nanoelectronics heat management.[717] Beyond thermal emission, also thermal transport can be significantly modified with phonon polaritons,[736,737] but has yet to be fully explored for anisotropic 2D materials, opening a new route to steer thermal dissipation and to build on-chip thermal diodes.

26.2.8 *Functional materials*

An emerging area is the integration of functional materials - such as ferromagnetic, ferroelectric and piezoelectric vdW materials - within 2D systems.[717] These materials may enable tunable polaritonic devices, with ferromagnetic vdW materials like $CrI_3$ and $VSe_2$ offering the potential for nonreciprocal devices controlled by external magnetic fields. Such behavior could lead to the development of optical isolators and circulators, with potential use in quantum communication and information processing. Recent experiments with CrSBr show that an intrinsic coupling of exciton polaritons to magnetic ordering is possible, offering a route to magnetic switching of polariton propagation (Sections 10 and 42).[170,294] Employing non-centrosymmetric materials moreover opens a route to enhance nonlinearities with hyperbolic polaritons, where phonon polaritons in quaternary oxides and excitons in ferroelectric $NbOI_2$ are promising candidates (Sections 14, 17, 17, and 20).[501,738]

In conclusion, optical anisotropy and hyperbolic polaritons in 2D vdW materials represent a booming research area with significant technological implications. Ongoing progress in material synthesis, polariton engineering and device integration is expected to drive important advances in light manipulation, nanophotonics, quantum technologies, sensing and thermal management.

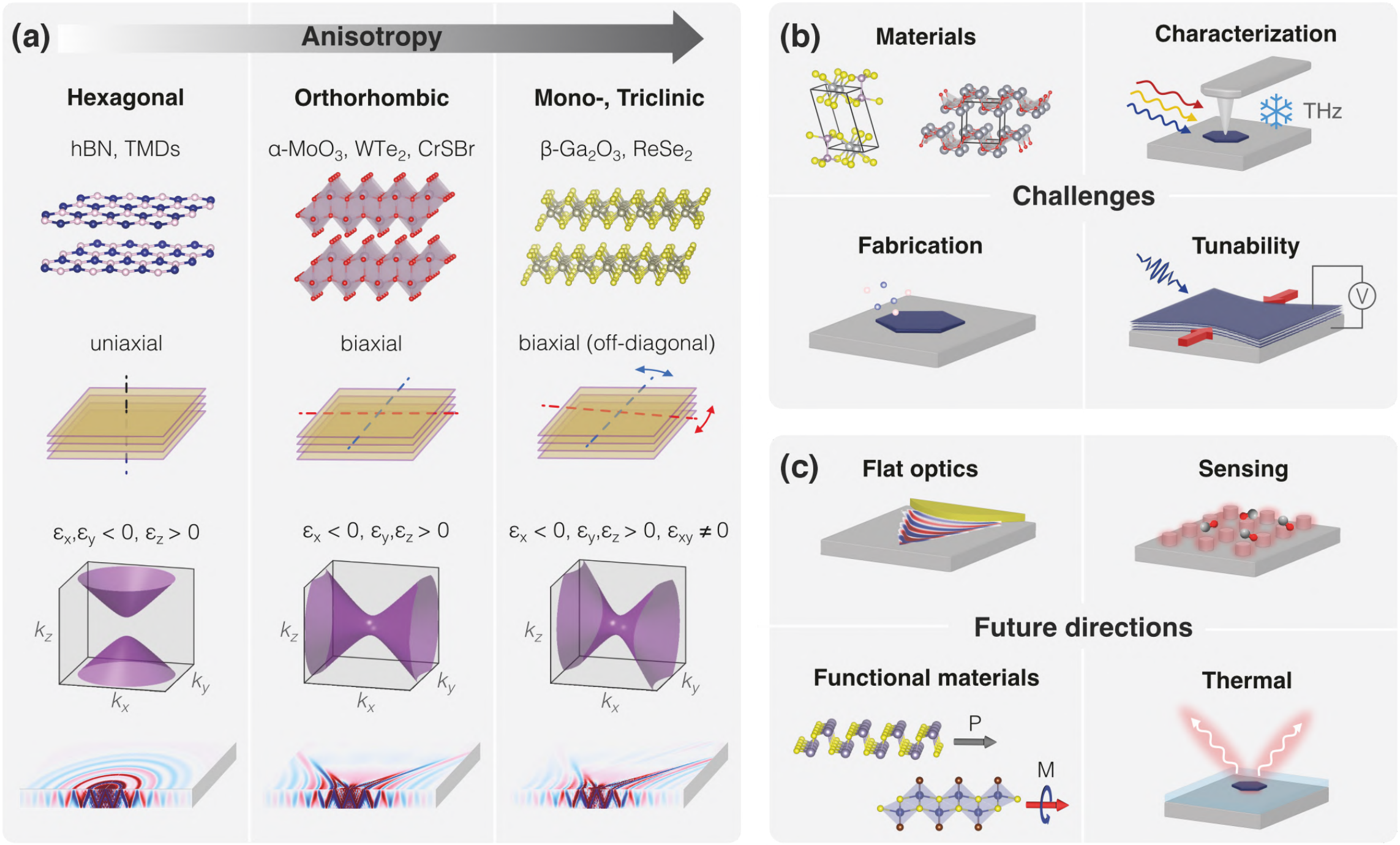

**Figure 26.** Role of crystalline anisotropy on polaritonic response and function. (a) Role of structural anisotropy in the optical properties of 2D vdW materials, including, from top to bottom, different crystal classes, optical axes leading to birefringence, hyperbolic isofrequency contours, and polariton propagation launched by a point dipole. (b,c) Challenges and future directions.

**Metasurfaces and Emerging Materials**

## 27. POLARITONIC METASURFACES / COUPLING 2D MATERIALS WITH PHOTONIC STRUCTURES

**Andrea Alù**[1,2,*]
[1]Photonics Initiative, Advanced Science Research Center, City University of New York, New York, NY 10031, USA
[2]Physics Program, Graduate Center, City University of New York, New York, NY 10016, USA
***Corresponding author.** Email: aalu@gc.cuny.edu

### 27.1 Current State of the Art

As discussed in other sections of this Roadmap, polaritons are quasiparticles that emerge from the strong coupling between light and material excitations. The ultrathin nature of 2D materials, combined with their strong material resonances, offers a plethora of opportunities in the context of polaritonic phenomena, stemming from excitons, phonons, magnons, plasmons and electronic transitions.[671] While a wide range of novel photonic phenomena have been explored in this context, broadly covered by the other sections of this Roadmap, recently it has been also realized that photonic engineering of polaritonic materials through nanostructuring offers a new playground for fundamental phenomena and applications. In addition, photonic engineering allows tailoring local density of states, field enhancement, quality factor, and modal engineering of the near-field light profiles that engage local material resonances, resulting in tailored polaritonic responses that are not accessible in the unstructured materials. In turn, polaritonic phenomena offer new opportunities to enhance the growing field of metasurfaces – ultrathin devices that exploit nanoscale engineered structures to tailor and control the optical wavefront in all its degrees of freedom. While conventional photonic metasurfaces are fundamentally limited in the way they interact with light due to the limited extent of light–matter interactions at the nanoscale in conventional materials, polaritonics offers a platform to overcome these limitations and realize metasurface responses that go beyond the conventional limits, in terms of field enhancement, nonlinearity, bandwidth, and other constraints dictated by passivity, linearity, and time invariance. Indeed, polaritonic metasurfaces have become in the past decade a blossoming field of research, offering opportunities for exciting interdisciplinary research and for technological breakthroughs.

For instance, polaritonic metasurfaces based on quantum engineered materials that support tailored electronic transitions have been shown to sustain giant optical nonlinearities,[515] orders of magnitude larger than those observed in natural crystals at the same frequencies, and free from phase matching constraints due to their inherently low profile. These have been used over the years to demonstrate efficient wave mixing and harmonic generation, as well as power limiting, emerging from ultrathin structures that leverage the optimized synergy between photonic engineering and tailored polaritonic responses. Similar approaches have been applied to a wide range of 2D materials, from boron nitride and other phononic materials, to transition metals and excitonic systems.[279,280739,740,741] In all these and many other demonstrations, enhanced polaritonic phenomena as well as enhanced metasurface functionalities have emerged by the optimal combination of material responses and photonic engineering at the nanoscale through nanostructuring.

Beyond numerical optimization of the metasurface geometries, the rational design of polaritonic metasurfaces through tailored broken symmetries has emerged as a powerful tool, leveraging both localized resonances with tailored broken symmetries[515,742] and delocalized nonlocal responses.[523] Particularly intriguing in this context is the opportunity that emerges from leveraging also the natural broken symmetry in the underlying 2D materials composing polaritonic metasurfaces,[716] which can support highly confined directional modes. In Section 25, we indeed discuss how twisting thin films of 2D materials can enable novel polaritonic phenomena, and it is then not surprising to realize that more sophisticated forms of photonic engineering, in the context of metasurface designs, can further enhance this control. Figure 27a, for instance, showcases a wide range of extreme polaritonic phenomena demonstrated in ultrathin planarized settings, obtained by structuring 2D vdW material interfaces,[741] demonstrating how the interplay of directional material resonances, nanostructuring in space, and metamaterial concepts, can be synergistically combined to demonstrate extreme control over nanoscale light. 2D polaritonic materials also naturally offer a very large index of refraction, large nonlinearities and tunability, all excellent features to augment photonic metasurfaces.

### 27.2 Challenges and Future Goals

While the field of polaritonic metasurfaces has been blossoming in the past decade - across various material platforms and frequency ranges - a few challenges have been hindering its full deployment and its impact beyond basic science explorations. First, directly structuring 2D materials may hinder their exceptional optical properties, as a function of the nanofabrication processes that may involve high temperatures or mechanical challenges. As an alternative, 2D materials can be coupled to photonic metasurfaces by simply stacking a thin film over a structured surface. This approach preserves the pristine properties of the polaritonic material but trades this benefit with the risk of weaker coupling with the metasurface. As an example, Figure 27b shows a topological polaritonic metasurface obtained by stacking a pristine 15nm film of boron nitride on a topological photonic crystal.[740] By optimizing the thickness of the polaritonic thin film, and tailoring the photonic crystal to support its nontrivial bandgap around the phonon resonance of boron nitride, it is possible to transfer the topological phase of the edge state living at the boundary between domain walls from light to the phonon polariton. The robustness inherited by the topological features of this mode are now observable in the far-field, and phonon vibrations in the thin film are guided around defects and sharp transitions due to the strong coupling to topological light. As shown in the right panel, the dispersion of the topological mode is strongly impacted by the phonon resonance, and as the mode gets more confined to the interface, for larger transverse momenta, the polaritonic nature of the topological mode emerges.

While this example features a phononic response clearly observable at room temperature, polaritonic metasurfaces involving excitonic responses often require operating at cryogenic temperatures, which enables stronger polaritonic responses. This clearly represents a challenge, which is typically accompanied by the fact that many 2D materials are poorly compatible with conventional optoelectronic platforms. A related challenge emerges when the 2D materials need to be exfoliated, which is often the case in the context of vdW materials. This challenge impacts repeatability and limits the available lateral extent of the polaritonic material.

The 2D nature of many polaritonic materials is an appealing feature for many applications, enabling strong coupling, large light–matter interactions and natural coupling with photonic structures, as in Figure 27, yet it can also represent a challenge due to the underlying symmetries. A 2D thin film naturally supports transverse electric responses, which are symmetric in the two half spaces above and below it. This implies fundamental limits on the efficiency of the optical response. In addition, the underlying crystal symmetry may prevent even-order nonlinear responses in 2D materials, which may be overcome selecting lower-symmetry crystal lattices, as discussed in another section of this Roadmap.

Another emerging challenge is the limited overall efficiency of the resulting polaritonic devices. While 2D materials offer giant refractive indexes and nonlinearity coefficients, these are typically a by-product of the tiny volumes involved in such planarized structures. As a result, the overall efficiency of polaritonic metasurfaces for linear and nonlinear phenomena tends to be limited, especially when operating at room temperature. Efforts to enable scalability, high-quality, and large-area fabrication are necessary to overcome this challenge.

### 27.3 Suggested Directions to Meet These Goals

The most appealing aspect of polaritonic metasurfaces is their inherently multi-physics operation, involving not only electromagnetic responses, but also other material responses, which can serve as reservoir for new degrees of freedom in photonic engineering. A by-product of this feature is that proper optimization of polaritonic metasurfaces inherently requires a co-optimization effort that goes beyond just optics and photonics. A good example is the polaritonic metasurfaces based on electronic transitions as in ref 515. Their optimal responses are achieved by optimizing the underlying electronic band structure of the multiple quantum well substrates while at the same time controlling the nanostructure design. This feature emerges also in the design of the topological polaritonic metasurface in Figure 27b,[740] where the optimization of the boron nitride thickness was crucial to ensure strong coupling with the photonic mode. This co-optimization requires new computational and design tools, which inherently span multiple disciplines – an interesting direction for future research, involving both theoretical and numerical efforts.

In parallel, efforts to improve large-area fabrication and growth of high-quality 2D materials will certainly enhance the performance of polaritonic metasurfaces, which may come hand-in-hand with progress in nanofabrication and structuring of these materials without affecting their polaritonic responses. Explorations to improve the material response and avoid the need for cryogenic temperatures will largely benefit the future of this research area.

In addition, the static nature of polaritonic metasurfaces, whose response is engrained in their spatial structure, represents an important challenge, which may be overcome by exploiting electric gating or optical pumping. Explorations in this context hold the promise to offer real-time tunability of the polaritonic metasurface responses. In this context, another promising direction is the use of magnetic materials, such as CrSBr, supporting polaritons. In a recent demonstration,[294] strong coupling between light and the excitonic response of CrSBr in optical cavities has resulted in giant tunability of the optical response through the applied magnetic bias. A transition from anti-ferromagnetic to ferromagnetic response, associated with large tuning of the optical properties and exciton resonance, was observed, ideally suited as a platform for magnetically tunable polaritonic metasurfaces and for the manipulation of magnons.

Finally, efforts to scale these devices, improve their reproducibility, and increase the compatibility of photonic systems are certainly needed to ensure a maximized impact of polaritonic metasurfaces for practical technologies, spanning from thermal management to sensing, imaging, wave mixing and more. The future of polaritonic metasurfaces is certainly vibrant, with important roles being played by the interplay and strong collaborative efforts among material scientists, physicists, chemists and optical engineers.

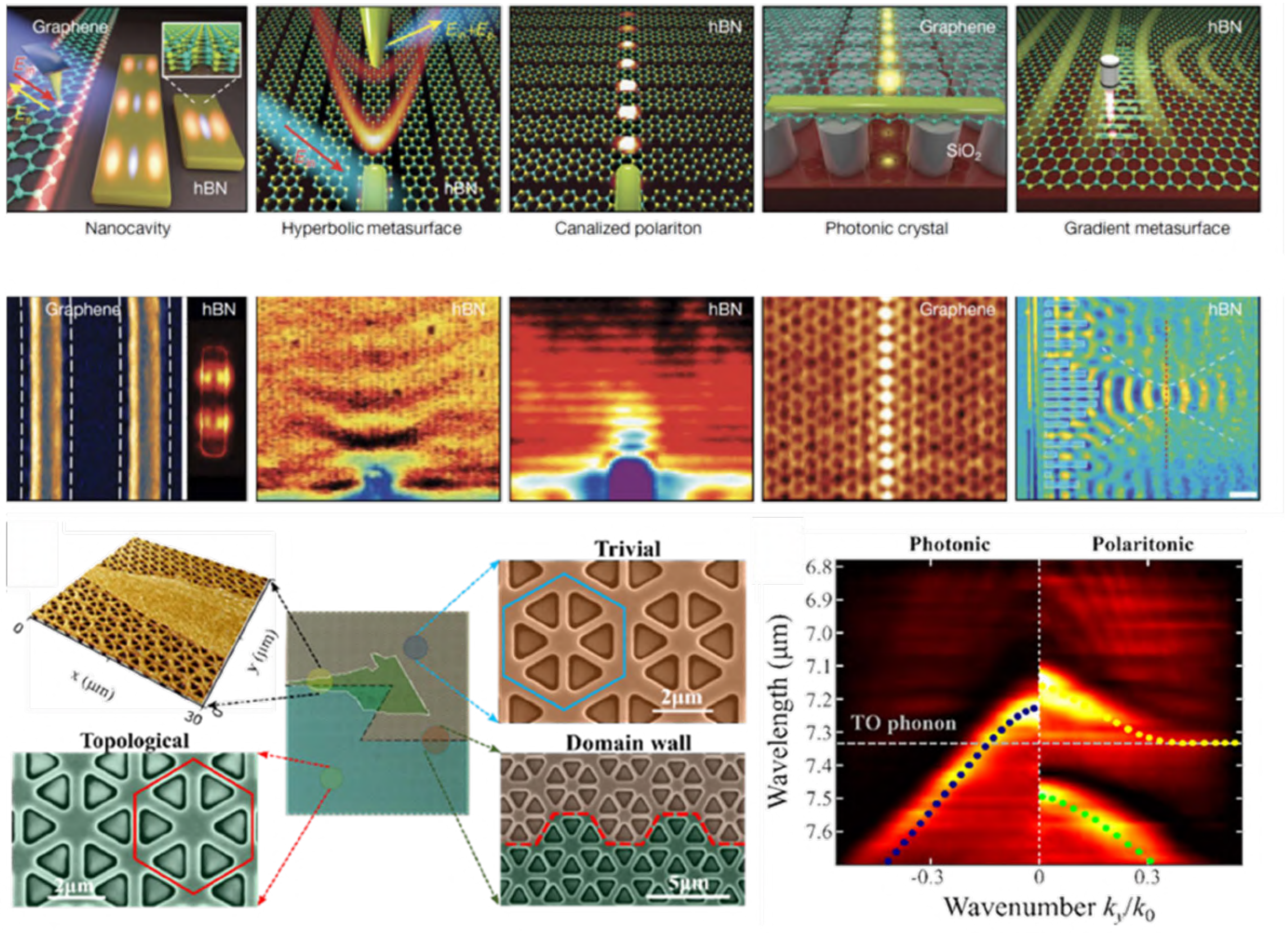

**Figure 27.** Polaritonic metasurfaces. (a) Polaritonic metasurfaces based on vdW interfaces to demonstrate a range of extreme polaritonic phenomena at the nanoscale. Adapted from ref 741 (Copyright 2021 Springer Nature). (b) Topological polaritonic metasurface obtained by loading a topological photonic crystal with a thin film of boron nitride. The right panel shows the effect of such loading: the topological edge state splits its dispersion within the bandgap of the photonic crystal due to the coupling with the phonon resonance of boron nitride. The topological phase is transferred to the polaritonic response. Adapted from ref 740 (Adapted from ref 6 (Copyright 2021 American Association for the Advancement of Science).

## 28. TRANSITION METAL DICHALCOGENIDES FOR PASSIVE AND ACTIVE METASURFACES

**Deep Jariwala,[1] Timur Shegai,[2] Jorik van de Groep[3,*]**
[1]Electrical and Systems Engineering, University of Pennsylvania, Philadelphia, Pennsylvania, USA
[2]Department of Physics, Chalmers University of Technology, Göteborg, Sweden
[3]Van der Waals-Zeeman Institute, Institute of Physics, University of Amsterdam, Amsterdam, The Netherlands
***Corresponding author.** Email: j.vandegroep@uva.nl

### 28.1 Introduction

Metasurfaces employ resonant light scattering in compact arrays of nanoscale structures to perform a collective optical function. Accurate control over the local scattering amplitude and phase enables (near-)arbitrary wavefront shaping. While initial metasurfaces employed plasmon resonances in metal nanoparticles, subsequent designs quickly moved to Mie-like resonances in dielectric and semiconductor nanostructures to mitigate the optical losses associated with the metallic light–matter interaction. Leveraging more advanced design concepts including propagation phase and geometric (Pancharatnam–Berry) phase, state-of-the-art metasurfaces now enable high-numerical aperture lensing, multi-color holography, beam steering, and even multifunctional flat optical elements. For dielectric metasurfaces, a high refractive index of the photonic material is of crucial relevance as it controls the scattering cross section and nanoscale confinement of light. As such, titanium dioxide ($TiO_2$, refractive index $n$~2.6) and silicon (Si, $n$~3.5) are common photonic materials for metasurfaces.

Over the past decade, TMDs have emerged as a new class of photonic materials for metasurfaces, for two reasons: (i) these 2D semiconductors exhibit remarkably high refractive indices ($n > 4$), which enable very strong light scattering (Figure 28a); (ii) their optical response is characterized by a very strong excitonic resonance in the red/near-IR spectral range (Fig. 1b). The steady-state formation and recombination of excitons give rise to a resonant light–matter interaction that is intrinsic to the material and, unlike plasmon and Mie resonances, this exitonic resonance occurs independent of the (nanoscale) geometry. In their monolayer limit, the exciton resonance energy and amplitude have been demonstrated to be very sensitive to external stimuli.[743] This opens new avenues for active metasurfaces where the optical function can be tuned post-fabrication using external parameters including electric/magnetic fields, strain, or optical pump fields.

Here, we outline the state-of-the-art in TMD based metasurfaces, highlight the key challenges in their development, and discuss potential pathways to address these challenges. Crucially, in this Roadmap we define *TMD metasurfaces* as those where the TMD functions as the *optically active material* of the metasurface (i.e., the principle medium for light–matter interactions and photon-field confinement), in contrast to decorating dielectric/metallic metasurfaces with a (monolayer) TMD.

## 28.2 State of the Art

### 28.2.1 *High-Index Metasurfaces*

The refractive index of many TMD materials reaches values > 4 in the visible and near-IR part of the spectrum, outperforming traditional materials like Si and GaAs (Figure 28a).[763] Moreover, due to their intrinsic vdW nature, TMDs are inherently anisotropic, resulting in birefringence values exceeding 1.5.[511] These optical properties have been experimentally measured by several methods including normal-incidence reflectivity, near-field optical microscopy,[511] and spectroscopic ellipsometry.[763] Such characteristics make TMDs exceptional candidates for passive high-index metasurfaces. Furthermore, optical anisotropy and the flexibility of integration with diverse substrates provide TMD metasurfaces with additional advantages over traditional Si and GaAs counterparts. Nanostructuring these materials enables the excitation of Mie resonances and related anapole states in TMD nanodisks.[744] Furthermore, several proof-of-principle experiments, including metasurfaces, waveguides, ring resonators, and photonic crystal defect cavities have also been reported recently.[745]

### 28.2.2 *Excitonic Polaritonic Self-Hybridization*

Depending on the spectral range and their overlap with excitons, the passive photonic modes discussed in the previous section could demonstrate polaritonic or pure photonic behavior. Because in the former case, the polaritonic behavior is demonstrated without the use of the external cavity, these polaritons are sometimes referred to as self-hybridized. In this scenario, the optical modes supported by the geometry of the material and its refractive index can strongly couple or hybridize with material excitations (excitons, phonons) within the very same material leading to the emergence of new quasiparticles—polaritons that inherit both light and matter character from their constituent parts. Such self-hybridization has been demonstrated in TMD films both above and below the light line,[517] in $WS_2$ nanodisks with anapole states,[744] 1D grating resonators on Au,[746] and more recently within passive $WS_2$ metasurfaces.[747] Self-Hybridization can also be observed with waveguide modes in monolayer TMDs when layered into a multi-quantum well (MQW) or a superlattice structure.[748]

### 28.2.3 *Optical Nonlinearity*

Due to the lack of inversion symmetry, monolayer $MoS_2$ is known to provide one of the highest second-order nonlinear coefficients ($10^{-7}$ m/V) in the near-IR part of the spectrum. At the same time, multilayer TMDs, which typically adopt AB stacking, result in a 2H-TMD configuration, where strong nonlinearity disappears in the bulk crystal. Recently, a new type of AA' stacked form of TMDs, particularly, 3R-$MoS_2$ and 3R-$WS_2$, has started gaining popularity. These materials combine high refractive index with substantial second-order nonlinearity, making them potentially useful in nonlinear nanophotonics. Specifically, giant second-harmonic generation (SHG),[491] piezophotovoltaic and piezoelectric effects, and spontaneous parametric down-conversion have been recently demonstrated using 3R-TMDs. Furthermore, structuring these materials into nanodisks enabled a combination of nanophotonic (anapole) and material (excitonic) resonances, which led to substantial (>$10^2$) enhancement of the SHG signal.[514] It is also recognized that 3R-$WS_2$ might offer an advantage over 3R-$MoS_2$ in terms of substantially smaller indirect band photoluminescence, which often complicates down-conversion measurements.[749]

28.2.4 *Exciton Resonance Tuning*

In their monolayer limit, TMDs exhibit a direct bandgap instead of the indirect bandgap that is associated with their bulk counterpart. Combined with reduced dielectric screening, the quantum confinement of the exciton wave function in the monolayer results in a few-hundred meV exciton binding energies.[750] This renders the exciton resonance stable, even at room temperature. On a fundamental level, tuning of this exciton resonance in monolayer TMDs has been explored using a wide range of external parameters,[743] including Stark and Zeeman tuning of the exciton energy using external electric and magnetic fields, respectively; Fermi-level tuning using solid-state gating or ion-liquid gating, which enables electrical control over the exciton decay channels, and thereby, the amplitude of the exciton oscillator strength; strain tuning using flexible substrates or suspended monolayers which induces large (red-)shifts in the exciton energy; and temperature tuning to control the exciton-phonon scattering rate, and thereby, the exciton linewidth. Of these tuning mechanisms, electrostatic doping has proved to be the most technologically relevant tuning mechanism because it can capitalize on the design and fabrication schemes of established CMOS technology.

28.2.5 *Active 2D Excitonic Metasurfaces*

While most studies focused on active tuning of the exciton photoluminescence, leveraging the strong and tunable resonant light–matter interaction offered by the steady-state excitation and recombination of excitons for coherent wavefront control has so far remained limited to a handful of initial demonstrations. The first works toward this end showed electrically tunable reflection and absorption in monolayer TMDs at cryogenic temperatures.[17,386] Since then, active 2D excitonic metasurfaces with tunable optical functions beyond simple reflection started to develop. These include the following: (i) Exciton resonance tuning of an atomically thin $WS_2$ lens using ion-liquid gating,[751] where the intensity in the focal point can be modulated with a depth of 33%. Tuning the optical function with temperature is studied in a follow-up work.[752] (ii) Excitonic beam steering in continuous layer $MoSe_2$ using nanopatterned gate electrodes to locally change the refractive index in the monolayer and induce an electrically controlled periodic grating.[753] (iii) Excitonic phase modulation using a superlattice or MQW structure of TMD monolayers that are biased through an oxide *via* a conducting substrate.[754]

## 28.3 Major Challenges, Future Goals, and Suggested Directions to Meet These Goals

28.3.1 *Scalability*

To enable the nanophotonic applications discussed in this section, it is essential to achieve wafer-scale production of TMD materials. In this respect, chemical vapor deposition (CVD) represents a promising method. However, CVD typically yields polycrystalline films, which introduce additional optical losses that can degrade device performance. For this reason, high-quality, single-crystalline multilayer TMDs are required to achieve optimal optical characteristics. These high-quality crystals are often produced by mechanical exfoliation from bulk crystals, typically grown through high-temperature physical processes. Recently, gold-assisted exfoliation methods have been demonstrated to selectively exfoliate the outer monolayer from a bulk TMD crystal.[755] This enables exfoliation of high-quality monolayers with active areas on the cm-scale. While this poses a tremendous opportunity for large-area TMD metasurfaces in an academic setting, the exfoliation technique is, however, not scalable, which limits its practical use in an industrial setting. Therefore, the challenge of synthesizing single-crystalline *multilayer* TMDs, suitable for nanophotonic applications, at the wafer scale remains significant. Recent advances in interfacial epitaxy growth methods[756] have made important steps toward addressing this issue. These methods offer the potential for wafer-scale, high-quality single crystals, making them highly relevant for practical high-index TMD nanophotonics applications. Rapid and encouraging progress in this field suggests that wafer-scale production of high-quality single-crystalline multilayer TMDs will likely be overcome soon, paving the way for innovative high-index nanophotonics and metasurface applications.

### 28.3.2 *Sub-Bandgap Nanophotonics*

TMDs are well-known for their high oscillator strength excitonic effects, which have led to numerous applications in optoelectronics and nanophotonics. However, these excitons also contribute to considerable optical absorption in the visible and near-IR spectrum. Beyond the bandgap, at longer wavelengths, there exists a broad spectral region of relatively low optical losses. This low-loss region is characterized by a high refractive index and optical anisotropy, making multilayer TMDs promising candidates for high-quality factor nanophotonic resonators, metasurfaces, and low-loss waveguides.[757] Currently, optical loss levels in this spectral range are determined primarily through ellipsometry and near-field measurements;[511,763] however, a more precise determination of these losses is required for developing integrated all-TMD nanophotonic elements.

### 28.3.3 *Selective Etching in Multilayer TMDs for Atomically Sharp Edges*

Fabrication of TMD metasurfaces is inherently associated with several technological challenges. In particular, it is often problematic to remove resist leftovers and avoid fabrication-induced defects and intolerances. To mitigate these issues, several anisotropic wet etching techniques were recently introduced.[747] These approaches enable the transformation of rough edges in nanostructured TMDs into self-limited atomically sharp zigzag edges, thereby providing control over the edge states of TMD metasurfaces and potentially reducing the scattering losses. This advancement offers significant advantages not only for nanophotonic applications but also for catalysis and sensing, where edge states play a critical role in governing the underlying physics. Nevertheless, the full potential of these ultrasharp edges for nanophotonic applications remains to be demonstrated.

### 28.3.4 *Efficiencies of Active 2D Metasurfaces*

Despite the uniquely strong tunability of excitons in monolayer TMDs, their applications in active metasurfaces face important challenges that hinder widespread development in dynamic wavefront shaping. The tunability of TMDs is most pronounced in their monolayer limit, which results in strong challenges associated with the optical efficiencies of 2D excitonic metasurfaces. In fact, all state-of-the-art demonstrations described above show large relative tunability but exhibit absolute efficiencies $<1\%$, limited by the atomic scale optical path length. Stronger light–matter interactions for the monolayer TMD are crucial to address this stringent limitation. Coupling monolayer TMDs to plasmonic or dielectric metasurfaces has been widely pursued to study strong coupling and exciton polaritons (see also Section 27 on polaritonic metasurfaces). Future work should focus on high-efficiency tunable metasurfaces where the material quality and tunability of the TMD remain intact while coupled to plasmonic or dielectric metasurfaces, for example by embedding the monolayer in a heterostructure with hBN within the metasurface.

### 28.3.5 *Tunability of Thicker Layers*

A promising approach to induce multilayer-like strong optical response with high efficiency yet maintain the direct bandgap nature and tunability of the monolayer, is to create MQWs and superlattices of 2D TMDs. This approach was initially attempted with exfoliated monolayers to realize quantum well-like electronic states.[758] However, it was soon realized that light-trapping and modulation can also be achieved in highly absorptive and high-index semiconductors such as monolayer TMDs when placed alternatively with insulating dielectric spacers such as hBN or other atomic-layer-deposition-grown oxides. The large mismatch in refractive indices and extinction coefficients of TMDs as compared to hBN or ALD insulating oxides in the visible means, one can observe near unity absorption at the exciton for just 5-7 monolayers of TMD stacked alternatively with an insulating spacer. Given the advances in CVD-grown materials over wafer scales this approach was achieved over $cm^2$ scales showing not only near-unity absorption at the $WS_2$ exciton but also exciton-polariton coupling and dispersion at oblique incidence angles where the incoming light gets trapped in a waveguide-like mode in the multilayer structure which further resonantly couples with the in-plane excitons.[748,] This MQW and superlattice approach with $WS_2$ has also been extended to demonstrate electrically tuned phase modulation of light in metasurfaces achieving full $2\pi$ phase modulation of light coming in at angles of 55 degrees away from normal.[754] While this approach works in principle, current demonstrations are limited to unpatterned multilayers, and MQW-like structures that do not exploit the geometric phase. Further, electrical tunability has been limited to a bottom gate connected concurrently with all $WS_2$ monolayers in the stack. To achieve bulk tunability individual electrical connections must be achieved to each TMD monolayer which is challenging from a fabrication and scaling perspective. Alternatively, use of patterned MQW or superlattice structures or even multilayer/bulk TMDs could be an alternative approach for electrochemical intercalation and tuning of their optical properties *via* insertion/de-insertion of alkali metal atoms or other organic ligands (Figure 28d). Such approaches to tuning have already been demonstrated on an individual flake scale by several groups.[759] However, extension of these concepts to optical metasurfaces and demonstration of repeated electrochemical cycling to tune the optical properties are yet to be made. In addition, it is expected that the tunability of an electrochemistry-based process that involves diffusion and intercalation will be inherently slow.

28.3.6 *Challenges of Exciton Tuning*

The excitonic nature of the resonance carries intrinsic limitations. First, exciton transitions are characterized by strong losses that can limit device efficiencies. By operating the metasurface at a frequency that is slightly detuned with respect to the exciton energy, losses can be decreased at the price of reduced tunability. Second, the operation wavelength is limited to the material's exciton energy and thus cannot be designed at arbitrary wavelengths. For applications in telecommunications and integrated photonics, it is of particular importance to achieve (excitonic) tunability for wavelengths around 1.5 microns. This could be potentially achieved with smaller-bandgap TMDs such as $MoTe_2$ and $ReS_2$ though their large-area synthesis with high optical quality is still in the early stages. The future of excitonic metasurfaces therefore strongly relies on explorative research in 2D material science to identify new semiconductor 2D materials.

## 28.4 Future Directions

While the above sections have already suggested a few forward-looking ideas for TMD-based metasurfaces, here we elaborate on some pressing needs and important directions that the community could take to push the concept of TMD-based metasurfaces into the practical domain. Primary among them is the ability to grow high crystalline quality layers with uniform and controlled thicknesses over wafer scales. While high-quality growth of monolayers over 8" wafers has been achieved, the growth of multilayers all oriented in-plane with high-crystalline quality remains far from accomplished due to the vdW nature of interplanar interactions which limits homogeneous nucleation of subsequent layers after the first layer of growth. Another major challenge regarding multilayers is the accurate measurement of optical losses and dielectric functions below the optical band edge (exciton resonance). This has been particularly tricky since most samples are small flakes or ultrathin films. One approach to address this challenge is to fabricate waveguides and ring resonator structures from TMD flakes or thick crystals to quantify the loss below the band edge. Recent attempts to fabricate such structures[760] and prior measurements of the full in-plane and out-of-plane dielectric function of $MoS_2$[761] have been a welcome step in this regard. Nonetheless, the determination of in-plane and out-of-plane dielectric functions of layered TMDs remains scarce and more efforts must be made to measure them and post them for community validation and further use in optical design. Similarly, efforts need to be made to measure and quantify such optical dielectric functions at low temperatures and under varying levels of electrostatic doping—a direction that has also been lacking.

Finally, one area where this community could truly lead to new innovations in metamaterials and metasurfaces is the use of epsilon-near-zero (ENZ) points emanating from the excitons. While these points are difficult to attain in TMDs at room temperature, they have been observed in $MoSe_2$[762] at cryogenic temperatures, thus raising the prospects of ENZ metamaterials and hyperbolic metasurfaces using excitons.

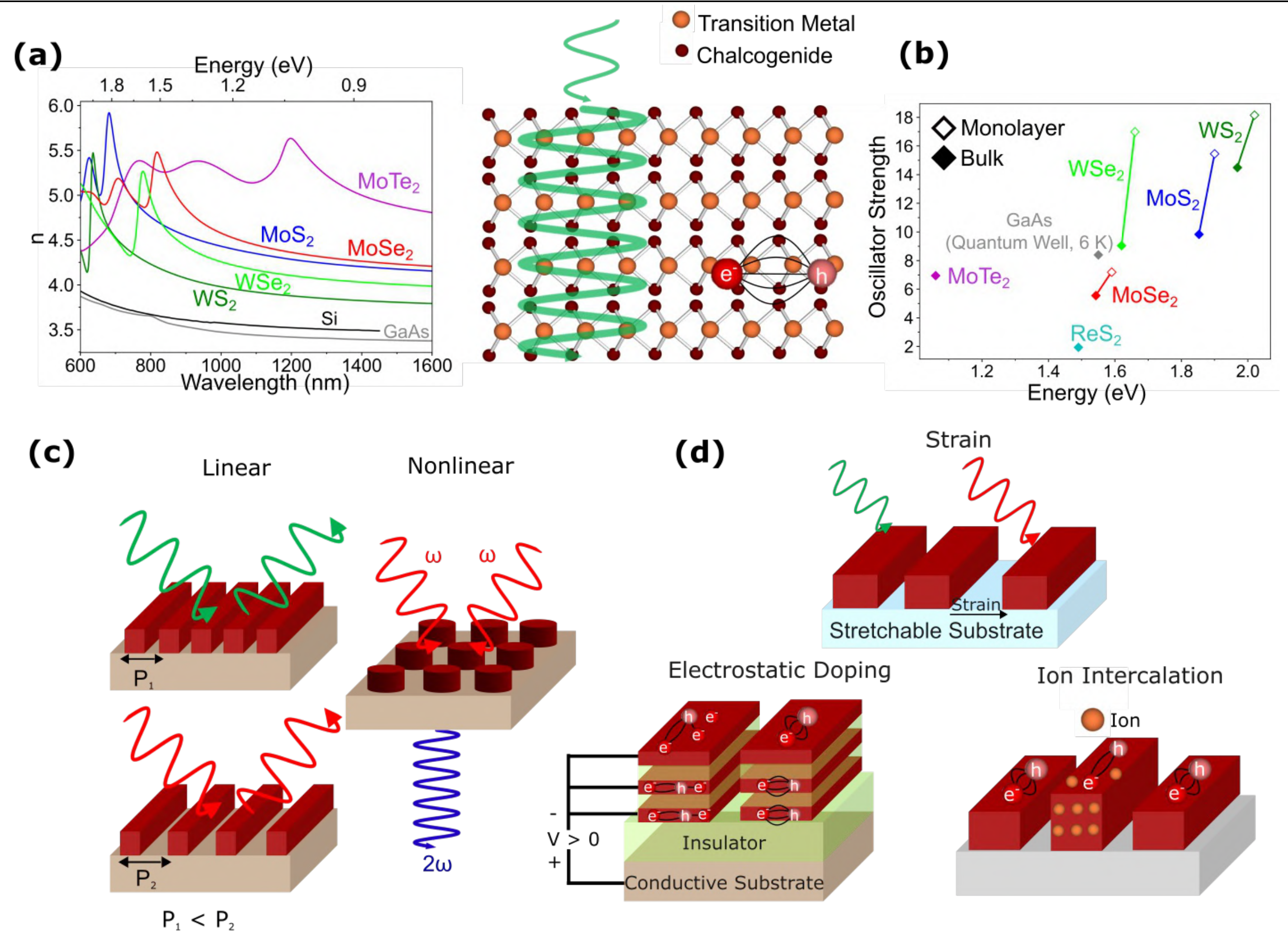

**Figure 28.** Comparison of TMD optical constants and schematic illustrations of TMD-based metasurfaces. (a) Comparison of the spectrally varying real parts of the refractive indices for the principal semiconducting TMDs in multilayer form together with Si and GaAs.[763,764,765] (b) Comparison of exciton oscillator strengths versus resonance energies of various monolayer and multilayer TMDs as measured at room temperature[517,744,745,766,767] A comparison of GaAs quantum wells at a temperature of 6 K is also provided.[768] (c) Schematic approaches to passive metasurfaces using bulk (multilayer) TMD active layers. Both linear and nonlinear approaches are depicted and discussed. (d) Schematic of approaches to active TMD-based metasurfaces through external parameters.

## 29. TRANSDIMENSIONAL MATERIALS AS A NEW PLATFORM FOR STRONGLY CORRELATED SYSTEMS

**Alexandra Boltasseva,[1,2] Igor V. Bondarev,[3] and Vladimir M. Shalaev[1,*]**
[1]Elmore Family School of Electrical and Computer Engineering, Purdue Quantum Science and Engineering Institute, and Birck Nanotechnology Center, West Lafayette, IN 47907, USA
[2]School of Materials Engineering, Purdue University, West Lafayette, IN 47907, USA
[3]Department of Mathematics & Physics, North Carolina Central University, Durham, NC 27707, USA
***Corresponding author.** Email: shalaev@purdue.edu

### 29.1 Status

Systems with strong electron–electron correlations, where the interactions between electrons are strong enough to dramatically influence their behavior, have long been in the focus of extensive studies across solid state physics, materials, and quantum science due to the unique properties that they enable. These include high-temperature superconductivity, unconventional magnetism, and several varieties of metal–insulator–transitions (MITs) related to remarkable physical phenomena, such as quantum- and disorder-related Anderson electron localization, Kondo effect, Wigner crystallization, and beyond.[338,339,769,770,771,772,773,774,775,776,777,778,779,780,781,782,783,784] These effects have been studied in strongly correlated materials, including insulators and metals, for electrons as well as for excitons and their complexes,[785,786,787,788,789] particularly in the low-dimensional regime in systems like semiconductor quantum wells, graphene, and TMDs.

One of the most interesting MIT phenomena is the electron Wigner crystal state formation[769] that has intrigued physicists since 1934.[338,339,770,771,772,773,774,775,776,777,778,779,780,781,782,783,784] In the Wigner crystal state, an electron gas crystallizes and forms a lattice of the electron density below some dimension-dependent critical value. Despite a large body of research on this topic, achieving and observing electron Wigner crystallization remains an outstanding challenge; it requires finding high-purity, high-quality materials that can offer a tailorable electronic response, sufficiently low electron densities, and low disorder. Thus far, signatures of Wigner crystals were observed in 2D electron gases under high magnetic fields,[770,771,772] bilayer TMD moiré superlattices (the so-called generalized Wigner crystals),[339,774,775,776] and only recently, some first microscopic images to prove charge excitations in 1D[773,777] and generalized 2D Wigner crystals were reported.[782,783] However, most Wigner crystal observations require low $T$, strong magnetic fields, or other external means, such as the moiré potential, making Wigner crystals challenging to observe in conventional materials and under reasonable operational regimes. Moreover, the generalized Wigner crystal state in bilayer moiré superlattices has little in common with the original Wigner's electron crystal concept as the *crystallization* there is achieved mostly due to the trapping of electrons by periodic moiré potential wells rather than being due to their Coulomb repulsion. To date, the genuine electron crystallization process has been observed indirectly in monolayers[338] and bilayers[784] of TMDs by monitoring exciton photoluminescence intensities, particularly, an extra resonance peak as a function of $T$ (~10 K) that could originate from the exciton Umklapp scattering by the 2D electron lattice formed below the Wigner crystal melting point.

As researchers are on the continuing lookout for new material systems to study strongly correlated phenomena, an exciting opportunity is offered by the so-called *transdimensional (TD) materials*[790] that have originally been proposed in the field of nanophotonics/plasmonics.[790,791,792,793] These ultrathin—between 2D and 3D—materials are expected to support strong electron–electron correlations and could potentially enable strongly correlated quantum phenomena including Wigner crystallization. Transdimensional plasmonic/metallic materials have thicknesses of only a few atomic layers and were shown to enable unprecedented tailorability of their optical response. This includes an unusually strong dependence on structural parameters such as thickness (number of atomic monolayers), composition (stoichiometry, doping), strain, and surface termination compared to conventional thin films, as well as an extreme sensitivity to external optical and electrical stimuli. Highly tailorable, quasi-2D designer-choice TD materials uniquely offer strong light confinement *via* highly localized surface plasmon excitations[793] and show remarkable tunability as well as novel quantum phenomena. Recently, epitaxially grown TD materials composed of transition metal nitrides (TMNs) such as TiN, ZrN, and HfN have been studied extensively[794] and showed a confinement-induced spatial nonlocal dependence of their plasma frequency as well as other related new physical phenomena.[795,796,797,798] However, while several electron-confinement-induced photonic effects in TD materials have been reported, until recently, TD materials have never been explored as a possible platform to study strongly correlated regimes (Figure 29). The first experimental evidence of the breakdown of plasmonic behavior and related MIT in HfN TD thin films was reported in a recent work.[798] The possibility of observing the electron confinement-induced MIT in metals is intriguing and could provide insights into the physical phenomena associated with strongly correlated systems in TD materials as well as new means to control them.

An ensemble of repulsively interacting particles (or quasiparticles) is expected to form a Wigner crystal lattice in a finite-thickness TD material film when its average potential interaction energy exceeds the average kinetic energy, that is,

$$\Gamma_0 = \frac{V_{KR}}{\langle E_{kin}\rangle} > 1, \tag{29.1}$$

as first formulated by Platzman and Fukuyama[799] and referred to as the PF ratio below. Here,

$$V_{\mathrm{KR}} = \frac{\pi e^2}{\epsilon_{\mathrm{film}}^{\mathrm{core}} d_{\mathrm{film}}}\left[\mathrm{H}_0\left(\frac{\epsilon_1+\epsilon_2}{\epsilon_{\mathrm{film}}^{\mathrm{core}}}\frac{\bar{\rho}}{d_{\mathrm{film}}}\right) - \mathrm{N}_0\left(\frac{\epsilon_1+\epsilon_2}{\epsilon_{\mathrm{film}}^{\mathrm{core}}}\frac{\bar{\rho}}{d_{\mathrm{film}}}\right)\right] \tag{29.2}$$

is the repulsive Keldysh–Rytova (KR) interaction potential[800] and $\langle E_{kin}\rangle$ is the mean electron kinetic energy per particle. The KR potential is a sum of the 0th-order Struve $\mathrm{H}_0$ and Neumann $\mathrm{N}_0$ (sometimes denoted as $\mathrm{Y}_0$) special functions and it represents the repulsive electrostatic potential interaction energy of a pair of electrons separated by the average in-plane distance $\bar{\rho}$ and confined vertically in the interior of the film of thickness $d$ with the background dielectric constant $\epsilon_{\rm film}^{\rm core}$, which is sandwiched in between a substrate and a superstrate with dielectric constants $\epsilon_1, \epsilon_2 < \epsilon_{\rm film}^{\rm core}$ (Figure 29). The KR potential energy in eq 29.2 indicates that the vertical electron confinement in optically dense ultrathin planar systems leads to an effective dimensionality reduction from 3D to 2D and replaces the $z$-coordinate dependence by the film thickness $d_{\rm film}$, which is now a parameter to represent the vertical size. The mean electron kinetic energy per particle at $T < T_F$ ($\sim 10^3 - 10^4$ K) can be obtained by integrating over the 2D Fermi surface (the circular zone of radius $k_F$ in the 2D reciprocal space). This gives $\hbar^2 \pi N_{2D}/(2m)$ where $N_{2D} = N_{3D}\, d_{\rm film.} = 1/(\pi\bar{\rho}^2)$ is the surface charge density while its volume counterpart is $N_{3D}$. Using the measured HfN film parameters $\epsilon_{\rm film}^{\rm core} = 5$, $\varepsilon_1 = 3$ (MgO substrate) and $\varepsilon_2 = 1$ (air),[798] from eqs 29.1 and 29.2 we obtain the PF ratio as a surface in the ($d_{film}$, $N_{2D}$) space (Figure 29). It can be observed that, for the 2-nm-thick HfN film (with $N_{2D} = 7.4\times10^{13}$ cm$^{-2}$ measured), a PF ratio $\Gamma_0 \sim 10$ can be achieved, which indicates the possibility of electron Wigner crystallization. In accordance with this, the 2-nm-thick film resistivity was observed to increase drastically with decreasing temperature, in contrast to what one would expect when lowering the temperature in conventional materials.

The observed effect could also occur due to the electron trapping by the random surface roughness potential of the film. However, this scenario would not explain the drastic resistivity increase with decreasing temperature. Clearly, more theoretical and experimental research is required in order to separate disorder/roughness-induced effects from the Winger crystallization signatures.

### 29.2 Challenges and Future Goals

While experimental progress in TD materials for nanophotonics has earlier been impeded by the challenges in producing atomically thin films of noble metals, plasmonic TMNs can easily be grown as epitaxial-quality films with thicknesses down to 1-2 nm (5-10 atomic layers).[796,797,798] Recently, the optical and electronic properties of TMDs have been studied both experimentally and theoretically; and the existing possibility to realize strongly correlated systems in TD materials is now ready to be explored. For example, to achieve the electron density required for the Wigner crystallization, static tailorability can first be used where the electron concentration is varied depending on the choice of material stoichiometry, deposition/annealing technique, and the film thickness. The dynamic tunability induced by electrical, optical and electro-optical stimuli should be also studied. Another avenue to tailor the optical properties of TD materials includes strain engineering. Below a critical thickness, an epitaxial thin film is expected to retain strain induced by the substrate. It has been theoretically demonstrated that by varying the in-plane lattice parameter of an ultrathin film its optical response can be tuned.[801] This can be achieved experimentally by growing strained ultrathin films on lattice-mismatched substrates. In conjunction with the thickness dependence of TD materials, strain engineering offers a novel way to tailor the optical response of plasmonic materials.

Since several physical effects could play a role in MITs observed in ultrathin plasmonic films, the experimental efforts should be closely aligned with the theoretical considerations for the Wigner crystallization phase diagram. This is crucial not only to guide experiments toward the realization of the strongly correlated state, but also to distinguish between Wigner crystallization and Anderson localization due to disorder-related formation of the random surface roughness potential. To distinguish between the two possibilities, one might use an in-plane magnetostatic field to reduce the number of the electron translational degrees of freedom from two (in-plane motion) to one (translational motion along the applied magnetic field direction), which would change drastically the Wigner crystallization picture whereas the Anderson localization process is not expected to change.

### 29.3 Suggested Directions to Meet These Goals

TD materials offer an interesting approach to exploring strongly correlated electronic systems. When compared to bulk metals, the screening in TD materials is greatly reduced, thus making the interactions between electrons dominant. Leveraging the possibility to control and tune the thickness-dependent plasma frequency of TMDs provides a *knob* not only to reduce their electron density but also to enhance the electron repulsive potential energy. Enhancing both factors at once could provide a viable route to achieve the Wigner crystallization state. The availability of various TMNs (TiN, ZrN, HfN, and others), their ability to grow as high-quality, ultrathin epitaxial films, and the sensitivity of the electron density to the material/structural/geometrical parameters provide a rich playground for the realization of strongly correlated electron systems in different regimes. The possibility of achieving Wigner crystallization in TD plasmonic materials is of great interest for photonic applications. When free electrons are crystallized, the thin film turns into a transparent dielectric. When the electron Wigner crystal melts, the material restores its metallic response. Exploring the conditions for electron Wigner crystallization opens up an entirely new avenue for the realization of optical modulation/switching and can be applied to develop tailorable/tunable photonic structures.

Importantly, a theory for electron Wigner crystallization in TD materials should be further developed based on the previous studies of 2D electron liquids.[799,802,803] Since the Wigner state is a crystal of the repulsively interacting particles with their average potential interaction energy exceeding the average kinetic energy, the specific PF ratio should be mapped for various materials and various thicknesses. The theoretical critical densities for observing the Wigner crystal as well as its phase diagram and melting curve should be studied for various TD materials and thickness regimes. The theory must also provide the guidance for experimentalists to distinguish between Wigner crystallization and Kosterlitz–Thouless transition in ultrathin TD plasmonic systems.[660] These efforts would be critical in both guiding experiments and proving the Wigner crystal formation. Equally importantly, the techniques to provide direct microscopic evidence for Wigner crystal formation, such as those utilizing scanning tunneling microscopy and other advanced microscopy techniques, should be applied to characterize strongly correlated electrons in TD materials.

## 29.4 Conclusion

Gaining fundamental insights into MITs in various TD materials and exploring the feasibility of the realization of Wigner crystal states potentially at room temperature and with no magnetic field applied represent a new direction in the field of strongly correlated systems. It is expected to bring critical fundamental insights into the physics of strongly correlated regimes and open applications that would eventually enable a new generation of tunable, reconfigurable, and multifunctional devices for nanophotonics, optoelectronics, and quantum technologies that are compact, ultrathin, and operate at low powers.

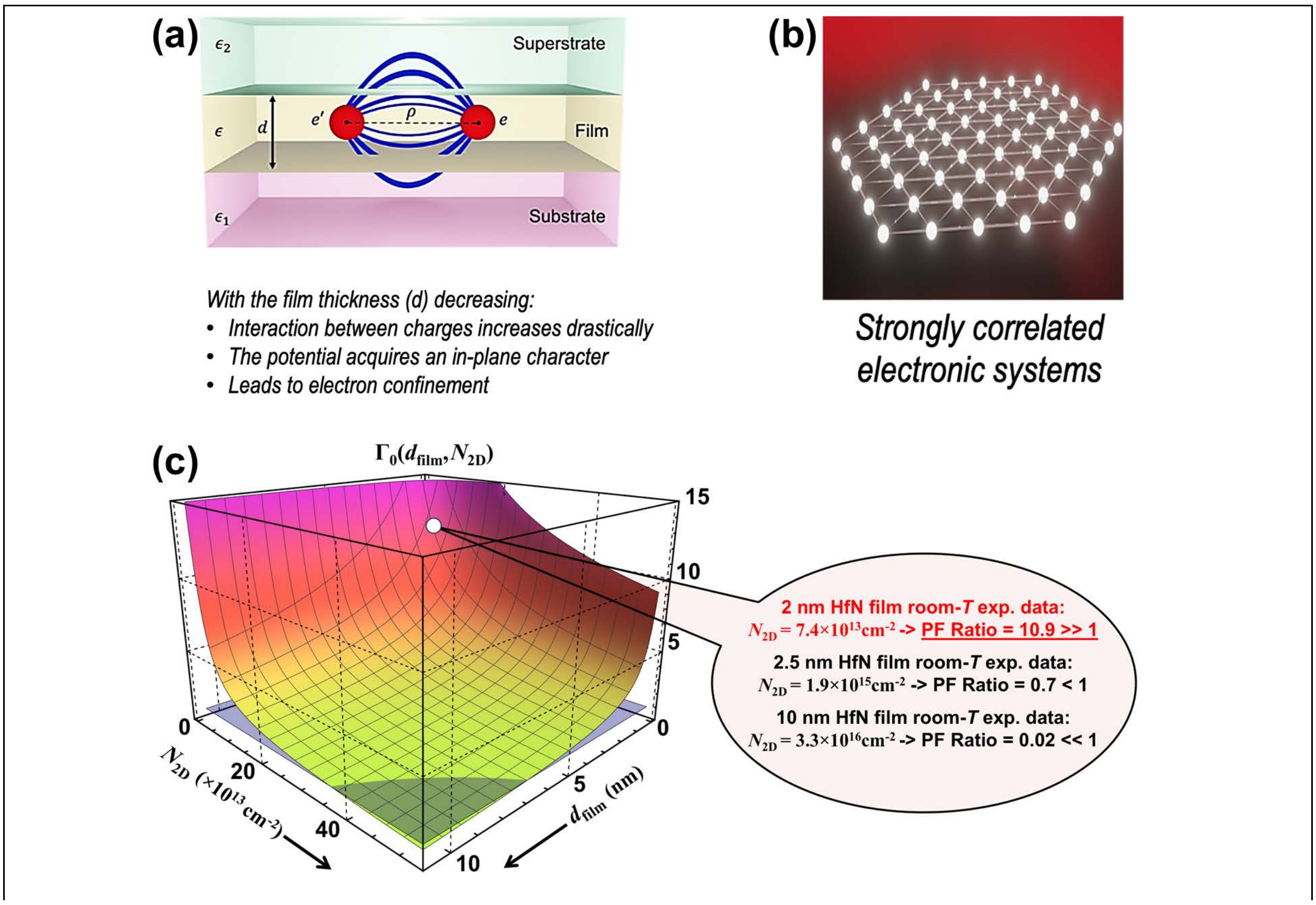

**Figure 29.** Strongly correlated electrons in transdimensional materials. (a) Electron confinement-induced plasmonic breakdown in a transdimensional material. We show a schematic of a plasmonic metal film sandwiched between a substrate and a superstrate. As the film thickness *d* decreases, the interaction potential among charges acquires an in-plane character and increases drastically, leading to electron confinement. (b) Schematic of Wigner crystallization when an interacting electron liquid has a sufficiently low electron density minimizing its total energy *via* crystallization into a quantum solid phase.[769] (c) Platzman–Fukuyama (PF) ratio given by eqs 29.1 and 29.2 plotted as a function of the surface charge density and film thickness with experimentally measured parameters of HfN TD films at room temperature.[798]

## 30. MXENES: EMERGING PHOTONIC MATERIALS

**Jeffrey Simon,[1] Colton Fruhling,[1] Vladimir M. Shalaev,[1] and Alexandra Boltasseva[1,2,*]**
[1]Elmore Family School of Electrical and Computer Engineering, Birck Nanotechnology Center and Purdue Quantum Science and Engineering Institute, Purdue University, West Lafayette, IN, 47907, USA
[2]School of Materials Engineering, Purdue University, West Lafayette, IN 47907, USA
***Corresponding author.** Email: aeb@purdue.edu

### 30.1 Introduction: Mxene Synthesis and Properties

MXenes are a family of 2D transition metal carbides, nitrides, and carbonitrides with the first member, $Ti_3C_2T_x$, (where $T_x$ denotes surface termination) being introduced in 2011.[804] They acquire their name from the combination of their general chemical formula $M_{n+1}X_nT_x$ and the suffix '*ene*' implying '*thin-film*'. Even though this discovery was overshadowed by the excitement toward graphene, which had won a Nobel prize just a year earlier, the materials and optics community swiftly began investigating the properties of MXenes. First, the prototypical MXenes $Ti_3C_2T_x$ and $Ti_3CNT_x$ were shown to be conductive,[805] and consequently, surface-plasmon properties in $Ti_3CNT_x$ were reported.[806] As a unique difference from graphene, the large interlayer spacing of stacked MXene flakes could allow for the independent polarization of different layers, thus enhancing control of plasmonic effects.[807] Another major discovery on titanium-based MXenes is their nonlinear optical properties and their application as passive saturable absorbing elements in near-IR ultrafast fiber lasers, both of which are now the subject of intense study.[808,809] The origin of these properties could relate to the density of states (DOS) or plasmonic nature.[810,811]

MXenes have the chemical formula $M_{n+1}X_nT_x$ and the flake structure (Figure 30) consists of $n + 1$ layers of early transition metal atoms (M) with layers of carbon or nitrogen (X) sandwiched in between. The number of layers can be in the $n = 1 - 4$ range, and $T_x$ is a surface termination group that commonly consists of =O, -OH, or -F, but can include many more elements (Figure 30). Over 60 different MXenes have been synthesized to date. Some include multiple transition metals combined to create both ordered atomic arrangements like $Mo_2TiC_2T_x$ and disordered arrangements such as high-entropy $TiVNbMoC_3T_x$.[805,812] The unprecedented tailorability of the MXene structure allows for the tuning of electrical and optical properties. Generally, MXenes with fewer layers $n$ are more likely to have insulating or semiconducting properties due to the fewer overlapping electronic bands. Conversely, MXenes with a higher $n$ number tend to have metallic behavior.[813] Theoretical results show that MXenes could even exhibit topological properties[814] and composition-controlled ferromagnetic and antiferromagnetic properties.[815]

Beyond the primary constituents, surface termination engineering can dramatically influence both the optical and electrical properties of MXenes. For example, the work function of MXenes can be tuned by the surface termination,[816] and *in situ* electrochemical manipulation of the surface termination can modulate the linear and nonlinear optical properties.[810,817,818]

The most studied MXene, $Ti_3C_2T_x$, has found many uses in optoelectronics. Its conductive properties allow for nanometer-thick films to be used as transparent conducting electrodes[813] and as a charge transfer layer for perovskite solar cells.[819] In the visible spectrum, MXenes generally exhibit large absorption due to interband transitions and, in combination with their metallic properties in the IR, can be employed as broadband absorbers.[813] Additionally, MXenes have been used for electromagnetic interference shielding in the THz, microwave, and RF spectral regions.[805] Most of these optical applications of MXenes, including photothermal, optical detection, and broadband absorption, focus on the lossy nature of the film form of MXenes.

## 30.2 Challenges and Goals

While losses in MXenes play a crucial role for some devices, they inhibit other applications. At the flake level, losses could occur from defects created in the MXene precursor or during the MXene synthesis process. With current technology, a solution-based process is used to synthesize MXene flakes from MAX phases *via* the removal of the A element. For example, $Ti_3C_2T_x$ is created from the MAX phase $Ti_3AlC_2$ when the Al layer is selectively etched with chemicals such as HF and delaminated. The result is a solution of multilayer and single-layer MXene flakes.[804] MXenes can be used either as single flakes, as is done during the exploration of surface plasmon properties, or as films, which are used for nonlinear applications.

To understand the loss mechanisms, researchers must explore how MXene films are made. Individual MXene flakes serve as the building blocks of optical films. At the film level, the optical permittivity is also expected to depend on the film morphology, surface charge, intercalants, and the interaction between flakes. Various techniques such as spin-casting, spray-coating, blade-coating, and dip-coating allow for deposition on a substrate and a free-standing film can be formed using vacuum filtration.[805] In the film, vdW forces cause overlapping flakes to adhere to one another. The relationship between the film morphology, which is different from each deposition technique, and the optical properties should be further investigated.

## 30.3 Applications Abound

The solution-based film creation process also leads to facile integration into many optical devices. For example, MXenes spray could be applied to an existing structure to enable electromagnetic interference shielding or a curved substrate could uniformly be coated for optical applications. Conductive MXenes are known to have an epsilon-near-zero (ENZ) point where the real part of the permittivity $\epsilon_1$ passes through zero, changing from a positive to negative sign.[820] This is especially exciting for dynamic and nonlinear optics as ENZ-hosting materials have been shown to exhibit large nonlinearities and novel dynamics.[821] Solid solutions or mixtures of multiple MXenes can be combined to tune the optical properties[822] and possibly enable the creation of films with tailorable optical properties such as the ENZ point. Further, anisotropic optical properties of MXenes have just started to be investigated.[823,824,825] Mixed MXene films consisting of one highly conductive and one lesser-conductive MXene have recently been showed to tune the ENZ point throughout the near-IR spectrum.[826] Further, engineered metamaterial structures could enable tailored optical properties including hyperbolic optical dispersion.

Additionally, because monolayer flakes have a thickness of approximately a single nanometer, incident light sees a single-species MXene film as an effective metamaterial composed of MXene and termination layers. In comparison to traditional vdW materials like graphene and TMDs, the interflake spacing for MXenes is larger due to the termination layers. This presents an exciting possibility to incorporate intercalants and make hybrid materials. The manufacturing process has allowed for molecules to be incorporated into MXenes to enhance nonlinear properties.[827] Intercalants used to improve electrode properties[828] could also be applied to control the optical properties for plasmonic applications.

## 30.4 Outlook

Since MXenes research is a developing field, some early experimental results may need to be reinterpreted. For example, MXene films can quickly grow oxides, which could alter electrical and nonlinear optical properties.[829] Additionally, $TiO_2$ oxide defects can result in photoluminescence.[830] It was shown that a standard fabrication method to produce $Ti_3C_2T_x$ MXene leads to oxygen defects at the titanium sites of the lattice.[831] With this coming to light, it is worthwhile to reinvestigate early optical studies. Currently, there are debates on dominate nonlinear mechanisms which could be re-examined.

The exploration of MXenes in the field of optics and photonics has unveiled a plethora of promising applications and unique properties. MXenes have already demonstrated significant potential in various domains, including optoelectronics, nonlinear optics, photothermal devices, broadband absorbers and electromagnetic interference shielding. The ability to tailor their structure and surface terminations offers unparalleled flexibility in tuning their optical and electrical properties, making them versatile materials for future technologies, including the expanding space of wearable electronics.[805] As research continues, the initial focus should be on understanding the fundamental mechanisms that govern the optical properties including losses in MXene films to expand capabilities in more traditional optical applications like lasing and optical modulation.[808] Therefore exploring new synthesis techniques[832,833] to minimize defects and other potential loss sources will be a key direction of research. Following from there, the development of hybrid materials and engineered metamaterials could pave the way for innovative applications in plasmonics and metamaterials, such as sensing.[834] The journey of MXenes in optics is just beginning, and with continued interdisciplinary collaboration, these materials could make a significant impact on optical technologies.

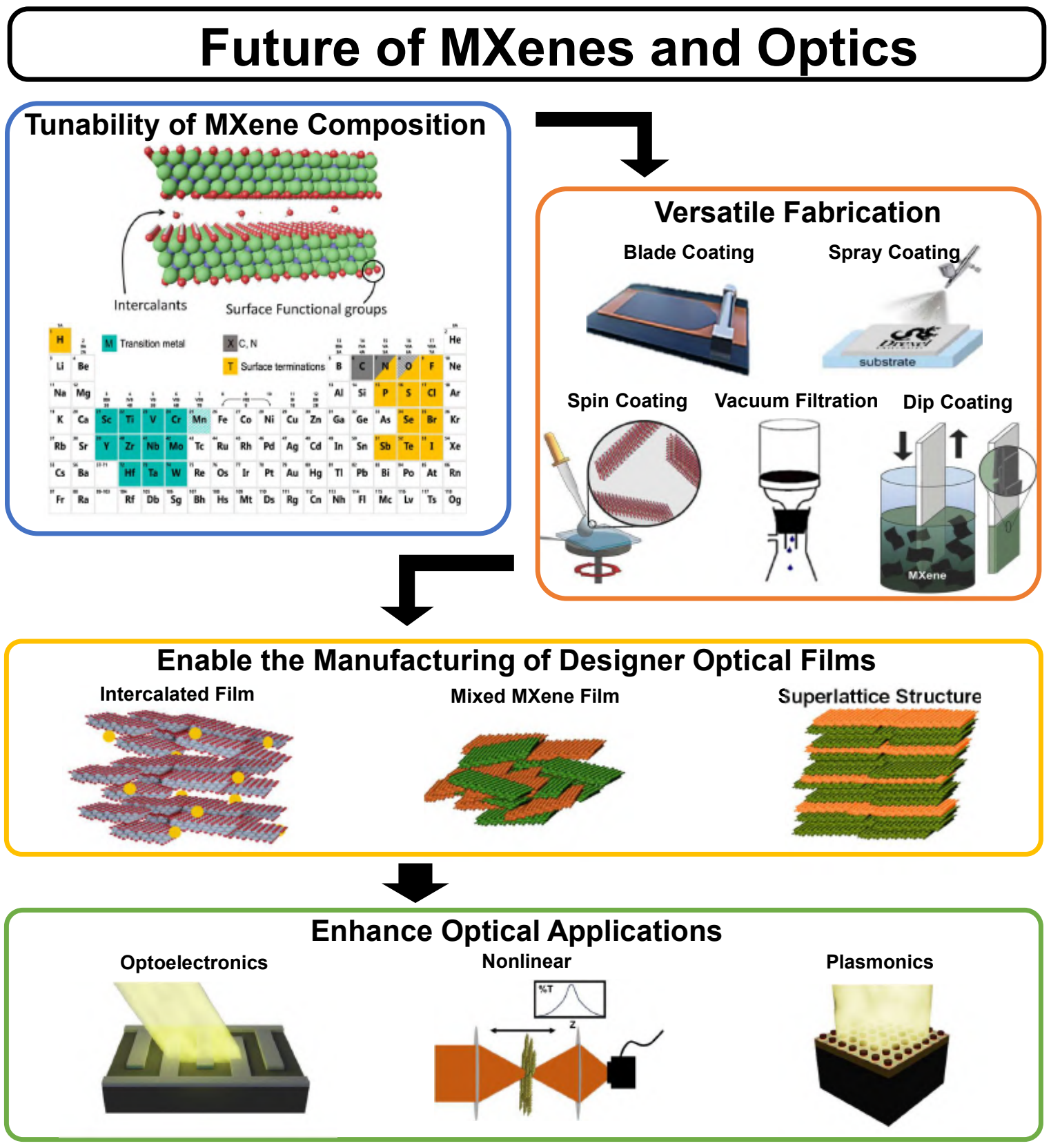

**Figure 30.** Pathways for the use of MXenes in optical applications. The tunability of MXene stoichiometry and versatile fabrication methods will allow for the manufacture of designer optical films to enhance current applications and pave the way for future uses. Periodic table adapted with permission from ref 835 (Copyright 2022 Springer Nature). Optical application and flake structure figures adapted from ref 836. Blade coating figure reprinted in part from ref 837 (Copyright 2024 American Chemical Society). Spray coating and vacuum filtration figures adapted from ref 838 (Copyright 2017 American Chemical Society). Dip coating image adapted with permission from ref 839 (Copyright 2018 Wiley). Spin coating figure reprinted in part from ref 840 (Copyright 2016 Wiley).

## 31. HOT ELECTRON GENERATION BY INELASTIC UV EXCITATION DOMINATES IN BLACK PHOSPHORUS

**Guangzhen Shen,[1] Dino Novko,[2] Shijing Tan,[1] Bing Wang,[1] and Hrvoje Petek[3,*]**
[1]Hefei National Research Center for Physical Sciences at the Microscale, New Cornerstone Science Laboratory, University of Science and Technology of China, Hefei, Anhui 230026, China
[2]Centre for Advanced Laser Techniques, Institute of Physics, 10000 Zagreb, Croatia
[3]Department of Physics and Astronomy and the IQ Initiative, University of Pittsburgh, Pittsburgh, Pennsylvania 15260, USA
***Corresponding author.** Email: petek@pitt.edu

### 31.1 Introduction

Two-dimensional materials hold promise for advanced photonics and electronics by providing ever greater miniaturization of electronic components and novel functionalities based on their optical and electronic properties and their application strategies. Black phosphorus (BP), a small bandgap semiconductor with prominent excitonic and highly anisotropic properties,[841] in particular, is interesting for its high carrier mobility[842] and *colossal* UV (>3 eV) photoresponsivity,[843] which give it intriguing photonic and electrical properties.

As its name suggests, black phosphorus is a highly photonically active material whose direct electronic transitions are supplemented (and in the UV region exceeded) by photoinduced inelastic electron scattering processes that augment its optical response on account of poor screening of optical fields that is common to 2D materials. This makes its excitonic properties susceptible to pressure, layer thickness, Floquet engineering, and parity symmetry tuning, as revealed by optical and photoelectron spectroscopies.[844,845,846]

By time-resolved multiphoton photoemission (TR-mPP) spectroscopy we investigate how the ultrafast photoresponse of BP changes between the visible and UV light excitation, and find that it transitions from conventional interband absorption, within the random phase approximation (RPA), to a higher-order beyond-RPA (b-RPA) response that has a dominant contribution from the inelastic electron–electron (*e*–*e*) scattering. We attribute this transition to the ineffective and dynamically modified screening of the Coulomb interaction that is a general property of 2D materials.[842,847,848]

The shrinking of electronic device dimensions presents challenges concerning the properties of electron quasiparticles in materials with reduced dimensionality. A distinctive intrinsic property of 2D materials that is likely to play a key role in any application is screening of the Coulomb interaction. The reduced screening of an electron by a free electron cloud and the strongly anisotropic dielectric functions that can gain large densities of states by band nesting and hyperbolic profiles can promote b-RPA interactions to compete with the direct interband optical responses.[847,848,849,850] The carrier scattering in 2D materials can be explored through transport properties or through their optical responses. Here, we elaborate on the ultrafast carrier scattering in bulk BP as revealed by energy-dependent rise of b-RPA features in its TR-mPP spectra.

In 1900, before the advent of quantum mechanics, Paul Drude proposed a model to explain the transport properties of electrons in metals, attributing resistivity to scattering of electrons by the relatively immobile ions, impurities, and defects. The concept of Drude absorption[851,852,853,854,855,856,857] has been extended to explain the frequency-dependent intraband optical absorption in alkali and noble metals.[858] Intraband absorption in metals is well documented when a photon generates a virtual electron–hole pair with a lifetime defined by energy–time uncertainty of detuning from resonant interband transitions, where the photon energy can only be absorbed by b-RPA scattering processes to a range of real states that satisfy energy and momentum conservation. Similarly, an electron promoted nonresonantly from the valence band to the conduction band of a semiconductor is in a virtual superposition state from which it must decay by scattering processes to energy- and momentum-conserving eigenstates.[848]

The contribution of *e*–*e* and electron–phonon (*e*–*ph*) scattering to intraband absorption in alkali and noble metals has been debated from 1960 to 1990 with respect to the frequency dependence and degree of energy exchange by virtual electron-scattering processes.[858,859,860] While *e-ph* scattering can dominate in the low-energy region (<1 eV), the electronic scattering involving multiple *e-ph* pair and plasmon excitations is likely to dominate at higher energies and appear as wings that decay from the Fermi level to higher energies.[857,860,861,862,863] This is broadly significant because hot electron distributions define the efficiency of processes such as hot-electron-induced chemistry. While hot electron and hole distributions are commonly assumed to have equal excitation probabilities of states between $E_F \pm \hbar\omega$ relative to the Fermi level $E_F$, with $\hbar\omega$ denoting the optical excitation energy, such distributions are not supported by a few experimental measurements of such distributions.[864] In fact, hot electron distributions have been observed to rise approximately exponentially from $E_F + \hbar\omega$ to $E_F$. Such distributions have been predicted by Hopfield for alkali metals[859,861] and have been measured in Ag(110) by TR-mPP.[864] If b-RPA scattering explains hot electron distributions in metals, it is also likely to explain hot electron distributions under ultrafast laser excitation in BP and similar 2D materials.

### 31.2 Results

Having elaborated on how b-RPA electronic scattering processes generate hot electron distributions in 3D metals, we examine the mPP spectra of layered BP as the excitation energy $\hbar\omega$ is tuned from visible to UV at ambient temperature under ultrahigh vacuum conditions. Contrary to optical spectroscopy, which records energy- and momentum-integrated optical transitions, mPP (Figure 31a) measures the energy- and momentum-resolved coherences and populations of transiently excited unoccupied states above $E_F$ that participate in multiphoton excitation leading to photoelectron emission.[864,865] Figure 31b shows representative angle-resolved mPP spectra of BP for selected $\hbar\omega$, which are fully explained by the known in-plane electronic bands from a DFT calculation (Figure 31c), and shown schematically in Figure 31d for the near-normal emission component together with the possible excitation pathways. The excitation with $\hbar\omega = 2.14$ eV is dominated by a near-resonant two-photon transition from valence band 1 (VB1) to conduction band 4 (CB4), with the angle-resolved spectrum displaying the negative band mass of CB4. The line profile taken at the Z point ($k_{||} = 0$ Å$^{-1}$) also shows a weak contribution from the first image potential state (IPS) that exists at the BP/vacuum interface. The signal associated with the near-resonant CB4 ← VB1 transition is far more intense than the signal at the work function edge, where the contribution from secondary processes (i.e., hot electrons) is expected to have its maximum intensity. Moving on to $\hbar\omega = 2.58$ eV excitation, we record CB1, CB4, and IPS features that are populated by one- or two-photon absorption and further photoemitted in overall three-photon excitation processes. It is notable, however, that, for this excitation energy, the work function signal has become dominant, indicating a strong contribution from secondary processes that generate a hot electron population in CB1. In this spectrum, all spectroscopic features of the conduction band involve three-photon absorption, so the relative intensities belong to the same photon order excitation. The VB1 signal, however, is excited by a nonresonant coherent two-photon absorption, where a fraction of the virtual intermediate state population scatters with the electrons remaining in VB1 to generate the intermediate-state hot electron population. Because both the populations of the virtual state generated by promotion of VB1 electrons by one photon and hot electrons population derived from it require one additional photon absorption to be photoemitted, we conclude that the virtual intermediate state scatters to form hot electrons with a high probability. A similar overlay of inelastic hot-electron generation and coherent interband mPP features has been observed in 2PP spectroscopy of Ag(110).[866] Upon further increasing to $\hbar\omega = 2.88$ eV excitation, we see that the coherent two-photon signal from VB1 is further diminished with respect to the incoherent signal associated with scattering of the VB1 + $\hbar\omega$ virtual state to form hot electrons. This shows that the b-RPA scattering increasingly dominates for UV excitation. The direct and indirect excitation processes from VB1 again involve one-photon absorption. The last spectrum for $\hbar\omega = 3.44$ eV shows that the hot electron distribution dominates over the vestige of coherent two-photon absorption from VB1. We note that IPS remains for all excitations as it is a coherent feature of the transient screening response of the BP/vacuum interface.[867] The primary hot electron distribution has an exponential energy dependence consistent with a Fermi–Dirac distribution with an effective electron temperature that increases with $\hbar\omega$ to ~1500 K at $\hbar\omega = 3.44$ eV. Figure 31e shows an interferometric two-pulse correlation measurement of 2PP signal as a function of the intermediate state energies. While above 1.5 eV the intermediate state electron lifetimes are nearly laser-pulse limited, below 1.5 eV one observes femtosecond timescale hot electron population decay with lifetimes that are quantified in Figure 31f. The hot electron population decays with an energy-dependent rate that can be interpreted by Fermi liquid theory.[868,869] While the hot electron distribution is weakly sample-temperature-dependent, the energy-dependent hot electron lifetimes tell that *e–e* scattering is its dominant excitation process.

## 31.3 Conclusion

The immediate significance of the reported hot electron generation in BP is that it explains its *colossal* photoresponsivity in the UV region.[843] This was attributed by Castro Neto and coworkers to a band nesting in the Γ-Z direction, a common feature in multilayer 2D materials. One would expect that such band nesting results in a high joint density of states for resonant optical transitions.[842,843,849] While this is a reasonable interpretation, our mPP spectra of BP show no evidence of interband transitions between nested bands. Instead, we argue for an alternative mechanism that can increase the density of states and dielectric field strengths of 2D materials. BP, like many other anisotropic 2D materials, has a dielectric function that is hyperbolic in different spectral regions.[870] In a 2D material, the hyperbolicity can be tuned by impurity or photodoping.[852] The ultrafast electronic excitation of BP dynamically performs such photodoping on a few femtosecond timescale. The hyperbolic dielectric functions introduce a large joint density of states for photon absorption and scattering, and can cause giant field enhancements in regions where the real part of the dielectric functions is near zero.[850] The hyperbolic dielectric properties are likely to be intrinsic to all 2D materials because of their anisotropic electronic structures giving them a metallic optical response in one direction with respect to the crystal plane, and semiconducting one in an orthogonal direction. In BP, and related materials with in-plane dielectric anisotropy we even anticipate quasi-1D excitonic, electronic, transport, and optic properties. These features are evident in the electronic[851,852,871,872] and THz[873,728] properties of BP as hyperbolic exciton polaritons and tunable plasmons.

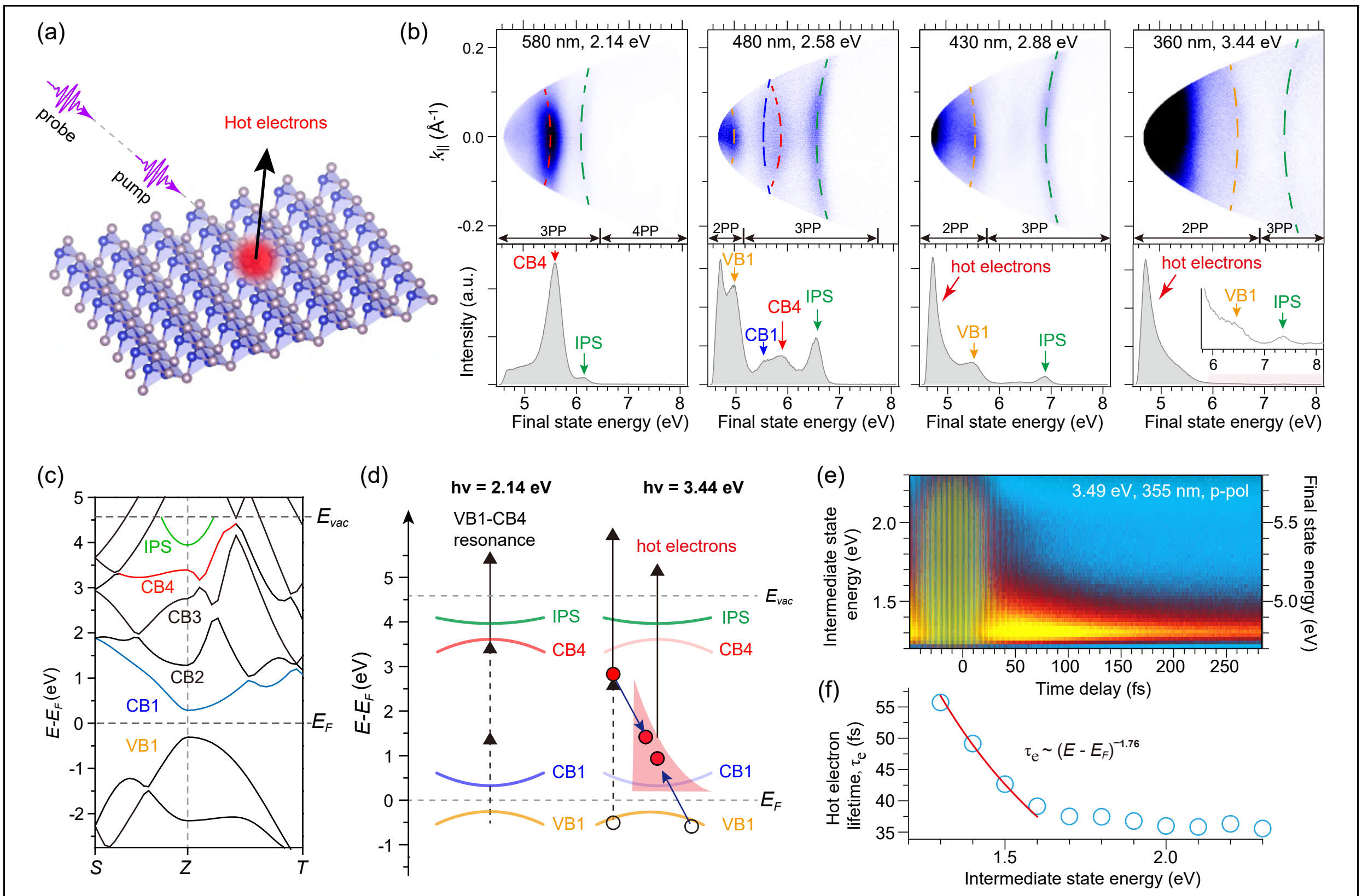

**Figure 31.** Distinct first-order interband and higher-order inelastic optical responses of BP from the visible to UV excitation regions. (a) Schematic excitation of mPP from BP, a representative 2D material. The frequency tunable pump pulse of ~30 fs duration generates interband and intraband (hot electron) excitations, and the identical delayed probe pulse records their population decay. (b) Representative mPP energy–momentum $E(k_{||})$ maps excited along the zigzag direction of BP with photon energies $\hbar\omega = 2.14$, 2.58, 2.88, and 3.44 eV, respectively. The dashed lines guide the band dispersions. The lower panels are the corresponding line profiles extracted at $k_{||} = 0$ Å$^{-1}$. The inset for $\hbar\omega = 3.44$ eV expands the mPP signal above 5.8 eV to show the remaining band features. The bottom arrows indicate the spectral regions that are excited by 2PP-4PP processes. (c) Calculated band structure of bulk BP with band dispersions along Z-S ($k_{AC}$) and Z-T ($k_{ZZ}$) directions. (d) Illustrative excitation diagram respectively showing band-to-band resonant excitation (left), and inelastic scattering of virtual states promoted from VB1 leading to the production and detection of hot electrons (right). (e) Interferometric two-pulse

correlation measurement of 2PP signal as a function of the intermediate state energy probing the BP polarization dephasing and hot electron population dynamics. (f) Extracted lifetime, $\tau_e$, of hot electrons as a function of the intermediate state energy. At an energy higher than 1.6 eV, the measurements are limited by the time resolution of ~35 fs at $\hbar\omega = 3.49$ eV. Below 1.6 eV, the data is fit to $\tau_e \sim (E - E_F)^{-1.76}$, which slightly deviates from the $\tau_e \sim (E - E_F)^{-2}$ behavior of a 3D Fermi liquid.

## 32. PLASMONICS IN FEW-ATOMIC-LAYER METALS: OPTICAL CONFINEMENT AND MODULATION

**Vahagn Mkhitaryan,[1,2,*] Renwen Yu,[3] Alejandro Manjavacas,[4] J. Enrique Ortega,[5,6] and F. Javier García de Abajo,[2,7]**
[1]Elmore Family School of Electrical and Computer Engineering, Birck Nanotechnology Center and Purdue Quantum Science and Engineering Institute, Purdue University, West Lafayette, IN 47907, USA
[2]ICFO-Institut de Ciencies Fotoniques, The Barcelona Institute of Science and Technology, 08860 Castelldefels, Barcelona, Spain
[3]Department of Electrical Engineering, Ginzton Laboratory, Stanford University, Stanford, CA, USA
[4]Instituto de Química Física Blas Cabrera (IQF), CSIC, 28006 Madrid, Spain
[5]Donostia International Physics Center, Paseo Manuel Lardizabal 4, 20018 Donostia-San Sebastián, Spain
[6]Departamento de Física Aplicada I, Universidad del País Vasco, 20018 San Sebastián, Spain
[7]ICREA-Institució Catalana de Recerca i Estudis Avançats, Passeig Lluís Companys 23, 08010 Barcelona, Spain
***Corresponding author.** Email: vmkhitar@purdue.edu

### 32.1 Introduction

Ultrathin metal structures down to single-atom thickness constitute a powerful platform to control light at the nanoscale.[792] Plasmons have been demonstrated in metallic systems with small dimensions, including zero-dimensional crystallites[874], 1D atomic metal wires (e.g., gold chains growth on the Si(557) vicinal surface[875]), and few-atomic-layer metal films (e.g., Ag(111) crystal layers grown on a silicon wafer[10,37]). Metal films have a much larger density of conduction electrons than graphene and doped 2D semiconductors, which makes them more difficult to modulate (i.e., less susceptible to external stimuli), although some possibilities have been formulated (see below). In exchange, they exhibit plasmons in the near-IR—a technologically appealing spectral region—and their optical properties are more robust.

Atomically thin noble metal nanoislands can support strong plasmon resonances in the near-IR spectral range. Figure 32a shows theory for the extinction cross section of a 20 nm nanodisk carved from a single monolayer of gold placed in vacuum,[792] as obtained by numerically solving the Maxwell equations with the frequency-dependent homogenous dielectric function for gold[876] and a thickness of 0.236 nm, corresponding to the separation between (111) atomic planes in the bulk metal. The extinction spectrum displays a pronounced plasmon resonance, yielding a peak that exceeds that of a gold nanosphere with the same diameter. The spectral position of the plasmon resonance and its response to near- and far-field excitation can be tuned by adjusting the aspect ratio of the nanostructure defined as the ratio of its diameter to thickness,[877] as illustrated in Figure 32b: for a fixed diameter, an increase in thickness blueshifts the plasmon resonance toward the visible range.[878] However, the increased number of charge carriers in thicker structures makes them harder to tune through, for example, electrical doping.

An alternative scheme to achieve tunable plasmons and dynamical light modulation is to exploit hybrid films consisting of an ultrathin metal film and a graphene sheet, as shown in Figure 32c. In such an ultrathin graphene-metal hybrid film, optical losses of the hybrid plasmon mode can be controlled by tuning the doping level of graphene and moving its Fermi energy $E_F$ relative to the neutrality point. In Figure 32c, the dispersion of the hybrid mode is illustrated through the reflectance for p polarization, plotted as a function of both the photon energy and wave vector. We find that, when the photon energy is larger than $2E_F$, there is a substantial quenching of the hybrid mode *via* coupling to interband transitions in graphene.[879] In contrast, for photon energies below $2E_F$, the plasmon dispersion remains almost intact (i.e., like that in the film without graphene).

### 32.2 State of the Art

Several techniques have been developed over the last decade for chemically synthesizing crystalline metallic nanostructures. Crystalline metallic nanoparticles, often synthesized *via* seed-mediated methods, are among the earliest achievements in this direction,[874] using small nanoparticles as nucleation sites for controlled growth by precisely managing reducing agents, temperature, and pH levels. This allows for the synthesis of nanoparticles with a large degree of control over morphology. Beyond simple geometries, these nanoparticles have served as templates for more complex structures,[880,881,882] including nanoplates[883] and ultrathin flakes.[545,884,885]

The synthesis of high-aspect-ratio structures like gold flakes represents an extension of these colloidal techniques. Gap-assisted methods, for instance, utilize a confined growth environment created by placing two substrates in proximity.[882] By introducing halide ions such as chloride and bromide, vertical growth is suppressed while lateral growth is promoted. This process yields gold flakes with lateral dimensions up to 250 μm and thicknesses around 10 nm. Template-assisted synthesis further refines this approach by using predefined scaffolds, such as layered hydroxides, to guide the deposition of gold atoms.[884,885] This method, which employs methyl orange as a confining agent to produce gold nanosheets with sub-nanometer thickness (as low as 0.47 nm), offers significantly less control over the lateral sizes and morphology of the films. Techniques for thinning these high-aspect-ratio flakes enable further advancements in precision and functionality. Atomic-level precision etching relies on chemical agents to remove material in a layer-by-layer fashion.[545] This approach allows one to achieve a thickness reduction down to single atomic layers while maintaining lateral dimensions of over 100 μm, ensuring smooth surfaces and uniformity, which are essential for quantum-confinement applications. In addition, transfer methods, such as graphene-inspired exfoliation, facilitate the integration of gold flakes onto diverse substrates.[886] In this method, gold films grown on copper are delaminated through chemical etching of the copper layer, thus ensuring the structural integrity of the transferred film.

A significant leap in scalability and quality has been achieved through ultrahigh vacuum (UHV) epitaxial growth techniques, as demonstrated in recent plasmonic studies of few-atomic-layer crystalline silver films[10,37] (Figure 32e-i) based on the silver-on-silicon platform.[887] These studies reveal that UHV conditions enable the growth of atomically thin, single-crystal films with smooth surfaces and low resistivity, reaching thicknesses as low as 8 monolayers (<2 nm) for silver films grown on atomically flat Si(111) substrates and extending across entire chips, far exceeding the lateral size limitations of chemically synthesized flakes. Furthermore, these films are compatible with advanced lithographic patterning techniques, including electron-beam lithography and reactive ion etching, allowing for the fabrication of plasmonic nanoribbon arrays with high crystalline quality.[10]

Prepatterned substrates have emerged as a strategy to enhance pattern quality:[37] by guiding the growth process, those substrates act as templates for the epitaxial deposition of silver, significantly minimizing damage often associated with post-patterning processes. This approach enables the fabrication of high-quality, atomically thin silver structures (Figure 32i) exhibiting larger quality factors and superior plasmonic performance than post-patterned structures. These films achieve both atomic-scale thickness and chip-scale lateral dimensions, providing a robust platform for high-performance photonic and quantum devices.

### 32.3 Challenges and Suggested Directions

The significantly reduced number of charge carriers in nanostructures consisting of an atomic layer of noble metal (compared to their 3D counterparts) makes these systems a promising platform for achieving efficient electrical light modulation. A practical implementation could involve a periodic array of identical nanostructures controlled by an electrical back gate. In this configuration, the average charge density in the layer of nanostructures is inversely proportional to the areal filling fraction occupied by the noble metal, as determined by capacitor theory for a fixed distance from the gate. Similar behavior is expected in systems comprising several atomic layers, although the degree of modulation is then reduced because the doping charge is distributed across the increased thickness.[878] An alternative approach could involve a continuous noble-metal atomic layer,[888] which also exhibits large electrical tunability of its propagating plasmons, and can be decorated with dielectric structures to bridge the light-plasmon momentum mismatch.

Building on the modulation of thin metal plasmons by varying the doping of an adjacent graphene layer (Figure 32c), we can simultaneously bridge the plasmon kinematic mismatch with the light cone (dashed line in Figure 32c) and realize efficient light-plasmon coupling by patterning the graphene-metal hybrid film into a periodic array of ribbons, as shown in Figure 32d. When examining the reflection spectra of such an array with undoped ($E_F = 0$) or highly doped ($E_F = 1$ eV) graphene, we observe a three-fold effect of doping: (1) the plasmon resonance peak is blue-shifted; (2) the magnitude of the resonance peak increases; and (3) the resonance linewidth becomes narrower for higher doping levels. The figure illustrates a modulation depth ≈ 36% at a photon energy around ≈ 0.9 eV. The doping level of graphene can be controlled through electrical gating at a relatively high speed depending on the design of the electronic circuit. As an alternative that can reach higher speeds, one could exploit ultrafast optical pumping to effectively dope graphene[18,563,889] through its ultrafast photothermal response (see Chapter 17).

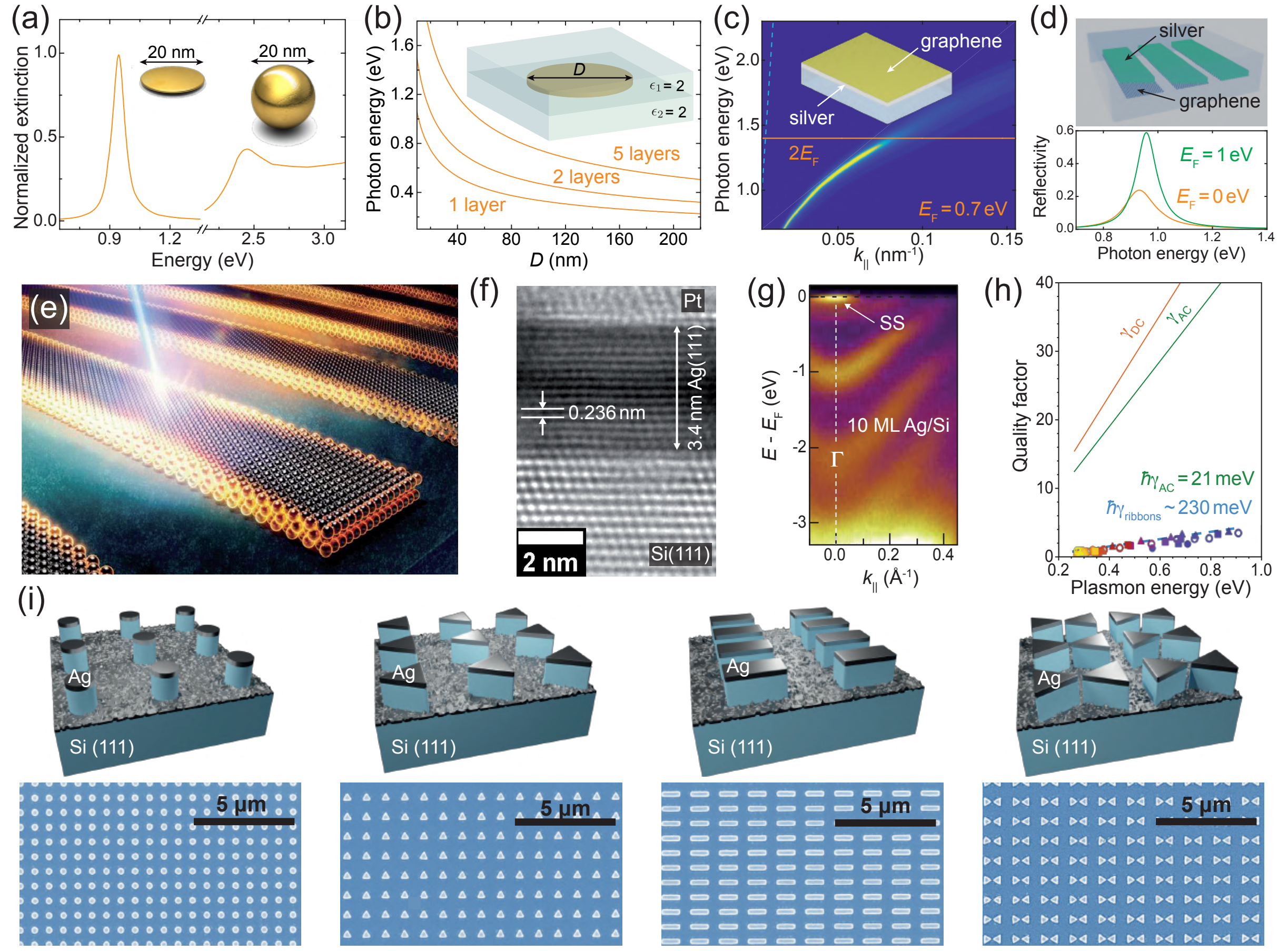

**Figure 32**. Plasmonics in ultrathin metal films. (a) A single-atomic-layer gold disk with (111) crystallographic orientation exhibits a sharp plasmon redshifted to a technologically appealing spectral region relative to a gold nanosphere with the same diameter. Adapted with permission from ref 792 (Copyright 2014 Springer Nature). (b) The plasmon energy of few-atomic-layer gold or silver disks scales roughly as $\sqrt{t/D}$ with the disk thickness $t$ and diameter $D$. Adapted with permission from ref 878 (Copyright 2015 Royal Society of Chemistry). (c,d) The plasmon band of an ultrathin silver layer (1 nm) deposited on graphene is quenched at plasmon energies exceeding the graphene optical gap $2E_F$, where $E_F$ is the graphene Fermi energy (c). This effect can be used to modulate the plasmons of a thin silver ribbon array, thus producing a radical change in light reflection depending on $E_F$ (d). Adapted with permission from ref 879 (Copyright 2016 Springer Nature). (e-h) Ultrathin crystalline silver ribbons can be prepared by epitaxial growth on silicon, followed by nanolithography (e). The resulting metal film exhibits high quality and wafer-scale crystallinity (f), reflected in the emergence of sharp electronic quantum-well states observed *via* angle-resolved photoemission spectroscopy (g) (i.e., the electron intensity as a function of parallel wave vector $k_\parallel$ and energy relative to the Fermi level, $E - E_F$). The quality factor (energy-to-width ratio, $\omega_p/\gamma$) of the obtained ribbon plasmons reaches ~4 (with a measured width $\hbar\gamma_{\rm ribbons} \approx 230$ meV), which is far from the large values predicted from the bulk Drude-model damping, as measured optically[876] ($\hbar\gamma_{\rm AC} \approx 21$ meV) or electrically ($\hbar\gamma_{\rm DC} \approx 24.6$ meV for a bulk conductivity of $6.30{\times}10^7$ S/m). This reduction in quality factor can be presumably ascribed to the effect of imperfections introduced during passivation of the film (with a silica coating) and nanolithography (h). Adapted from ref 10 (Copyright 2019 American Chemical Society). (i) Prepatterning of silicon followed by epitaxial deposition of silver crystalline films leads to higher quality structures with great

flexibility to pattern different morphologies, as shown in the upper sketches and the lower secondary-electron-microscope images. The measured quality factor of the bowties reaches ~10. Adapted with permission from ref 37 (Copyright 2024 Wiley).

**Applications: Integrated Photonics**

## 33. INTEGRATED PHOTONICS WITH 2D MATERIALS

**Xu Cheng[1], Hao Hong,[2] Ruijuan Tian,[1] Dong Mao,[3] Dries Van Thourhout,[4] Xuetao Gan,[3] Kaihui Liu,[2] and Zhipei Sun[1,*]**

[1]QTF Centre of Excellence, Department of Electronics and Nanoengineering, Aalto University, Espoo, Finland

[2]State Key Laboratory for Mesoscopic Physics, Frontiers Science Center for Nano-optoelectronics, School of Physics, Peking University, Beijing, China

[3]School of Physical Science and Technology, Northwestern Polytechnical University, Xi'an, China

[4]Photonics Research Group, Department of Information Technology, Ghent University–imec, Gent, Belgium

***Corresponding author.** Email: zhipei.sun@aalto.fi

### 33.2 Introduction

Integrated photonics is reshaping modern technology by enabling low-cost, compact, energy-efficient, and multifunctional optical systems. To expand the potential of integrated photonics, 2D materials have emerged as a powerful tool mainly due to their unique optical and electronic properties. These ultrathin materials are highly versatile, offering strong light–matter interaction, tunable optical characteristics, and compatibility with existing photonic integration technologies. They can function on-chip devices like light sources, modulators, and photodetectors, making them ideal for the next generation of integrated photonics. In addition, 2D materials cover a completed material regime across insulators like hBN, semiconductors such as molybdenum disulfide ($MoS_2$) to semimetal graphene, and superconductors such as $NbSe_2$, thus providing a broad range of passive and active photonic and optoelectronic function possibilities. This combination puts 2D materials at the forefront of integrated photonics, offering exciting opportunities to advance research and applications in areas such as optical communication and computing, sensing and imaging, and quantum technologies.

### 33.2 Current State of the Art

Over the past two decades, integrating 2D materials, primarily focusing on monolayer and few-layer 2D materials, with waveguides for photonics has shown two key advantages: (1) It can significantly enhance the device performance by improving light–matter interaction along the waveguide structures while preserving the intrinsic waveguide advantages (e.g., low transmission loss, dispersion and broad bandwidth); (2) It can enable new functional or multifunctional optoelectronic devices by flexibly combining different 2D materials with various waveguides. The integration of 2D materials thus opens up exciting possibilities for integrated photonics. In the following sections, we will concentrate on two primary types of 2D materials integrated optical waveguide structures: optical fibers and on-chip waveguides,[507,890] while the integration of polariton-based (including plasmon, phonon, and exciton polaritons) waveguides is addressed in separate sections of this Roadmap.

#### 33.2.1 *Two-Dimensional Materials Integrated Optical Fiber Devices*

Optical fibers play a crucial role in long-distance communication, sensing applications, and biomedical technology due to their low attenuation, customizable optical properties, flexibility, and well-established manufacturing processes. These make them the current most widespread waveguide for practical applications. The integration of 2D materials with optical fibers has mainly evolved through three main stages: starting with the fiber facet, progressing to the fiber surface, and incorporating them into the interior of the fiber (Figure 33a).

Fiber facet integration. This approach offers a convenient and compact platform for directly coupling light with 2D materials or their polymer composites, typically sandwiched between two fiber facets. By leveraging this integration method, the exceptional properties of 2D materials and their heterostructures have been utilized to create high-performance fiber-based photonic and optoelectronic devices, including lasers, modulators, and photodetectors. Among these applications, a key research focus is the integration of 2D materials between the fiber facets in a fiber-ring cavity to exploit their ultrafast saturable absorption capabilities for generating ultrafast optical pulses.[891] The demonstrated performance of the ultrafast lasers includes down to sub-100 fs pulse duration with the operation wavelengths covering from ~1 to 2 μm.

Fiber surface integration. This method involves transferring various 2D materials onto the surface of nanofibers or side-polished fibers, significantly enhancing light–matter interaction along the fiber, in contrast to the fiber facet integration method. Using this approach, numerous linear and nonlinear optical effects have been achieved in the hybrid fibers, including anisotropic absorption, saturable absorption, four-wave mixing, difference frequency generation, and sum frequency generation.[892] These advancements open the door to a range of applications, such as polarizers with an extinction ratio of up to 27 dB,[893] ultrafast all-optical modulators (with a response time at ~ ps scale),[474] ultrasensitive gas detection (down to sub-ppm),[894] high-speed logic gates, all-fiber phase-modulators, and continuous-wave nonlinear optical converters.[892]

Fiber interior integration. This route was initially demonstrated by filling 2D material dispersions into microstructured fibers, and has now been revolutionized by advancements *via* the chemical vapor deposition method. This deposition technique enables the controlled growth of 2D materials within the air-holes of fibers, creating high-quality hybrid fibers that can extend up to tens of centimeters. This approach allows significantly stronger light–matter interaction compared to the surface or facet integration methods, thanks to the enhanced length and considerate overlap between optical modes and 2D materials. This integration also unlocks new possibilities for photonic and optoelectronic applications, such as graphene-integrated photonic crystal fibers for broadband (~1150–1600 nm) electro-optic modulators with impressive modulation depths (~20 dB $cm^{-1}$ at 1550 nm),[484] semiconducting 2D materials in hollow-core fibers for enhanced second- and third-harmonic generation by ~300 times.[506] Recent advancements also improve supercontinuum generation efficiency, exhibiting a 70% reduction of the threshold power for spectra with one octave spanning and distinct layer-dependent behavior only working for 2D materials thinner than five atomic layers.[895] These breakthroughs broaden the application areas by meeting the demand of large-scale production, paving the way for highly integrated all-fiber devices.

33.2.2 *Two-Dimensional Materials Integrated on-Chip Waveguide Devices.*

In contrast to optical fibers, on-chip waveguides offer greater compactness, making them more suitable for photonic chips, and enhanced flexibility in structural design, enabling diverse engineering of the optical field. Within this framework, different 2D material integration methods are also utilized for a wide range of applications (Figure 33b).

Waveguide surface integration. This integration method, which involves straightforwardly transferring or depositing 2D materials onto the surfaces of various waveguides, is widely adopted. Its popularity stems primarily from its simplicity and compatibility with complementary metal-oxide semiconductor (CMOS) technology. To date, a variety of typical 2D materials have been explored, including graphene, TMDs, black phosphorus, gallium selenide and various 2D material heterostructures, such as moiré structures. The demonstrated waveguides include but are not limited to channel waveguides, photonic crystals and microresonators, which enable photonic and optoelectronic applications in both linear and nonlinear optical regimes. In the linear optical domain, 2D materials significantly enhance integration functions, enabling high-performance devices such as electro-optic modulators (over 15 dB modulation depth and 30 GHz bandwidth operating at telecommunication wavelengths),[896] photodetectors (over 100 GHz bandwidth and 100 A $W^{-1}$ responsivity separately),[897,898] polarizers (~20 dB extinction ratio),[899] and absorbers (~85% unpolarized absorptivity covering almost the entire solar spectrum of 300-2500 nm with graphene metamaterials).[900] In the nonlinear optical domain, 2D materials enlarge the optical responses of hybrid waveguides, addressing the limitations of traditional bulk materials with relatively weak intrinsic nonlinearities. For example, 2D materials introduce second-order nonlinearities, such as second-harmonic generation and the Pockels effect, into silicon-based materials that naturally lack these properties due to their crystal centrosymmetry. Additionally, they enhance third-order processes, including four-wave mixing (>9 dB conversion efficiency improvement),[901] supercontinuum generation (>300 times enhancement of the nonlinear coefficient),[902] and all-optical modulation (with switching energy and time of ~35 fJ and 260 fs, respectively).[903] Furthermore, the gate-tunability of 2D materials (including graphene and TMDs) introduces an additional degree of freedom for refractive index (including real and imaginary parts) adjustments for novel optoelectronic applications (e.g., shifting the frequency comb lines from ~2.3 to 7.2 THz, and facilitating efficient tuning respectively).[485,904] These contributions greatly expand the functional capabilities of waveguide-based photonic systems.

Waveguide interior integration. Integrating 2D materials directly within waveguides presents a promising strategy for enhancing light–matter interactions by maximizing the overlap between the optical modes and material. This approach shows significant potential for applications such as polarizers, modulators, and photodetectors. However, its adoption remains limited due to the involved fabrication complexities, the associated device losses, and the already-achieved relatively satisfactory light–matter interaction through the waveguide surface integration. Currently, the waveguide interior integration is primarily explored in theoretical studies.[905] Nonetheless, this method merits further investigation, especially as advancements in hybrid waveguide fabrication techniques and evolved complex 2D material integration (e.g., hybrid 3D integration).

### 33.3 Challenges and Future Research Directions

Integrated photonics with 2D materials has demonstrated great potential in many aspects and has been extensively studied and achieved massive advancements in the past years. However, its direct implementation as a replacement for current commercial devices is still challenging. To fully realize the potential of 2D materials in photonics and optoelectronics for practical applications, several challenges and areas for improvement must still be addressed, thereby enhancing its competitiveness in the near future.

2D materials-based integrated platforms. 2D materials can feature atomically smooth surfaces to minimize scattering losses at interfaces, high-refractive index (for example, $MoS_2$ ~4.2 at near-IR region),[906] large transparent window (for example, hBN for extending into the deep UV band),[907] negligible interlayer electronic coupling, and monolayer-like excitonic behavior in bulk form. Thus, they have recently emerged as a promising material platform for integrated photonics (Figure 33c) instead of integrating with conventional platforms (e.g., silicon). For example, given their large optical nonlinearity, 2D material-based waveguides, in principle, can achieve high efficiency of nonlinear optical conversion toward new functions (e.g., rhombohedral $MoS_2$ waveguide for broadly tunable second-harmonic generation).[491] Further, a versatile template-assisted method enables the efficient fabrication of various perovskite waveguides with predefined geometries and the observation of their edge lasing, which can also apply to many other 2D materials. [908] Sequentially, with an increasing number of new vdW crystals identified and invented, it is foreseeable that these emerging materials hold the complete potential to develop into new integrated photonic platforms.

New 2D materials for integration. In the realm of searching for novel 2D materials for integrated photonics, there are three potential directions. First, there is an interest in expanding the material library by investigating and synthesizing new 2D materials beyond the most common 2D materials. This expansion can encompass materials like borophene, silicene, germanene, antimonene, 2D oxides, and 2D nitrides, as well as synthetic materials not found in nature, such as Janus structural materials and heterostructures, for integrated photonic demonstrations. Second, it is important to explore novel 2D materials exhibiting unique physical properties, such as superconductors for single-photon detection, highly nonlinear materials for entangled photon pair generation, and 2D topological, ferroelectric, and magnetic materials for integrated spintronics, quantum devices, and other advanced integrated photonic technologies. Finally, further exploration of new organic 2D materials that offer tunable electronic and optical properties and can be easily integrated with existing photonic platforms and sustainable integration systems is also warranted for future research.

Novel structural engineering. In addition to searching for new materials, structural engineering plays a pivotal role in achieving high-performance integrated photonic devices. This approach mainly encompasses two key aspects: stack engineering of 2D materials and the development of innovative waveguide structures. Regarding the stacked 2D materials and their homostructures in the former, the twisted bilayers or multilayers of 2D materials stacked in two or three dimensions can not only generate novel physical phenomena, for example, moiré patterns and superconductivity but also facilitate the phase manipulation of the optical field, enhancing nonlinear optical processes such as second- and third-harmonic generation for frequency conversion,[486] modulation, and signal processing.[493] Further, the lateral or vertical heterostructures consisting of different 2D materials enable tailored bandgaps and electronic properties, adjustable *via* twist angles, creating wavelength-specific photonic devices and integrating quantum photonic functions (e.g., sources, memory, detectors) with classical photonic integrated circuits. For waveguide structures, optimized waveguide designs significantly enhance linear and nonlinear processes. For instance, a better phase-matching design in hollow-core fibers for hybrid $MoS_2$ integration can boost nonlinear conversion efficiency to rival bulk crystals.[506] Similarly, careful structure and dispersion tuning in 2D materials integrated on-chip waveguides can improve nonlinear optical performance, supporting further studies like high-order nonlinearities, soliton dynamics, and frequency combs.

Innovative device function exploration. Beyond the novel material/structural engineering considerations outlined above, another pivotal approach that necessitates enhancement in current demonstrations is external (thermal, acoustic, optical, electric/magnetic, mechanical, etc.) modulation for innovation functionalities previously unattainable with conventional materials. For instance, temperature impacts the phonon populations in 2D materials, influencing their interaction with electronic states and affecting light scattering and absorption processes. Understanding and leveraging these interactions can drive innovations in optoelectronic and photonic devices, including new thermal management and phonon-assisted photonic devices. Additionally, acoustic waves can introduce periodic strain fields, which can alter the electronic band and crystal structures, affecting optical transitions and enabling dynamic control over light propagation, respectively. This capability can be harnessed for dynamic modulation of optical responses in both linear and nonlinear regimes for new devices such as optical modulators. Furthermore, novel magnetic vdW materials have impressive Verdet constants, rendering them highly responsive to external magnetic fields. The application of a magnetic field can induce significant Faraday effects, enabling the manipulation of light polarization and offering potential applications in high-performance magneto-optic isolators, circulators, and sensors.

Hybrid integration advantages. The term *hybrid integration* broadly encompasses the combination of various well-established waveguide platforms, diverse active materials, and multifunctional electro-optical systems (Figure 33d). To date, most 2D material-integrated photonics have been demonstrated on traditional material platforms, such as silica fibers and silicon-based on-chip waveguides. However, emerging material platforms and technologies - such as silicon-core and chalcogenide glass fibers, lithium niobate, lithium tantalate and III–V semiconductors photonic platforms, and laser-written waveguides in glasses, polymers, and sapphires - are gaining interest for hybrid 2D material integration, presenting significant opportunities to enhance performance and expand the functional scope of photonic devices. Integrating various 2D materials with these platforms can further enrich their wide-ranging properties and functions, and allow for extensive hybrid integration, enabling both passive and active photonics and optoelectronics. For example, insulators like hBN for wideband waveguides, semiconducting 2D materials for lasers, modulators, and detectors, ferroelectric 2D materials for modulators, metallic 2D materials for contacts, and superconducting 2D materials for single-photon detectors fully support highly specialized or integrated applications. Further, graphene and semiconducting 2D materials, in particular, exhibit outstanding electronic properties, making them also ideal for applications such as transistors, receivers, memories, sensors,[909] and beyond (e.g., in-memory optical computers). In this regard, 2D materials hold the potential to fully realize true electro-optical hybrid integration, paving the way for seamless on-chip convergence of communications, sensors, computers, electronic, photonic and optoelectronic systems into a single platform for next-generation technologies.

Large-scale and high-quality 2D materials fabrication. One of the most pressing challenges in the field of 2D materials for integrated photonics is associated with their large-scale and high-quality preparation. Currently, the existing research is predominantly confined to lab-level studies at the micrometer scale, relying heavily on the well-known mechanical exfoliation method and manual transfer techniques. Therefore, scalable synthesis or integration methods, such as chemical vapor deposition, molecular beam epitaxy, liquid-phase exfoliation, and atomic layer deposition, hold promise to be developed for their large-scale production with high-quality *via* precise control over parameters (e.g., thickness, uniformity, crystallinity, composition, and stacking) on wafer scales. For 2D materials integrated waveguides, only few materials like graphene and $MoS_2$ have been realized in waveguides on a large scale. In general, future advancements should prioritize the production or transfer of high-quality and uniform 2D materials and the expansion of their diversity deposited throughout the entire waveguide area. Key improvements include the fabrication of single-crystal materials with pristine surfaces, ensuring consistent layer thickness with minimal defects, or achieving seamless coverage over curved and edged surfaces. Further, the development of techniques for creating on-demand vertical and lateral heterostructures is essential to enable new functionalities and enhance performance by leveraging the unique properties of each layer. Additionally, the ability to facilitate selective area and pattern growth for directly synthesizing 2D materials on a waveguide is paramount for both single hybrid integrated waveguides and the integration of multiple 2D materials with diverse functionalities. On 2D materials-based integrated platforms, the recent successful implementation of both interfacial epitaxy and bevel-edge epitaxy techniques to produce large-scale and single-crystal vdW materials (up to 15,000 layers) provides a significant opportunity to establish a truly integrated photonics platform.[756]

Future applications. With the advancement of high-performance devices achieved through the diverse integration of 2D materials with waveguides and the implementation of varied fabrication strategies for high-quality hybrid integration, the field of integrated photonics with 2D materials demonstrates significant potential. For instance, the exceptionally high nonlinear responses, combined with the distinctive peaks and scattering properties-such as Raman and Brillouin effects-of 2D materials, compared to those of conventional silicon, make it particularly advantageous for advancing optical fiber sensing technologies. Additionally, the demonstrated capability for high-speed data processing and communication, along with the complete integration of devices on a single chip, is expected to yield more compact and efficient optical interconnects within data centers and 6G networks. And owing to their exceptional mechanical flexibility, 2D materials can be seamlessly integrated onto nonplanar substrates for the development of flexible and wearable integrated photonics, as well as for the creation of lightweight, and high-performance optical components essential for augmented/virtual reality displays and headsets, thereby enhancing user experiences. Furthermore, given that vdW stacked 2D materials possess exceptional features enabling the precise manipulation and control of quantum states at the nanoscale, they hold significant potential in the realms of quantum technologies, encompassing communication, computing, sensing, and networking.

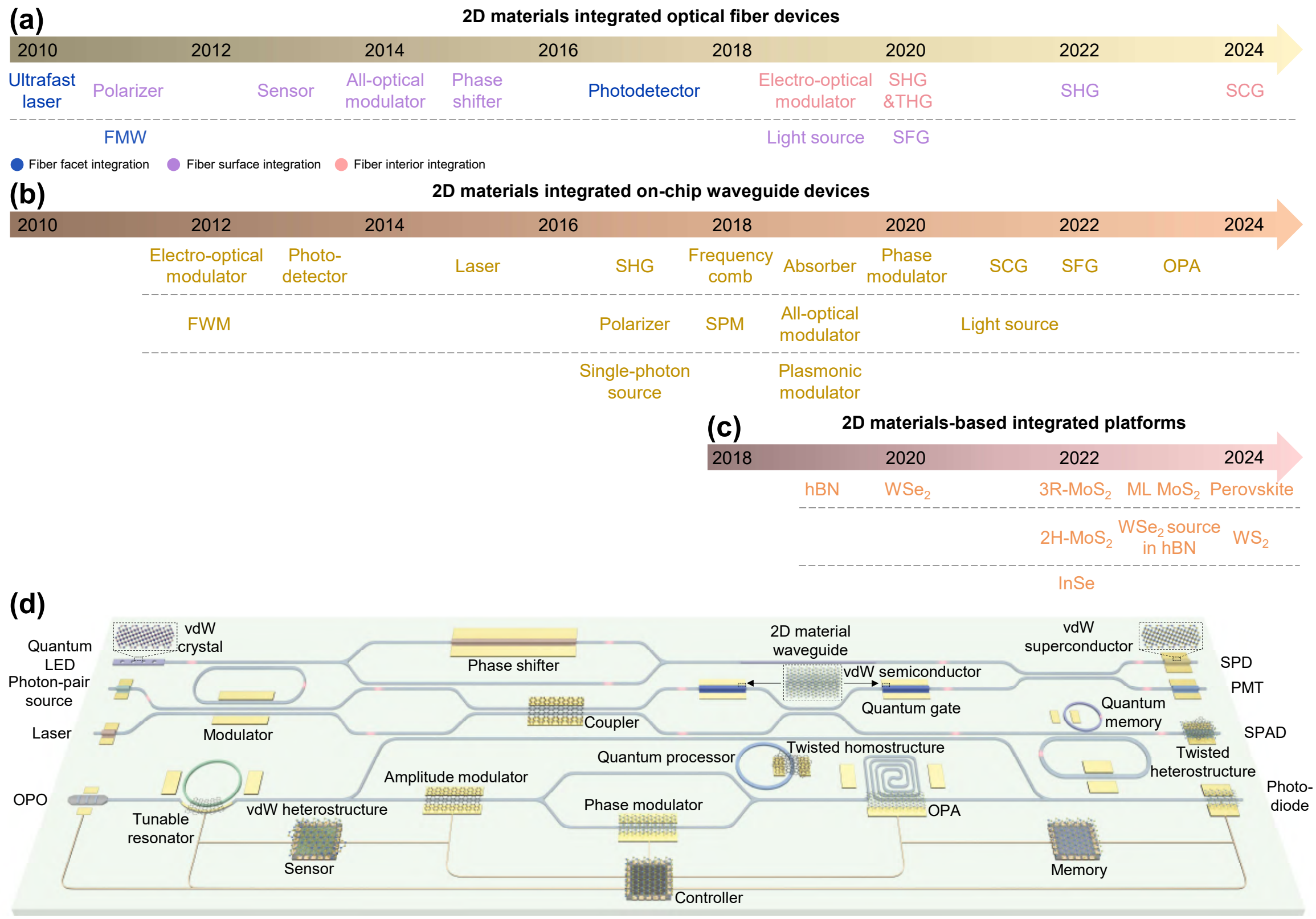

**Figure 33.** Two-dimensional materials for integrated photonics. (a-c) Selected demonstrations of 2D materials integrated optical fiber devices (a), on-chip waveguide devices (b), and novel 2D materials-based integrated platforms (c) over the past years. (d) Schematic illustration of future hybrid integration, providing a vision of the combination of various waveguide platforms, active materials, and their engineered structures, along with multifunctional optoelectronic interconnection systems, aimed at developing the next generation of highly integrated photonics.

## 34. OPTOELECTRONIC INTERCONNECTION OF POLARITONS IN 2D MATERIALS

**Qing Dai**[1,2,3,*]
[1]School of Materials Science and Engineering, Shanghai Jiao Tong University, Shanghai 200240, P. R. China
[2]CAS Key Laboratory of Nanophotonic Materials and Devices, CAS Key Laboratory of Standardization and Measurement for Nanotechnology, National Center for Nanoscience and Technology, Beijing 100190, P. R. China
[3]Center of Materials Science and Optoelectronics Engineering, University of Chinese Academy of Sciences, Beijing 100049, P. R. China
***Corresponding author.** Email: daiqing@sjtu.edu.cn

### 34.1 Current State of the Art

Electronic chips are facing dual challenges related to computational power and energy consumption as they approach their physical limits. To address these issues, three technical pathways have been proposed for advancing next-generation information technologies: More Moore, More-than-Moore, and Beyond Moore.[910] The essence of the Beyond Moore approach is optoelectronic integration, which incorporates photonic functionalities into integrated circuits. This integration harnesses the rapid information transmission capabilities of photons while capitalizing on the significant advantages of photonic matrix computations, greatly enhancing chip performance. The primary research focus is on achieving high integration of optoelectronic devices through heterogeneous integration and advanced packaging solutions. Various schemes for optoelectronic integration have been proposed, including silicon-based photonics that are compatible with complementary metal-oxide-semiconductor (CMOS) processes. These schemes leverage the plasma dispersion effects of silicon;[911] lithium niobate photonics, which utilizes the nonlinear electro-optic effects of lithium niobate crystals;[912] and wide-bandgap III–V semiconductor materials, which harness quantum well confinement effects and offer advantages in source devices.[913] Additionally, there are platforms based on micro-sized metallic plasmonics.[914]

### 34.2 Challenges and Future Goals

Despite significant advancements in transmission speed and photonic computation, the integration of optoelectronic devices remains limited by the optical diffraction limit. Key photonic devices, such as modulators, typically range from millimeter to hundred-micron scales, creating a substantial size mismatch—over three orders of magnitude—compared to nanoscale electronic devices. This severe disparity necessitates additional specialized interfacing components, which act as *bridges* to connect the two domains. These components generally include modules for digital signal processing, serial-to-parallel conversion, amplification, and driving, and therefore, their power consumption can account for more than two-thirds of the entire system, presenting a significant bottleneck for the large-scale adoption of optoelectronic integration technologies.

To address these challenges, there is an urgent need to significantly miniaturize optical devices, transitioning from microscale to nanoscale optoelectronic integration. This advancement is essential for achieving size and performance compatibility with integrated circuits. A critical requirement is the development of electro-optic modulators that are compact, highly efficient, and low in power consumption. Consequently, the search for innovative materials that can drastically reduce the size of electro-optic modulators while substantially improving key performance parameters, such as modulation bandwidth, modulation speed, half-wave voltage, and overall modulator size, has become a primary focus for advancing next-generation optoelectronic device integration.

### 34.3 Suggested Directions to Meet These Goals

When light strikes a specific material surface, photons excite charge carriers within the material, resulting in collective oscillations. These oscillations give rise to special electromagnetic modes known as polaritons. As unique quasiparticles, polaritons exhibit both the wave-like characteristics of light and the particle-like properties of matter. Depending on the coupling between the incident light and various charge carriers in the material, a diverse array of polaritons can be formed, including plasmon polaritons, phonon polaritons, and exciton polaritons.[63]

In contrast to traditional 3D materials like gold and silver, 2D materials possess unique characteristics stemming from their atomic-scale thickness, distinct band structures, and vdW hetero-stacking. These features enable the formation of polaritons with enhanced light field confinement, a broad bandwidth response, long propagation distances, and the potential for electrical dynamic control.[718] These diverse polaritonic modes can break the diffraction limit, compressing the wavelength of light into the nanoscale, thus facilitating new breakthroughs in the design of optical functionalities at the nanoscale.

Compared to electrons and photons, polaritons offer four distinct advantages as carriers for on-chip information processing and the development of novel electro-optic modulators (Figure 34): (1) Polariton modes in 2D materials can be more tightly localized at the material surface, even achieving sub-diffraction-limit light field confinement down to atomic scales, which is conducive to optoelectronic integration at the nanoscale; (2) They allow for multidimensional control over amplitude, phase, polarization, and angular momentum, significantly enhancing bandwidth for optoelectronic interconnection and information processing through multidimensional multiplexing; (3) The *semi-light/semi-matter* characteristics of polaritons can overcome the limitations of electrical control typically associated with photons as bosons; (4) The hybrid coupling and strong interactions between different types of polaritons provide the potential for efficient signal processing and the realization of photonic matrix computations.[741]

Recently, polaritons in 2D materials, represented by graphene plasmonics, have shown significant potential in constructing nanoscale high-performance modulators and photonic computing chips, leading to the development of a series of nano-polariton photonic devices.[9] In terms of optical-field mode confinement, methods utilizing acoustic graphene plasmons,[118] whispering-gallery phonon polaritons,[118] and gap-confined modes on gold substrates have demonstrated wavelength confinement capabilities exceeding hundreds of times.[56] For polariton transmission, mechanisms of losses due to phonon scattering, electron scattering, and defect scattering have been studied, confirming enhancements in the quality factor and propagation distance of polaritons through heterojunction structures, low-temperature environments,[5] isotope purification,[915] and suspended substrates.[38] Furthermore, directed and diffractionless propagation properties can be achieved through in-plane asymmetric polariton modes hybridization, such as with twisted α-$MoO_3$,[44,698] and $MoO_3$/graphene heterostructures.[916] In terms of modulation, efficient multidimensional control methods have been developed,[718] along with the fabrication of nanoscale electro-optic modulators.[917] These polaritonic electro-optic devices cannot only control the phase of mode transmission[918] but also achieve functionality for positive and negative refraction.[194] For detection, plasmon-assisted resonant techniques can achieve wide-band and high-sensitivity optoelectronic detection performance.[919,920,921]

However, current research primarily focuses on the materials supporting polaritons, polariton performance tuning, characterization methods, and discrete functional polaritonic devices. There is yet to be a consensus on establishing nanoscales for optoelectronic interconnections driven by polaritons as a main research direction. This lack of consensus arises from the need to overcome a series of physical, material, and process challenges to achieve this ambitious research goal. These challenges include addressing inherent losses in polaritonic modes; achieving good performance in a spectral ragion that covers the communication frequency bands; developing multiple essential photonic devices for on-chip applications, particularly various electro-optic modulators, and further advancing polaritonic devices with nonlinear modulation; overcoming challenges in heterogeneous integration across materials and scales; and achieving functionalities for polariton storage and polariton matrix computation.

In this context, future research should focus on interdisciplinary collaboration, integrated optics, electronics, materials science, and physics to tackle these complex foundational scientific problems. In response to the demand for large-scale integration, high speed, and low energy consumption in on-chip optoelectronic interconnections and optical computing, the development of polaritonic optoelectronic interconnection frameworks is imperative. The research goal can be divided into four stages: principle verification, device fabrication, architectural design, and chip computation. The first stage aims to clarify the fundamental physics of polaritons, develop materials and fabrication methods for polaritons, and establish characterization methods for their optoelectronic responses, with significant progress already achieved in this area in recent decades. The second stage focuses on fabricating fundamental polaritonic components, such as modulators (Mach-Zehnder (MZM), resonant ring (RRM), and in-phase quadrature (IQM)) and switching devices (high-frequency electro-optic switches, ultrafast optical switches, thermal-optic switches, and optical transistors), where notable advancements have already been made. The third stage aims to realize polaritonic photonic circuits and optoelectronic interconnection modules, with each circuit containing polaritonic light sources, waveguides, modulators, switches, transistors, and detectors, ensuring efficiency in device cascading. Currently, much of this work remains at the theoretical and simulation stages, with few experimental results. The fourth stage seeks to achieve the packaging of polaritonic photonic chips and the validation of optical matrix computations. In terms of processes, the development roadmap of electronic chip technology based on 2D materials, as exemplified by “The Roadmap of 2D Materials and Devices Toward Chips”,[922] can be referenced to facilitate collaboration between academia and industry.

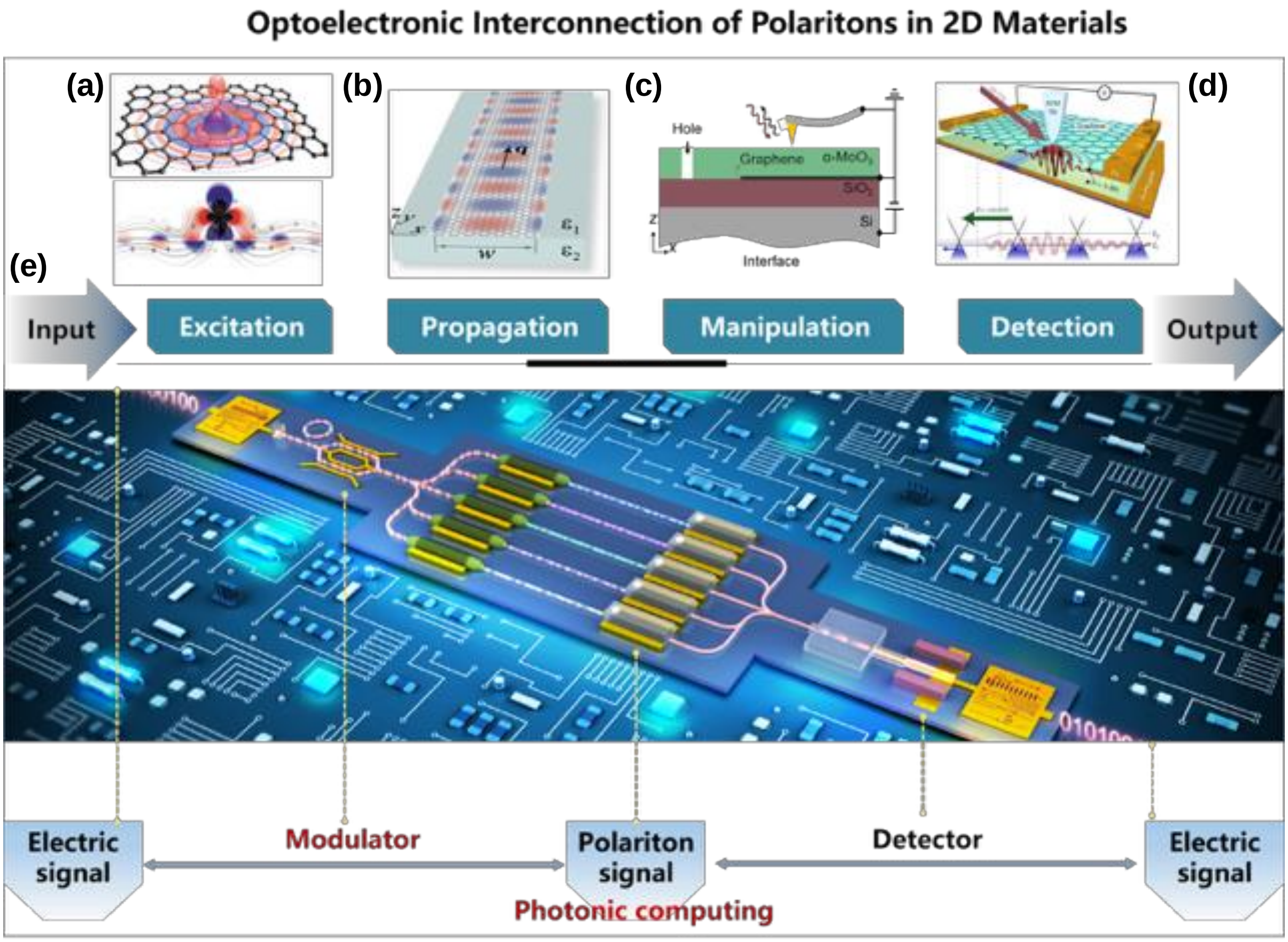

**Figure 34.** Optoelectronic interconnection of polaritons in 2D materials. (a) Interaction between a dipole emitter and plasmons in doped homogeneous graphene (upper panel) and near-field generated by a perpendicular dipole positioned 10 nm from doped graphene (lower panel). Reprinted from ref 9 (Copyright 2011 American Chemical Society). (b) Edge and waveguide THz surface plasmon modes in graphene microribbons. Reprinted from ref 923 (Copyright 2011 American Physical Society). (c) Gate-tunable topological polaritons in vdW heterostructures with high spatial confinement. Reprinted with permission from ref 194. (Copyright 2023 AAAS). (d) Thermoelectric detection and imaging of propagating graphene plasmons. Continuous-wave laser light is scattered by a movable metalized AFM tip, which excites plasmons in the hBN-graphene-hBN heterostructure. Reprinted with permission from ref 924 (Copyright 2016-Springer Nature). (e) Schematic illustration of the optoelectronic interconnection of polaritons in 2D materials featuring several key components arranged from left to right: the module for converting electrical signals into polariton signals, the polariton waveguide for signal transmission, the polariton signal modulation module, the polariton photonic computing module, and the polariton detection module. Each modulator is individually controlled by an electronic chip to encode the optical signals. Additionally, a silicon waveguide connects the modulators to the photodetector and can be extended and adapted for fiber optics, enabling long-distance signal transmission.

## 35. EFFECTIVE RESPONSE OF QUANTUM MATERIALS DRIVEN BY OPTICAL NEAR FIELDS

**Aaron Sternbach,[1,2,*] You Zhou,[2,3] and Mohammad Hafezi[4]**
[1]Department of Physics, University of Maryland, College Park, MD, USA
[2]Maryland Quantum Materials Center, University of Maryland, College Park, MD, USA
[3]Department of Materials Science and Engineering, University of Maryland, College Park, MD 20742, USA
[4]Joint Quantum Institute (JQI), University of Maryland, College Park, MD 20742, USA
***Corresponding author.** Email: ajs01123@umd.edu

The unification of 2D materials, sub-diffraction-limited photonics, and light-induced states is creating opportunities for material design. Emerging possibilities are rooted in the exquisite properties of 2D materials. These compounds often possess exceptionally clean interfaces and sharp lateral boundaries, and they are amenable to experimental control. Moreover, 2D materials can host polaritons—hybrid collective modes that possess characteristics of both light and matter.[112,671] Polaritons often display nonintuitive properties, including the confinement of radiation to sub-diffraction-limited length scales and diffractionless propagation, that defy conventional limits of far-field radiation. Polaritons in 2D materials have garnered attention for their potential to produce cornerstones of photonics in next-generation technologies including miniaturized devices, optical modulators, and superlenses. Numerous proposals have emerged suggesting novel paradigms to tailor the properties of quantum materials with structured radiation. In this section, we discuss proposed prospects for polariton modes to deliver unique properties, experimental results, and suggestions to advance this emerging line of research. We discuss three topics of importance for controlling quantum materials with near fields: (1) Engineering near fields, or 'nano-light' (Figure 35a); (2) Using strong near fields to drive quantum materials (Figure 35b); (3) Passively mediating quantum states with near fields in cavities (Figure 35c).

### 35.1 Engineering Nano-Light with 2D Materials

Polaritons are produced when dipole-active resonances, including the collective motion of lattice vibrations, electronic charges, or spins, hybridize with IR radiation.[671] Extensive research has demonstrated exceptional control over the near-field radiation patterns produced by polaritons. In a single sheet of 2D graphene, plasmon interference can produce polaritonic standing waves with sub-diffraction-limited confinement. Twisted bilayers can naturally form nanoscale photonic crystal structures that derive from their moiré superlattices (Figure 35a).[43,171] Polaritons can also propagate along ray-like trajectories with ultraslow group velocity within highly anisotropic layered compounds known as hyperbolic materials.[169] When 2D materials are stacked, hybrid polaritons with properties that are distinct from the constituent layers can form in hetero- or homo-bicrystals.[44,45,194] Hybrid polaritons often display avoided crossings where the frequency–momentum dispersion of the unhybridized crystals intersect and form large strong coupling gaps.[45,112] The directional propagation of hyperbolic polaritons can be tuned in bicrystals by changing the relative twist angle of the two stacked hyperbolic crystals. Polaritonic radiation can also be focused and steered in circulating loops or spirals *via* negative or all-angle refraction.[45,53,194] Moreover, polariton trajectories are amenable to *in situ* control by a variety of experimental stimuli including electrostatic gating, magnetic fields,[183] and photoexcitation.[112] Thus, there are many ways to confine, steer, and structure programmable polaritons below the diffraction limit in 2D materials.

Within the last decade many ways and means to engineer nano-light have been established. Yet, several opportunities remain just beyond the horizon of front-leading experiments. For instance, the polaritons that have been visualized to date, including plasmons, phonon polaritons, and exciton polaritons, are all governed by the dispersion of the permittivity in the material. Systems where the permeability governs the electrodynamics of polaritons could lead to magnetic near fields,[925] which remain largely unexplored. Investigations of polaritons in superconductors, correlated materials, and topological systems[112] are in their infancy. These systems are candidates to host polaritons with desirable properties for controlling quantum materials, including longer propagation lengths and stronger modulation of nanoscale periodic potentials created by polaritonic standing waves. Moreover, properties could be generally useful for subwavelength optical control including beam routing and nonlinear optics below the diffraction limit for next-generation nanophotonics. A first goal is the development of advanced near-field probes that can be used to study polaritons and expand the toolset to control nano-light.

Next, we address suggestions to characterize polaritons that could enable novel functionalities. The THz frequency range is appropriate to examine low-energy collective excitations in 2D materials.[926] Terahertz dipole-active resonances, including magnetic resonances, orbital transitions, and superconducting plasma resonances, could produce polaritons with exceptional properties. Yet, THz near-field spectroscopy remains challenging. Further advancements are required to reach the highest possible spatial resolution and perform quantitatively accurate near-field measurements across the entire THz frequency range. Even higher spatial resolution could be attained with electron microscopies. Emerging electron-assisted optical methods including photon-induced near-field electron microscopy (PINEM),[927] THz-STM,[220] and NOTE[191] spectroscopies are promising. The determination of the local optical response function at the smallest length scales will require further developments.

## 35.2 Nano-Light Can Be Used To Control Quantum Materials

Experimental results demonstrating that polaritons can be used to control quantum materials are emerging. Two distinct regimes, classified as driven and passive, can be identified. In driven systems, large numbers of photons are used to perturb quantum materials. In passive systems, the interactions between quantum materials and vacuum radiation fields are considered.

### 35.2.1 *Quantum Materials Driven By Optical Near Fields*

Strong pump lasers can be used to perturb charge, spin, lattice, and orbital degrees of freedom in quantum materials. While most experiments have been conducted in the far-field regime, optical near fields can also be used to control quantum materials. Enhanced electric fields from localized plasmons were used to induce an insulator-to-metal transition in thin films of vanadium dioxide.[928] Propagating polaritons were used to reverse the ferroelectric polarization of $LiNbO_3$ outside of the photoexcitation volume.[929] These experiments suggested a highly nonlinear process involving propagating photoexcitation, near super-luminal domain wall motion, and re-emission of perturbative radiation, all of which could play a role in the observed dynamics.

In a 2D $MoSe_2$ monolayer, subwavelength optical lattices have been employed to create a spatially varying Stark shift (Figure 35b).[939] In plasmonic structures engineered to exhibit hyperbolic dispersion, surface-plasmon polaritons (SPPs) can feature diffractionless propagation,[930] allowing for the excitation of SPP modes tightly confined within a spatial region as small as ~100 nm. When these modes were optically excited, the resulting field was shown to induce a spatially varying AC Stark shift[931] in excitons within a $MoSe_2$ monolayer, effectively creating a subwavelength periodic potential. Importantly, the SPP-driven AC Stark shift is significantly more power-efficient than conventional far-field approaches, making this method a promising platform for exploring optically induced phenomena at sub-diffraction-limited length scales.

Inspiring atomic and molecular optics experiments have successfully demonstrated effects including trapping, cooling, and creation of condensates in optical lattices. In condensed matter systems, spatially uniform drives have been proposed—and partially demonstrated—to yield phenomena including Bloch-Floquet states and nonthermal light-induced phase transitions.[932,933] With spatially structured driving radiation, one can conceive, for example, imprinting an optical lattice for itinerant excitons and electrons to engineer band structures[934] and light-induced fractional quantum Hall phases.[935] The bandwidth of the emergent effective bands is, however, directly related to the periodicity of the structured radiation. Requirements for tight optical lattice periods demand the creation of sub-diffraction-limited photonic superlattices to realize several proposed effects. Motivated by these proposals, the second goal of investigating the properties of quantum materials that are driven by structured radiation below the diffraction limit or polaritonic lattices is suggested.

To meet the goal of accessing states that are driven by structured radiation, the optical response function of the polaritonic host materials must be known at the relevant momenta. Thus, adequate characterization tools must be developed and utilized to determine the properties of materials hosting tightly confined polaritonic superlattices. Experimental observables must also be time-resolved and local, at least within the driving pulse and region of space where the transient potential is uniform. Evaluating spectral features that change within the probe pulse duration, which is common in Floquet engineering, could require a generalized understanding of reflectivity and transmission in certain cases.[936]

### 35.2.2 *Mediating Quantum States in Passive Near-Field Cavities*

The possibility of mediating quantum states of matter with cavity modes has gained considerable attention both theoretically and experimentally. Experiments addressing the impact of polaritons on condensed matter systems are emerging. Changes in the metal–insulator transition temperature of 1T-$TaS_2$ were documented within a Fabry–Pérot cavity.[937] Modifications of phase transitions have been witnessed in quantum materials proximate to plasmons, including enhancements of ferromagnetism.[73] Pioneering experiments have recently supported the notion that strong gradients of the electric near fields could enhance fractional quantum Hall conductivity in bulk GaAs (Figure 35c).[940]

Interest in passive cavities is largely driven by the desire to Floquet-engineer solids. The question of whether the effects that have been witnessed in cavities are unambiguously related to Floquet states generated by vacuum fluctuations remains open. Further work is required to establish general design protocols to produce cavity-mediated states and assess their origins. A third goal is to establish physical observables that can yield unambiguous insight into the physics of cavity-mediated states and rigorously test the agreement between theory and experiment.[78,938]

Even in the absence of a radiation source, passive cavities could change the properties of quantum systems. Unambiguous observables with explicit connections between theory and experiment are desired. The Purcell modification of emission in the polaritonic near field is one example, where the dynamics of optical excitations is not yet fully understood. The collective properties of charge carriers can also be modified by cavities. In a gaped system, the relevant coupling strength is quantified by a strong coupling gap relative to the system's decay rates. The description of strong coupling in a gapless sea of itinerant fermions is subject to subtleties. Recent proposals have suggested that enhancements of the effective mass,[938] or equivalently, changes of the Fermi velocity,[78] could reveal cavity-induced changes of Fermi liquids. Two-dimensional materials are likely to play a central role, as these can be placed precisely within engineered near-field environments.

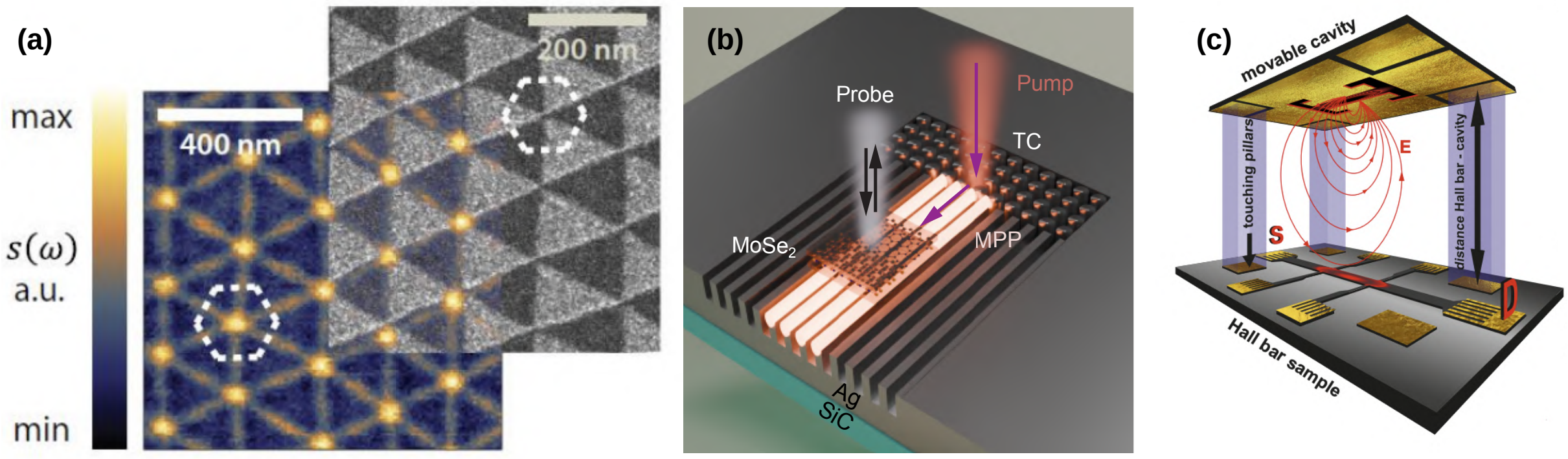

**Figure 35.** Controlling quantum materials with optical near fields. (a) Producing structured near-field radiation can be accomplished with polaritons, including sub-diffraciton-limited superlattices of IR radiation in moiré-patterned twisted bilayer graphene. Adapted from ref 171 (Copyright 2018 American Association for the Advancement of Science). (b) A 2D material driven by a periodic potential of plasmons with subwavelength periodicity. Adapted from ref 939 (arXiv). (c) Near fields interacting with a Hall bar of GaAs. Adapted from ref 940 (arXiv).

**Applications: Light Emission and Detection**

## 36. SINGLE-PHOTON EMITTERS

**Dmitrii Litvinov,[1,2] Magdalena Grzeszczyk,[2,3] Kostya S. Novoselov,[1,2] Maciej Koperski[1,2,*]**
[1]Department of Materials Science and Engineering, National University of Singapore, 117575, Singapore
[2]Institute for Functional Intelligent Materials, National University of Singapore, 117544, Singapore
[3]Center for Quantum Nanoscience, Institute for Basic Science (IBS), Seoul 03760, Republic of Korea
***Corresponding author.** Email: msemaci@nus.edu.sg

### 36.1 Introduction

Single-photon emitters (SPEs) in 2D materials originate from quantized energy levels formed by intrinsic imperfections within the crystal structure, as well as through engineering methods such as doping, defect creation, or local strain. These emitters can also be induced by potential traps that confine excitons at the nanoscale. For example, in hBN, point defects like nitrogen or boron vacancies, along with substitutional impurities such as carbon, disrupt lattice symmetry and create mid-gap states that facilitate precise optical transitions across the UV to near-IR spectrum.[256,941] Quantum dots also represent a promising mechanism for the generation of single photons. These nanoscale regions can occur naturally in 2D materials[942,943,944,945] or be engineered *via* fabrication techniques, effectively confining excitons.[946,947] Beyond intrinsic defects and strain-induced traps, functionalization has also been explored to modify optical properties in 2D materials. While single-photon emission has been observed in functionalized carbon-based nanomaterials such as graphene quantum dots and nanotubes,[948,949,950] its realization in functionalized extended 2D flakes, including TMDs and graphene, remains an ongoing challenge.[364] Additionally, 2D materials can be assembled into heterostructures, generating periodic potential variations that trap carriers leading to quantized excitonic states conducive to single-photon emission.[951,952,953]

To fully exploit the potential of SPEs in 2D materials, it is essential to create them in a deterministic and controllable manner. Such control allows for the tailoring of emission energy and enhances reproducibility and stability, which is crucial for practical applications. Moreover, achieving operational stability at ambient conditions is a key goal, as it facilitates the integration of SPEs into quantum photonic chips. These advancements include coupling to waveguides and resonators and enabling coherent control of qubits for applications in quantum communication and computation.

### 36.2 Current State of the Art

Techniques like strain engineering, nanopatterning, and moiré superlattices are used to control and enhance the emission properties of SPEs, allowing for their on-demand creation. Mechanical strain has proven effective for modulation of the local bandgaps and exciton confinement in 2D materials. This approach has been successfully demonstrated in monolayer TMDs such as $WSe_2$,[954] $WS_2$,[392] $MoSe_2$,[946,947] and $MoTe_2$,[24] where improved light–matter interactions and stabilized photon emission were achieved. A fabrication process using strain engineering is illustrated in Figure 36a, where a PDMS-assisted transfer onto a prepatterned $Si/SiO_2$ substrate creates localized deformations visible in dark-field optical microscopy image in Figure 36b.

Another approach to deterministically create point defects involves site-selective electron or ion beam exposure of 2D materials.[390,955,956,957] Compared to strain engineering, the resolution of irradiation is limited by the resolution of the beam,[390] making it well suited to structure 2D materials at the nanometer scale. In irradiated hBN samples, additional thermal annealing in an oxygen atmosphere is often necessary. This method has enabled the deterministic creation of emitters with a precision better than 50 nm and a creation yield exceeding 35%.[957]

Stacking 2D materials with slightly mismatched lattice constants creates moiré superlattices—periodic potential variations that trap excitons, leading to enhanced single-photon emission. These superlattices enable the realization of ordered arrays of quantum emitters, improving both the uniformity and tunability of SPEs. For example, spectrally tunable quantum emission from interlayer excitons trapped in moiré superlattices has been proposed for TMD heterobilayers[9] and realized in heterostructures such as $WSe_2/MoSe_2$.[958]

Integrating SPEs with photonic structures such as optical cavities and waveguides enhances photon emission rates through the Purcell effect. This embedding improves emission efficiency and enables effective photon routing, which supports the development of scalable quantum networks and precise, on-demand photon delivery. Recent advancements in this area include $WSe_2$ SPEs coupled to high-refractive index GaP nanoantennas demonstrating enhancement of the radiative rates and the reduction of the nonradiative decay rates with $10^2$ - $10^4$ enhancement of the PL intensity.[403] Another example shows the coupling of SPEs in $WSe_2$ with plasmonic cavities, resulting in a significant increase in quantum yield from 1% to 65%, as well as a substantial reduction in photon lifetime, with a Purcell factor of up to 551.[402] The schematic of a representative plasmonic Au nanocube cavity array is depicted in Figure 36d. Figure 36e presents the PL intensity versus excitation power for both uncoupled and coupled $WSe_2$ emitters, highlighting the enhanced emission when integrated with the cavity structures. Additionally, a photonic waveguide can simultaneously induce strain-localized SPEs and couple them into a waveguide mode. This was demonstrated for the $WSe_2$ monolayer, where autocorrelation measurements were obtained from the top and through the waveguide output,[411] schematically depicted in Figure 36c. Progress has also been made in integrating hBN emitters with photonic structures. For instance, tunable cavities on PDMS substrates have achieved a Purcell factor of 4.07, a reduction in linewidth from 5.76 nm to 0.224 nm, and a twofold improvement in $g^{(2)}(0)$ values.[959] In another study a monolithic cavity with a quality factor of 5,000 demonstrated a Purcell factor of 15, resulting in a 10-fold increase in emission intensity and a shortened emitter lifetime.[960] The coupling of hBN SPEs to waveguides has also seen notable progress. Monolithic integration of red emitters, confirmed by finite-difference time-domain (FDTD) simulations, has shown higher efficiency than hybrid configurations. Experimental data has demonstrated successful coupling with an efficiency of 0.032 and the observation of photon antibunching.[961] Further studies have shown the coupling of red emitters to dielectric SiN waveguides[962] and the top-down integration of blue emitters into photonic platforms,[963] marking significant steps toward scalable quantum photonic systems.

Furthermore, the spin states inherent in 2D materials like TMDs and hBN can be utilized for coherent control of qubits. Techniques like optically detected magnetic resonance (ODMR) allow for the manipulation and readout of spin states *via* photon emission. Hexagonal boron nitride has emerged as a standout platform with room-temperature ODMR arising from a magnetic dipole transition between the spin sublevels of the ground state.[964] The zero-field splitting is around 3.47 GHz, which is typical for spin-1 systems related to boron vacancy ($V_B^-$). This is demonstrated in Figure 36f and the evolution of the ODMR spectrum with a magnetic field applied parallel to the sample is shown in Figure 36g. The room-temperature spin coherence time ($T_2$) was measured to be 18 μs in this material.[965] In TMD monolayers, strong spin-valley coupling arises from the lack of inversion symmetry and significant spin–orbit interactions, leading to spin splitting in the conduction and valence bands in the K and K' valleys. For example, in $WSe_2$, external magnetic fields have been employed to manipulate both spin and valley states, enabling coherent control that can be observed through Shubnikov–de Haas oscillations.[966] Quantum coherence for single moiré excitons within a twisted $WSe_2/MoSe_2$ heterobilayer has revealed additional beating patterns that visualize the presence of quantum coupling between them.[967] These findings showcase the potential of valley locking and advanced quantum systems in 2D materials, emphasizing their capacity to manipulate multiple quantum states.

### 36.3 Challenges, Future Goals, and Directions to Meet Goals

Despite substantial progress in the creation of SPEs, significant challenges still remain. A major obstacle for the implementation of such SPE in real-life applications is achieving controlled and reproducible emission at specific energy levels.[968] While strain-engineering techniques have shown promise in the deterministic placement of emitters, scaling these methods continues to be a challenge. Ion or electron beam irradiation offers higher precision for defect placement across larger areas, but maintaining consistency in emitter behavior remains an issue. Additionally, controlling the mesoscopic environment to minimize variability between emitters is essential.[969] Identifying and eliminating decoherence channels, along with developing high-fidelity, scalable pumping schemes—such as electrically driven optical emission and selective excitation methods that allow for greater control over emission properties—are important for optimizing emitter performance, improving single-photon purity, and achieving practical applications. Accurate characterization of emitters is also vital for understanding their electronic and optical properties. Techniques such as scanning tunneling microscopy (STM)[970] and atomic force microscopy (AFM), integrated with optical methods, have been effectively employed to gain deeper insights into these systems.

For example, AFM-based nanoindentation has been applied to locally deform $WSe_2$ monolayers, creating SPEs with a high degree of purity, evidenced by values of the second-order correlation $g^{(2)}(0)$ values as low as 0.02. The combination of AFM topography and photoluminescence mapping revealed that the SPEs were not located at the center of the nanoindents but instead formed around the edges, indicating a complex interaction between strain-induced potential wells and exciton localization.[401]

Another notable example involves apertureless scanning near-field optical microscopy (SNOM), which has been used to resolve nanoscale exciton localization in $WSe_2$ monolayers. This technique revealed that localized excitons formed doughnut-shaped distributions around nanobubble edges, with localization lengths as small as 10 nm.[206] The AFM topography and corresponding integrated PL intensity map of these $WSe_2$ nanobubbles are shown in Figure 36h,i.

Additionally, tip-enhanced photoluminescence (TEPL) has been proven instrumental in achieving high-resolution optical imaging of quantum emitters. A study utilizing adaptive TEPL achieved a $4.4 \times 10^4$-fold enhancement in photoluminescence in $WSe_2$ monolayers through feedback-based optimization of the spatial phase mask, leading to a spatial resolution of approximately 14 nm.[238]

An electrically driven photon emission from defects in monolayer $WS_2$, employing a plasmonic scanning tunneling tip to induce optical transitions *via* sequential inelastic tunneling was also demonstrated. The system's schematic is presented in Figure 36j. This technique allows for tuning emission by adjusting the applied bias voltage and mapping photon emission from individual defects as illustrated by the photon map of a sulfur vacancy in Figure 36k.[180] This method, known as STM luminescence (STML), has been applied in studies of metallic surfaces and molecular systems, surpassing the diffraction limit by two orders of magnitude.[236,971] However, the universal application of such measurements is limited by the stability and spatial resolution, which are typically sufficient only at cryogenic temperatures. Moreover, the requirement of conductive substrates provides additional limitations, as hybridization effects may arise between commonly used plasmonic noble metal substrates and semiconducting samples modifying and quenching the intrinsic optical emission.[972] Epitaxial graphene grown on SiC has been shown to be an effective substrate preserving the native TMD band structure.[973] Recent work demonstrated that inserting a few-layer graphene flake between the TMD monolayer and the metallic substrate significantly weakened interfacial coupling. This setup successfully preserved STML signals at negative sample voltages, enabling the observation of excitonic features from various local structures within a $MoSe_2$ monolayer.[215] A similar approach was employed for electrically driven photon emission from defects in monolayer $WS_2$, where the samples were prepared by growing TMD islands on epitaxial graphene on SiC using chemical vapor deposition.[180]

### 36.4 Outlook and Future Directions

A key focus moving forward will be the on-chip integration of SPEs into photonic circuits, such as waveguides, resonators, and cavities. The Purcell effect, which enhances photon emission rates by coupling emitters to optical cavities, will support the integration of SPEs into scalable quantum photonic networks. Additionally, developing electrically driven SPEs[974,975] compatible with CMOS technology is important for creating practical quantum applications.

While many current systems rely on cryogenic cooling to maintain quantum coherence, recent success in achieving room-temperature operation of SPEs in materials such as hBN and TMDs is encouraging. Future research should focus on engineering defects that remain stable under ambient conditions, reducing the need for complex cooling.

Addressing decoherence, which limits the performance of quantum systems, is also important. Increasing the coherence time of excitons or spin states by minimizing interactions with environmental factors is key for quantum information processing. Approaches such as stacking 2D heterostructures, applying encapsulation, and material engineering could help to reduce decoherence channels.

Nanoscale imaging and spectroscopy techniques, including STM, AFM, SNOM, and TEPL, will continue to provide valuable insights into quantum states at the atomic level. Enhancing surface quality—through cleaner fabrication techniques, encapsulation, and ultrahigh vacuum environments—can improve the precision of optical and electrical measurements. Integrating high-efficiency detectors, such as superconducting nanowire single-photon detectors (SNSPDs), into STM/AFM setups will also aid simultaneous electrical and optical measurements. These detectors, known for their low dark count rates, high timing resolution, and low noise, can improve photon detection from low-light emitters and boost the signal-to-noise ratio in correlated quantum readout. This is particularly beneficial for detecting weak photon emissions from individual defects or quantum dots.

In addition, advances in spin-resolved techniques, such as Electron Spin Resonance (ESR)-STM, will help map and manipulate spin states within quantum emitters. 2D materials are also emerging as promising platforms for generating entangled photon pairs. The use of interlayer excitons in heterostructures offers a promising mechanism for photon entanglement, which is important for secure quantum cryptography and advancing quantum networks.

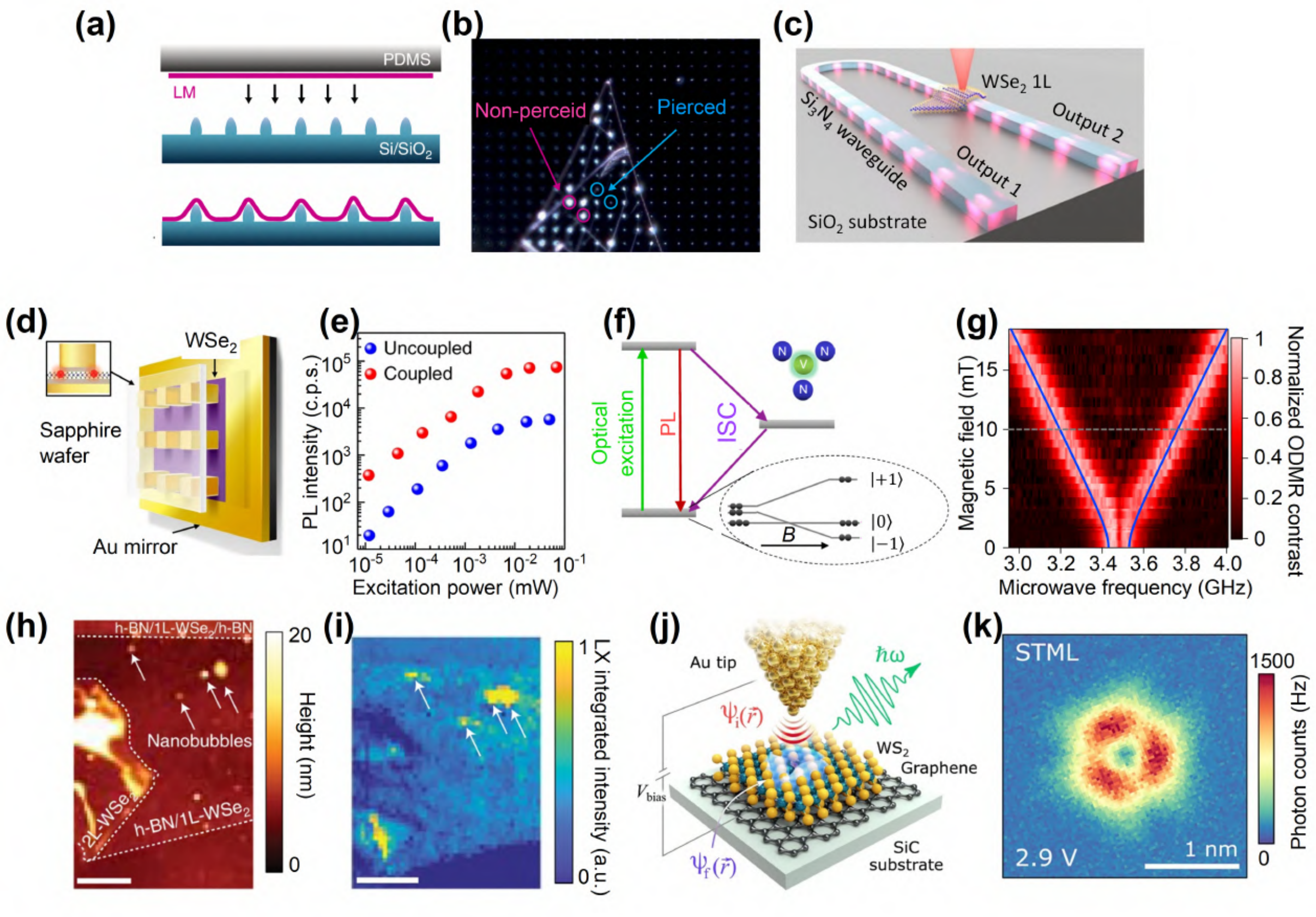

**Figure 36.** Single-photon emitters in 2D materials. (a) Schematic illustrating strain engineering *via* PDMS-assisted transfer of a monolayer TMD onto a prepatterned $Si/SiO_2$ substrate, creating localized deformations that form quantum emitters. (b) Dark-field optical microscopy image showing pierced and non-pierced defects in a 2D material lattice. The image covers an area of 170 µm by 210 µm. (c) Illustration of a monolayer $WSe_2$ integrated with a $Si_3N_4$ waveguide for photon routing, featuring two output channels. This configuration allows excitation either through the waveguide or a standard confocal setup. (d) Cross-sectional schematic of a $WSe_2$ monolayer on a gold mirror and sapphire substrate, enhancing photon emission. (e) Plot of PL intensity versus excitation power for uncoupled and coupled $WSe_2$ emitters, demonstrating enhanced emission when coupled to optical structures. (f) Simplified energy-level diagram for the negatively charged boron vacancy in hBN, illustrating ground, excited, and metastable states. (g) Evolution of the ODMR signal with an external magnetic field applied parallel to the hexagonal c axis (B ∥ c) of hBN. (h) AFM topography of hBN/$WSe_2$ nanobubbles, showing height variations. (i,h) Integrated PL intensity map of $WSe_2$ nanobubbles (i) over the 1.5–1.6 eV energy range corresponding to localized excitons (LX), correlated with the topography in panel (h). The scale bar indicates 500 nm length. (j) Schematic of scanning tunneling microscopy (STM)-induced photon emission from a monolayer $WS_2$ on graphene and a SiC substrate, where inelastic tunneling induces light emission. (k) Spectrally integrated photon map of a sulfur vacancy in $WS_2$, acquired with a constant current of I = 20 nA and a bias voltage of V = 2.9 V. Corresponding photon counts are displayed. Panels (a,b) are adapted from ref 392 (Copyright 2017 Springer Nature); panel (c) is adapted from ref 411 (Copyright 2021 American Chemical Society); panels (d,e) are adapted from ref 402 (Copyright 2018 Springer Nature); panels (f,g) are adapted from ref 965 (Copyright 2020 Springer Nature); panels (h,i) are adapted from ref 206 (Copyright 2020 Springer Nature); panels (j,k) are adapted from ref 180 (Copyright 2020 American Association for the Advancement of Science).

## 37. BAND-STRUCTURE ENGINEERING FOR LIGHT-EMITTING ELECTRON TUNNELING DEVICES

**Sotirios Papadopoulos[1] and Lukas Novotny[1,*]**
[1]Photonics Laboratory, ETH Zurich, Zurich 8093, Switzerland
***Corresponding author.** Email: lukas.novotny@ethz.ch

### 37.1 Introduction

The possibility of stacking 2D materials into heterostructures with twist-angle control provides a unique capability for band-structure engineering. In particular, band-structure engineering can be exploited to control electron tunneling between individual layers and the coupling between electrons and material excitations, such as phonons and excitons, by taking advantage of the conservation of both energy and (in-plane) momentum.

Here, we briefly present the progress in the understanding of tunnel junctions made of 2D heterostructures and discuss their use in light-emitting devices, electronic circuitry, and optical spectroscopy.

### 37.2 Current State of the Art

One of the first realizations of vertical transport light-emitting diodes with 2D materials, showcasing band-structure engineering, came in 2015 by Withers *et al.*[976] In this study, a $WSe_2$ monolayer was sandwiched between two thin hBN layers, acting as tunnel barriers, and two outer graphene layers that acted as transparent electrodes. Such a quantum well (QW) configuration is illustrated in Figure 37a. If the applied bias potential is higher than the electronic bandgap of $WSe_2$, electrons and holes can tunnel to the conduction and valence bands of $WSe_2$, respectively. The incorporation of tunnel barriers increases the probability of the injected carriers forming excitons that decay radiatively. This device structure achieves an external quantum efficiency (EQE) of 1% and can be improved by multiple QWs in series. Designing efficient on-chip optical sources requires an in-depth understanding of electronic transitions and optical modes in such structures.

In a follow-up study on single-QW devices,[977] an important observation was made. Excitons were created in the TMD with voltages lower than the electronic bandgap with an onset voltage that matched the optical bandgap. This observation suggested mechanisms beyond charge injection that contribute to the creation of excitons in the monolayer TMD. Similar studies, but with heterobilayer TMDs in place of monolayer TMDs, observed interlayer-exciton electroluminescence.[977,332] Furthermore, in the work of Binder *et al.*,[977] upconverted electroluminescence from intralayer excitons was reported. This observation was interpreted as Auger scattering of interlayer excitons and subsequent formation of intralayer ones.

Subsequent studies introduced alternative interpretations, like resonance energy transfer.[978,979] Pommier *et al.*,[979] by studying STM-induced electroluminescence from TMDs, confirmed the creation of excitons and their radiative recombination with an onset voltage close to the optical bandgap of the TMD and they proposed resonance energy transfer as the mechanism responsible for sub-bandgap electroluminescence.

Papadopoulos *et al.*[980] demonstrated electrically excited light emission from graphene–hBN–graphene tunnel junctions with a TMD placed outside of the tunnel barrier. In this configuration, the TMD is not part of the electronic pathway and is only optically coupled to the tunnel junction. By separating the TMD with a hBN spacer of variable thickness from the tunnel junction, one could study the emission intensity as a function of separation and provide evidence for a Foerster-type energy transfer mechanism. Electroluminescence in this case is a result of inelastic electron tunneling assisted by exciton creation. In fact, the coupling to excitons appears to be 10,000 times stronger than the coupling to photons, making energy transfer an attractive candidate for electrical excitation of luminescence.

Moreover, it was shown that the energy transfer mechanism is also responsible for electroluminescence upconversion. As reported in the work of Shan *et al.* [981] on TMD coupled tunnel junctions, emission at the exciton energy has been observed for bias voltages as low as half the optical bandgap, a process that involves excitation *via* two electrons. Follow-up studies have provided further evidence for upconversion *via* energy transfer in tunnel junctions.[982]

The logic of the studies discussed so far is based on energy conservation, but the role of momentum conservation is being ignored. While in localized tunnel junctions, such as in point contacts, momentum conservation is largely relaxed, this is not so for 2D interfaces with monoatomic flatness. In fact, energy and in-plane momentum conservation give rise to distinct resonances in tunnel junctions based on 2D materials and allow for the selective coupling to photons, phonons, plasmons, and excitons, as illustrated in Figure 37c.

Energy and momentum-conserving tunneling is referred to as resonant tunneling and is best exemplified for a pair of twisted and separated graphene layers. Twisting gives rise to Dirac cone separation in reciprocal space. For twist angles for which the Dirac cones overlap at the K points one observes a negative differential resistance peak in the current-to-voltage curves.[983] This peak indicates a resonance (energy and momentum conservation) in the elastic tunneling process. Resonant inelastic tunneling is also possible. Photon-assisted electron tunneling becomes resonant for twist angles near 0°, giving rise to a narrowing of the photon emission spectrum and an increase of the external quantum efficiency (EQE).[984]

For graphene–gold tunnel junctions there is a momentum mismatch of 17 $nm^{-1}$ between the Γ point in Au and the K point in graphene. Early STM studies suggested that this momentum gap is bridged through phonon-assisted tunneling processes.[985] When the graphene–gold tunnel junction is coupled to a TMD, as reported by Wang *et al.*,[986] one observes distinct conductance peaks associated with exciton-assisted electron tunneling, meaning that the momentum gap is bridged by coupling to indirect excitons.

The aforementioned studies demonstrate that energy and momentum engineering in tunnel junctions can be exploited for tuning the electrical and optical properties of 2D tunnel junctions, and hence for optoelectronic applications.

### 37.3 Future Goals, Challenges, and Suggested Directions

Tunneling between graphene electrodes has the benefit of atomically flat interfaces (little disorder) and partial transparency (outcoupling of optical radiation). Resonances in elastic and inelastic electron tunneling not only depend on the graphene–graphene twist angle but also on the structure of the tunnel barrier and the outer environment of the tunnel junction. For light emitting tunnel junctions one aims at maximizing the photon-assisted inelastic tunneling process.

As shown in Figure 37b, by introducing a twist angle between the two graphene layers we can address interactions that involve at minimum the momenta of $\Delta\mathbf{K}_1$ and $\Delta\mathbf{K}_2$. The bias voltage applied to the tunnel junction provides an energy offset $eV_b$. Figure 37c illustrates the corresponding energy–momentum landscape. The two Dirac cones are projected onto planes at $z = 0$ and $z = d$. Transitions between the two cones have to account for the energy offset $eV_b$ and the momentum difference $\Delta\mathbf{K}$. The energy and momentum difference can be provided by excitations (modes) such as phonon, photons, plasmons, or excitons. For example, TMDs support several momentum-direct and indirect excitons and by coupling TMDs to tunnel junctions one can selectively excite these excitons.

For light-emitting devices, one can design the tunnel junction so as to excite efficiently photons or direct excitons ($\theta = 0°$, $\Delta\mathbf{K} = 0$). However, for voltages in the range of exciton energies (> 1.6 V) the background current is rising, thereby making the inelastic tunneling process less probable and hence decreasing the EQE. One possibility to suppress the background current at these energies is to use sufficiently high twist angles, such that momentum-direct transitions are less likely. Another possibility is to excite electron–hole pairs with finite momenta inside the conduction and valence band at the K point (Figure 37d) that are allowed to relax (by coupling to phonons) to ultimately form direct excitons. Thus, the momentum gap is bridged by low-energy momentum non-conserving interactions.

To establish momentum conservation the twist angle between graphene electrodes has to be precisely controlled and angle disorder throughout the heterostructure lowers the reproducibility during fabrication and affects the quality of a device. The quantum twisting microscope can alleviate such problems during the design stage, as demonstrated by the Ilani group.[987] With the addition of optical access for allowing electroluminescence spectroscopy the entire angular spectrum can be addressed, thereby accelerating the exploration process of angle-tuned light emission. We note that not only the twist angle between graphene electrodes matters but also the crystallographic orientation of the TMD coupled to the tunnel junction. This is an area of ongoing exploration and it highlights the richness of control parameters involved in the design of simple single-QW structures (Figure 37a).

As pointed out above, the quantum efficiency of a light-emitting tunnel junction depends on the ratio of inelastic to elastic electron tunneling.[988] Elastic tunneling is responsible for a background current that is not coupled to any electromagnetic modes of the system. One way to suppress this background current is to use electrodes with no electronic states of low energy into which electrons can tunnel. Therefore, semiconducting materials like TMDs can be of interest as electrical electrodes. In fact, tunnel junctions with TMD electrodes have been shown to depend on the twist angle,[989] which provides control through momentum engineering as in the case of graphene electrodes. A major challenge, however, is the efficient injection of electrons and holes into TMD electrodes since Schottky barriers are formed at the TMD-metal contact regions. A possible solution is to control charge injection by gating.[92] Gate voltages not only control charge injection but also the Fermi level of a TMD which can strongly enhance or suppress radiative recombination of excitons.[366] All these parameters have to be fine-adjusted to achieve efficient exciton emission.

In conclusion, band-structure engineering of tunnel junctions offers numerous possibilities for device innovation through layer combinations, twist angles, and bandgap selection of 2D semiconductors. However, realizing these designs requires extensive experimentation to address feasibility and to uncover new physical phenomena within these heterostructures.

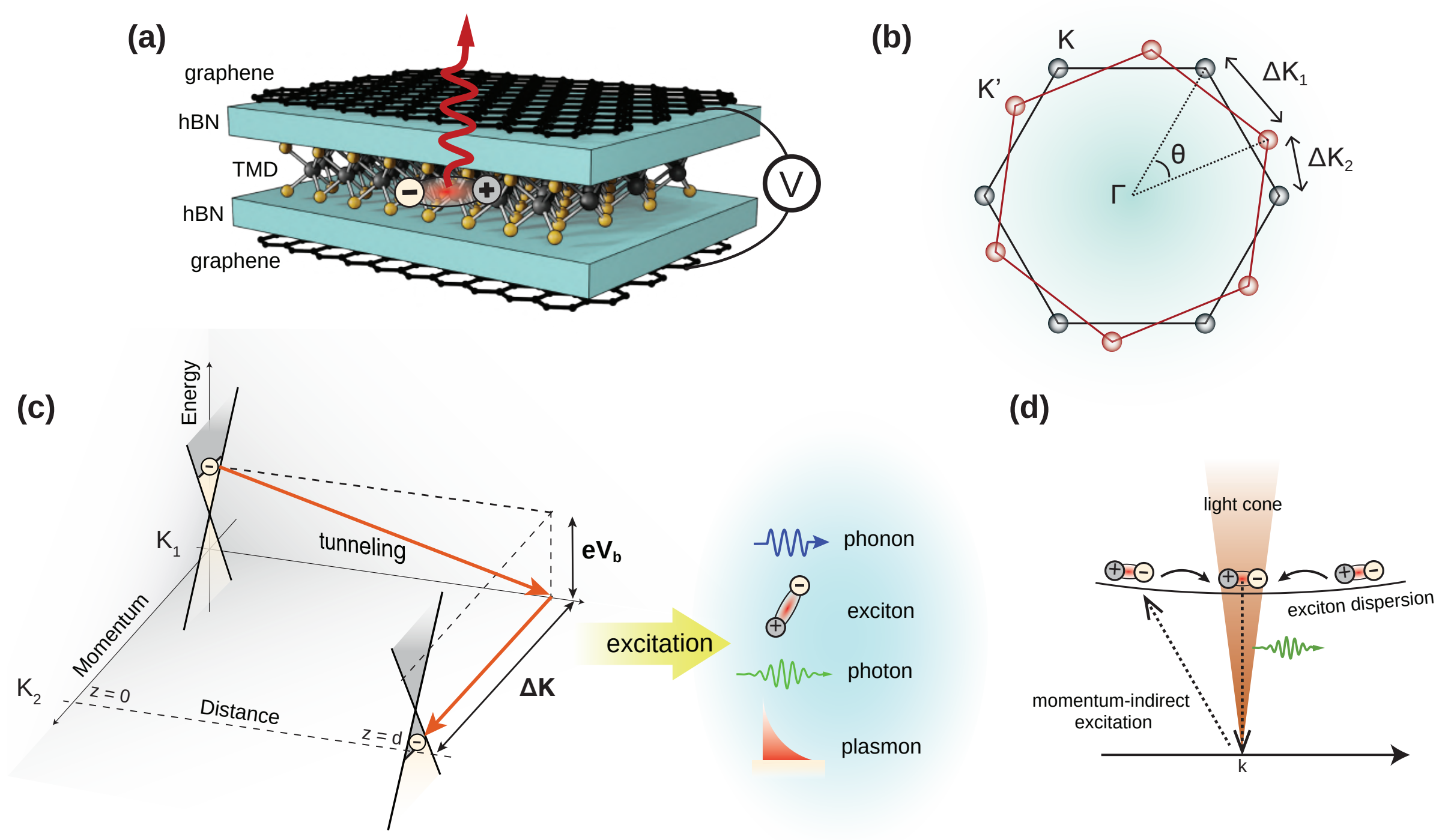

**Figure 37**. Inelastic electron tunneling between graphene monolayers. (a) Schematic of an all 2D quantum-well type light-emitting device. Charge carriers are injected in the device through the two graphene electrodes. Light emission is facilitated by exciton generation and their subsequent radiative decay. (b) First Brillouin zones (1BZ) of two graphene layers overlapping with a twist angle θ. This leads to a momentum mismatch between the two layers of $\Delta\boldsymbol{K}_1$ and $\Delta\boldsymbol{K}_2$ for neighboring K points. (c) A 3D expansion of a band-diagram to accommodate for both energy and momentum conservation, representing an inelastic tunneling process between two graphene electrodes. The three axes represent energy, momentum, and distance along which tunneling occurs. The initial and final states of the tunneling electron are separated by an energy $eV_b$ and momentum $\Delta\boldsymbol{K}$. The inelastic tunneling process occurs by exciting modes (e.g., phonons, photons, excitons, plasmons) that can fulfill energy and momentum conservation. (d) Momentum-indirect excitation of excitons can lead to direct exciton formation by phononic relaxation at the light cone.

# 38. PROSPECTS IN FAR-INFRARED PHOTONIC DEVICES AND UNCONVENTIONAL LIGHT SOURCES

**Leonardo Viti[1] and Miriam Serena Vitiello[1,*]**
[1]NEST, CNR-Istituto Nanoscienze and Scuola Normale Superiore, Piazza San Silvestro 12, Pisa 56127, Italy
***Corresponding author.** Email: miriam.vitiello@sns.it

## 38.1 Introduction

Dynamical phenomena in 2D materials and vdW heterostructures typically occur on a timescale of picoseconds, which corresponds to frequencies that fall in the THz range (i.e., in the far-IR). Consequently, in recent years, 2D materials, and particularly graphene, attracted wide interest as new material platforms in photonics for engineering THz-frequency devices and integrated systems, including photodetectors;[990,991,992,993,994] amplitude, phase, frequency, and polarization modulators;[995,996,997] plasmonic devices;[85] and ultrafast lasers.[891,998] These extraordinary advances are a consequence of some key properties of graphene: i) The extremely high electron mobility (300,000 $cm^2/Vs$), which means a very efficient electrical conductivity; this property is particularly important for applications in the THz bandwidth, where fast response times are often required. ii) Absorption of electromagnetic radiation over a wide range of frequencies, including the 1–10 THz window, which allows graphene to be used for devising broadband sensors, modulators, and detectors for THz waves. iii) The optical properties of graphene can be tuned using external stimuli such as electrical bias or chemical doping, which can be useful for tunable THz frequency devices. Graphene also shows ultrafast electron cooling dynamics with quasi-instantaneous (~100 fs) carrier–carrier thermalization and THz-rate (~ps) cooling timescales to the driving electric fields.[999] The combination of wide tunability and ultrafast dynamics allows for dynamic control over the response to THz radiation, as well as provide graphene with a strong, nonlinear, ultrafast, and adjustable response to THz frequency radiation. Furthermore, due to the recent availability of industrial production processes and the compatibility with a wide range of substrates, if placed on a chip with flat optical circuits, 2D material nanodevices can allow maximal interaction with light, prospecting a wealth of applications in transformational optics.[1000]

## 38.2 Current State of the Art

One of the most vibrant research fields in the last decade has certainly been 2D material-based light detectors, with the quest of nano-receivers that are highly sensitive (i.e., have a low noise-equivalent power, NEP) and simultaneously operate at room temperature (RT), exhibit a fast photoresponse, display a broad dynamic range (between the lowest and highest measurable incident light power), operate over a broad range of THz frequencies, are possibly suitable for upscaling to multi-pixel architectures, and can be eventually on-chip integrated with modulators.[1001] Photo-thermoelectric detection at RT has been demonstrated with high-quality hBN-encapsulated graphene field effect transistors (GFETs)[993,1002,1003] or scalable graphene-based architectures[994,1004] reaching NEP~100 $pWHz^{-1/2}$ and response times of a few nanoseconds, or even with layered black phosphorus[1005] and topological insulators.[1006]

Recently, graphene has been also innovatively exploited for photonic integration in semiconductor heterostructure lasers, either being integrated on-chip, as a micro-thermometer, for monitoring laser cooling,[1007] or as nonlinear elements for inducing phenomena that cannot be spontaneously sustained in quantum cascade lasers (QCLs).[1008] As an example, the back-facet integration of graphene in the Fabry–Pérot cavity of a QCL proved to be a key ingredient to induce intensity-dependent losses that promoted a stable frequency-modulated comb regime.[1009] Harmonic frequency combs of predefined order have been also demonstrated to be spontaneously generated in a QCL[1010] by locally patterning equidistant graphene islands on the top QCL waveguides.[1010] Furthermore, being graphene a very efficient THz saturable absorber,[1011] it has been embedded intracavity to promote self-starting pulse formation (4 ps long) in a THz QCL,[998] a long-dreamed goal since the initial demonstration of QCLs. Finally, inserting multilayer graphene in a random THz laser, mode-locking was realized.[1012] These achievements provided the first convincing experimental demonstration that, even if technologically challenging, graphene integration in well-known solid-state platforms can lead to a concrete technological revolution in photonics.

Graphene also exhibits a remarkably high third-order THz susceptibility $\chi^{(3)}\sim10^{-9}$ $m^2/V^2$,[532,1013] which is 7 orders of magnitude larger than that reported in a semiconductor heterostructure laser.[1009] Such a huge $\chi^{(3)}$ has been exploited to demonstrate, by frequency upconversion, high-harmonic generation (HHG), up to the 7th harmonic using moderate fields (~10 kV/cm) at frequencies $\leq$ 2.2 THz[1014,1015] even in grating–graphene metamaterials.[1016] Terahertz third-harmonic generation has also been reported in topological insulators (TIs)[1017,1018] and Dirac semimetals[1019,1020] and more recently, proved at high THz frequencies (9.63 THz) in QCL-pumped split-ring resonator arrays embedding single-layer graphene sheets.[1021]

### 38.3 Challenges and Future Goals

One of the current challenges in THz photonics is the engineering of THz photodetectors providing quantum-enhanced sensitivity and high saturation intensities (mW), which are of major interest in astronomy, quantum sensing, and quantum information. Traditional radiation detectors, such as superconducting hot-electron bolometers (HEB), superconducting transition-edge sensors,[1022] and kinetic inductance detectors[1023] operating at cryogenic (sub-Kelvin) temperatures, have been prevailing for decades. These technologies are now reaching the photon noise limit required in astronomy, with a noise equivalent power NEP~$10^{-20}$ W $Hz^{-1/2}$.[1024] THz photon counting[1025] has been also demonstrated, albeit in a narrow band around 1.5 THz, using quantum capacitance detectors.[1026] Despite these advances, suitable solid-state devices for detection of squeezed THz frequency light are still missing due to major technological challenges.

None of the state-of-the-art THz detectors developed so far are indeed fully adapted to quantum technology applications. Squeezing measurements across the THz require NEPs ~0.1 pW/$Hz^{1/2}$, electrical bandwidths above tens of MHz, and large dynamic ranges. Commercial HEBs have either a low electrical bandwidth or a modest dynamic range that limits the usable THz power to a few µW, which, considering the required ~0.1 pW/$Hz^{1/2}$ NEP, is not sufficient to detect sub-shot noise features. Also, HEBs show extremely low saturation powers (~10 pW). Interestingly, 2D materials may provide a valuable solution in this respect. In particular, photodetectors based on graphene show linear response for the expected power levels in the mW range,[993,1003] and excellent performances at room temperature (NEP ~100 pW/$Hz^{1/2}$, sub-ns response time, responsivities ~100 V/W).[993] Recently, photodetectors based on single-layer graphene (SLG) and bilayer graphene (BLG) have shown large dynamic ranges[992,993,1002] (up to four orders of magnitude) due to their high saturation currents and efficient electron cooling dynamics (electron-optical phonon scattering time ~1 ps)[1027]. Additionally, BLG, which is a gapless semiconductor, allows bandgap opening either through interaction with a substrate[1028] or by applying an out-of-plane electric field (Figure 38a) controlled by external double gates.[1029,1030] Exploiting this effect, BLG tunnel field-effect transistors (TFETs) have been demonstrated to operate as extremely sensitive receivers (NEP ~ 30-300 × $10^{-15}$ $WHz^{-1/2}$) at frequencies of 0.1 THz in single-top-gated[1031] or top p–n junction[1032] configurations, hence potentially representing a fitting platform for multiphoton quantum detectors in the THz.

Novel and brilliant opportunities can also raise from the integration of large area, wafer-scalable 2D and vdW crystals with artificial semiconductor heterostructures. Current advanced nanofabrication technologies, combined with new resonator concepts, have recently facilitated the confinement and handling of electrons and photons with an extraordinary degree of control. As an example, microcavities,[1033] either plasmonic[1034] or high-quality-factor optical ring resonators, able to manipulate and confine light in small volumes and at preselected frequencies, have expanded the functionalities of photonic devices. They have enabled a dramatic size reduction from laboratory scale to chip-level of optical filters, modulators, frequency converters, and frequency combs (FCs), mostly in the visible and the near-IR spectral domains. [1035,1036] The THz frequency range, despite being of utmost interest for chemical and biological sensing, communications, and imaging, does not yet fully benefit from integrated photonic solutions. Novel possibilities promise to open if 2D materials are embedded in semiconductor heterostructure resonators to engineer a fully novel generation of light sources. To provide a few examples, graphene can be a relevant building block for devising a miniaturized solid-state platform for the generation of nonclassical squeezed states of THz frequency light, or for devising innovative nonlinear electrically pumped light sources in frequency domains where solid-state narrowband light sources do not exist.

### 38.4 Suggested Directions to Meet These Goals

Combining SLG and BLG with hBN (Figure 38b) in a TFET architecture can promise a rectifying capability of the junction (Figure 38c), orders of magnitude larger than that achieved in a standard FET, which translates in significantly higher sensitivities. The concept is frequency scalable (from 0.1 THz to 10 THz) across the entire sub-THz and THz ranges. TFETs can hence offer concrete perspectives for detecting nonclassical states of THz light that could be for instance generated by miniaturized quantum sources such as THz QCL frequency combs. This can open exciting opportunities for developing THz quantum platforms all based on miniaturized solid-state devices, which can also be integrated on-chip.

Graphene can also potentially allow one to develop a miniaturized solid-state platform for the generation of nonclassical squeezed states of THz frequency light. High-order $n > 5$ harmonic frequency combs with quantum-correlated sideband modes and a custom-tailored output spectrum can be for instance sculptured by embedding on-chip integrated local graphene frequency filters (Figure 38d) that suppress non-harmonic modes, preserving defined high-order harmonic modes.

The recent demonstration of third-harmonic frequency upconversion in graphene at high THz frequency (9.63 THz)[1021] anticipates major advances in the field of innovative nonlinear light sources. As an example, novel integrated solutions for light generation in the *Reststrahlen* band of conventional III–V semiconductors (6-12 THz), where optical photon absorption hinders band-structure engineering for light emission, can be conceived. A spectrally narrowband technology to entirely access this range does not exist. Available technologies (thermal sources commonly adopted in Fourier-transform IR (FTIR) spectroscopy, spintronic emitters, or time-domain spectroscopy (TDS) systems, which rely on III–V photoconductive switches) suffer from poor performance or are cumbersome and lack stability (as, for example, difference-frequency generation in gas lasers (e.g., $NH_3$)). Distributed feedback Bragg grating cavities in double-metal QCLs are particularly suited for graphene intracavity integration, stemming from the possibility of lithographically engineering gratings of shape, pitches, and depth designed from scratch and tunable on purpose. Although technologically challenging, engineering graphene plasmons in a specific fashion in a ribbon grating superimposed to a distributed-feedback-laser pattern on a QCL resonator can provide the field enhancement needed for harmonic generation (HG) and can offer a valuable solution to engineer a monolithic, electrically pumped solid-state source in the 6.5–12.0 THz range. The patterned ribbons can allow optimizing the nonlinearity, since the HG efficiency depends on the carrier concentration, which can be locally tuned, for example, by gating. Importantly, unlike in other spectral domains, only moderate THz fields (< 40 kV/cm) are required to induce HG in graphene,[1014] comparable to those available from THz QCLs when electromagnetic radiation is confined, for instance, *via* plasmonic effects. This represents a long-dreamed frontier in THz photonics.

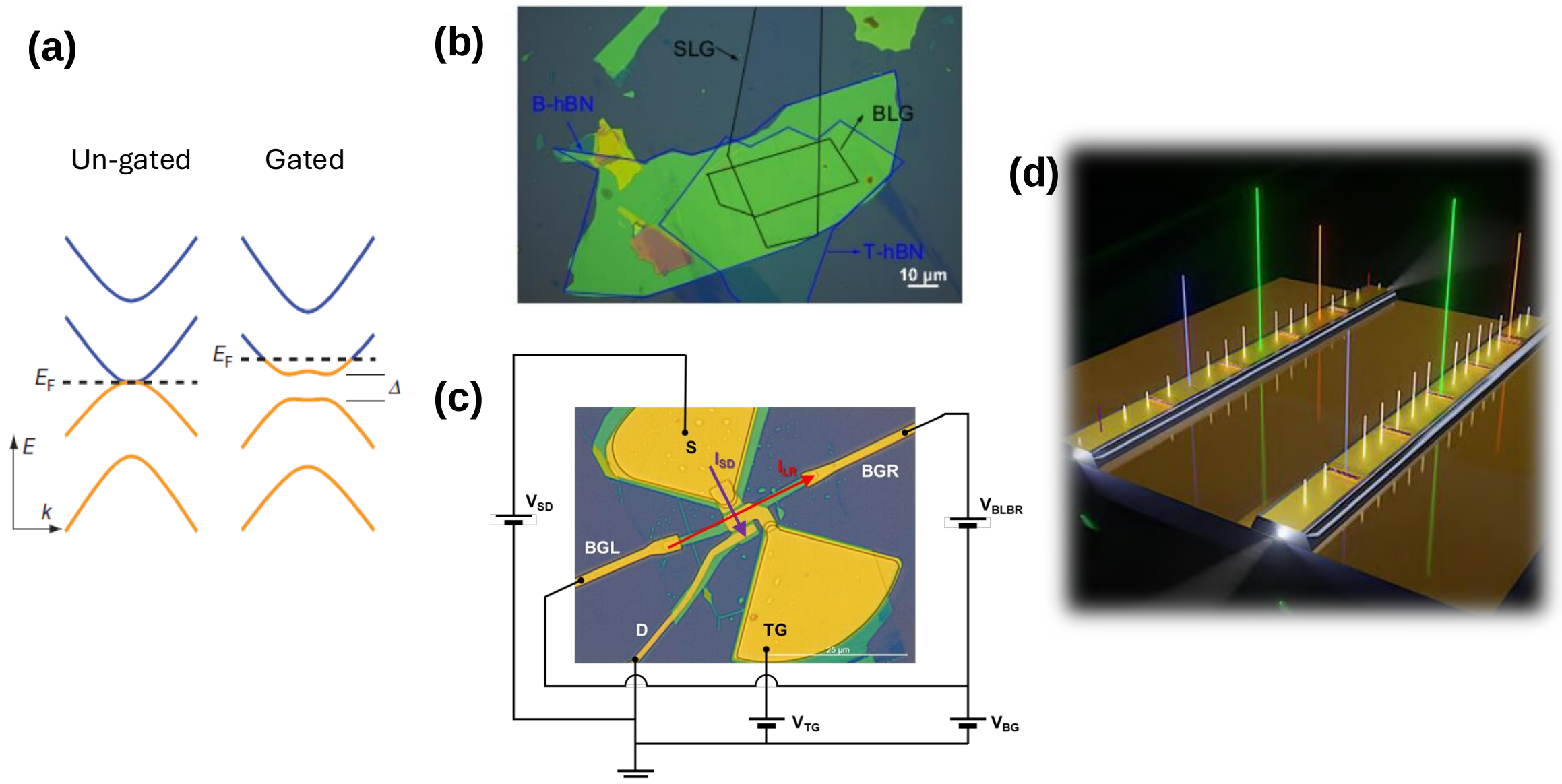

**Figure 38.** Graphene-integrated THz frequency devices (a) Schematic representation of a bandgap opening in bilayer graphene (BLG) induced by an out-of-plane electric field. Adapted from ref 1030 (Copyright 2009 Springer

Nature). (b) Optical microscope image of an assembled vdW heterostructure comprising a top-hBN flake, BLG, and a bottom hBN flake on a large-area single-layer graphene flake. (c) Schematic of a dual back-gate (BGL, BGR) tunneling field-effect transistor (TFET) with an on-chip planar bowtie antenna connecting the source (S) and top-gate (TG) electrodes of the TFET. The drain and S contacts connect the heterostructure of panel (b). (d) Schematic of 5$^{th}$- and 6$^{th}$-order harmonic frequency-comb QCLs in a wire cavity with top graphene-etched trenches acting as frequency filters.

## 39. 2D MATERIALS FOR SOLAR-BLIND ULTRAVIOLET PHOTON SENSING

**Nathan D. Cottam,[1] Benjamin T. Dewes,[1] Oleg Makarovsky,[1] and Amalia Patanè[1,*]**
[1]School of Physics and Astronomy, University of Nottingham, Nottingham, NG7 2RD, UK
***Corresponding author.** Email: Amalia.Patane@nottingham.ac.uk

### 39.1 Current State of the Art

The UV-C spectral range (100–280 nm) has garnered significant interest in surveillance, communication, and defense technologies due to inherently minimal interference from natural solar radiation, which is mostly absorbed by the upper atmospheric ozone layer at these wavelengths.[1037] Thus, the UV-C spectrum is commonly referred to as a solar-blind region. Additionally, UV-C sensors can be employed in flame sensing, the detection of biological and chemical agents, and enhanced imaging in low-visibility environments (Figure 39a).[1038] These applications require devices with high sensitivity, signal-to-noise ratio, frequency bandwidth, and spectral selectivity. To date, commercial photomultiplier tubes (PMTs) based on vacuum glass tubes represent the best sensing technology. However, PMTs are bulky and operate at high voltage (>1 kV). High-gain photodetectors, such as avalanche photodiodes (APDs), unipolar field-effect phototransistors (FEPTs), and bipolar heterojunction phototransistors (HPTs) based on wide bandgap semiconductors (e.g., $Ga_2O_3$, $Mg_xZn_{1-x}O$, SiC, and III-nitrides) are alternatives to PMTs. They are compact and energy-efficient, but require further advances in material quality and fabrication processes.[1039,1040]

The 21$^{st}$ century boom of atomically thin 2D semiconductors (2DSEM) has led to new platforms for energy-efficient solar-blind photodetectors (Figure 39b) that exploit photovoltaic[1041,1042] and photoconductive[1043] effects for the conversion of UV-C light into an electrical signal. Photodetectors with enhanced responsivity can be achieved by photogating. In these devices, a high gain in the electrical response can be realized by charge traps, whereby one of the photogenerated carriers is trapped, producing an additional electric field that increases the channel conductance. Due to the large surface-to-volume ratio and reduced screening in 2DSEM, impurities and defects can cause a large gain and responsivity, but tend to increase the carrier lifetime, thus limiting the frequency bandwidth. Examples include 2D metal chalcogenides with strong absorption in the UV-C (e.g., GaS[1044] and GaSe[1045]), all-inorganic 2D perovskites (e.g., $Sr_2Nb_3O_{10}$[1046]), and graphene functionalized with UV-C absorbing layers.[1047] A number of wide-bandgap 2DSEM are currently under investigation for UV-C spectral selectivity, such as $NiPS_3$, $GeSe_2$, and $GaPS_4$,[1043,1048,1049] but they all tend to suffer from a slow response. Monolayer hBN is also a promising candidate due to its direct bandgap (~ 6 eV) and large band-edge absorption coefficient ($7.5 \times 10^5$ cm$^{-1}$). Liu *et al.* reported on hBN devices with a weak responsivity $R = 0.1$ mA/W at 225 nm and slow response time (~ 0.1s).[1050] The responsivity can be increased through localized surface plasmon resonances in Al nanoparticles on the hBN surface but remains small.[1051] Another distinct advantage of 2DSEM over traditional materials is their compatibility with flexible substrates. Veeralingam *et al.* reported a cost-effective solid-state reaction synthesis of hBN nanosheets onto a flexible Cu(111) foil to form a metal-semiconductor Schottky junction with high responsivity ($R > 5$ A/W at 210 nm), but slow response (~ 0.2s).[503]

One of many directions to cementing a place for 2DSEM in modern optoelectronics is by integration with other systems, such as $Ga_2O_3$, with an absorption cut-off in the UV-C region.[1052] Chen *et al.* fabricated a graphene/$Ga_2O_3$ Schottky junction on an Al/Si/Al substrate for position-sensitive-detection in the UV-C range.[1053] A 25-nanometre-thick film of $Ga_2O_3$ was grown by plasma-enhanced chemical vapour deposition (CVD) and a graphene film was transferred on $Ga_2O_3$. This thin junction exhibits photovoltaic response with $R = 48.5$ mA/W at 250 nm and a strong 250 nm to 400 nm rejection ratio of $R_{250}/R_{400} = 5 \times 10^3$. A fast (~ µs) response time was also achieved making it suitable for fast data transfer in UV-C communications. Another advantageous property of 2D materials is the relative ease in the formation of heterojunctions through mechanical exfoliation and stacking, as shown by Zeng *et al.*, who reported on a $MoTe_2$/Ta:$\beta$-$Ga_2O_3$ p–n junction with $R = 358.9$ A/W at 254 nm, ms response time, and a rejection ratio $R_{260}/R_{400} \sim 10^3$.[1054] The device could operate in photovoltaic mode, although with a lower responsivity of 9.2 mA/W. Finally, the scalable conversion of gallium-based 2D layers (e.g., GaS and GaSe) into an oxide offers an alternative route to the fabrication of ultrathin UV-C sensors. Using high-quality GaSe grown by molecular beam epitaxy (MBE), Cottam *et al.* demonstrated the complete conversion of nanometer-thick layers of GaSe into $\beta$-$Ga_2O_3$ through oxidation and crystalization at $T > 600$ ºC.[1055] The oxidized material was employed in solar-blind photodetectors with $R = 2.1$ mA/W and a response time of ~ 10 ms. In summary, the recent literature demonstrates that 2D UV-C photodetectors can be fabricated by various techniques using hybrid heterostructures or wide-band 2DSEM, but require further development toward improved performance and scalability (Figure 39c).

**39.2 Challenges and Future Goals**

There are numerous challenges in the development of high-performance solar-blind sensors based on 2DSEM. Traditionally, a large proportion of devices based on 2DSEM utilize top-down fabrication techniques, such as mechanical exfoliation and stacking. This results in high-quality 2D materials with an extra degree of freedom in heterostructure stacking for solar-blind detectors. However, mechanical exfoliation of vdW bulk crystals generally produces only small-area flakes, which is incompatible with the large-scale deployment required for these 2D systems to reach their full potential. Bottom-up approaches to material growth *via* epitaxy offer a route to the scalable growth of 2DSEM, but require highly technical and costly equipment as well as vast expertise in the operation and optimization of growth parameters. Whilst wafer-scale growth of 2D materials is a maturing field, the challenge remains in the large-scale transfer of grown layers onto technologically relevant substrates to unlock the versatility of this approach. Strain engineering, defect engineering, and dopant implantation offer the opportunity to modify the optoelectronic properties of the grown layers, but these techniques still require considerable investigation.

Many 2D materials suffer from oxidation and degradation under ambient conditions,[1056] which presents a significant obstacle to their deployment in relevant applications. 2DSEM can be influenced by interactions at the material–air boundary, most notably with $O_2$ and $H_2O$. This oxidation and degradation in ambient conditions can reduce the performance of a device over time. Thus, technological improvements in surface capping materials and processes that do not impinge on device performance are required. However, introducing a protective, UV-C transparent coating that does not negatively influence the properties of the underlying 2D layer (e.g., carrier concentration and mobility) brings additional challenges.

The performance of 2D photodetectors requires the careful characterization of many key performance parameters and one challenge is the inconsistency in the reporting of such parameters within the community. Specific detectivity is particularly useful as it enables a comparison of diverse device geometries. However, different interpretations of how to calculate the noise, optical power density, and response time of a photodetector can lead to an overestimation of the specific detectivity.[1057] Furthermore, for solar-blind 2D photodetectors, the testing conditions are affected by the limited availability of UV-C laser sources; in particular, the optical power, wavelength, and sensor operating voltage vary considerably throughout the literature. Thus, direct comparison of photodetectors that incorporate different materials, structures, and photosensing mechanisms is challenging.

**39.3 Suggested Directions to Meet These Goals**

Despite increasing work on UV photodetectors based on 2D materials,[1058] deployment for applications in the UV-C range are still limited. Future progress and innovative solutions for UV-C specific spectral selectivity require a shift toward novel approaches to synthesis and advanced integration technologies. State-of-the-art manufacturing technologies for precise engineering of 2DSEM, such as for the wide bandgap hBN,[1059] are needed to overcome the reliance on exfoliated layers. Advanced solution-based exfoliation methods can offer scalability but can compromise crystal quality; most importantly, physical properties associated with specific compositions, crystal structures, doping, and size are difficult to control by exfoliation. Other techniques, such as MBE, CVD, and metal-organic CVD (MOCVD) are less common due to the high entry-level costs of the necessary infrastructure and/or the specialized requirements for operational expertise. The fabrication of single and arrays of photodetectors for UV imaging also presents significant challenges. Bespoke fabrication strategies are needed to minimize process-induced material damage and flexibility in the integration of ultrathin and flexible single-crystalline 2D membranes on different platforms, such as with silicon-on-insulator (SOI) waveguides for photonics and Si-CMOS-based read-out circuits.[1060] In particular, new approaches are required to optimize light–matter interaction in the deep UV. Localized surface plasmons can be used to enhance the photon absorbance of 2D materials, but current approaches work mostly in the near-UV and visible ranges.[1061] Finally, as the growth and fabrication of 2D materials advance, new regimes of photodetection can be expected. The reduction in thickness of 2D materials to a few layers or the combination of 2D materials to form new 2D–2D layer structures, such as moiré superlattices,[1062] can dramatically impact electronic properties, leading to new forms of charge, magnetic, and superconducting order, driven by weakly screened electron correlations. These phenomena could enable new concepts for fast and sensitive photon sensors *via* the manipulation of carrier correlations. Other device technologies are phototransistors based on 2D ferroelectrics for self-powered applications and polarization-sensitive detection.[1063] In summary, we can expect a myriad of further developments in this field by advances in the science and technologies of 2D materials.

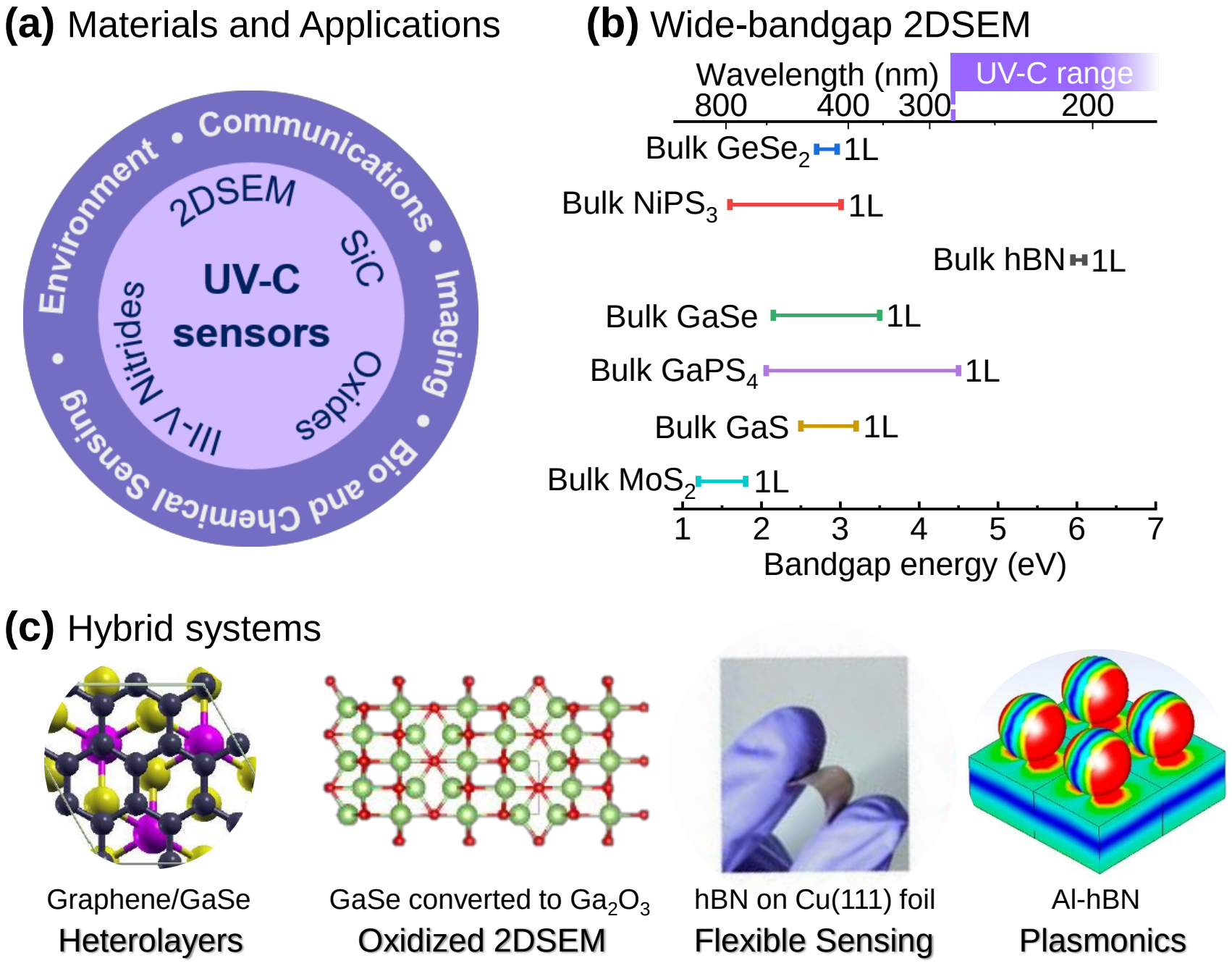

**Figure 39.** Materials for UV-C photonics. (a) Graphic of current technologies for UV-C sensing: materials and applications. (b) A diagram of the bulk and monolayer (1L) bandgap energies of 2D semiconductor materials relevant for UV-C photodetection. (c) Examples of different hybrid systems used for UV-C sensing. Images adapted with permission from (left) Bradford *et al.*[1047] (Copyright 2024 Wiley-VCH), (middle-left) Cottam *et al.*[1055] (Copyright 2024 American Chemical Society), (middle-right) Veeralingam *et al.*[503] (Copyright 2021 American Chemical Society), and (right) Kaushik *et al.*[1051] (Copyright 2022 American Chemical Society).

# 40. BLACK PHOSPHORUS AND ITS ALLOY FOR INFRARED SENSING

**Yihao Song,[1] Mingyang Cai,[1] Jiazhen Chen,[1] Doron Naveh,[2] Houk Jang,[3] Suji Park,[3] and Fengnian Xia[1,*]**
[1]Department of Electrical and Computer Engineering, Yale University, New Haven, CT 06511, USA
[2]Faculty of Engineering and Bar-Ilan Institute for Nanotechnology and Advanced Materials, Bar-Ilan University, 52900, Ramat-Gan, Israel
[3]Center for Functional Nanomaterials, Brookhaven National Laboratory, Upton, NY 11973, USA
***Corresponding author.** Email: fengnian.xia@yale.edu

## 40.1 Bandgap Tunability

### 40.1.1 *Physical Properties of Black Phosphorus (BP)*

Black phosphorus (BP) is the most stable allotrope of phosphorus,[1064,1065] and it has a puckered honeycomb lattice as shown in Figure 40a. The atomic layers bonded by weak vdW force allow for the mechanic exfoliation down to monolayer limit.[1066,1067,1068,1069,1070] Black phosphorus is a promising layered material for IR sensing, imaging and spectroscopy due to its high carrier mobility, a moderate bandgap, and wide tunability by electric field or strain. It is a direct bandgap semiconductor, and the bandgap varies from around 0.3 eV in the bulk (> 10 nm) to 2.0 eV in a monolayer due to enhanced quantum confinement.[1071] Due to the crystal-symmetry reduction arising from the puckered honeycomb lattice, BP exhibits highly anisotropic electronic and optical properties.[1067,1069,1070] The effective mass and carrier mobility are different along *x* (armchair) and *y* (zigzag) directions.[1070] A hole mobility above 1,000 $cm^2V^{-1}s^{-1}$ was reported in 15-nm thick BP film at 120 K[1069], and a higher hole mobility above 5,000 $cm^2V^{-1}s^{-1}$ was observed in few-layer BP encapsulated in hBN layers even at room temperature.[1072] This high mobility makes BP a promising material for field-effect transistors[1066] (FET).

The strong in-plane anisotropy due to the crystal structure of BP induces angular dependent optical absorption and emission,[1073] which can be exploited for polarization-sensitive photodetectors[1074] or light-emitting diodes.[1075] Photoluminescence (PL), Raman spectroscopy, DC conductance, and Hall mobility measurements have been leveraged to determine the lattice orientation of the BP flake.[1069] A previous study on excitons in monolayer BP showed that a linearly polarized emission was observed regardless of the excitation laser polarization.[1076] A light-emitting diode consisting of BP/$MoS_2$ p–n junctions that show strongly polarized emission within mid-IR range has been demonstrated by Wang *et al.*[1075]

### 40.1.2 *Bandgap Tunability*

As discussed above, the BP bandgap can be effectively tuned by film thicknesses[1077] to cover the spectral range from near-IR to mid-IR range up to ~ 3.7 µm. First-principle calculations based on density function theory (DFT) have predicted that monolayer BP can switch from a direct bandgap semiconductor to an indirect semiconductor, and eventually to a metal with strain applied on x-direction.[1078] Kim *et al* experimentally confirmed the strain tunability of BP bandgap from 0.22 to 0.53 eV.[1079] An external electric field can also continuously modulate the bandgap of a 10 nm thick BP film from around 300 meV to below 50 meV,[1080] as illustrated in Figure 40b. Such tuning can significantly extend the wavelength coverage of BP based optoelectronic devices. Leveraging this dynamic tuning method, a 5 nm BP photodetector with photoresponse up to 7.7 µm was demonstrated.[1081] Additionally, in black arsenic-phosphorus (b-AsP) alloys, the bandgap can be tuned from 0.33 to 0.15 eV by varying the molecular fraction of arsenic.[1082] The tunability of BP bandgap by different approaches greatly extends its operational wavelength and provides possible pathways for more advanced optoelectronic applications.

### 40.1.3 *Black Phosphorus in Infrared Sensing and Spectroscopy*

The operational wavelength range of BP photodetectors spans from UV to mid-IR with a cutoff wavelength of ~4 µm.[872,1081,1083,1084,1085] The high carrier mobility and fast carrier dynamics enable ultrafast[1086] and THz[1005] photodetection. Integration of BP with silicon photonic platforms, including waveguides[1087] and photonic crystal cavities,[1088] can significantly enhance the interaction of light and BP. Silicon-integrated avalanche BP photodetectors have been demonstrated at wavelengths of around 1.55 µm with responsivity up to 125 A/W.[1089] Its optical anisotropy has also been utilized for polarization-sensitive photodetection covering a wavelength range of 400–3750 nm[1074] with a polarization contrast of 288 at 1450 nm.[1085] Interestingly, the photothermal effect in BP has been utilized to acquire THz images.[1090]

Beyond imaging, the tunable properties of BP photodetectors[1081] have been utilized for spectroscopy applications. A double-gated BP FET tunable by an electric field can function as a spectrometer (Figure 40c), which is capable of measuring the mid-IR spectrum in the range of 3.5–9.5 µm[1091] by leveraging the regression algorithm in the interpretation of the high-dimensional photoresponse. Moreover, BP photodetectors can also be designed to possess characters of memristors, where the photocurrent is dependent on the previous operational conditions. Such characteristics have enabled the implementation of a programable image sensor capable of performing cognitive operations such as edge detection.[1092]

40.1.4 *Research Progress on Arsenic Phosphorus*

As discussed above, the relative arsenic concentration in b-AsP can be varied to tune its bandgap. When the molecular fraction of arsenic changes from 0 to 0.83, the bandgap of b-AsP varies from around 0.33 to 0.15 eV. Thus, b-$As_{0.83}P_{0.17}$ photodetectors can cover a wavelength range of up to 8.27 µm.[1082] The polarization-sensitive absorption characteristics inherent to BP are preserved in b-AsP, as confirmed experimentally across various molecular fractions[1082]. Amani *et al.* have demonstrated a b-AsP photodetector operational in the mid-IR range with a detectivity up to $2.4 \times 10^{10}$ cm $Hz^{-1/2}$ $W^{-1}$,[1093] which is an order of magnitude better than commercial detectors at room temperature. A room temperature mid-IR photodetector based on b-AsP/$MoS_2$ heterojunctions with high detectivity (> $4.9 \times 10^{9}$ cm $Hz^{-1/2}$ $W^{-1}$) and suppressed dark current with noise equivalent power (NEP) below 0.24 pW $Hz^{-1/2}$ was also reported in a spectral range of 3 to 5 µm.[1094]

**40.2 Material Challenges and Recent Progress**

40.2.1 *Stability*

One major obstacle to scaling and commercializing BP devices is the instability of the material under ambient environmental conditions. As a result, passivation and encapsulation are necessary for practical applications of BP[1095]. In particular, the lone pair electrons in P atoms favor reaction with oxygen.[1096] The oxidation mechanism can be summarized as follows:[1095] first, the surface P atoms are oxidized to form phosphorus oxide ($P_xO_y$); then the oxide reacts with the moisture to form phosphoric acid on the surface. Experimentally, the BP oxidation can be observed as the formation of small bumps shortly after exfoliation followed by coalescent into large droplets, as shown in Figure 40d. A quantitative photo-oxidation investigation shows that the oxidation rate is linearly proportional to both the oxygen concentration and the light intensity, and oxidation increases significantly from thin film to monolayer.[1097]

Hence, to prevent the degradation of BP devices, isolating BP from water and oxygen are of paramount importance. Various encapsulation methods have been reported and can be summarized into three categories: top encapsulation,[1098] full encapsulation,[1099] and inert gas encapsulation.[1096] Top encapsulation only involves one cover layer on top of the exfoliated BP flake. The top layer can be atomic-layer-deposition oxide, 2D insulator hBN, or common sealing polymer such as parylene. This method is simple and improves the device stability as shown in Figure 40e, but the results are not optimal, and the substrate used for BP device fabrication should be hydrophilic to reduce water diffusion.[1100] Full encapsulation involves using both top and bottom cover layers, usually by exfoliated hBN flakes or oxide grown by atomic layer deposition. Full-hBN-encapsulated BP devices have shown stability over months under ambient conditions and already exhibit great potential for applications such as photodetection and spectroscopy.[1091,1101] Inert gas chamber has also been utilized for the packaging for BP devices. This is usually combined with the previous two methods to achieve superior passivation.[1102] Notably, Higashitarumizu *et al.*[1096] have tested the lifetime of fully encapsulated BP-$MoS_2$ LED with $Al_2O_3$ in nitrogen chamber. Under a high driving current density of ~75 A/$cm^2$ and a temperature of up to 140°C, the extrapolated half-life time for their devices reaches ~15,000 hours.

40.2.2 *Large Scale Production*

To fully unleash the potential of BP in IR detection, it is essential to develop methods for wafer-scale synthesis of high-quality crystalline BP films. Due to the high chemical potential barrier, utilizing high temperature and high pressure at several gigapascals (GPa) was a dominant way to turn bulk red phosphorus or white phosphorus into large-scale BP crystals.[1064] Li *et al.* reported successful synthesis of BP thin film at 1.5 GPa pressure on millimeter-scale sapphire substrates with a carrier mobility of around 160 $cm^2 V^{-1} s^{-1}$ and a grain size of 40–70 µm.[1103] Moreover, Higashitarumizu *et al.* reported the successful conversion of red phosphorus thin film into centimeter-scale BP film (~ 18 nm thick) on mica by applying 8 GPa pressure at room temperature and thermal annealing.[1104] Wu *et al.* demonstrated successful growth of few-layers BP at centimeter-scale on mica by utilizing pulsed laser deposition. No visible grain boundaries across the film have been found under polarized Raman mapping.[1105] The number of BP layers can also be tuned by adjusting the laser pulse during the growth.

Furthermore, with the introduction of mineralizer assistants such as Sn, $SnI_4$, and $I_2$, the formation energy of the BP phase can be reduced and conversion can be realized at high temperature without involving high pressure.[1106] Chen *et al.* have shown that, by using silica sand as the diffusion matrix, as shown in Figure 40f, $P_4$ pressure can be lowered for controllable growth of single crystalline BP thin films at sub-centimeter-scale with mineralizer.[1106] This strategy has also been applied to synthesize large-scale b-$As_xP_{1-x}$ alloy.[1106] Moreover, lateral growth of BP thin films has been demonstrated on silicon substrate starting with $Au_3SnP_7$ flakes.[1107]

40.2.3 *Reduced Light–Matter Interaction*

Another challenge relates to the thin-film nature of the materials utilized in device realizations. This property allows for flexibility in device integration and reconfigurability. However, for vertically incident light excitation, light–matter interaction is reduced and the efficiency of such thin-film devices can be low. As discussed in Section 40.3.2, integration of such thin-film materials with IR waveguides could offer a low-loss solution and represents a direction for future research.

## 40.3 Suggested Future Directions

40.3.1 *Encapsulation and Large-Scale Device Fabrication with Q-Press*

Reliable encapsulation and heterostructure stacking are key steps for the realization of large arrays of BP and b-AsP optoelectronic devices. However, manual mechanical exfoliation and stacking processes are labor-intensive and time-consuming, and the reproducibility is poor. To overcome these obstacles, recently the community is developing a new automatic process suitable for manufacturing named Quantum Material Press (Q-Press).[1108] Q-Press comprises three core modules—exfoliator, cataloger, and stacker—integrated with peripheral semiconductor processing units like a reactive ion etcher, a physical vapor deposition (PVD) system, and a high-temperature annealer within a glovebox filled with inert gas. These components are connected by robotic sample transfer systems, making process automation possible.

Each core module addresses specific challenges in the manual fabrication process. The mechanized exfoliator (Figure 40g) enables reliable, recipe-based exfoliation with precise control over critical variables that can affect the exfoliation result, such as time, pressure, temperature, and speed. The automated cataloger pairs custom analysis software with an automated optical microscope to perform rapid identification and categorization of flakes based on thickness, size, and shape. Lastly, the robotic stacker equipped with six-axis movement capabilities—including translation and rotation—performs linear motions with sub-micrometer precision to eliminate bubbles in the stacking process. These modules can be utilized to enable the fabrication of large arrays of devices consisting of a wide range of 2D materials and heterostructures including black phosphorus and b-AsP. Since synthesized large-area BP and b-AsP thin films are already available, in the future these thin films can be utilized in Q-Press systems for the manufacturing of photonic devices in large quantities.

40.3.2 *Mid- and Long-Wavelength Infrared Waveguide Integration*

As discussed in Section 40.2.3, another key challenge in thin film BP and b-AsP optoelectronic devices is the reduced light–matter interaction. To effectively tune the electronic properties of BP or b-AsP, the thickness of these narrow-gap semiconductors should be small (less than 20 nm) to minimize screening effects,[1080] leading to reduced light–matter interactions for vertically incidence light. A classic approach to resolve this issue is through waveguide integration.[1087,1109,1110] In fact, classic semiconductors such as silicon and germanium can cover spectral ranges of 3 to 5 µm and 8 to 12 µm, respectively.[1111]

Moreover, thin-film black phosphorus has been previously integrated with a chalcogenide glass waveguide platform for the detection of mid-IR light.[1112] In the future, the use of these prefabricated waveguide platforms in the Q-Press system will enable the realization of complex mid- and long-wavelength photonic integrated circuits through their integration with thin BP or b-AsP films. Mid-IR light emitters, modulators, waveguides, and detectors can be monolithically integrated within a unified platform for various applications such as sensing, imaging, and surveillance. Integration of thin BP or b-AsP films with other photonic platforms such as photonic crystal cavities and plasmonic structures can also enhance their light–matter interaction properties.

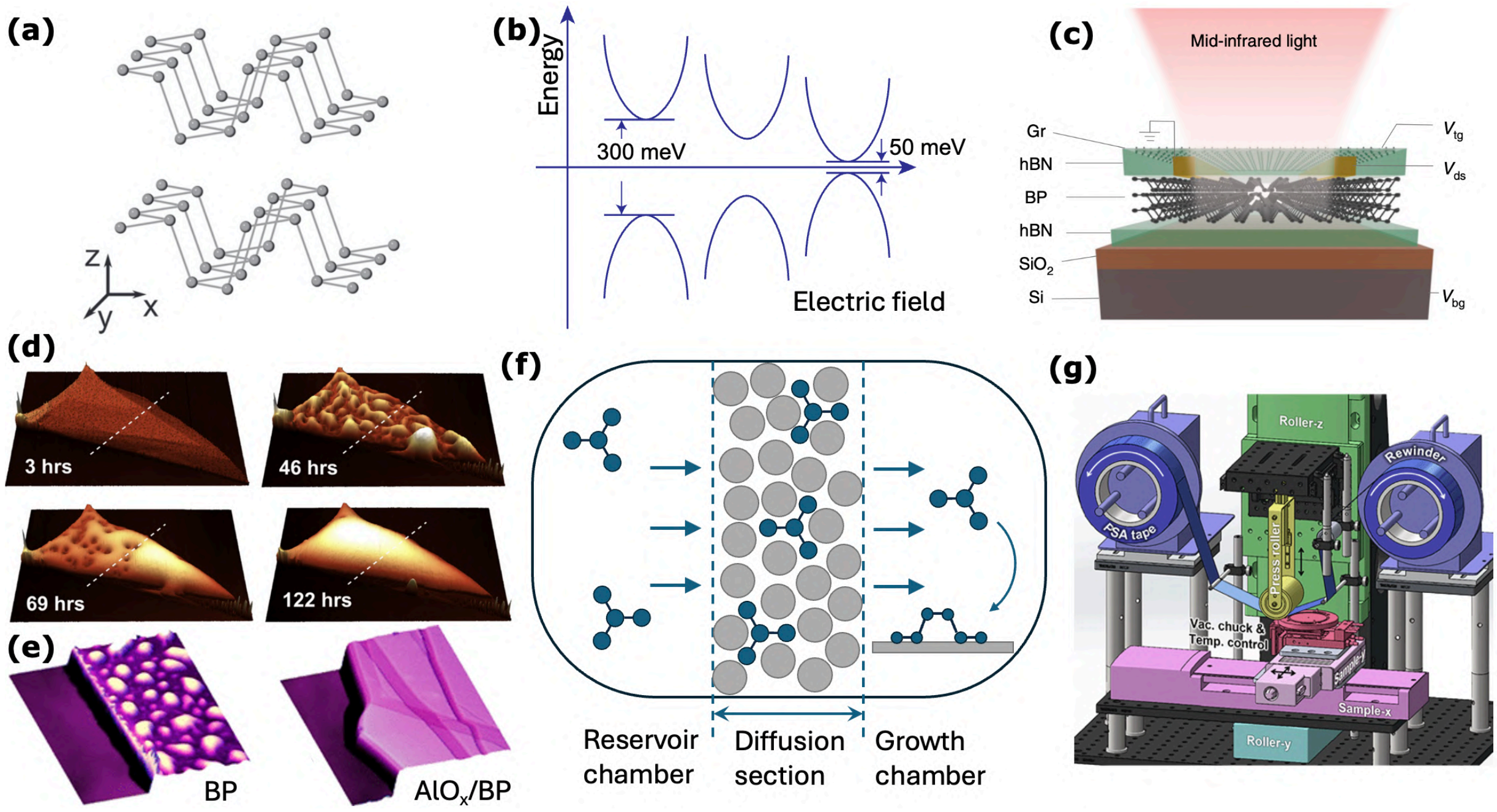

**Figure 40.** BP properties and its applications in IR sensing. (a) Schematic of black phosphorus crystalline structure. Reproduced from ref 103 (Copyright 2014 IOP Publishing). (b) Illustration of bandgap tunability by vertical external electric field in BP thin film (~10 nm). (c) Schematic of a hBN encapsulated BP photodetector in dual-gate configuration. Reproduced from ref 1091 (Copyright 2017 Springer Nature). (d) Atomic force microscopy (AFM) images of a BP flake after exposed to ambient environment for different times. Reproduced from ref 1095 (Copyright 2015 IOP Publishing). (e) AFM images of BP flake exposed to air without (left) and with (right) aluminum oxide ($AlO_x$) encapsulation. Reproduced from ref 1100 (Copyright 2014 American Chemical Society). (f) The schematic growth setup to achieve a low $P_4$ pressure for BP and b-AsP synthesis. The diffusion section is filled with silica sand. Adapted from ref 1106 (Copyright 2023 Springer Nature). (g) 3D schematic view of the exfoliator of Q-Press, consisting of a sample stage, a press roller, a rewinder, and a vacuum chuck with a heater.

**Applications: Disruptive Directions**

# 41. 2D MATERIALS FOR PHOTONIC QUANTUM INFORMATION PROCESSING

**Lee A. Rozema,[1,*] Philipp K. Jenke,[1,2] Josip Bajo,[1,2] Benjamin Braun,[1,2] and Philip Walther[1,2,3]**
[1]Faculty of Physics, Vienna Center for Quantum Science and Technology (VCQ), University of Vienna, Boltzmanngasse 5, 1090 Vienna, Austria
[2]Faculty of Physics & Vienna Doctoral School in Physics, University of Vienna, Boltzmanngasse 5, 1090 Vienna, Austria
[3]Research Network Quantum Aspects of Space Time (TURIS) & Christian Doppler Laboratory for Photonic Quantum Computer, University of Vienna, Boltzmanngasse 5, 1090 Vienna, Austria
***Corresponding author.** Email: lee.rozema@univie.ac.at

## 41.1 Current State of the Art

Quantum protocols are anticipated to pave the way for new technological advances, but many essential components remain imperfect. 2D materials are emerging as promising solutions to several of these challenges. Photonic quantum information requires three main components: sources of single photons, mechanisms to manipulate these photons (preferably with nonlinear interactions), and efficient photon detectors. Here, we focus on how 2D materials may enhance photonic quantum information, particularly in relation to photon sources and photon manipulation. Three main research directions are highlighted: defect-based single-photon emitters (SPEs) (Figure 41a), spontaneous parametric photon pair production in vdW crystals (Figure 41b), and the use of plasmons to manipulate quantum information (Figure 41c).

41.1.1 *Single Emitters*

Traditionally, photonic quantum information has relied on spontaneous parametric down-conversion (SPDC) to produce photon pairs. SPDC sources face limitations because the emission is fundamentally nondeterministic. Thus, researchers have increasingly turned their attention to SPEs in solid-state platforms, such as III–V semiconductor quantum dots and NV centers. While InGaAs quantum dots offer excellent performance in the near-IR range, they have not reached the same level of performance at other wavelengths, and their operation is confined to cryogenic temperatures.

The discovery of defect-based SPEs in 2D materials opens new avenues for quantum information processing, with monolayer TMDs and monolayer or multilayer hBN currently being some of the most promising. These materials can potentially operate at room temperature[1113] and offer tunability beyond existing platforms.[1114] SPEs have been observed in TMDs like $MoS_2$, $WS_2$, $MoSe_2$, $WSe_2$, and $MoTe_2$. These TMD-based emitters cover a wide range of wavelengths—from the visible to the telecom C and O bands—making them suitable for integration with other quantum systems such as quantum memories or repeaters. While the central wavelength is determined by the precise defect, TMD emitters can achieve wavelength tunability through mechanical and electric strain, surface acoustic waves, and electric field engineering. While most TMD emitters currently require temperatures below 30 K to operate, emission at 150 K has been demonstrated.[13] SPEs in $WSe_2$ at cryogenic temperatures have even allowed quantum key distribution (QKD) to be performed,[1115] highlighting their potential in practical quantum communication.

While TMD single-photon emission has been limited to 150K, hBN has achieved room-temperature operation. With a large bandgap of 6 eV, hBN emitters can generate photons in the visible spectrum and achieve emission rates in the MHz range, competitive with leading III–V quantum dots. In hBN, the defects can be deterministically placed using methods such as thermal annealing, electron beam irradiation, and strain engineering. Compatibility with integrated photonics has been shown and quantum protocols using hBN SPEs have been demonstrated.[1116] Moreover, the spin degree of freedom in hBN shows quantum coherence even at room temperature, making it a potential platform for spin qubits, quantum memories, and quantum sensors.

41.1.2 *Spontaneous Parametric Down-Conversion*

Although deterministic single-photon emission has advantages, non-deterministic methods such as SPDC or spontaneous four-wave mixing remain essential for generating multi-photon entangled states. SPDC sources continue to play a key role in scalable photonic quantum computing, scattershot boson sampling, and quantum communication. Additionally, since photon pairs are generated in these processes, two-photon entanglement can be directly created across various degrees of freedom, such as orbital angular momentum or frequency.

In the context of 2D materials, SPDC in ultrathin vdW materials, particularly TMDs, holds significant potential,[526] and for example, one of the key advantages is the relaxed phase-matching condition, where the two-photon state is defined primarily by energy conservation. This allows for photon-pair generation across a wide spectral range, resulting in continuous-variable energy–time entanglement.[1117] The nonlinear response in these materials also enables direct control of the quantum polarization state, without the requirement to consider phase-matching. This feature has been utilized to create photon pairs with tunable, separable polarization states in thin films, such as 400-nm-thick GaP[1118] and 46-nm-thick $NbOCl_2$.[489] TMDs can directly produce maximally entangled Bell states.[492] Unlike most bulk polarization-entanglement sources, TMDs do not require compensation crystals to correct temporal walk-off, making them highly attractive for real-world applications and integrated quantum systems.

Though efficiency is currently modest, nonlinear coefficients in 2D materials are up to 1,000 times higher than those in conventional nonlinear crystals, offering promising opportunities for future developments.[471] Techniques such as periodic poling have been shown to increase photon-pair rates by factors of up to 20 through quasi-phase-matching in $MoS_2$.[490] Other approaches, such as cavity enhancement and coupling to intrinsic resonances (e.g., excitons or plasmons) remain to be explored.

#### 41.1.3 *Plasmonics*

Manipulating quantum information encoded in photons is a key challenge for scalable photonic quantum computing. Single-photon states are relatively easy to manipulate using linear optical devices such as waveplates, beamsplitters, and phase shifters, and they can also be readily created in integrated settings. However, the resulting structures are still relatively large, and further miniaturization would be desirable. Furthermore, even in bulk settings, two-photon interactions, which are necessary for two-qubit quantum logic gates, require nonlinear effects. Surface-plasmon polaritons in 2D materials provide an innovative way to encode quantum information in subwavelength structures[1119] with intrinsic nonlinearities.[1120]

Plasmons, which confine electromagnetic fields beyond the diffraction limit, enhance light–matter interactions significantly. Quantum plasmonics—a field that explores light–matter interactions *via* collective electron excitations—has recently gained traction. Photons and plasmons both exhibit bosonic behavior and demonstrate wave-particle duality.[1121] In metal plasmonics, it has been shown that quantum properties such as entanglement and indistinguishability are preserved during the conversion from photons to plasmons and vice versa.[1122] Analogously to photonic systems, integrated linear plasmonic devices and superconducting nanowire detectors have also been developed for metallic plasmons. Although promising, Ohmic losses and mode distortion have limited metallic quantum plasmonics. 2D materials, which exist as metals or semimetals, can address these challenges, hosting long-lived plasmons[5] and providing tunability through chemical or electronic doping. Graphene plasmons, for example, have demonstrated long lifetimes and tunable plasmon resonances.[85] Other platforms, such as silicene and germanene, still need to prove their potential for plasmonic applications.

An exciting aspect of 2D plasmonics is the possibility of enhanced nonlinear interactions.[530] Graphene plasmons can provide field confinement that is significantly stronger than in metallic plasmons, amplifying weak nonlinearities. These interactions hold immense promise for highly efficient frequency conversion and ultrastrong nonlinear light–matter interactions beyond their metallic counterparts. It has even been proposed that the nonlinearities of graphene could be used for efficient quantum plasmonic logic gates or as sources of single photons[1120] or plasmons.[1123] Recently, ultrathin crystalline noble metal films demonstrated significant plasmon loss reduction and suggested strong nonlinearities, holding potential for a comeback of metal plasmonics for efficient second-harmonic generation or SPDC.[544]

### 41.2 Challenges and Future Goals

Despite their promise, the integration of 2D materials into photonic quantum systems faces important challenges. We shall now identify these for each of the three research directions.

#### 41.2.1 *Single Emitters*

Although SPEs in TMDs and hBN are competitive in terms of single-photon purity and emission rates, multiphoton quantum protocols require photon indistinguishability. This comes at two levels: first, successively emitted photons from the same SPE must be identical and, second, photons from different SPEs should also be indistinguishable. Indistinguishability can be measured through the Hong-Ou-Mandel (HOM) interference visibility. III–V quantum dots have achieved nearly perfect HOM visibility between successive photons from the same source, but indistinguishability between different quantum dots remains a challenge.

In 2D materials, HOM visibility has remained low. Drawer *et al.* reported 2% visibility in $WSe_2$,[436] likely due to spectral diffusion caused by environmental factors. Fournier *et al.* achieved ~50% visibility in hBN.[1124] Another hurdle in the creation of multiphoton states is the direct generation of entanglement. In quantum dots, photon entanglement can be generated *via* biexciton–exciton cascades. A similar cascade has been identified in $WSe_2$-based defects, but entanglement has yet to be demonstrated.[1125] Scaling beyond two-photon states has not yet been explored in 2D materials.

Integrating 2D material-based SPEs into photonic structures can enhance the quality of the emitted photons, as observed in $WSe_2$, wherein the emission properties have been improved by coupling to different structures and resonances.

41.2.2 *Spontaneous Parametric Down-Conversion*

Current photon-pair sources based on 2D materials are primarily limited by low conversion efficiency, with rates of about 10 Hz. Photonic protocols typically require rates in the kHz or MHz range. Techniques such as quasi-phase-matching, multi-pass systems, and cavity-based enhancement, which have been used in bulk crystals,[1126] could help boost rates. Moreover, while standard bulk nonlinear crystals do not typically take advantage of any material resonances, 2D materials possess exciton transitions and plasmonic resonances, and can be readily integrated with metasurfaces, all of which could provide significant enhancements to the conversion efficiency.[1127]

Another challenge relates to achieving an efficient collection due to the high divergence angle of the emitted photons, which results from the relaxed phase-matching. This can cause coupling losses, while internal reflections within the material can further reduce efficiency. One approach is to introduce cavities or metasurfaces to enhance directional emission,[1128] which could also enhance the emission at specific frequencies. Nonparametric background noise, which lowers the purity of two-photon states, can be addressed by post-processing techniques such as temporal distillation.

41.2.3 *Plasmonics*

Implementing two-qubit plasmonic logic gates is particularly challenging due to demanding fabrication addressing the required high-quality plasmonic structures. This has limited most experimental work in graphene to the mid-IR. However, resonant graphene nanostructures have been recently achieved in the near-IR. While promising, these methods rely on chemical etching, resulting in randomly formed structures. Deterministic nanofabrication of graphene nanostructures that are resonant in the visible or near-IR has yet to be realized, though heterostructures may offer a solution.[22] Other 2D platforms, such as buckled honeycomb materials (e.g., silicene), also have potential, but further research is required to prove their suitability as a competing plasmonic platform.

Finally, efficiently coupling photons from free space to highly confined plasmonic modes remains a challenge, but classical light has been coupled to graphene plasmons with 94% efficiency.[1044] Further development of these methods could enable similar success at the single-photon level.

### 41.3 Suggested Directions to Meet These Goals

Photonic quantum information processing is a relatively mature and advanced field. Thus, if 2D materials are to be integrated into quantum experiments, they will have to compete with well-established systems. For example, while 2D materials show promise as photodetectors,[1129] superconducting nanowire detectors are arguably nearly perfect, achieving close-to-ideal efficiency. A remaining challenge lies in scaling up their production. This makes it unlikely for 2D materials to make a large impact on single-photon detection unless qualitatively better performance is demonstrated, such as efficient detection at new wavelength regimes or operation at higher temperatures. However, the situation for single-photon sources and nonlinear manipulation is different. For these applications, quantum photonic technology is being developed in various platforms to improve efficiency and feasibility. Here, 2D materials could provide substantial breakthroughs that would speed up the advent of photonic quantum information processing.

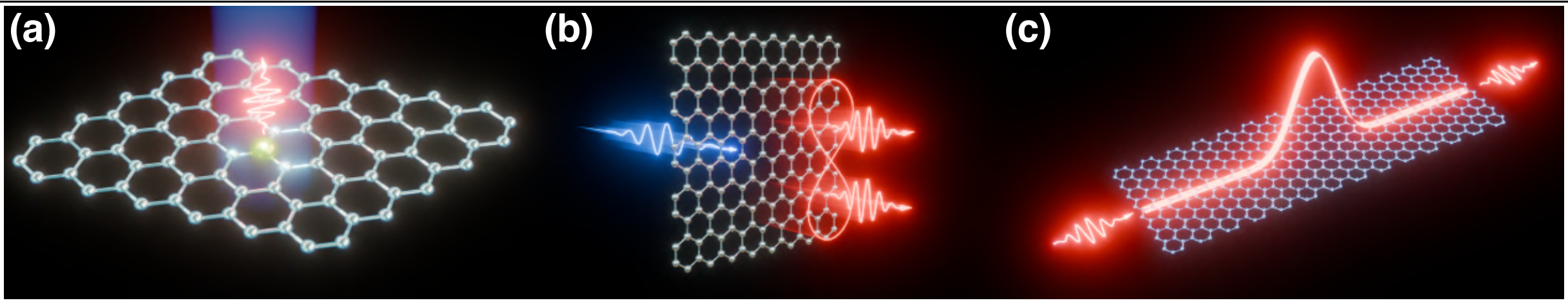

**Figure 41.** 2D materials have the potential to revolutionize photonic quantum technologies. They are already making significant inroads in the development of (a) single-photon emitters, (b) spontaneous parametric down-conversion, and (c) nonlinear plasmonics, with potential applications in quantum logic gates.

## 42. 2D MAGNETIC PHOTONICS

**Kenneth S. Burch,[1,*] Liuyan Zhao,[2] Vinod M. Menon,[3,4] Xiaodong Xu[5,6]**
[1]Department of Physics, Boston College, Chestnut Hill, MA, USA
[2]Department of Physics, University of Michigan, Ann Arbor, MI, USA
[3]Department of Physics, City College of New York, New York, NY, USA
[4]Department of Physics, The Graduate Center, City University of New York, New York, NY, USA
[5]Department of Physics, University of Washington, Seattle, WA 98195, USA
[6]Department of Materials Science and Engineering, University of Washington, Seattle, WA 98195, USA
***Corresponding author.** Email: burchke@bc.edu

In this section, we describe the potential of combining magnetic 2D atomic crystals (2D magnets) with photonic structures to study novel states, answer long-standing questions, and create new devices. For example, despite decades-old predictions of nontrivial quantum correlations and entanglement in 2D frustrated magnets, these have yet to be directly observed.[1130,1131] Furthermore, 2D magnets offer a unique and widely explored ability to tune the magnetic properties through strain,[1132] gating,[1133,1134] and heterostructures.[1130,1135] Thus, as shown in Figure 42a, 2D magnets can be employed for tunable on-chip photonic sources of helical light. However, the controls of 2D magnets employed to date are rather blunt tools that do not finely tailor the interactions. Nonetheless, 2D magnets possess the potential for strong light–matter interactions resulting from placement in cavities and their inherent large excitonic features. Thus, photonic structures offer exciting possibilities to probe, control, and incorporate 2D magnets into novel device architectures for future quantum applications: light sources, detectors, memories, transduction, entanglement generation, and control.

### 42.1 Future Devices

Magnetism plays a significant role in memory and computing applications. Compared to their 3D counterparts, certain 2D magnets offer new opportunities *via* their strong light–matter interactions. 2D magnetic semiconductors, such as $CrI_3$ and CrSBr, host excitons strongly coupled to the interlayer magnetic order. For instance, the magnetic state-dependent helical optical selection rules of excitons in $CrI_3$ produces spin-polarized photovoltaic effects,[1136] using spin-filtering in magnetic tunnel junction (MTJ) devices[1137,1138] whose giant photo-magneto current is tunable by electrical bias. In CrSBr, the spin-dependent interlayer exchange interaction produces a unique coupling between excitons and magnons, whose energies differ by five orders of magnitude.[295,1136] By integrating a 2D magnet with a photonic cavity, magnon-exciton polaritons have also been realized.[294] This photon–magnon coupling enables the optical detection of coherent spin waves for potential applications in hybrid quantum magnonics and quantum transduction.

Expanding beyond these initial demonstrations are possibilities relying on the control of the interlayer exchange *via* electrical means[1133,1134] or mechanical strain.[1132] An example device concept is shown in Figure 42b, an electrical write–optical read multilevel memory. The device concept follows the recent demonstration of non-volatile electric-field control of magnetization of CrSBr[1139] with the added role of the light emission from the bulk polaritons acting as the read-out mechanism. Furthermore, the multiple polariton states in bulk CrSBr provide wavelength multiplexing for free. The last example utilizes the flexibilities of 2D materials to create artificial structures on demand *via* vdW layer assembly.[1135]

## 42.2 New Probes

The probes traditionally employed to study the ground state and collective excitations of 3D bulk magnets, namely elastic and inelastic scattering, are exceedingly challenging in 2D magnets. Here, the atomic thickness with lateral dimensions below 10μm, along with small cross sections, makes these approaches nearly impossible. While optical signals from 2D magnets are expected to be at least an order of magnitude weaker than those from their 3D counterparts, since the early days of 2D atomic crystals, optical techniques have been central to detecting magnetic order and excitations[576,577,1140] (Figure 42c). Over the past few years, improving the optical detection sensitivity in studying 2D magnetism has been actively pursued. To this end, many have exploited optical resonances strongly coupled to the magnetism to boost the optical signal. For example, it has been demonstrated that the magneto-optical effects in 2D $CrI_3$ at the charge-transfer excitation wavelength (633 nm) are *giant*.[576] Additionally, optical sensing layers have been employed. For example, TMD monolayers have been interfaced with $CrI_3$ to enhance magneto-optical signals.[1141] Third, enhanced sensitivity has been achieved using cavities.[1142,1143]

Nonetheless, these techniques are also limited by the long wavelength and small photon momentum. As such, they are not sensitive to the translational symmetry at the lattice scale and, hence, only probe the $q = 0$ magnetic excitations along with the point and not the space group. This makes antiferromagnetic phases belonging to the gray magnetic point groups typically invisible to optics. To this end, the focus turned to magnetic circular/linear dichroism (MCD/MLD) or birefringence (MCB/MLB). Here, MCD/B detects the broken time reversal while MLD/B is sensitive to the broken rotational symmetry due to the magnetic order parameter. As such, investigations of 2D ferromagnetism (FM) relied on MCB and/or MCD.[576,577] In addition, MCD detected the layered AFM with a compensated magnetization, for example, even-layer $MnBi_2Te_4$,[1144] and the noncollinear spins in moiré magnets.[1135] Due to changes in hopping upon entering the magnetically ordered state, MLD/B has probed zigzag AFM[1145] and imaged their magnetic domains. As MLD/B is proportional to the square of the Néel vector, it is effectively capable of probing magnetic fluctuations and has been applied to probe Potts nematicity.[1146] In addition, the onset of various magnetic orders and magneto-elastic coupling can be observed *via* the phonons. For example, Raman scattering revealed 2D intralayer AFM orders *via* the appearance of zone-boundary phonons folded by the magnetic wave vectors,[1146,1147,1148] tracking 2D FM orders through phonon symmetry reduction[1148] and resolving 2D interlayer AFM orders *via* magnetism-assisted phonon scattering.[1142]

To expand the range of magnetic orders probed, second- and third-harmonic generation (SHG/THG) were applied due to their finer resolution into the point groups. SHG has been applied in several cases to capture the inversion-symmetry-breaking AFMs[583,585,1149] as the non-centrosymmetric magnetic phases turn on the leading-order electric dipole (ED) SHG. Electric quadrupole (EQ) SHG has been pushed to resolve inversion symmetric phases in 3D bulk materials[1150] and in 3D CrSBr to separate surface and bulk AFMs.[1151] While cavities could enhance the low signal levels, SHG is forbidden under normal incidence with an out-of-plane two-fold rotational symmetry. This suggests that future efforts to exploit well-tuned optical structures to enhance the THG efficiency could address the symmetry limitations of SHG and the inherent low signal levels.

Beyond static responses, magnetic fluctuations and collective excitations in 2D magnets provide the strength of various anisotropy and exchange terms for loss mechanisms, detect hidden orders, and exploit novel excitations for spintronics, quantum transduction, and computing. For example, time-resolved MCB and reflectivity captured the in-phase and out-of-phase magnons in 2D layered AFMs.[295,1141] In addition, scanning time-resolved reflectivity microscopy measured magnon propagation in CrSBr flakes.[295] Real-time MCD imaging captured the critical slow-down of spin fluctuations and the divergent correlation length near the critical temperature for a Heisenberg FM monolayer.[1152] Time-resolved SHG has been used to detect and separate various electro-magnons in vdW multiferroics.[1149] Alternatively, inelastic light scattering spectroscopy provides the energy, lifetime, statistics, and symmetry of the excitations. Magneto-Raman spectroscopy has been used to show zone-center single magnons in 2D and vdW layered AFM,[1143] zone-boundary two magnons in 2D AFM,[1143,1147] quasi-elastic scattering in 2D AFM,[1146,1147] and fractional excitations.[1140]

Recent developments in measuring single-photon correlations *via* Raman scattering, combined with on-chip photonics, could provide new insights into the nontrivial classical and quantum correlations in 2D magnets. Specifically, consider the Stokes-antiStokes (SaS) process shown in Figure 42d. Here, a *write* photon scatters, producing a Stokes (S) photon and a magnon. This magnon is annihilated *via* an anti-Stokes (aS) process with a second pump photon. To ensure the S and aS photons originate from the same mode, one measures the $g^{(2)}$ correlation between them using single photon detectors. This S–aS process has been successfully demonstrated in water, diamond, graphene, and twisted bilayer graphene,[1153] where the size of the quantum correlations induced by the mode on the S and aS photons were evaluated using the Clauser–Horne–Shimony–Holt (CHSH) form of the Bell inequality.[1154] Going forward, we envision employing the S–aS process to probe individual magnons directly in 2D materials. Such studies could open the door to using 2D magnets in future quantum memories and quantum photonic devices. Furthermore, by studying their quantum dephasing as 2D magnets are tuned through different magnetic orders, one could gain insights into the evolution of their strong fluctuations in producing novel phenomena.

To explore novel spatial correlations in these systems, consider a 2D magnet placed on a series of resonant photonic cavities to allow for local excitation and probing of the S–aS process (see Figure 42d right). First, one can measure how the CHSH parameter is modified as a function of the distance between the read and write beam on length scales relevant to 2D magnets. Next, one could explore the nonlocal quantum correlations between magnons excited in different physical locations and times. While such correlations are expected even in classical systems, an open question is to what extent they exist in 2D systems where strong quantum fluctuations destroy the long-range order. Lastly, since the correlations could be tuned by local gates, strain, or magnetic fields, this could provide a significant opportunity for generating and controlling local entanglement or quantum memories in photonic settings.

### 42.3 Photonic Control of Magnetism

For 2D magnets, the intrinsic magnetic anisotropy and relative strength of the exchange terms are crucial for determining their ground states and the emergence of novel quasiparticles. Noting that the anisotropy and specific forms of exchange are highly sensitive to the local lattice symmetry, it seems highly promising to manipulate 2D magnets by excitation of specific phonon modes. Indeed, recent advances in nonlinear phononic pumping have demonstrated the ability to control magnetism by exciting specific optical modes[1155] or through Floquet renormalization.[571] Nonetheless, such studies are limited by the need for large-amplitude electric fields and are constrained to zone centers and optical modes. Thus, future efforts could explore the combination of optomechanics and photonics to allow for fine-tuned dynamic control of 2D magnets. For example, as shown in Figure 42e, placing a 2D material on a patterned substrate can create a cavity for both acoustic modes and light. One could dynamically excite the strain using piezoelectric substrates while using the optical cavity to monitor the acoustic modes and magnetic properties simultaneously. Alternatively, one can attempt to dynamically destroy long-range order or manipulate specific wave vectors *via* magnonic cavities. Using the same setup but with a magnetic substrate (e.g., yttrium-iron-garne), we envision coherent optics exciting and the cavity focusing magnetic excitations into the 2D material for direct control of the magnetism *via* finite $(q, \omega)$ excitations. We note such structures would also be helpful in future quantum transduction efforts or as ultrasensitive probes of magnetism.

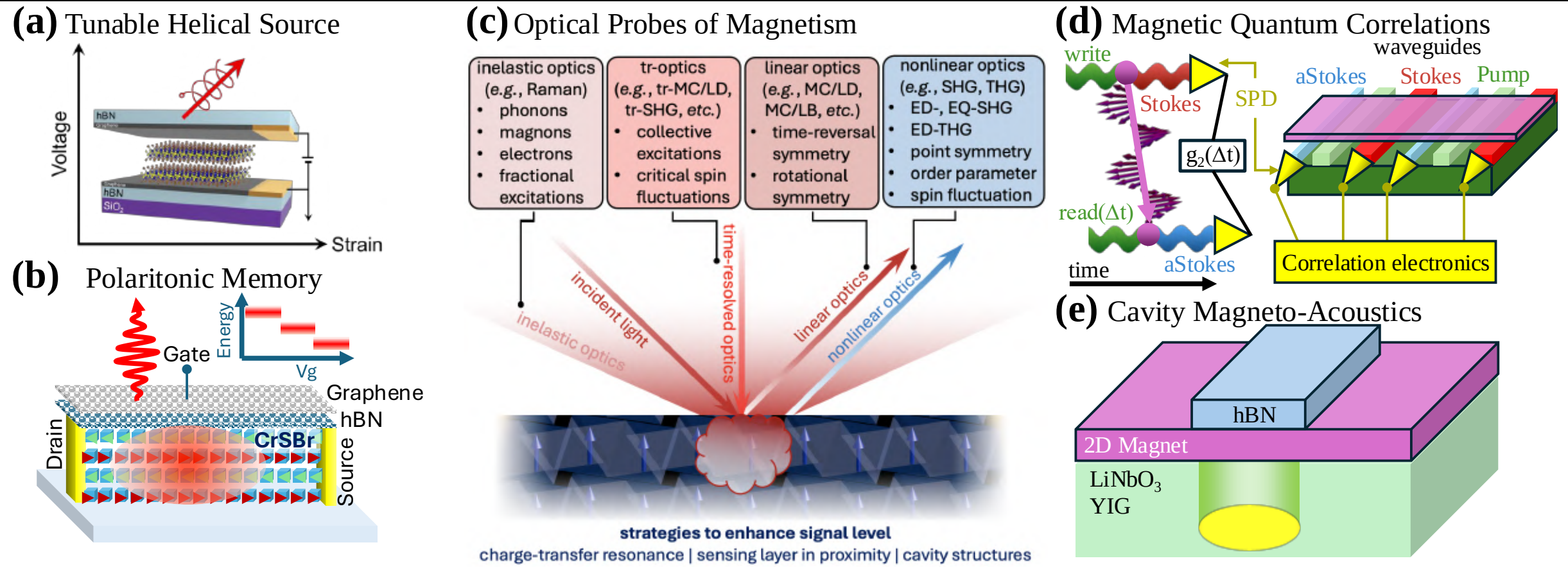

**Figure 42.** Photonics meets magnetism in 2D. (a) Exploiting the electrical and strain control of 2D magnets for on-chip sources of helical light needed for protected quantum communications. (b) Self-hybridized polaritons optically read out the multilevel magnetization.[1151] (c) Various means by which optics is employed to probe the magnetic properties of 2D atomic crystals. (d) Stokes/anti-Stokes scattering, allowing for the detection of single magnetic excitations and their quantum correlations. Incorporation into photonic structures provides nonlocal corrections. (e) Combining optical, acoustic, and magnonic cavities to enable fine-tuned control of 2D magnets toward novel ground states.

**Conclusion**

## 43. CONCLUSION

**F. Javier García de Abajo,[1,2,*] D. N. Basov,[3] and Frank H. L. Koppens[1,2]**
[1]ICFO-Institut de Ciencies Fotoniques, The Barcelona Institute of Science and Technology, 08860 Castelldefels, Barcelona, Spain
[2]ICREA-Institució Catalana de Recerca i Estudis Avançats, Passeig Lluís Companys 23, 08010 Barcelona, Spain
[3]Department of Physics, Columbia University, 1150 Amsterdam Avenue, New York, New York 10027, United States
***Corresponding author.** Email: javier.garciadeabajo@nanophotonics.es

From the standpoint of the complexity inhering in biological systems and their astounding degree of dynamical evolution, interrelations, and functionalities, materials science is still in its early infancy. Moving from bulk to surfaces (famously described by Pauli as a devil's invention) has kept chemists and physicists engaged for decades. Once the domain of surface science, the combination of exfoliation, stacking, and twisting has now emerged as a powerful alternative for creating atomically controlled structures. In-plane complexity in 2D materials represents a natural extension of this evolution, enabling the integration of increasingly complex structures, dynamics, and functionalities in atomically thin structures—a small step in the complexity ladder. This evolution is already unfolding in the field of photonics.

We are witnessing a shift in our understanding of material response functions and light–matter interactions, driven by concepts from topology and quantum geometry. Quantum properties of light and materials are now being combined in an effort to identify intrinsically joint quantum states. Additionally, hyperbolic media are broadening our understanding of nanoscale photonics, previously focused on field confinement, and now extending this concept to include propagating rays, opening new possibilities for light manipulation at the nanoscale.

The wide range of topics covered in this Roadmap underscores the transformative impact of 2D materials on photonics. These materials not only provide better opportunities for exploring and practically exploiting several previously known effects, but also facilitate the discovery of new phenomena and the development of innovative applications. These include 2D polaritonics (Section 2–6), excitons with exceptional properties (Sections 10–13), substantial advances in nonlinear (Sections 14–20) and magnetic (Section 42) nanophotonics, insights into chirality and moiré systems (Sections 21–26), new uses of metasurfaces and emerging materials (Section 27–32), and applications in integrated photonics (Sections 33–35) and light emission/detection (Section 36–41). Tip-based nanoscopies (Section 7–8) play a crucial role in driving these discoveries, while electron microscopy also provide valuable insights (Section 9). However, it remains impractical to compile a fully comprehensive collection of topics, and certain important areas such as, for example, optical biosensing, neuromorphic computing, and photocatalysis, are still underrepresented. These fields can benefit from the high surface-to-volume ratio of 2D materials, enabling them to progress toward bioinspired applications with increased complexity.

We should continue to focus on designing materials with enhanced nonlinearities, an area where 2D crystals have generated great expectations. In parallel, the strong, tunable excitons featured by 2D semiconductors pave the way for future applications in active nanophotonics. Additionally, controlling the creation and placement of atomic defects is crucial to harness their potential in integrated photonics, where they can serve as elements with precisely defined light emission, absorption, and scattering properties. From a fabrication perspective, the pursuit of higher-quality materials remains ongoing. Atomically resolved images of 2D materials often reveal a high density of defects, and therefore, improve synthesis methods are essential. These methods must also be scalable to enable massively parallelized fabrication, and should include standardized metrology to characterize these materials and design well-defined and reproducible manufacturing protocols.

Advanced first-principles simulations should be adapted to facilitate scientific exploration in these areas and provide deeper insights into experimental findings, particularly those arising unexpectedly from undesired disorder and adsorbates, as well as from surface-chemical modifications and engineered atomic defects in a more controlled manner.

We conclude by extending our gratitude to the contributors who have expertly summarized the state of the art in this dynamic field—an extraordinary outcome of the significant collective efforts devoted so far to researching photonics of (and with) 2D materials. They have identified future objectives and generously outlined strategies to achieve them. This rapidly expanding area of research promises breakthroughs in ultrafast, quantum, and nonlinear optics, as well as in nanophotonics, light sources and detectors, optical sensing, optomechanics, thermal management, and other emerging areas yet to be fully defined. We believe that this effort will ultimately bring substantial benefits to society as a whole, and therefore, we are excited to present this Roadmap, hoping it will guide future endeavors in photonics and pave the way for groundbreaking research in the years to come.

## ▪ LIST OF SELECTED ACRONYMS

| | |
|---|---|
| 1D, 2D, 3D | one-, two-, three-dimensional |
| IR | infrared |
| mid-IR | mid-infrared |
| near-IR | near-infrared |
| THz | terahertz |
| TMD | transition metal dichalcogenide |
| UV | ultraviolet |
| vdW | van der Waals |

## ▪ AUTHOR CONTRIBUTIONS

Authors listed at the beginning of each section are responsible for the content of that section. The corresponding author in each section should be the person to be contacted if questions arise regarding the content of that section.

## ▪ ACKNOWLEDGMENTS

This work has been supported in part by the founding sources listed next (grand numbers in parentheses). Section 2: European Research Council (ERC) (101141220 QUEFES); Spanish Ministry of Science, Innovation, and Universities (MICIN) (CEX2019-000910-S); Fundaciós Cellex and Mir-Puig (FCMP). Section 3: Gordon and Betty Moore Foundation (GBMF) (12212); FCMP; Generalitat de Catalunya and European Social Fund—FEDER. Section 4: Italian Ministry of University and Research (PRIN 2020 2020JLZ52N); European Union (EU) (101131579, 873028). Section 5: Portuguese Foundation for Science and Technology (UIDB/04650/2020, COMPETE 2020, PORTUGAL 2020, FEDER, PTDC/FIS-MAC/2045/2021); Independent Research Fund Denmark (IRFD) (2032-00045B); Danish National Research Foundation (DNRF) (165). Section 6: Office of Naval Research (ONR) (N00014231270). Section 7: U.S. Department of Energy, Office of Science, Basic Energy Sciences (DOE) (DE-SC0019443). Section 8: Catalan CERCA; MICIN (SEV-2017-0706, PID2022-140845OB-C63, CNS2022-135658, TED2021-132388B-C41, PID2021-125309OA-100, TED2021-129416A-I00); EU (PRTR-C17.I1); Comunidad de Madrid (TEC-2024/TEC-459, SINMOLMAT-CM); CSIC (20226AT011); CSIC and Czech Academy of Sciences (CAS) (BILAT23033/CSIC-24-10); CAS (GAČR 22-18718S). Section 9: French National Agency for Research (ANR) (ANR-20-CE42-0020). Section 10: F.D.: Emmy Noether Program (ENP) (534078167); H.D.: GBMF (10694), ONR (N00014-21-1-2770); C.S.: German Research Society (DFG) (SPP 2244 project SCHN1376 14.2); V.M.M.: GBMF (12764), ONR (W911NF-22-1-0091). Section 11: T.D.: DFG (426726249, DE 2749/2-1, DE 2749/2-2); A.C.: DFG (314695032 (SFB 1277 B05), 417590517 (SFB 1415 B11), 390858490 (EXC 2147 ct.qmat)), ERC (101001764 CoulENGINE); K.S.T.: Novo Nordisk Foundation Challenge Programme 2021 (BIOMAG NNF21OC0066526); K.S.T.: VILLUM FONDEN (37789). Section 12: Y.A.: Air Force Office of Scientific Research (AFOSR) (FA9550-23-1-0375), GBMF (12246); V.K.S. and M.C.H.: Northwestern University (NSF DMR-2308691); M.T.: AFOSR (FA9550-23-1-0447), National Science Foundation (NSF) (DMREF 2323296). Section 13: X.C., T.F.H., and L.Y.: DOE (Q-NEXT DE-AC02-76SF00515). Section 14: EU (IR0000016, ID D2B8D520, CUP B53C22001750006, 101130384, HORIZONEIC-2023-PATHFINDEROPEN-01, QUONDENSATE, M4C2 - 2022LA3TJ8 – CUP D53D23002280006, MSCA-2021 101066108, NQSTI PE00000023-q-ANTHEM-CUP H43C22000870001). Section 15: DFG (Li 580/16-1, DE 3578/3-1, SFB1372/2-Sig01, RTG1885, INST 184/163-1, INST 184/164-1); Niedersächsisches Ministerium für Wissenschaft und Kultur and Volkswagen Foundation (SMART, DyNano, Wissenschaftsraum ElLiKo). Section 16: Research Council of Finland (RSF) (353364, 360411, 359009, 365686, 367808, 320167, PREIN); EU (MSCA-RISE-872049 IPN-Bio, 101155089, ALTOS); ERC (834742). Section 17: DOE (DE-SC0019443). Section 18: IRFD (0165-00051B); DNRF (165); ERC (101141220-QUEFES). Section 19: DFG (CRC 1375 NOA, 398816777 C4, IRTG 2675 'Meta-Active' 437527638 A4); Federal Ministry for Education and Research (BMBF) (16KIS1792 SiNNER). J.W.: DFG (ENP 503985532, CRC 1277 - 314695032 A03, RTG 2905 - 502572516). Section 20: National Key Research and Development Program of China (NKRDPC) (2024YFA1409800); National Natural Science Foundation of China (NNSFC) (12034003). Section 21: Ministry of Education Singapore (MES) (MOE- T2EP50222-0011, MOE- MOET32023-0003). Section 22: EU (851780); Hellen Diller Quantum Center; I.K.: GBMF (11473); S.T.: Israeli Academy of Science and Humanities (Yad Hanadiv Foundation Rothschild), Israel Council for Higher Education (VATAT-Quantum), Technion (Viterbi). Section 23: DFG; BMBF; Vector-Stiftung; Zeiss-Stiftung; EU. Section 24: MICIN (PID2023-146461NB-I00, CEX2023-001316-M). Section 25: J.D.: Beijing Natural Science Foundation (Z240005), NNSFC (12474027); A.A.: ONR, Simons Foundation (SF); J.D.C.: ONR MURI (N0001-23-1-2567); A.Y.N.: MICIN (PID2023-147676NB-I00), Basque Department of Education (PIBA-2023-1-0007); P.A.-G.: ERC (101044461 TWISTOPTICS), MICIN (PID2022-141304NB-I00). Section 26: A.Y.N., P.A.G., and L.M.-M.: MICIN (PID2023-147676NB-I00, PID2022-141304NB-I00, PID2023-148359NB-C); L.M.-M.: Aragon Government (Q-MAD);. P.A.-G.: ERC (101044461, TWISTOPTICS); A.A.: ONR (N00014-2212448);. J.D.C and T.L.: ONR MURI (N0001-23-1-2567). Section 27: Department of Defense; NSF; SF. Section 28: J.v.d.G.: Dutch Science Foundation (VI.Vidi.203.027), ERC (101116984); T.S.: 2D-TECH VINNOVA Competence Center (Ref. 2019-00068); D.J.: ONR (N00014-23-1-203). Section 29: I.V.B.: U.S. Army Research Office (W911NF2310206); A.B. and V.M.S.: DOE (DE-SC0017717); AFOSR (FA9550-20-1-0124). Section 30: Purdue team: ONR (N00014-21-1-2026); J.S.: NSF (Graduate Research Fellowship DGE-1842166). Section 31: H.P.: NSF (CHE-2102601); S.T.: CAS (YSBR-054). Section 32: ERC (101141220 QUEFES); MCINN (CEX2019-000910-S, PID2022-137569NB-C42). Section 33: RCF (353364, 360411, 359009, 365686, 367808, Flagship Programme 320167 PREIN); EU (MSCA-RISE-872049 IPN-Bio); Jane and Aatos Erkko Foundation; Technology Industries of Finland Centennial Foundation (Future Makers 2022); ERC (834742); NNSFC (12374167, 12422406); NKRDPC (2022YFA1403500). Section 34: NNSFC (51925203, U2032206, 52072083, 51972072, 2021YFA1201500). Section 35: Y.Z.: DOE (DE-SC-0022885). Section 36: MES (EDUN C-33-18-279-V12, I-FIM, MOE-T2EP50122-0012); AFOSR and ONR (FA8655-21-1-7026); MAT-GDT Program at A*STAR (M24N4b0034). Section 37: Swiss National Science Foundation (200021-227862). Section 38: ERC (101157731 Terascan); EU (EXTREME IR 944735, Graphene Flagship core 3, 101070546, PE0000023-NQSTI. Section 39: Defense Science and Technology Laboratory and Engineering and Physical Sciences Research Council (EP/T019018/1, EP/X524967/1). Section 40: ONR; NSF; Government of Israel. Section 41: P.W.: AFOSR (FA8655-20-1-7030 PhoQuGraph, FA8655-23-1-7063 TIQI); Austrian Science Fund (FWF 10.55776/F71); Austrian Federal Ministry of Labour and Economy; Christian Doppler Research Association; L.A.R.: Erwin Schrödinger Center for Quantum Science & Technology (ESQ Discovery). Section 42: K.S.B.: NSF (EPMD EPMD-2211334), DOE (DE-SC0018675); L.Z.: DOE (DE-

SC0024145), NSF CAREER (DMR-174774); V.M.M.: DOE (DE-SC0025302); X.X.: DOE (DE-SC0018171), AFOSR MURI (FA9550-19-1-0390). L.C. and M.P. thank D. Basov, P. Jarillo-Herrero, F. Koppens, and I. Torre for many useful discussions. N.M.R.P. acknowledges the hospitality of the POLIMA Center. The authors of Section 22 thank Qinghui Yan for fruitful discussions.

## Note

Views, findings, conclusions, and recommendations expressed in this Roadmap are those of the authors and do not necessarily reflect the views of the funding agencies.

## ▪ REFERENCES